\documentclass[aps,pre,
showkeys,
longbibliography,
superscriptaddress,
reprint,                %% preprint, draft, 
twocolumn,              %% onecolumn, 
nofootinbib,
floatfix]{revtex4-2}

\usepackage[T1]{fontenc}
\usepackage[utf8]{inputenc}
\usepackage{amsmath}
\usepackage{amsfonts}
\usepackage{graphicx}
\usepackage[labelformat=simple]{subcaption}

\usepackage[colorlinks=true,hyperfootnotes=true,breaklinks=true]{hyperref}
\usepackage[capitalise]{cleveref}

\begin{document}

\title{Universality of opinions disappearing in sociophysical models of opinion dynamics: From initial multitude of opinions to ultimate consensus}

\author{Maciej Wo{\l}oszyn}
\thanks{\href{https://orcid.org/0000-0001-9896-1018}{0000-0001-9896-1018}}
\affiliation{AGH University, Faculty of Physics and Applied Computer Science, al.~Mickiewicza~30, 30-059 Krak\'ow, Poland}

\author{Tomasz Mas{\l}yk} 
\thanks{\href{https://orcid.org/0000-0001-9901-5859}{0000-0001-9901-5859}}
\affiliation{AGH University, Faculty of Humanities, al.~Mickiewicza~30, 30-059 Krak\'ow, Poland}

\author{Szymon Paj\k{a}k}
\affiliation{AGH University, Faculty of Physics and Applied Computer Science, al.~Mickiewicza~30, 30-059 Krak\'ow, Poland}

\author{Krzysztof Malarz}
\thanks{\href{https://orcid.org/0000-0001-9980-0363}{0000-0001-9980-0363}}
\email{malarz@agh.edu.pl}
\affiliation{AGH University, Faculty of Physics and Applied Computer Science, al.~Mickiewicza~30, 30-059 Krak\'ow, Poland}

\begin{abstract}
It is also intriguing what the dynamics of such processes is.
To address those problems, we performed computer simulations using well-established models of social opinion formation, namely the voter, Sznajd, and Latan\'e models.
We investigated opinion dynamics in cases where the initial number of opinions is very large, equal to the number of actors (the voter and Latan\'e models) or when every second actor has their own opinion (Sznajd model), with some variations on the update schemes, lattice topologies, effective ranges of interaction, and information noise levels.
For all considered models, the number of opinions assumed by the actors is finally almost always reduced to only one.
However, while the voter and Latan'e models exhibit a power-law time decrease in the number of opinions, the Sznajd model follows a complex three-stage behavior.
We also demonstrated that the mean/median time of reaching the consensus scales with system size according to a power law for voter and Sznajd models, while for the Latan\'e model this increase is even faster.
Our results show that in the studied models the consensus is possible, provided that a long enough and model-dependent time to reach this state is available.
\end{abstract}

\date{May 6, 2024}


\maketitle

%% ###############################################################
\section{Introduction}
%% ###############################################################

Sociophysics \cite{Galam_2012,Stauffer_2013,Sen_2014,Schweitzer_Sociophysics,Matjaz_2019,Jusup_2022} uses tools and methods of statistical physics to explore issues arising in sociology.
The modern trends of sociophysics \cite{SI:Trends_Sociophysics,da_Luz_2023} in its 40-year history \cite{SI:SergeGalam70} still indicate the dynamics of opinion \cite{Castellano-2009,Galam_2006,Weisbuch_2006,Stauffer_2013,Sobkowicz_2019,Grabisch_2020,Jusup_2022,Ellero_2023} as a core topic in a field.
In principle, sociophysical models reveal either consensus of opinions or population polarization. 
However, the second effect may be treated as an artificial one, as very often only two opinions available in the system are initially assumed.
Not surprisingly, for such an initial condition, the path into consensus is widely open and system polarization is naturally assumed at the very beginning of the system modeling.
In contrast, some attempts to model the availability of multiple opinions (at least at the beginning of the simulation) were also considered \cite{ISI:000226629700050,*Rodrigues_2005,*Kulakowski2010,*Ozturk_2013,*1902.03454,*Martins_2020,*Zubillaga_2022,*Li_2022,*Doniec_2022,*000409112600017,*000316891200004,*000432967700004,*Mobilia_2023}.

Very recently \citeauthor{Maslyk_2023}~\cite{Maslyk_2023}---based on \citeauthor{2211.04183}~\cite{2211.04183} and \citeauthor{2002.05451}~\cite{2002.05451} papers---showed how the initial diversity of opinions influences the possibility of reaching unanimity of opinions.
Thus, in this paper, we decided to check the final outcome of computer simulations for three sociophysical models of opinion formation.
Namely, we apply the voter, Sznajd and Latan\'e models also when initially the number $K$ of opinions available in the system is comparable with the system size $N$.
The degree of complication of the rules of these models subsequently increases from the simplest (toy model rules for the voter model \cite{Holley_1975,Liggett_1999}) via an intermediate level of complexity (for the Sznajd model \cite{Sznajd-2000}) to quite sophisticated (in the case of Latan\'e model of social impact \cite{Latane-1981}). 

Moreover, to make our studies as general as possible, various aspects of model implementation are considered. For example: 
\begin{itemize}
    \item in the voter model various strategies of update scheme; 
    \item in the Sznajd model various lattice topologies/neighborhoods;
    \item in the Latan\'e model various effective ranges of interaction and noise level.
\end{itemize}

Our studies concentrate on the ultimate number of opinions and their temporal evolution---including characteristics of distribution of times of reaching the final state, and this distribution themselves.
The principal aim of the studies presented here is to verify the possibility of reaching consensus when the initial number of opinions in the system is huge, i.e., it is at the order of magnitude of the number of actors in the system.

The paper is organized as follows. The rest of this section is devoted to the presentation of sociological aspects of the considered models.
In \Cref{sec:methods}, we provide a detailed description of the implementation of the models mentioned above.
\Cref{sec:results} is devoted to the presentation of the results obtained from the simulation, while \Cref{sec:discussion} contains their discussion.
The paper is closed with \Cref{sec:conclusions} which contains the conclusions.
In \Cref{app:Snapshots} snapshots of the system evolution are presented.

%% ###############################################################
%% ###############################################################
\section{\label{sec:methods}Methods}
%% ###############################################################
%% ###############################################################

In this Section we provide the technical details on direct implementation of the considered models.
Independently of the model considered: 
\begin{itemize}
    \item each actor occupying the site $i$ (from the $N$ sites available in the system) is characterized at the time $t$ by its opinion $\lambda_i(t)$ from the finite set $\{\Lambda_1, \cdots, \Lambda_K\}$ (from the $K$ opinions available in the system);
    \item  single Monte Carlo steps (MCS) described in \Cref{subsec:Voter,subsec:sznajd,subsec:latane} are repeated until the consensus state, with only one common opinion in the whole system, is achieved---or the maximum simulation time passes. 
\end{itemize}

%% ###############################################################
\subsection{\label{subsec:Voter}Voter model}
%% ###############################################################

In the voter model \cite{Holley_1975,Liggett_1999} (see, for example, References~\onlinecite{PhysRevE.100.022304,Doniec_2022,Mobilia_2023,PhysRevE.109.024312} for recent papers) an actor mimics the opinion of their randomly selected neighbor from their neighborhood.
When the number of such opinions is finite (restricted), the selection of the most popular in the neighborhoods becomes the most probable, and thus sociologically the conformity theory \cite{Cialdini_1984,Morgan_2012} becomes the most suitable strategy to describe the underlying principles behind these model rules.

Accepting or rejecting a given opinion can be derived from social pressures that lead individuals to conform to the group to avoid rejection, gain acceptance, or avoid conflict. 
Individuals succumb to social conformism and thus adapt their beliefs, attitudes, or behavior to social norms and expectations. 

Conformism can manifest itself in different ways and cover different areas of life, such as beliefs, values, lifestyle, or behavior. 
Several factors can be identified that influence social conformity. 
The first is the pressure of the group. The need to belong and be accepted prompts individuals to conform to the norms of the group \cite{Asch_1956}. 
Individuals not only seek to remain part of their own group or to win the favor of outside groups, but also strive to remain members of a society as a larger whole. 
Then they are subjected to the influence of social norms. 
The society in the primary or secondary socialization process exerts pressure on individuals to adopt certain social norms \cite{Deutsch_1955}. 

Another factor is the pressure of authority. 
Authorities can pressure individuals to conform to their expectations using social legitimacy and thus power and control \cite{Milgram_1963,Blass_1999}. 

Conformist attitudes also form in situations of social uncertainty. 
In such a situation, when an individual does not know how to behave, other people's behavior will provide a model of appropriate behavior. 
The adoption of such behavior will be an adaptation mechanism in situations of coping with uncertainty \cite{Cialdini_2004}. 

In some cases, the influence of minority groups may be a factor in the formation of conformist attitudes. 
This is the case when the minority group has coherent views, its members are confident in the rightness of their reasons and show determination to defend them, are a representative of the majority, or receive support from outside (from authorities, organizations, or other social groups) \cite{Moscovici-Nemeth_1974}.

In the voter model, actors take their opinion in the next time step ($t+1$) based on the opinion of a randomly selected neighbor at the time $t$.
For this model we consider three update schemes: sequential, synchronous, and random (asynchronous).

In the case of sequential order, actors are investigated one-by-one in typewriter order, and their opinions are updated immediately. We assume that one MCS takes $N$ such inspections.

In the random (asynchronous) scheme, we select the investigated actor randomly and one MCS takes $N$ such draws.

In the synchronous version of the update scheme we deal technically with two matrices of opinions (for times $t$ and $t+1$) and the single MCS means updating all $N$ matrix elements in the matrix corresponding to time $t+1$ based on the matrix for time $t$.

%% ###############################################################
\subsection{\label{subsec:sznajd}Sznajd model}
%% ###############################################################

The Sznajd model \cite{Sznajd-2000,ISI:000166141900014,Sznajd-2005a} (see References~\onlinecite{Sznajd-2005,Sznajd-Sznajd-Sznajd} for a review) assumes that individuals who have congruent beliefs reinforce each other's beliefs through mutual confirmation. If two neighboring individuals have identical beliefs to begin with, their beliefs will be reinforced as a result of the interaction. Then the confirmation of the beliefs occurs. 

When two individuals have different beliefs, it is assumed that they can come to an agreement as a result of the interaction. This mechanism can be understood as a kind of social negotiation in which individuals adjust their initial beliefs toward a common position. What we have here is a negotiation of different beliefs. Through interactions between neighboring individuals, a common belief is gradually developed throughout the community. Despite initial differences, the process of confirming beliefs and negotiating leads to the situation where the whole group approaches consensus. The development of consensus occurs through social interactions that are essential in the formation of collective beliefs. Group dynamics, including confirmation and negotiation between individuals, is a key element in the formation of shared opinions \cite{Asch1955,Milgram1969}.

We consider the Sznajd model on three different lattices: square, triangular, and honeycomb, each with periodic boundary conditions.
In the first case, two neighborhoods are used: von Neumann and Moore, both for deciding which individuals may form a pair possibly convincing others and for finding the set of those others who are influenced when the pair is in agreement (i.e. those who are neighbors of at least one of the individuals in the pair).
It means that when the von Neumann neighborhood is applied, each individual can form a pair with one of four neighbors, and such a pair may convince six other individuals; the Moore neighborhood allows for eight possibilities of forming a pair by a given individual and then convincing ten or twelve other individuals.
The triangular lattice has those numbers between the values given above, as it offers six possible pairs including the selected individual, and then influencing eight other individuals.
Choosing the honeycomb lattice results in the lowest numbers, with only three possible pairs that include the initially selected individual, and four individuals influenced by the pair.
In all cases, each of the individuals, $i=1, \ldots, N$, may have an opinion $\lambda_i \in \{\Lambda_1, \ldots, \Lambda_K\}$.

Regardless of the underlying lattice, the method of simulation is generally the same.
The initial state is randomly generated, with each individual given one of the allowed opinions with uniform probability.
Then, it is verified if there exists at least one pair of neighbors with the same opinion.
If not, a new random initial state is generated until success, since in the other case the system cannot change its configuration at all.
The probability that the initial state cannot be accepted due to such verification is equal to $(N/(N-2))^{N M/2}$ (typically 0.2--5.0\%), where $M=3$, $4$, $6$, $8$ is the number of nearest neighbors for the honeycomb, square (with the von Neumann neighborhood), triangular, and square (with the Moore neighborhood) lattice, respectively.

During one MCS of the simulation, a pair of neighboring individuals is randomly chosen $N$ times.
After the pair is selected, and only if both individuals $i$ and $j$ in the pair have the same opinion, $\lambda_i = \lambda_j$, all the individuals influenced by the pair change their opinion one level towards the opinion of the pair.

%% ###############################################################
\subsection{\label{subsec:latane}Latan\'e model}
%% ###############################################################

\citeauthor{Latane-1981} \cite{Latane-1981} maintained that individuals are subject to social influence rendered by the presence of others or by their actions.
This effect is the result of various social forces operating in a social force field. 
When an individual is not a member of a group, the influence of the group on the individual increases with the strength, immediacy, and number of people in the group. 
There is then a multiplication of influence. 

When an individual is a member of a group, there is a diffusion or distribution of influence, so that an increase in the size, strength, or directness of the group results in a reduction in the influence of external forces on individual group members. 
The comparison with other group members and the adjustment of one's own behavior to that of other group members are aimed at maintaining a satisfactory self-assessment. This evaluation is compromised when the majority of group members begin to share opinions that differ from those of the individual. 

The strength and therefore the relevance, power, validity, or intensity of a given source are related to the socio-economic status of the group members, their age, and previous relationships with the individual or anticipated power relationships in the future.
Directness implies proximity in space or time and the absence of barriers to communication. 
The number, on the other hand, depends on the size of the group in question \cite{Darley1968,Latane-1976,Latane-1981,Latane-1981a}.

The computerized version of the Latan\'e model \cite{Nowak-1990} is based on the theory of social impact (see \cite{ARCPIX253} for a review of computational efforts for the modeling of such systems).
In this model, we assume two steering parameters: the effective range of interactions $\alpha$ and the level of information noise (social temperature) $T$.

In case of the presence of information noise ($T>0$), we directly follow the methodology presented in Reference~\onlinecite{Maslyk_2023}, while for the deterministic version of the algorithm (the absence of information noise, $T=0$) we apply the strategy presented in Reference~\onlinecite{2002.05451}.

As stated by \citeauthor{Maslyk_2023} \cite{Maslyk_2023} ``the social impact $\mathcal{I}_{i,k}(t)$ exerted in time $t$ on an actor $i$ by all actors who share opinions $\Lambda_k$ is calculated as
%% ================================================================
\begin{subequations}
\label{eq:szamrej}
%% ----------------------------------------------------------------
\begin{equation}
\label{eq:szamrej_uni}
\mathcal{I}_{i,k}(t) = \sum_{j=1}^{N}{\frac{4s_j}{g(d_{i,j})} \cdot \delta(\Lambda_k, \lambda_j(t)) \cdot \delta(\lambda_j(t),\lambda_i(t))}
\end{equation}
%% ----------------------------------------------------------------
or
%% ----------------------------------------------------------------
\begin{equation}
\label{eq:szamrej_sum_diff}
\mathcal{I}_{i,k}(t) = \sum_{j=1}^{N}{\frac{4p_j}{g(d_{i,j})} \cdot \delta(\Lambda_k, \lambda_j(t)) \cdot [1-\delta(\lambda_j(t),\lambda_i(t))}], 
\end{equation}
\end{subequations}
%% ----------------------------------------------------------------
where $s_j$ is $j$-th actor supportiveness,
$p_j$ is $j$-th actor persuasiveness,
$d_{i,j}$ stands for Euclidean distance between actors $i$ and $j$,
$g(\cdot)$ is an arbitrary distance scaling function,
and Kronecker delta $\delta(x,y)=0$ when $x\ne y$ and $\delta(x,y)=1$ when $x=y$.''
[\dots] ``To ensure a lower impact on the opinions of actors from a more distant neighbor, the distance scaling function $g(\cdot)$ must be an increasing function of its argument.
Here, we assume that
%% ----------------------------------------------------------------
\begin{equation}
\label{eq:fg}
g(x) = 1+x^\alpha,
\end{equation}
%% ----------------------------------------------------------------
where the exponent $\alpha$ is a model control parameter while the first addition component ensures finite self-supportiveness.''
[\dots] ``The parameters supportiveness $s_i$ and persuasiveness $p_i$ describe $i$-th actor intensity of interaction with actors sharing his/her opinions or with believers in opposite opinions, respectively.''
Here, we decided to use random values of $p_i$ and $s_i$ %%(as in Ref.~\onlinecite{1902.03454})
because whether it is easier to stick to our opinion or change it depends on numerous factors, such as the social context, emotions, beliefs, authorities, and persuasion strategies.

For the probabilistic version of the algorithm social impact \eqref{eq:szamrej} implies probabilities
%% ================================================================
\begin{subequations}
    \label{eq:probability_eq}
%% ----------------------------------------------------------------
    \begin{equation}
    p_{i,k}(t) =
    \begin{cases}
    0 & \iff \mathcal I_{i,k}=0,\\
    \exp\left(\frac{\mathcal{I}_{i,k}(t)}{T} \right) & \iff \mathcal I_{i,k}>0,
    \end{cases}
    \label{eq:probability_p}
    \end{equation}
%% ----------------------------------------------------------------
that actor $i$ adopts opinion $\Lambda_k$.

Probabilities \eqref{eq:probability_p} require proper normalization
%% ----------------------------------------------------------------
    \begin{equation}
    \label{eq:probability_P}
    P_{i,k}(t) = \frac{p_{i,k}(t)}{\sum^K_{j=1} p_{i,j}(t)}.
    \end{equation}    
%% ----------------------------------------------------------------
\end{subequations}
%% ================================================================

\citeauthor{2002.05451} \cite{2002.05451} say that 
``If $T=0$, then a lack of noise is assumed, and the actor $i$ adopts an opinion $\Lambda_k$ that has the most {impact} on it:
\begin{equation}
\begin{split}
\label{eq:T=0}
\lambda_i(t+1)=\Lambda_k \\ 
\iff I_{i,k}(t)=\max(I_{i,1}(t), I_{i,2}(t),\cdots,I_{i,K}(t)),
\end{split}
\end{equation}
where $k$ is the label of this opinion which believers exert the largest social impact on $i$-th actor and $I_{i,k}$ are the social influence on actor $i$ exerted by actors sharing opinion $\Lambda_k$.''

A MCS step is finalized when each actor induces his/her opinion for time ($t+1$) based on the opinions of all other actors in time $t$.

%% ###############################################################
%% ###############################################################
\section{\label{sec:results}Results}
%% ###############################################################
%% ###############################################################

Initially, the number of opinions $n_\text{o}(t=0)=K$ and $K=N$, $N$, and $N/2$ for the voter, Latan\'e and Sznajd model, respectively.

%% ###############################################################
\subsection{Ultimate number of opinions}
%% ###############################################################

%% ===============================================================
\begin{figure*}[htbp]
%% ---------------------------------------------------------------
\begin{subfigure}[t]{0.30\textwidth}
\caption{\label{subfig:VSSE_no_vs_t}} %%voter model, sequential update}
\includegraphics[width=.99\textwidth]{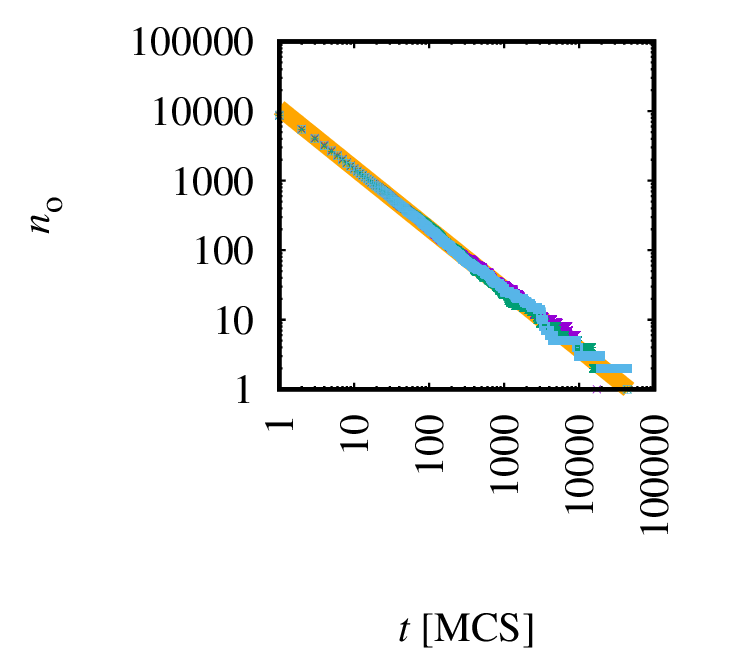}
\end{subfigure}
%% Final set of parameters            Asymptotic Standard Error
%% =======================            ==========================
%% cvsse           = 11115            +/- 20.39        (0.1835%)
%% bvsse           = -0.867533        +/- 0.000509     (0.05867%)
\hfill %% ---------------------------------------------------------------
\begin{subfigure}[t]{0.30\textwidth}
\caption{\label{subfig:VSRA_no_vs_t}} %%voter model, random update}
\includegraphics[width=.99\textwidth]{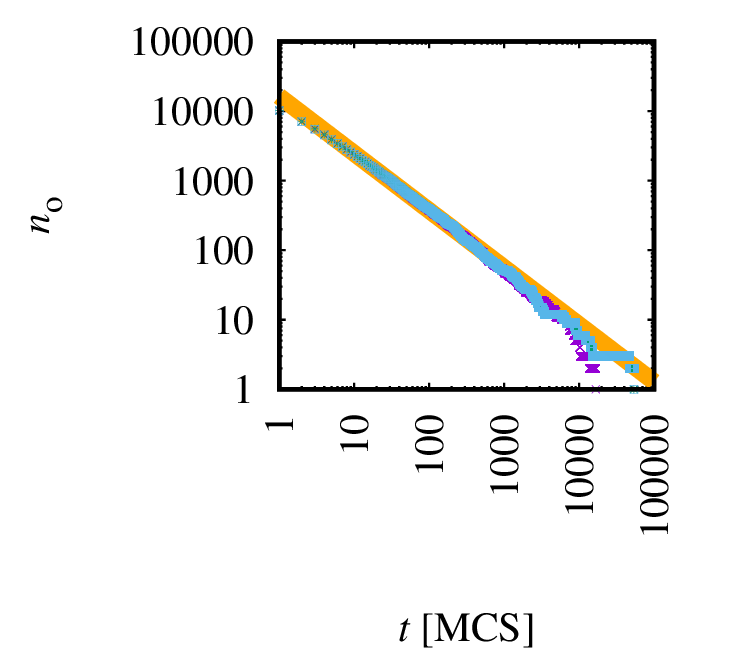}
\end{subfigure}
%% Final set of parameters            Asymptotic Standard Error
%% =======================            ==========================
%% cvsra           = 16344.7          +/- 31.93        (0.1953%)
%% bvsra           = -0.821022        +/- 0.0005246    (0.0639%)
\hfill %% ---------------------------------------------------------------
\begin{subfigure}[t]{0.30\textwidth}
\caption{\label{subfig:VSSY_no_vs_t}} %%voter model, synchronous update}
\includegraphics[width=.99\textwidth]{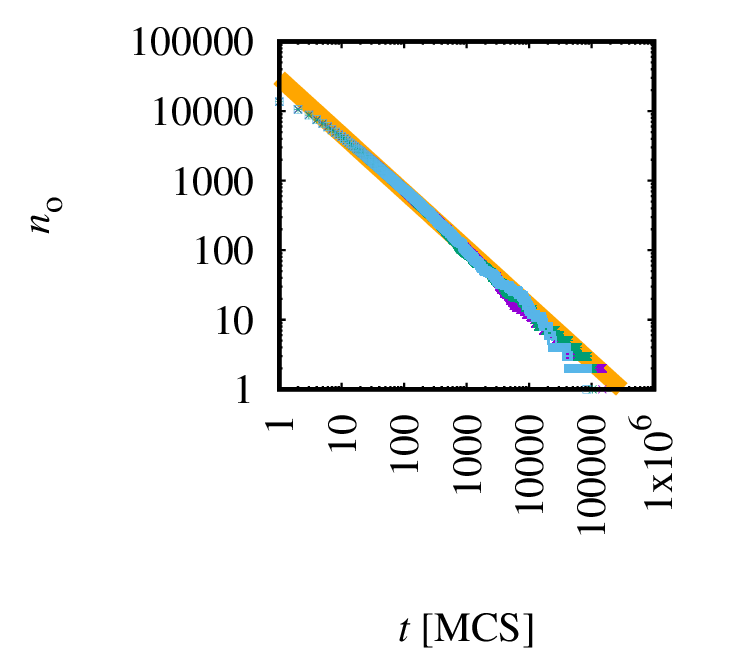}
\end{subfigure}
%% Final set of parameters            Asymptotic Standard Error
%% =======================            ==========================
%% cvssy           = 29974.6          +/- 25.55        (0.08525%)
%% bvssy           = -0.816631        +/- 0.0002141    (0.02622%)
%% ---------------------------------------------------------------
\begin{subfigure}[t]{0.48\textwidth}
\caption{\label{subfig:SZ_VN_no_vs_t}} %%Sznajd model, square (von Neumann)}
\includegraphics[width=.99\textwidth]{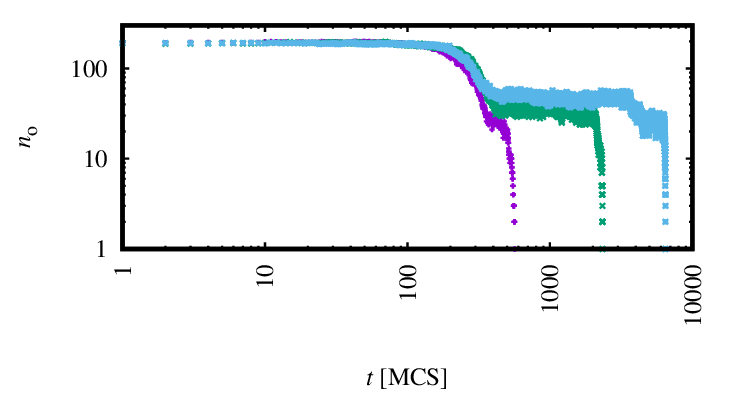}
\end{subfigure}
%% ---------------------------------------------------------------
\begin{subfigure}[t]{0.48\textwidth}
\caption{\label{subfig:SZ_M_no_vs_t}} %%Sznajd model, square (Moore)}
\includegraphics[width=.99\textwidth]{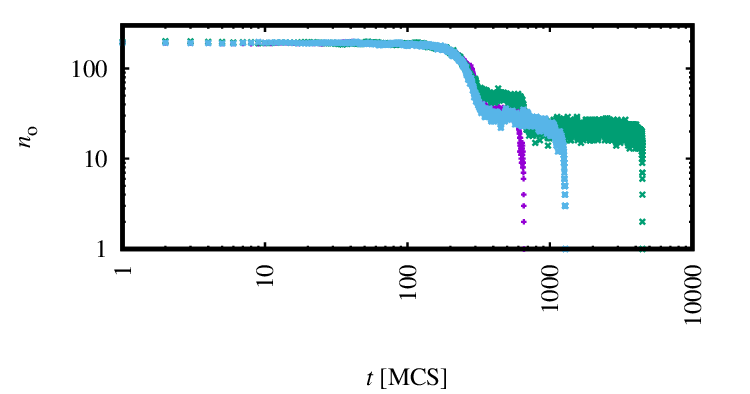}
\end{subfigure}
%% ---------------------------------------------------------------
\begin{subfigure}[t]{0.48\textwidth}
\caption{\label{subfig:SZ_TR_no_vs_t}} %%Sznajd model, triangular}
\includegraphics[width=.99\textwidth]{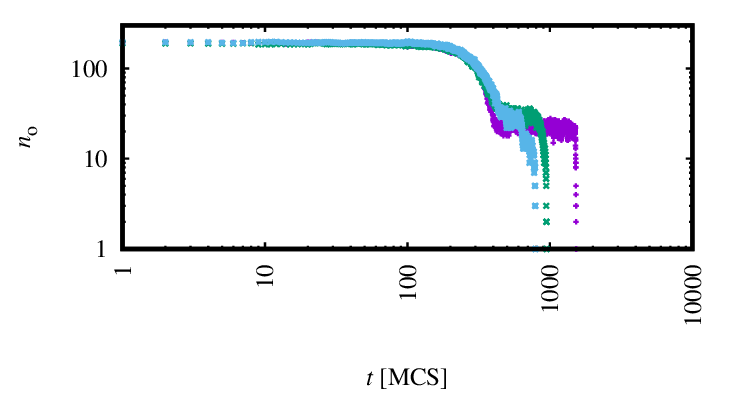}
\end{subfigure}
%% ---------------------------------------------------------------
\begin{subfigure}[t]{0.48\textwidth}
\caption{\label{subfig:SZ_HC_no_vs_t}} %%Sznajd model, honeycomb}
\includegraphics[width=.99\textwidth]{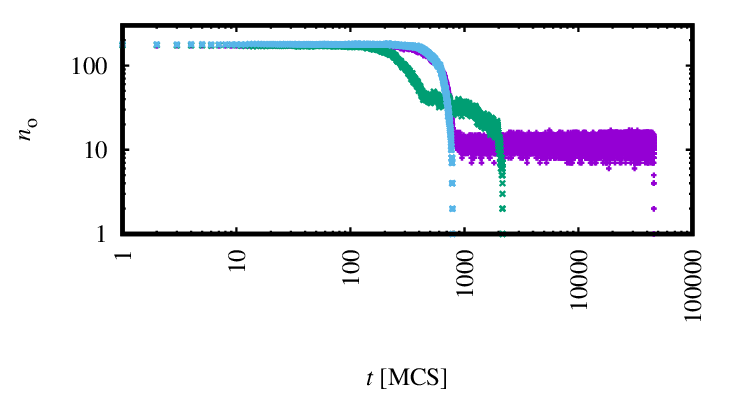}
\end{subfigure}
%% ---------------------------------------------------------------
\begin{subfigure}[t]{0.30\textwidth}
\caption{\label{subfig:L21a20T125_no_vs_t}} %%Latan\'e model, $\alpha=2, T=1.25$}
\includegraphics[width=.99\textwidth]{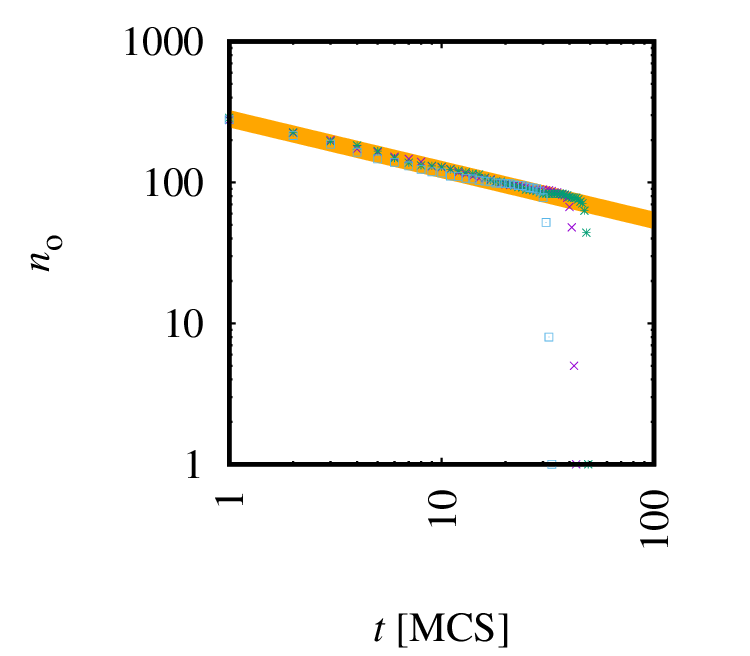}
\end{subfigure}
%% Final set of parameters            Asymptotic Standard Error
%% =======================            ==========================
%% cL21a20T125     = 283.152          +/- 3.359        (1.186%)
%% bL21a20T125     = -0.360404        +/- 0.008751     (2.428%)
\hfill %% ---------------------------------------------------------------
\begin{subfigure}[t]{0.30\textwidth}
\caption{\label{subfig:L21a20T20_no_vs_t}} %%Latan\'e model, $\alpha=2, T=2$}
\includegraphics[width=.99\textwidth]{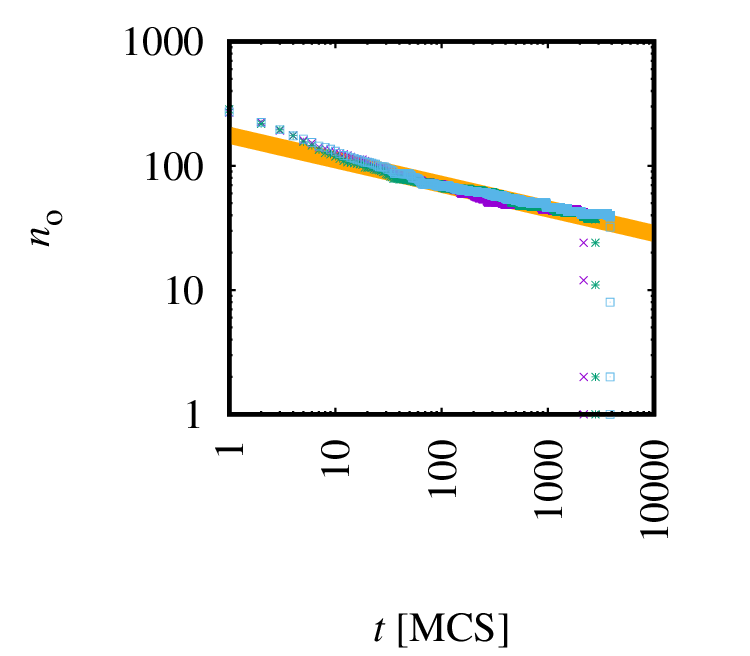}
\end{subfigure}
\hfill %% ---------------------------------------------------------------
\begin{subfigure}[t]{0.30\textwidth}
\caption{\label{subfig:L21a3T15_no_vs_t}} %%Latan\'e model, $\alpha=3, T=1.5$}
\includegraphics[width=.99\textwidth]{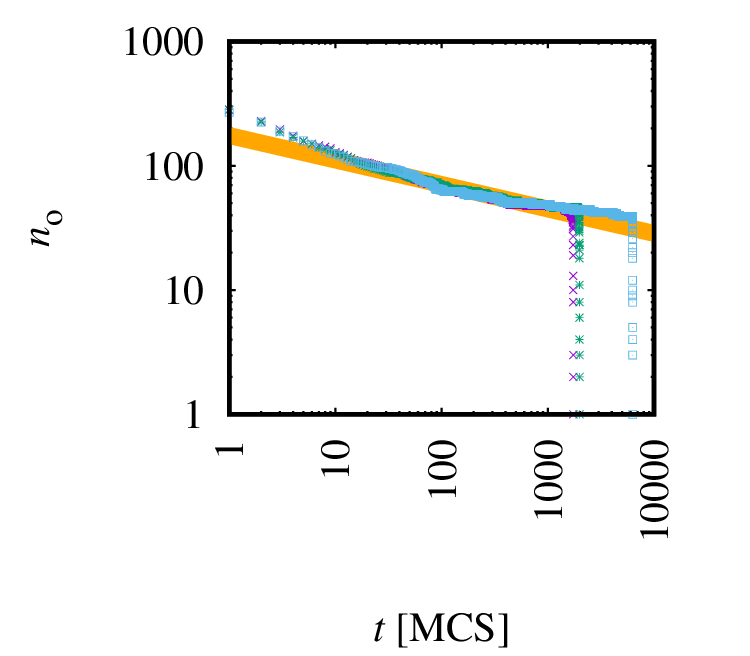}
\end{subfigure}
%% ---------------------------------------------------------------
\caption{\label{fig:no_vs_t}Examples of $n_\text{o}(t)$ evolution for voter model ($L=141$) for 
\subref{subfig:VSSE_no_vs_t} sequential, 
\subref{subfig:VSRA_no_vs_t} random and 
\subref{subfig:VSSY_no_vs_t} synchronous update; 
Sznajd model for 
\subref{subfig:SZ_VN_no_vs_t} square (with von Neumann neighborhood, $L=21$), \subref{subfig:SZ_M_no_vs_t} square (with Moore neighborhood, $L=21$), \subref{subfig:SZ_TR_no_vs_t} triangular ($L=21$), \subref{subfig:SZ_HC_no_vs_t} honeycomb lattice ($L=20$)
and Latan\'e model ($L=21$) for 
\subref{subfig:L21a20T20_no_vs_t} $\alpha=2$, $T=2$, 
\subref{subfig:L21a20T125_no_vs_t} $\alpha=2$, $T=1.25$, 
\subref{subfig:L21a3T15_no_vs_t} $\alpha=3$, $T=1.5$.
The orange lines in Figures \subref{subfig:VSSE_no_vs_t}--\subref{subfig:VSSY_no_vs_t} and \subref{subfig:L21a20T20_no_vs_t}--\subref{subfig:L21a3T15_no_vs_t} are the power-law fits.}
%% ---------------------------------------------------------------
\end{figure*}
%% ===============================================================

In \Cref{fig:no_vs_t} examples of $n_\text{o}(t)$ evolution for
the voter [\Cref{subfig:VSSE_no_vs_t,subfig:VSRA_no_vs_t}],
Sznajd [\Cref{subfig:SZ_VN_no_vs_t,subfig:SZ_M_no_vs_t,subfig:SZ_TR_no_vs_t,subfig:SZ_HC_no_vs_t}], 
and Latan\'e [\Cref{subfig:L21a20T125_no_vs_t,subfig:L21a20T20_no_vs_t,subfig:L21a3T15_no_vs_t}],
models are presented.
The ultimate number $n_\text{o}^\text{u}=\langle n_\text{o}(t\to\infty)\rangle$ of opinions for various models of the dynamics of opinion is collected in \Cref{tab:disc}.
The symbol $\langle\cdots\rangle$ represents an average over $\mathcal R$ simulations.

%% ===============================================================
\begin{figure}[htbp]
\includegraphics[width=.49\textwidth]{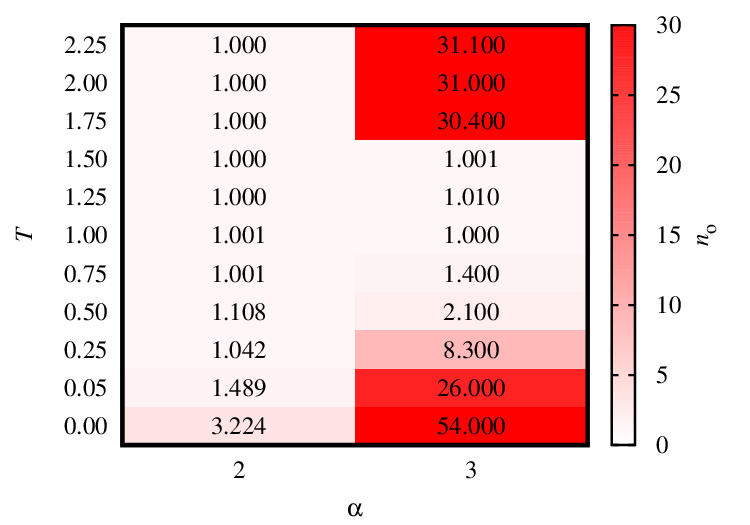}
\caption{\label{fig:Latane_no}Final (averaged over $\mathcal R=1000$ simulations) number of opinions $n^{\text{u}}_{\text{o}}$ versus $(\alpha, T)$ for Latan\'e model (see also Figure~1(a) in Reference~\onlinecite{Maslyk_2023} but for worse statistics and shorter time of system observation). If consensus is not reached, then $n^{\text{u}}
_{\text{o}}$ is measured at $t_\text{max}=10^6$ MCS for $\alpha=2$ and at $t_\text{max}=10^5$ MCS for $\alpha=3$}
\end{figure}
%% ===============================================================

%% ===============================================================
\subsubsection{Voter model}
%% ===============================================================

In \Cref{app:Snapshots} in \Cref{fig:V_snap} snapshots for the system evolution with the voter model are presented.
They provide examples of simulations performed for all three types of opinions updates:
synchronous (odd $L=21$) [\Cref{subfig:V_sync_t=0000,subfig:V_sync_t=0010,subfig:V_sync_t=0100,subfig:V_sync_t=1000,subfig:V_sync_t=end}];
synchronous (even $L=20$) [\Cref{subfig:V_even_t=0000,subfig:V_even_t=0010,subfig:V_even_t=0100,subfig:V_even_t=1000,subfig:V_even_t=end}];
sequential [\Cref{subfig:V_seql_t=0000,subfig:V_seql_t=0010,subfig:V_seql_t=0100,subfig:V_seql_t=1000,subfig:V_seql_t=end}]; 
and random [\Cref{subfig:V_rand_t=0000,subfig:V_rand_t=0010,subfig:V_rand_t=0100,subfig:V_rand_t=1000,subfig:V_rand_t=end}].

\Cref{subfig:VSSE_no_vs_t,subfig:VSRA_no_vs_t,subfig:VSSY_no_vs_t} show examples of $n_\text{o}(t)$ evolution for the voter model ($L=141$) for sequential, random, and asynchronous updates, respectively.
For this model---independently of the assumed system update scheme---a roughly power-law decay of the number of opinions is observed.
This decay follows $n_\text{o}(t)\propto t^{-\gamma}$ with the exponent $\gamma\approx 0.86753(51)$, $0.82102(52)$ and $0.81663(21)$ for a sequential, random, and synchronous opinion update scheme, respectively.
The least-squares method fits are marked in \Cref{subfig:VSSE_no_vs_t,subfig:VSRA_no_vs_t,subfig:VSSY_no_vs_t} by orange solid lines.

%% ===============================================================
\subsubsection{Sznajd model}
%% ===============================================================

In \Cref{fig:S_snap} in \Cref{app:Snapshots} snapshots of the system evolution with the Sznajd model are presented.
They provide examples of simulations performed for all four types of lattices, with $N=L^2$, and $L=21$ for the triangular lattice 
[\Cref{subfig:SZ_triangular_t=0,subfig:SZ_triangular_t=100,subfig:SZ_triangular_t=200,subfig:SZ_triangular_t=400,subfig:SZ_triangular_t=end}]
and both varieties of the square lattice (with von Neumann [\Cref{subfig:SZ_neumann_t=0,subfig:SZ_neumann_t=100,subfig:SZ_neumann_t=200,subfig:SZ_neumann_t=400,subfig:SZ_neumann_t=end}] and Moore neighborhood [\Cref{subfig:SZ_moore_t=0,subfig:SZ_moore_t=100,subfig:SZ_moore_t=200,subfig:SZ_moore_t=400,subfig:SZ_moore_t=end}] or $L=20$ for the honeycomb lattice [\Cref{subfig:SZ_honeycomb_t=0,subfig:SZ_honeycomb_t=100,subfig:SZ_honeycomb_t=200,subfig:SZ_honeycomb_t=400,subfig:SZ_honeycomb_t=end}] requiring an even value of $L$.
In this model, the initially agreeable pairs start to convince their neighbors, which forms one or more expanding clusters of the same opinion and leads to creation of several domains. 
Eventually, one opinion overwhelms the entire population after $\tau$ steps.

The number of opinions found in the system is shown as a function of time in \Cref{subfig:SZ_VN_no_vs_t,subfig:SZ_M_no_vs_t,subfig:SZ_TR_no_vs_t,subfig:SZ_HC_no_vs_t}.
The process of reaching one opinion takes place in several steps, from the initial phase of slow change to the abrupt change towards the final stable state.
Even from this limited number of examples we can see that the relaxation time $\tau$ may vary considerably from one sample to another, depending as well on the type of the lattice.
For this reason a more detailed study of $\tau$ is provided in \Cref{subsec:relaxation}.

%% ===============================================================
\subsubsection{Latan\'e model}
%% ===============================================================

In \Cref{app:Snapshots} snapshots of the evolution of the system with the Latan\'e model for various effective ranges of interaction $2\le\alpha\le 6$ 
[$\alpha=2$: \Cref{subfig:L_a20T10_t=0000,subfig:L_a20T10_t=0005,subfig:L_a20T10_t=0010,subfig:L_a20T10_t=0050,subfig:L_a20T10_t=end}, 
 $\alpha=3$: \Cref{subfig:L_a30T10_t=0000,subfig:L_a30T10_t=0005,subfig:L_a30T10_t=0010,subfig:L_a30T10_t=0050,subfig:L_a30T10_t=end}, 
 $\alpha=5$: \Cref{subfig:L_a50T10_t=0000,subfig:L_a50T10_t=0100,subfig:L_a50T10_t=1000,subfig:L_a50T10_t=5000,subfig:L_a50T10_t=end}, 
 $\alpha=6$: \Cref{subfig:L_a60T10_t=0000,subfig:L_a60T10_t=5e3,subfig:L_a60T10_t=1e4,subfig:L_a60T10_t=5e4,subfig:L_a60T10_t=end}] and various social temperatures $T$ (for $T=0$ in \Cref{fig:L_snap_T0} and for $T=1$ in \Cref{fig:L_snap_T1}) are presented. 

\Cref{fig:Latane_no} shows final (averaged over $\mathcal R=1000$ simulations) number of opinions $n^{\text{u}}_{\text{o}}$ versus $(\alpha, T)$ for Latan\'e model. 
The rows for $T>2$ and for the columns for $\alpha>3$ are presented in Figure~1(a) in Reference~\onlinecite{Maslyk_2023}.
The ambiguous role of information noise (social temperature) $T$ can be induced from this plot: the finite $T>0$ but small value of information noise promotes unanimity of opinion, while further increasing $T$ destroys consensus and promotes opinion randomness as elaborated in References~\onlinecite{Maslyk_2023,2002.05451}.
The figure mentioned above allows us to identify sets of $(\alpha,T)$ parameters, where reaching consensus is possible (and the time to reach the final state of the system is unambiguously defined).
For a phase diagram [but for ten times worse statistics ($\mathcal R=10^2$) and the order of magnitude shorter time of observation $t_{\text{max}}=10^5$] in the Latan\'e model see Figure 3 in Reference~\onlinecite{Maslyk_2023}.
As one can deduce from the phase diagram mentioned above, both: the effective range of interaction ($\alpha$) and the level of information noise ($T$) influence the possibility of reaching consensus.
As we can see in \Cref{fig:L_snap_T0} deterministic version of Latan\'e model ($T=0$) successfully prevents the consensus to be reached.
The low level of information noise ($T=1$, \Cref{fig:L_snap_T1}) allows for reaching consensus when the effective range of the interaction of the actors is high ($\alpha<4$, \Cref{subfig:L_a20T10_t=0000,subfig:L_a30T10_t=0000,subfig:L_a20T10_t=0005,subfig:L_a30T10_t=0005,subfig:L_a20T10_t=0010,subfig:L_a30T10_t=0010,subfig:L_a20T10_t=0050,subfig:L_a30T10_t=0050,subfig:L_a20T10_t=end,subfig:L_a30T10_t=end}) but for its lower value [$\alpha>4$, \Cref{subfig:L_a50T10_t=0000,subfig:L_a60T10_t=0000,subfig:L_a50T10_t=0100,subfig:L_a60T10_t=5e3,subfig:L_a50T10_t=1000,subfig:L_a60T10_t=1e4,subfig:L_a50T10_t=5000,subfig:L_a60T10_t=5e4,subfig:L_a50T10_t=end,subfig:L_a60T10_t=end}] not quite (at least after waiting $t_{\text{max}}=10^5$ MCS).

In \Cref{subfig:L21a20T125_no_vs_t,subfig:L21a20T20_no_vs_t,subfig:L21a3T15_no_vs_t} selected examples of the time evolution of the number of opinions $n_{\text{o}}(t)$ for $L=21$, $\alpha=2$, $T=1.25$, and $\alpha=2$, $T=2$, and $\alpha=3$, $T=1.5$, respectively.
The system evolution is rather closer to the behavior of the voter model [\Cref{subfig:VSRA_no_vs_t,subfig:VSSE_no_vs_t,subfig:VSSY_no_vs_t}] than to the Sznajd model [\Cref{subfig:SZ_VN_no_vs_t,subfig:SZ_M_no_vs_t,subfig:SZ_HC_no_vs_t,subfig:SZ_TR_no_vs_t}].
For $\alpha=2$ and $T=1.25$, the initial decrease of $n_\text{o}(t)$ follows power-law $n_\text{o}(t)\propto t^{-\gamma}$ with exponent $\gamma\approx 0.3604(88)$.
For other sets of parameters ($\alpha,T$), exact power-law decay of the number of opinions is absent, and a much more rapid decrease of the number of opinions at the end of the evolution of the system is observed.
On the other hand, long intermediate states observed for the Sznajd model are also absent.

%% ###############################################################
\subsection{\label{subsec:relaxation}Relaxation time}
%% ###############################################################

%% ===============================================================
\subsubsection{Average/median time of reaching consensus}
%% ===============================================================

\begin{figure}[htbp]
%% ===============================================================
\begin{subfigure}[t]{0.49\textwidth}
\caption{\label{subfig:tau_vs_N_V}} %%voter, $\mathcal R=10^3$}
\includegraphics[width=.95\textwidth]{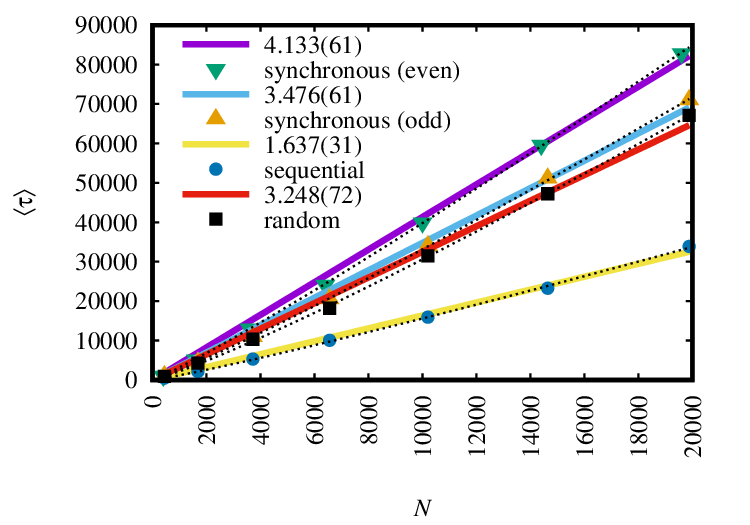}
\end{subfigure}
%% ---------------------------------------------------------------
\begin{subfigure}[t]{0.49\textwidth}
\caption{\label{subfig:tau_vs_N_S}} %%Sznajd, $\mathcal R=100$ ($\mathcal R=10^5$ for honeycomb)}
\includegraphics[width=.95\textwidth]{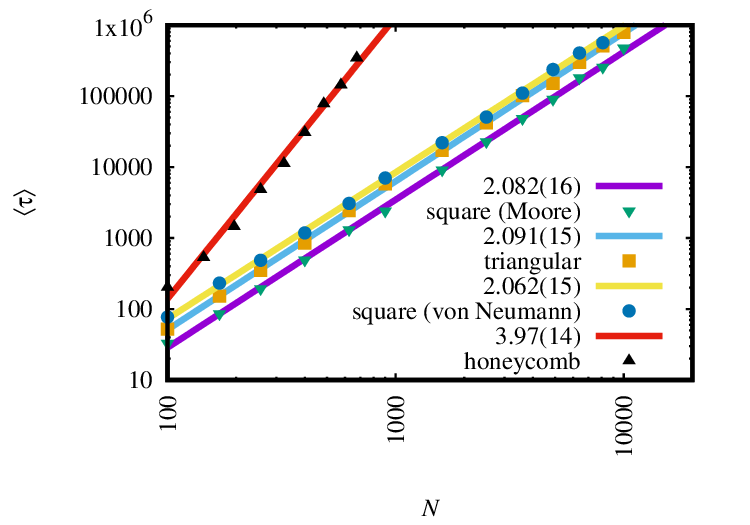}
\end{subfigure}
%% ---------------------------------------------------------------
\begin{subfigure}[t]{0.49\textwidth}
\caption{\label{subfig:tau_vs_N_L}} %%Latan\'e, $\mathcal R=10^3$, median}
\includegraphics[width=.95\textwidth]{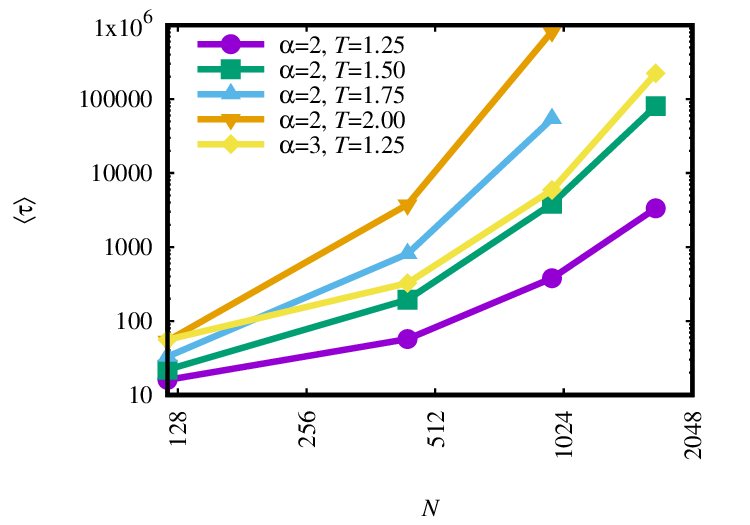}
\end{subfigure}
%% ---------------------------------------------------------------
\caption{\label{fig:tau_vs_N}Time $\langle\tau\rangle$ of reaching consensus (averaged over $\mathcal R$ simulations, measured in MCS) versus system size $N$ for various models: 
\subref{subfig:tau_vs_N_V} voter ($\mathcal R=10^3$), with linear fit $\langle\tau\rangle=bN$, coefficients $b$ are displayed,
\subref{subfig:tau_vs_N_S} Sznajd ($\mathcal R=100$ for square and triangular but $\mathcal R=10^5$ for honeycomb lattice), with power-law fit $\langle\tau\rangle=cN^\beta$, exponents $\beta$ are displayed,
and \subref{subfig:tau_vs_N_L} Latan\'e ($\mathcal R=10^3$), faster than power-law}
%% ===============================================================
\end{figure}

\Cref{fig:tau_vs_N} presents times $\langle\tau\rangle$ of reaching consensus versus system size $N$ for various models:
voter [\Cref{subfig:tau_vs_N_V}];
Sznajd [\Cref{subfig:tau_vs_N_S}];
and Latan\'e [\Cref{subfig:tau_vs_N_L}] model.
The symbol $\langle\cdots\rangle$ stands for a mean value in $\mathcal R$ simulations for the votermodel, but for the median for the Sznajd and Latan\'e models.
The power-law fit parameters $\langle\tau\rangle=cN^\beta$ are presented in \Cref{tab:disc}.

%% ===============================================================
\begin{table}[bp]
\caption{\label{tab:disc}Mean (for voter model) and median (for Sznajd and Latan\'e models) time $\langle\tau\rangle = cN^\beta$ of reaching of the final sate of the system and the ultimate number $n_\text{o}^\text{u}=\langle n_\text{o}(t\to\infty)\rangle$ of opinions for the voter, Sznajd and Latan\'e model of opinion dynamics}
\begin{ruledtabular} 
\begin{tabular}{llrrr}
model    & details & $\beta$ & $c$ & $n_\text{o}^\text{u}$ \\
\hline %% --------------------------------------------------------
voter    & synchronous\footnote{odd $L$}    & $1.1135(76)$ & $1.165(86)$ & $1$\\
%% async1 = 3.47586 +/- 0.0613   (1.764%)
%% csync1 = 1.1649  +/- 0.0857   (7.357%)
%% bsync1 = 1.11352 +/- 0.007575 (0.6803%)
         & synchronous\footnote{even $L$}   & $1.0942(60)$ & $1.671(97)$ & $2$\\
%% async2 = 4.13271 +/- 0.06081  (1.471%)       
%% csync2 = 1.67087 +/- 0.09679  (5.793%)
%% bsync2 = 1.09419 +/- 0.005977 (0.5462%)
         & sequential               & $1.115(20)$ & $0.54(11)$ & $1$\\
%% aseql = 1.63656  +/- 0.03072 (1.877%)
%% cseql = 0.539206 +/- 0.1069  (19.83%)
%% bseql = 1.11529  +/- 0.02042 (1.831%)
         & random                   & $1.144(13)$ & $0.81(10)$ & $1$\\
%% arand = 3.24841  +/- 0.07164 (2.205%)          
%% crand = 0.814296 +/- 0.1     (12.28%)
%% brand = 1.14365  +/- 0.01264 (1.105%)
\hline %% --------------------------------------------------------
Sznajd   & square\footnote{von Neumann neighborhood} & $2.062(15)$ & $0.00547(61)$ & $1$\\
         & square\footnote{Moore neighborhood}       & $2.082(16)$ & $0.00196(23)$ & $1$\\
         & triangular                   & $2.091(15)$ & $0.00337(35)$ & $1$\\
         & honeycomb                    & $3.97(14)$  & $0.16(13) \times 10^{-5}$ & $1$\\
\hline %% --------------------------------------------------------
Latan\'e  & $\alpha=2$, $T=1.25$ &  &  &  $1$\\
          & $\alpha=2$, $T=1.5$  &  &  &  $1$\\
          & $\alpha=2$, $T=1.75$ &  &  &  $1$\\
          & $\alpha=2$, $T=2$    &  &  &  $1$\\
          & $\alpha=3$, $T=1.25$ &  &  & $1.010$\footnote{in ten cases, at $t=10^6$, consensus were not reached}\\
          & $\alpha=3$, $T=1.5$  &  &  & $1.001$\footnote{in one case, at $t=10^6$, consensus was not reached}\\
\end{tabular}
\end{ruledtabular}
\end{table}
%% ===============================================================

%% ---------------------------------------------------------------
\paragraph{Voter model}
%% ---------------------------------------------------------------

In \Cref{subfig:tau_vs_N_V} the system size $N$ dependence of the median time $\langle\tau\rangle$ to reach the final state is presented.
These dependencies again follow a power law with exponents:
$1.1135(76)$ for synchronous and odd $L$;
$1.0942(60)$ for synchronous and even $L$\footnote{here the final state corresponds to (rather artificial) chess-board pattern instead of consensus};
$1.115(20)$ sequential;
$1.144(13)$ random update schemes.
The exponents are quite close to unity, so linear fits for these dependencies $\langle \tau\rangle=(bN+\text{const})$ were also performed.
These linear fits are shown as solid lines in \Cref{subfig:tau_vs_N_V} and their slopes $b$ are indicated there as well.

%% ---------------------------------------------------------------
\paragraph{Sznajd model}
%% ---------------------------------------------------------------
The relaxation times observed in the Sznajd model vary a lot even in relatively small number of experiments, and to have a better statistics in this case we have to average all the results over much larger number of simulations, $\mathcal R = 10^4$ or even $\mathcal R = 10^5$ if needed.
Since the observed values of $\tau$ are quite often very distant from the average value of $\tau$, and it is also not rare that even very long simulation times are still not adequate to reach $n_{\text{o}}=1$, for this model we present the results in terms of the median $\tau$ instead of its average value.
\Cref{subfig:tau_vs_N_S} illustrates the observed power-law dependence of the median $\tau$ on the number of nodes $N$, $\langle\tau\rangle \propto N^\beta$, with exponents $\beta$ given in detail in \Cref{tab:disc}.

%% ---------------------------------------------------------------
\paragraph{Latan\'e model}
%% ---------------------------------------------------------------

In \Cref{subfig:tau_vs_N_L} the dependence of the mean time $\langle\tau\rangle$ to reach consensus (averaged over $\mathcal R$ simulations) versus the size $N$ of the system is presented.
In contrast to the voter and the Sznajd model, here the growth $\langle\tau\rangle$ vs. $N$ is faster than the power-law observed for other models considered independently on the noise level ($T$) and the range of interaction ($\alpha$)---at least for which unambiguously the consensus takes place.

%% ===============================================================
\subsubsection{Probability distribution function of times of reaching consensus}
%% ===============================================================

\begin{figure}[htbp]
%% ===============================================================
\begin{subfigure}[t]{0.49\textwidth}
\caption{\label{subfig:P_vs_tau_V}} %%voter, $N=141^2$, $\mathcal R=10^3$}
\includegraphics[width=.95\textwidth]{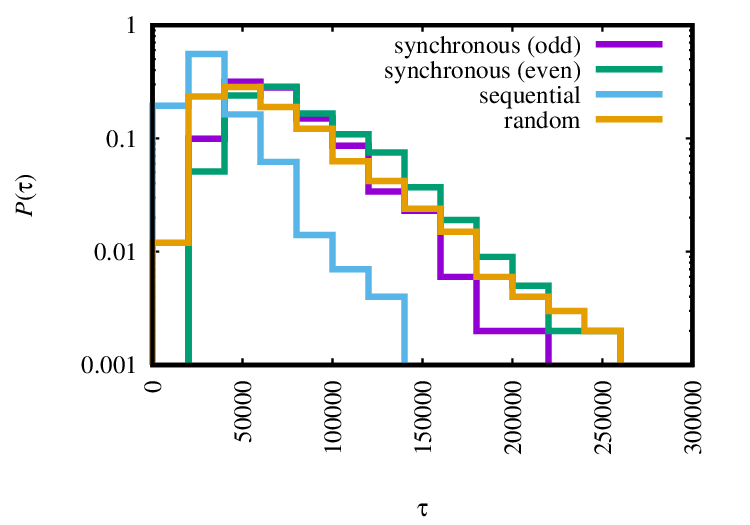}
\end{subfigure}
%% ---------------------------------------------------------------
\begin{subfigure}[t]{0.49\textwidth}
\caption{\label{subfig:P_vs_tau_S}} %%Sznajd, $N=20^2$, $\mathcal R=10^4$}
\includegraphics[width=.95\textwidth]{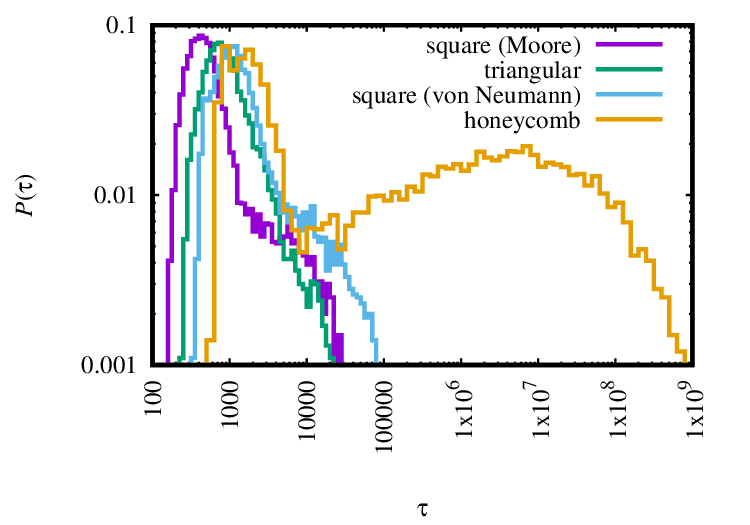}
\end{subfigure}
%% ---------------------------------------------------------------
\begin{subfigure}[t]{0.49\textwidth}
\caption{\label{subfig:P_vs_tau_L}} %%Latan\'e, $N=21^2$, $\mathcal R=10^3$}
\includegraphics[width=.95\textwidth]{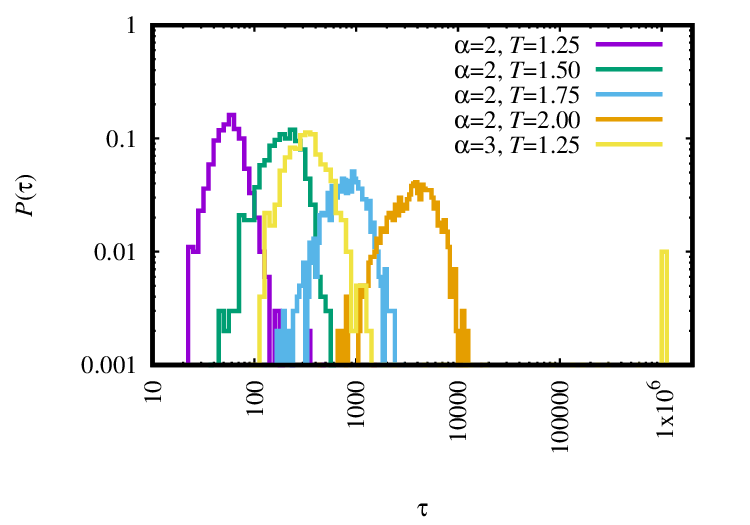}
\end{subfigure}
%% ---------------------------------------------------------------
\caption{\label{fig:P_vs_tau}Probability distribution function $P$ of times $\tau$ of reaching consensus (gathered from $\mathcal R$ simulations, measured in MCS) for various models: 
\subref{subfig:P_vs_tau_V} voter ($N=141^2$, $\mathcal R=10^3$), 
\subref{subfig:P_vs_tau_S} Sznajd ($N=20^2$, $\mathcal R=10^4$),
\subref{subfig:P_vs_tau_L} Latan\'e ($N=21^2$, $\mathcal R=10^3$).
In the latter, for $\alpha=3$ and $T=1.25$ in ten cases, consensus was not reached, resulting with artificial presence $P(10^6)=0.01$ after reaching the assumed $t_\text{max}=10^6$ MCS}
%% ---------------------------------------------------------------
\end{figure}

\Cref{fig:P_vs_tau} shows probability distribution function $P$ of times $\tau$ to reach consensus (gathered from $\mathcal R$ simulations) for various models:
voter [\Cref{subfig:P_vs_tau_V}]; 
Sznajd [\Cref{subfig:P_vs_tau_S}];
and Latan\'e [\Cref{subfig:P_vs_tau_L}].

%% ---------------------------------------------------------------
\paragraph{Voter model}
%% ---------------------------------------------------------------

In the \Cref{subfig:P_vs_tau_V} probability distribution function $P$ of times $\tau$
to reach consensus (for the size of the system $N=141^2$ and averaged over the $\mathcal R=10^3$ simulations) is presented.
Data are gathered in bins of length 20000.

%% ---------------------------------------------------------------
\paragraph{Sznajd model}
%% ---------------------------------------------------------------
The probability distribution function of relaxation times in \Cref{subfig:P_vs_tau_S} was obtained for all lattices used in the Sznajd model simulations.
Its long tails confirm that very long times needed to reach the final state of $n_{\text{o}}=1$ are observed quite often (we note that the log-scale is also on the horizontal axis here).
This property is most evident for the honeycomb lattice, where the probability distribution function becomes clearly bimodal and the observed values of $\tau$ spread over many orders of magnitude, in agreement with previous remarks in \Cref{subsec:relaxation}.

%% ---------------------------------------------------------------
\paragraph{Latan\'e model}
%% ---------------------------------------------------------------

In \Cref{subfig:P_vs_tau_L} probability distribution function $P(\tau)$ for $N=21^2$ averaged over $\mathcal R=10^3$ simulations is presented.
The binning is performed in logarithmic fashion.
The function shape is similar to that observed for the voter model and far away from this for the bimodal shape for the Sznajd model.

%% ===============================================================
%% ===============================================================

%% ===============================================================
%% ===============================================================

%% ###############################################################
\section{\label{sec:discussion}Discussion}
%% ###############################################################

%% ###############################################################
\subsection{Voter model}
%% ###############################################################

For the voter model---with the simplest model rules among models considered here---the time evolution of the number $n_\text{o}(t)$ of opinions follows power-law decay. 
This is most spectacular for sequential updating as presented in \Cref{subfig:VSSE_no_vs_t}.

The mean time $\tau$ to reach consensus increases almost linearly with the size of the system $N$ [see \Cref{subfig:tau_vs_N_V}].
This increase is the slowest for the sequential update scheme and the most rapid for the synchronous scheme with an even linear system size.

Histograms [see \Cref{subfig:P_vs_tau_V}] for the voter model do not show unexpected behaviors.
The histogram is the most well-knit for the sequential update scheme, as expected based on the dependence of $\langle\tau\rangle$ vs. $N$ [see \Cref{subfig:tau_vs_N_V}].

%% ###############################################################
\subsection{Sznajd model}
%% ###############################################################

The complexity of rules of the Sznajd model is in between the toy rules of the voter model and the extreme complexity of the Latan\'e model.

\Cref{subfig:SZ_VN_no_vs_t,subfig:SZ_M_no_vs_t,subfig:SZ_TR_no_vs_t,subfig:SZ_HC_no_vs_t} shows that unlike the voter model and the Latan\'e model, for the Sznajd model we can observe several stages of the dynamics leading to $n_{\text{o}} = 1$.
Initially, the number of opinions remains virtually unchanged when the first clusters begin to emerge. These clusters are rather small at this stage, with a lot of unchanged, random parts still unaffected.
Then, the clusters grow rapidly and, when no more random areas are present, the evolution slows down: the number of clusters gradually decreases together with $n_{\text{o}}$.
Only after much longer times, one of the clusters prevails, and the final stable state is achieved after time $\tau$.

For the square lattice with the Moore neighborhood, the triangular lattice and the square lattice with the von Neumann neighborhood, the values of the exponent $\beta$ in the power law $\langle\tau\rangle=cN^\beta$ are very similar [see \Cref{subfig:tau_vs_N_S}], in fact equal within the range of the estimated standard errors, even though the number of nearest neighbors is not the same but equals eight, six and four, respectively.
However, a further decrease of the number of nearest neighbors, to three in the case of the honeycomb lattice, changes the exponent $\beta$ almost by a factor of two, rapidly increasing the time needed to reach the final state with only one opinion.

The distribution of the times needed to achieve consensus, shown in \Cref{subfig:P_vs_tau_S}, results from the details of the initial (random) distribution of opinions.
Sometimes, it allows one opinion to quickly overcome all other alternatives, while in other cases two or more competing opinions remain in the system for a long time before one of them finally wins.

%% ###############################################################
\subsection{Latan\'e model}
%% ###############################################################

The Latan\'e model poses the most complex rules among the ones considered here.
For this model and in the region of low $\alpha$ (wide effective range of interaction) and low but finite information noise $T$ [see \Cref{subfig:L21a20T125_no_vs_t} for $\alpha=2$ and $T=1.25$, \Cref{subfig:L21a20T20_no_vs_t} for $\alpha=2$ and $T=2$, \Cref{subfig:L21a3T15_no_vs_t} for $\alpha=3$ and $T=1.5$] initial more or less power-law decay of $n_\text{o}(t)$ is similar to this observed for the voter model, finalized with abrupt (however, on the logarithmic time scale) decease of opinions toward consensus.

In contrast to the voter and Sznajd models, the increase in the median time $\langle\tau\rangle$ to reach consensus [see \Cref{subfig:tau_vs_N_L}] is much faster than according to $N$ to any power.

The histograms of the times to reach consensus [\Cref{subfig:P_vs_tau_L}] for Latan\'e model exhibit much more similarity to those for the voter than to the Sznajd model, i.e., they do not exhibit bimodal characteristics.

%% ###############################################################
\section{\label{sec:conclusions}Conclusions}
%% ###############################################################

The voter model, the Sznajd model, and the Latan\'e model are three different abstract models that attempt to describe social processes such as opinion formation in a population. There are certain common features between them that, at the same time, emphasize the importance of various social phenomena.

All three models assume that agents (whether voters, individuals in the Sznajd model or participants in the Latan\'e model) make decisions based on their own preferences, beliefs, or experiences. These decisions are usually represented by a change in opinion or behavior.
In each of these models, agents interact with each other, which may lead to a change in their opinions or behavior. These interactions can be direct (e.g. discussions between voters in the voter model), neighborhood-based (in the Sznajd model) or relating to the perceptions of others (in the Latan\'e model).

All three models lead to some emergent behavior in the population that results from interactions between agents. For example, in the voter model there may be an emergence of a dominant opinion in the population, in the Sznajd model there may be a consensus of opinion in neighborhood groups, and in the Latan\'e model there may be social effects such as the spiral of silence effect \cite{Noelle-Neumann_1984}.

All three models may contain elements of randomness that affect the dynamics of change in the population. 
Random factors may include random choices of agents to interact, random events that influence opinions or behavior, and other factors that introduce uncertainty into the process.
Both the Sznajd model and the Latan\'e model are based on a network structure in the population in which agents are connected in a particular way. 
In the Sznajd model, neighboring agents pair up, while in the Latan\'e model there is a network of connections through which information is transmitted. The voter model can also be extended to include the network structure in the interactions between voters.

Examples of opinion alignment and reduction can be found in the contemporary world dominated by electronic media. In social media, people succumb to the influence of the majority opinion or more influential individuals in the network, known as influencers. This leads to the formation of information bubbles, where opinions become more homogeneous \cite{Eady_2019}. Additionally, these effects can be reinforced due to homophily, which is the tendency to form relationships and bonds with individuals who are similar to oneself in terms of characteristics such as age, gender, race, socioeconomic status, political views, interests, or values \cite{Lin_2008}. Creating social groups based on demographic similarities promotes opinion alignment and reduction. A negative consequence here may be an increase in social segregation, deepening differences between social groups, and reinforcing social conformity. The diversity of opinions may also undergo a positive reduction due to cross-cutting exposure \cite{Goldman_2011}. This phenomenon refers to situations where individuals with different political or social views live or work in the same physical or social environment. Interactions between people with different beliefs can lead to mutual understanding. Cross-cutting exposure is crucial for the functioning of a democratic society because it enables the exchange of views, public debate, and thus the alignment of common norms or values.

In conclusion, in this paper, the universality of opinions disappearing in sociophysical models of opinion dynamics is studied.
Universality is ensured by studying three different sociophysical models of opinion formation with various details of the implementation of the model rules. 
These models themselves ensure universality of studied behaviors as they use various
\begin{itemize}
\item range of interaction (with the nearest neighbors interactions for voter and Sznajd models in contrast to Latan\'e model---where everybody interacts with everybody);
\item in-flow (voter and Latan\'e models) and out-flow (Sznajd model) transfer of opinions;
\end{itemize}
but also various:
\begin{itemize}
\item updates schemes (random, sequential, asynchronous) for a voter model;
\item lattice topology (square, triangular, honeycomb) for a Sznajd model;
\item information noise levels ($T$) and effective range of interactions ($\alpha$) for a Latan\'e model.
\end{itemize}

For all the analyzed models the maximum initial diversity of opinions is assumed, which means that the number of available opinions is of the order of the number of actors.
In all the considered cases \emph{unanimity of opinion is reached}.
The only observed exception is that of the artificial chess board-like pattern for the even linear size of the system in the voter model and some sets of $(\alpha,T)$ parameters for the Latan\'e model---see~\Cref{fig:Latane_no} for values far different from unity.
For the Latan\'e model---with not too high social temperature ($T=1$) but effectively short range of interaction ($\alpha>4$)---the society polarization ($\alpha=5$, \Cref{subfig:L_a50T10_t=end}) or finally observed three opinions ($\alpha=6$, \Cref{subfig:L_a60T10_t=end}) are possible. As it has been shown in \Cref{fig:Latane_no} for effectively large range of interaction ($\alpha=3$) but very low ($T<1$) and high enough ($T>1.5$) social temperature the consensus is hardly reachable (although for various reasons in low and high temperatures). These issues for Latan\'e model were also discussed earlier in Refs.~\cite{2211.04183,Maslyk_2023}, also in terms of signatures of self-organized criticality observed in the system~\cite{2002.05451}.

The results presented here provide evidence that consensus is achievable in the long term, either for many initial opinions or when only two of them exist at the beginning.  One can say that at least for the considered models, it is finally possible to remove any doubts on that matter, central in the 40-year tradition of sociophysics.  However, this generally optimistic conclusion has to be taken with some caution.  Not only the time required to reach the consensus may be relatively long, but we are also limited by factors which are not included in the models, for example external influence or deliberate actions aiming to prevent reaching an agreement in opinion.

%% ################################################################
\appendix
\section{\label{app:Snapshots}Snapshots from system evolution}
%% ################################################################

\begin{figure*}[htbp]
%% ---------------------------------------------------------------
\begin{subfigure}[t]{0.23\textwidth}
\caption{\label{subfig:V_sync_t=0000}synchronous, odd $L$}
\includegraphics[trim={0mm 4mm 12mm 31mm},clip,width=.99\textwidth]{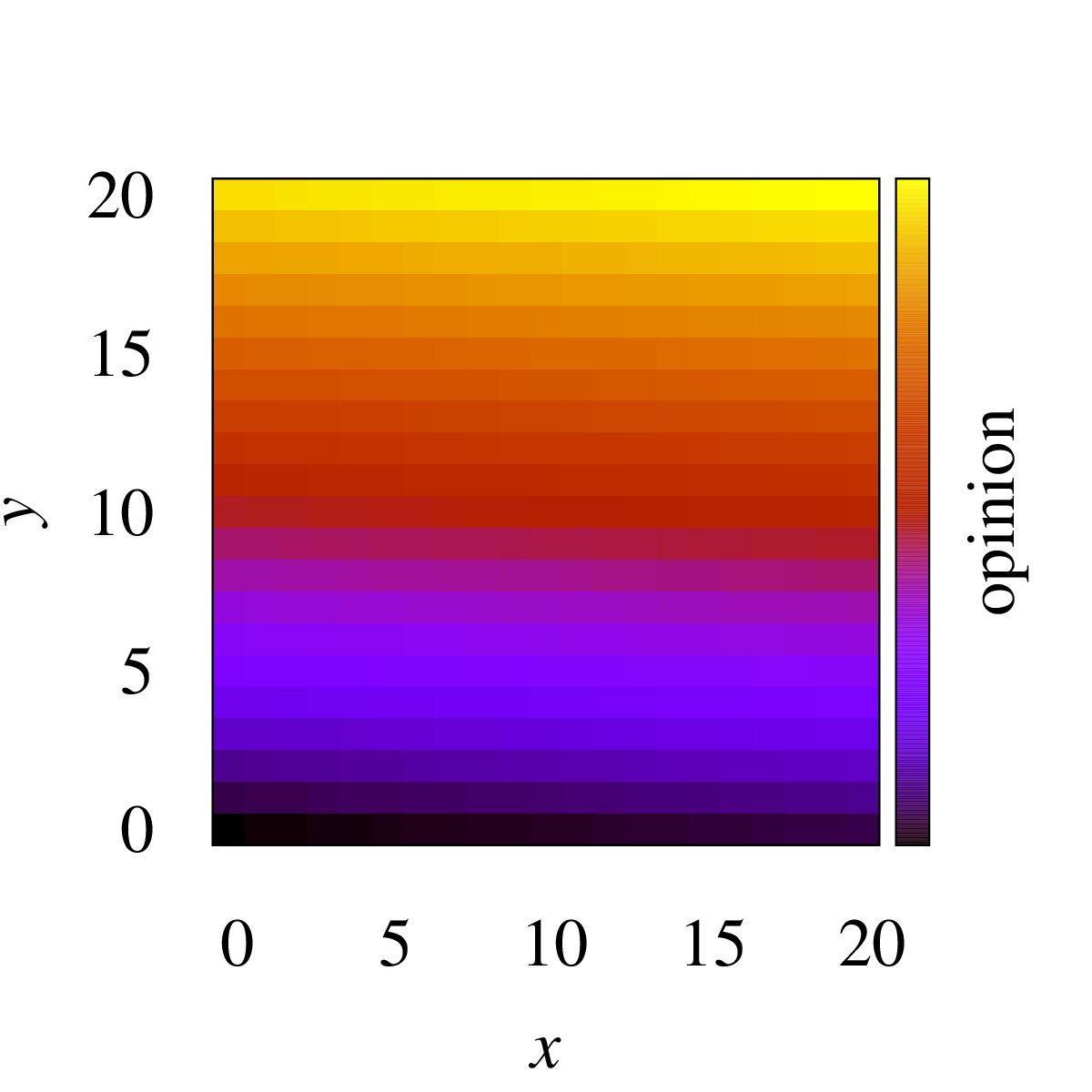}
\end{subfigure}
\hfill%% ---------------------------------------------------------------
\begin{subfigure}[t]{0.23\textwidth}
\caption{\label{subfig:V_even_t=0000}synchronous, even $L$}
\includegraphics[trim={0mm 4mm 12mm 31mm},clip,width=.99\textwidth]{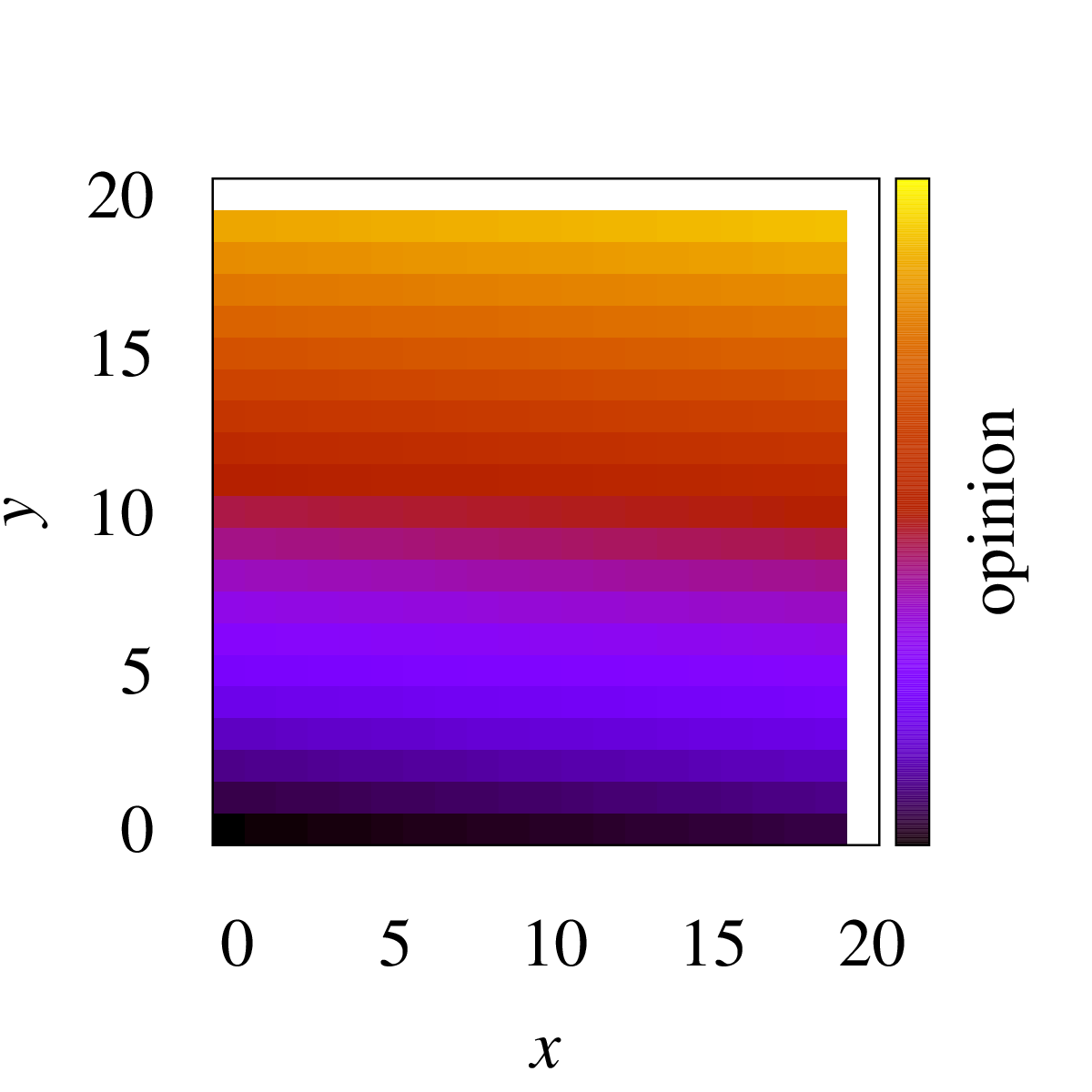}
\end{subfigure}
\hfill%% ---------------------------------------------------------------
\begin{subfigure}[t]{0.23\textwidth}
\caption{\label{subfig:V_seql_t=0000}sequential}
\includegraphics[trim={0mm 4mm 12mm 31mm},clip,width=.99\textwidth]{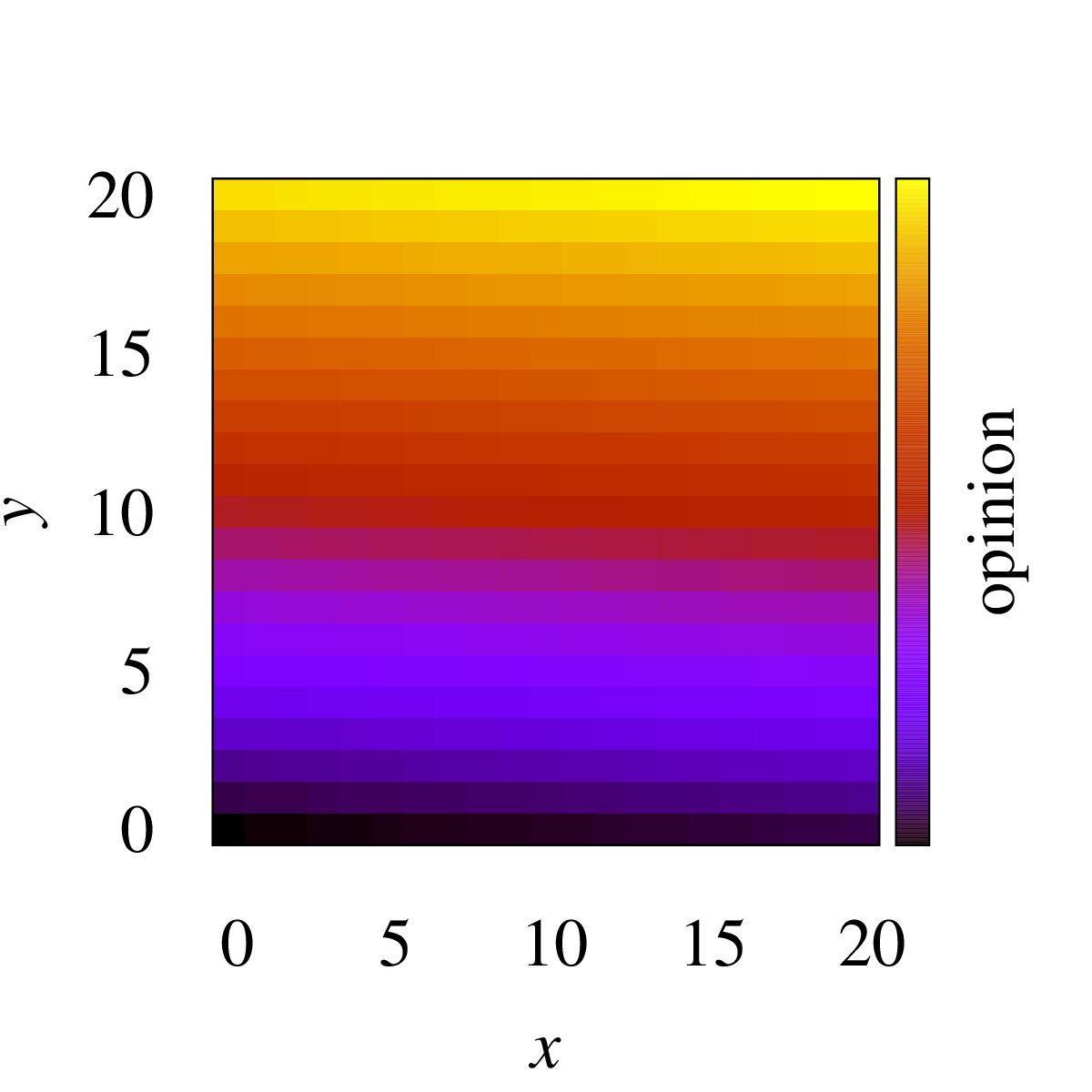}
\end{subfigure}
\hfill%% ---------------------------------------------------------------
\begin{subfigure}[t]{0.23\textwidth}
\caption{\label{subfig:V_rand_t=0000}random}
\includegraphics[trim={0mm 4mm 12mm 31mm},clip,width=.99\textwidth]{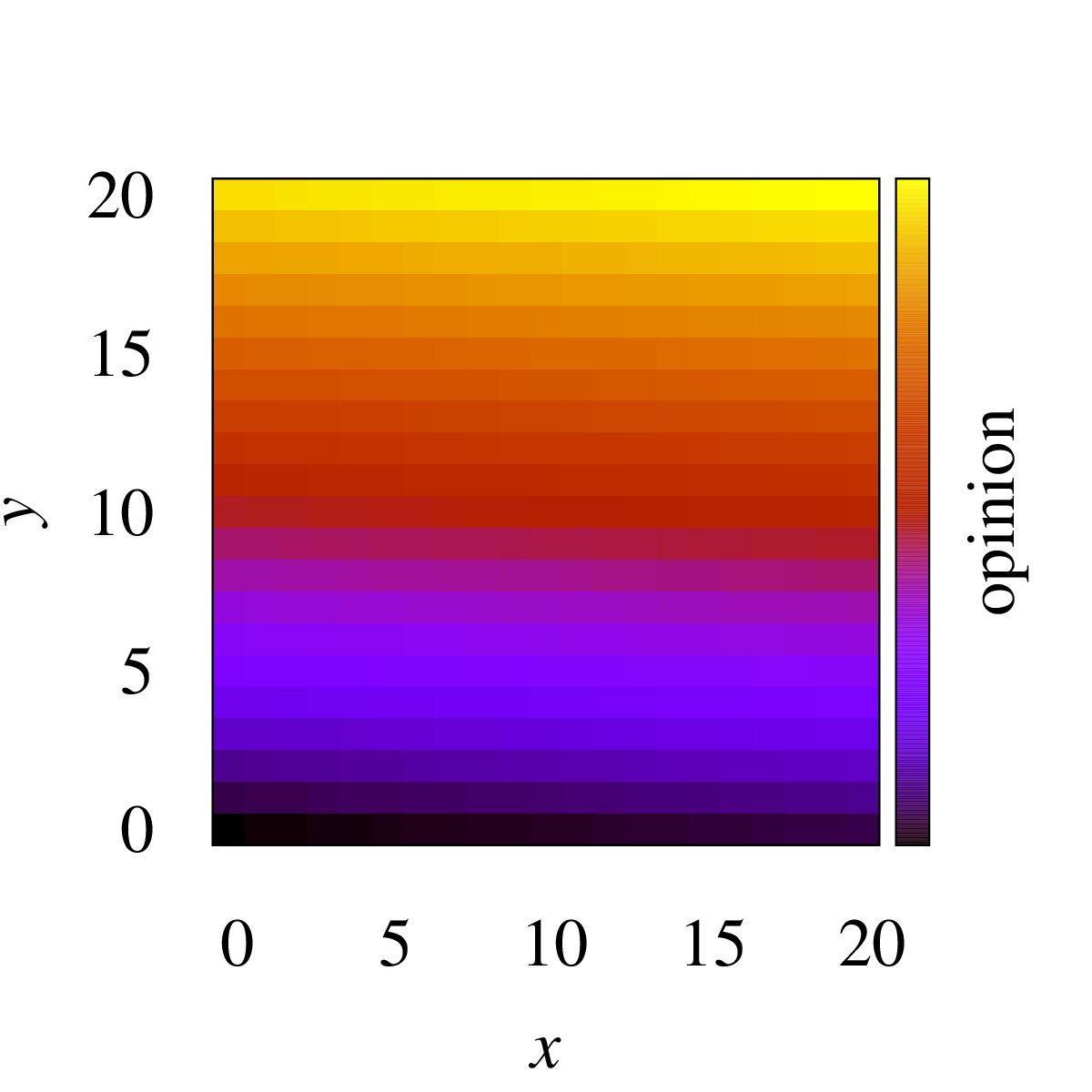}
\end{subfigure}
%% ---------------------------------------------------------------
%% ---------------------------------------------------------------
\begin{subfigure}[t]{0.23\textwidth}
\caption{\label{subfig:V_sync_t=0010}$t=10$}
\includegraphics[trim={0mm 4mm 12mm 31mm},clip,width=.99\textwidth]{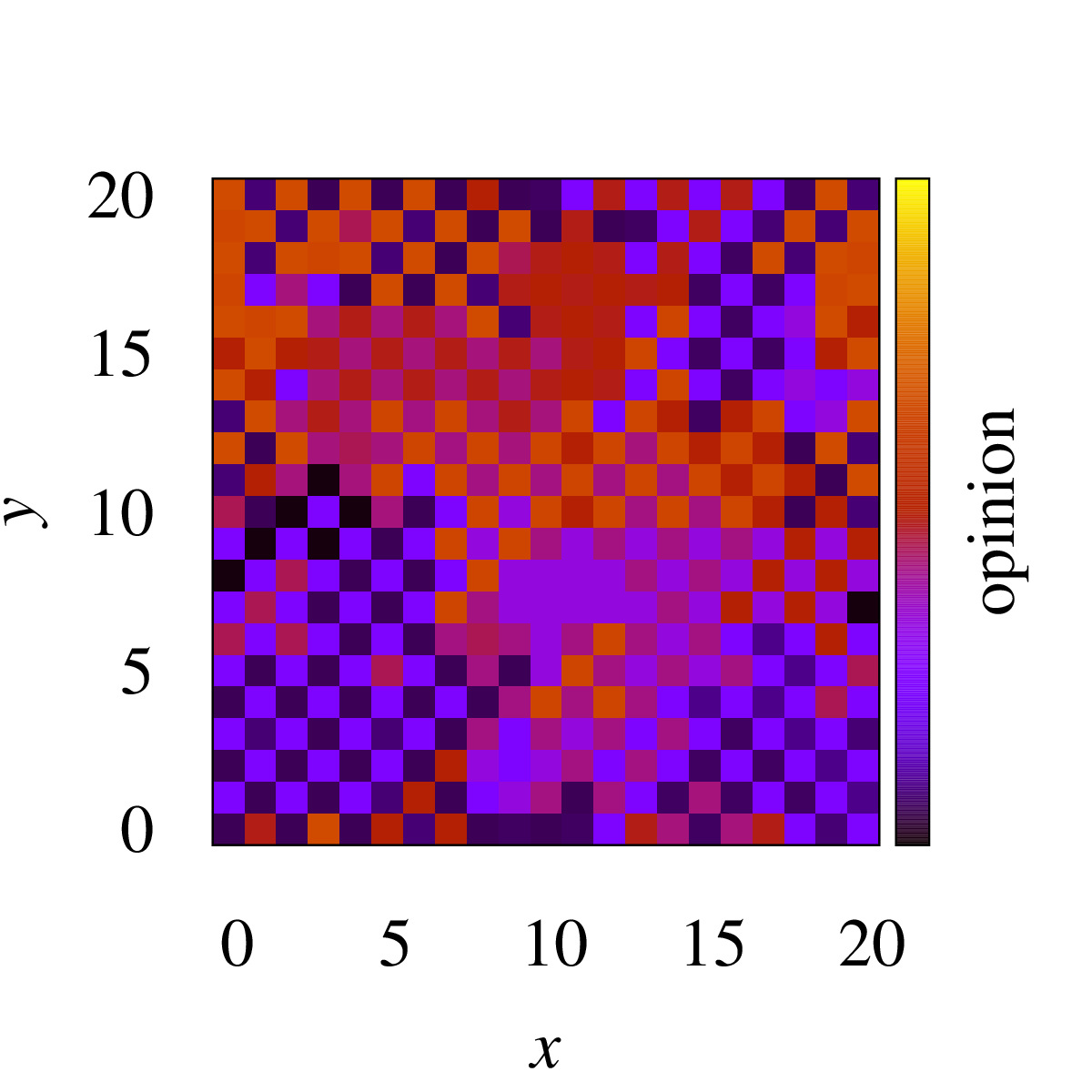}
\end{subfigure}
\hfill%% ---------------------------------------------------------------
\begin{subfigure}[t]{0.23\textwidth}
\caption{\label{subfig:V_even_t=0010}}
\includegraphics[trim={0mm 4mm 12mm 31mm},clip,width=.99\textwidth]{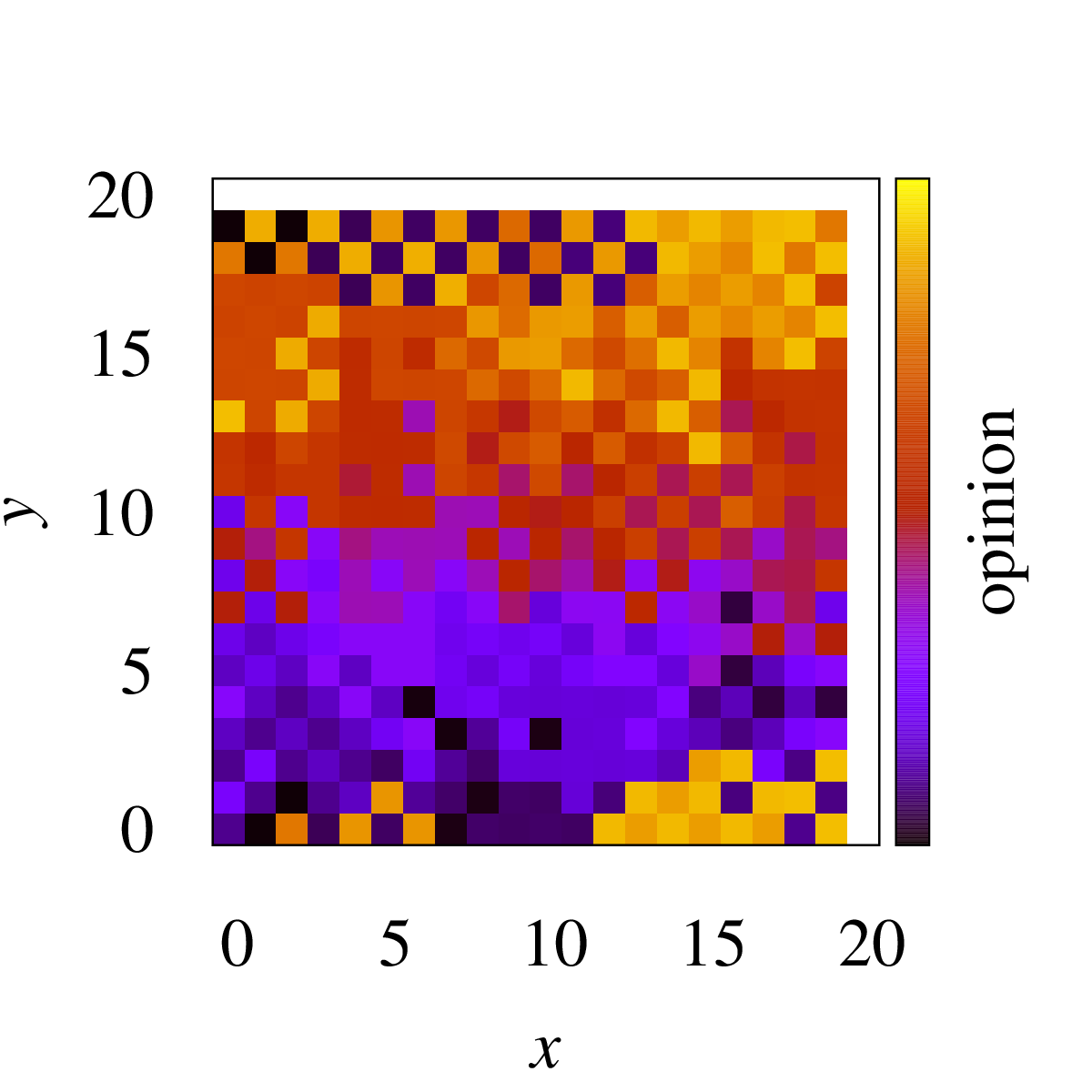}
\end{subfigure}
\hfill%% ---------------------------------------------------------------
\begin{subfigure}[t]{0.23\textwidth}
\caption{\label{subfig:V_seql_t=0010}}
\includegraphics[trim={0mm 4mm 12mm 31mm},clip,width=.99\textwidth]{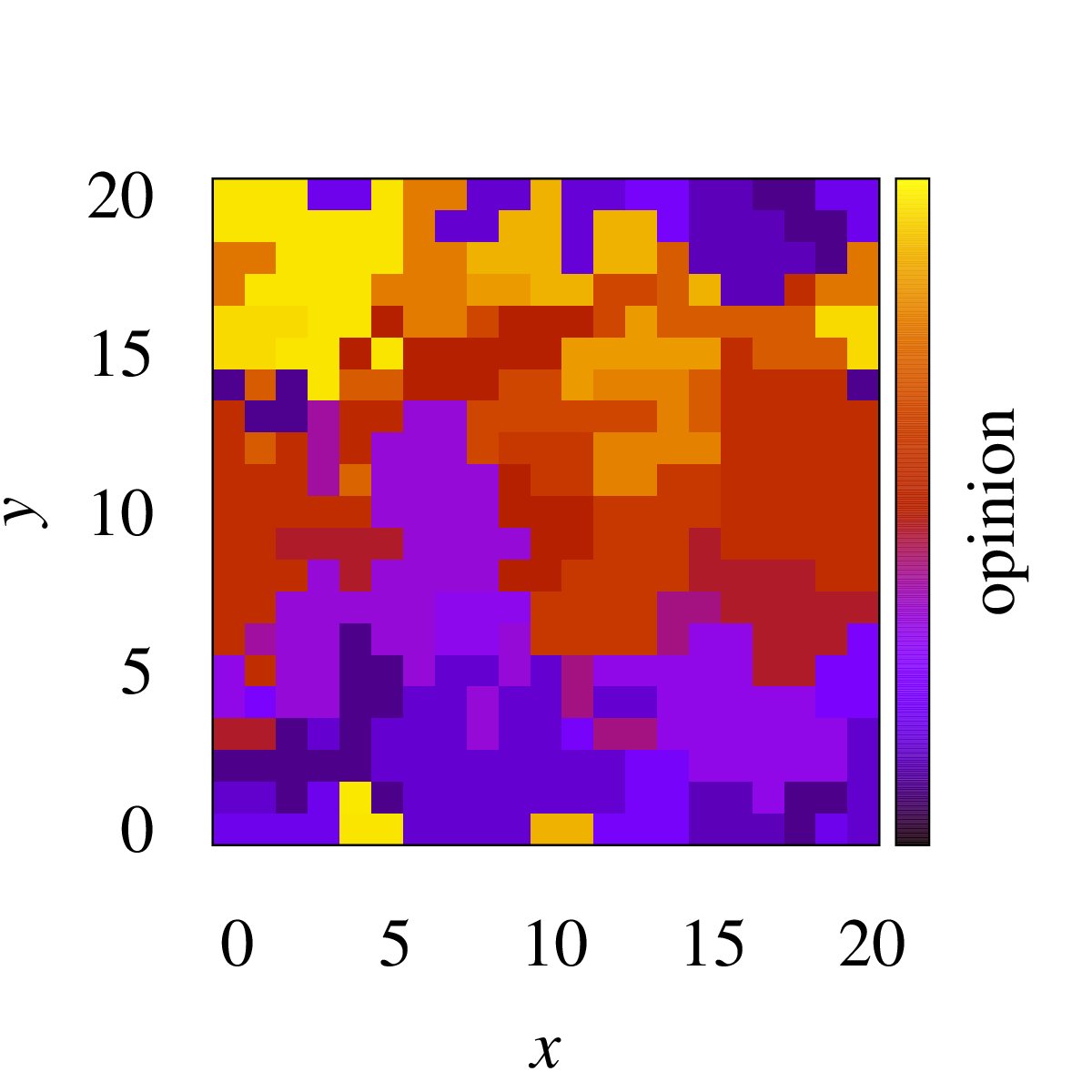}
\end{subfigure}
\hfill%% ---------------------------------------------------------------
\begin{subfigure}[t]{0.23\textwidth}
\caption{\label{subfig:V_rand_t=0010}}
\includegraphics[trim={0mm 4mm 12mm 31mm},clip,width=.99\textwidth]{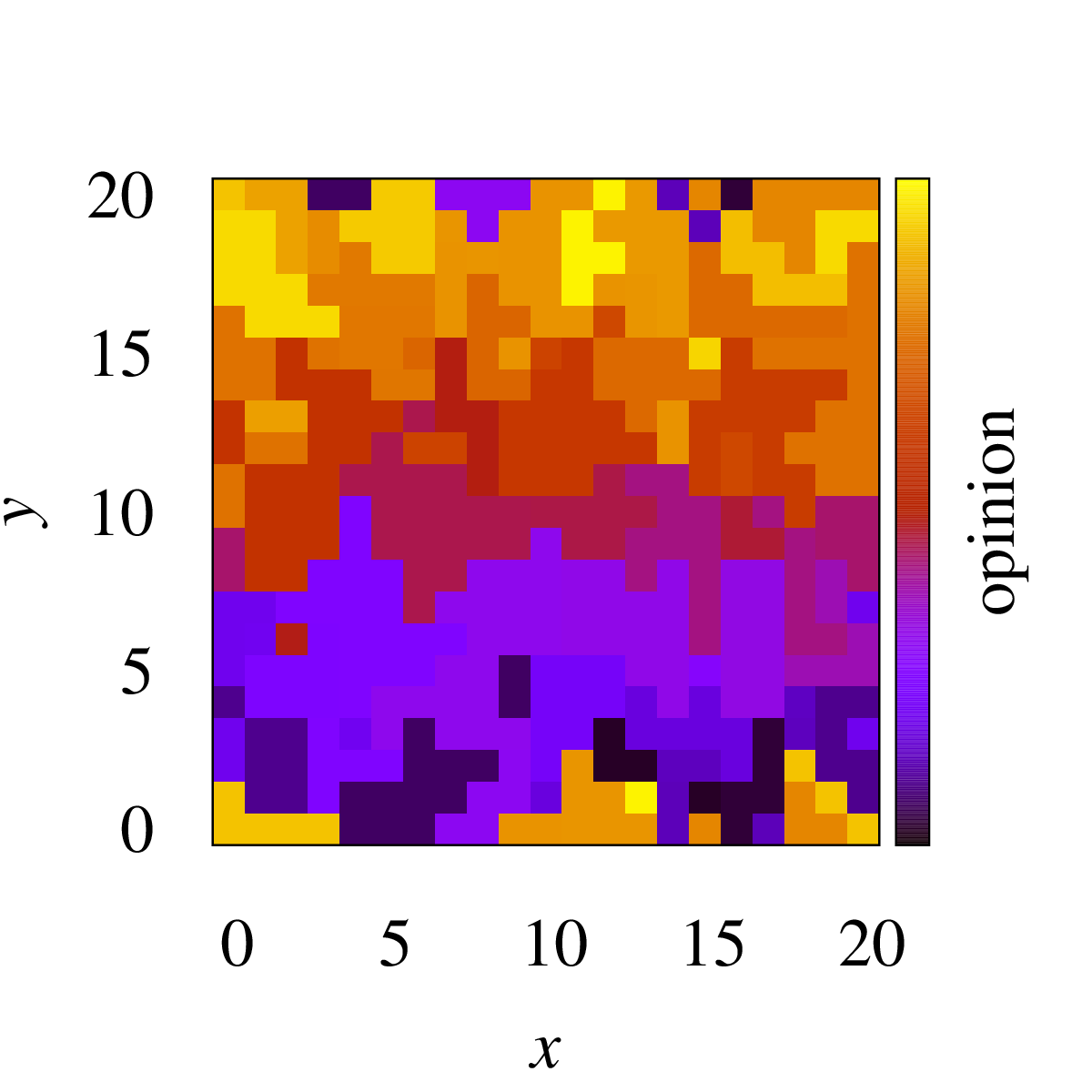}
\end{subfigure}
%% ---------------------------------------------------------------
%% ---------------------------------------------------------------
\begin{subfigure}[t]{0.23\textwidth}
\caption{\label{subfig:V_sync_t=0100}$t=10^2$}
\includegraphics[trim={0mm 4mm 12mm 31mm},clip,width=.99\textwidth]{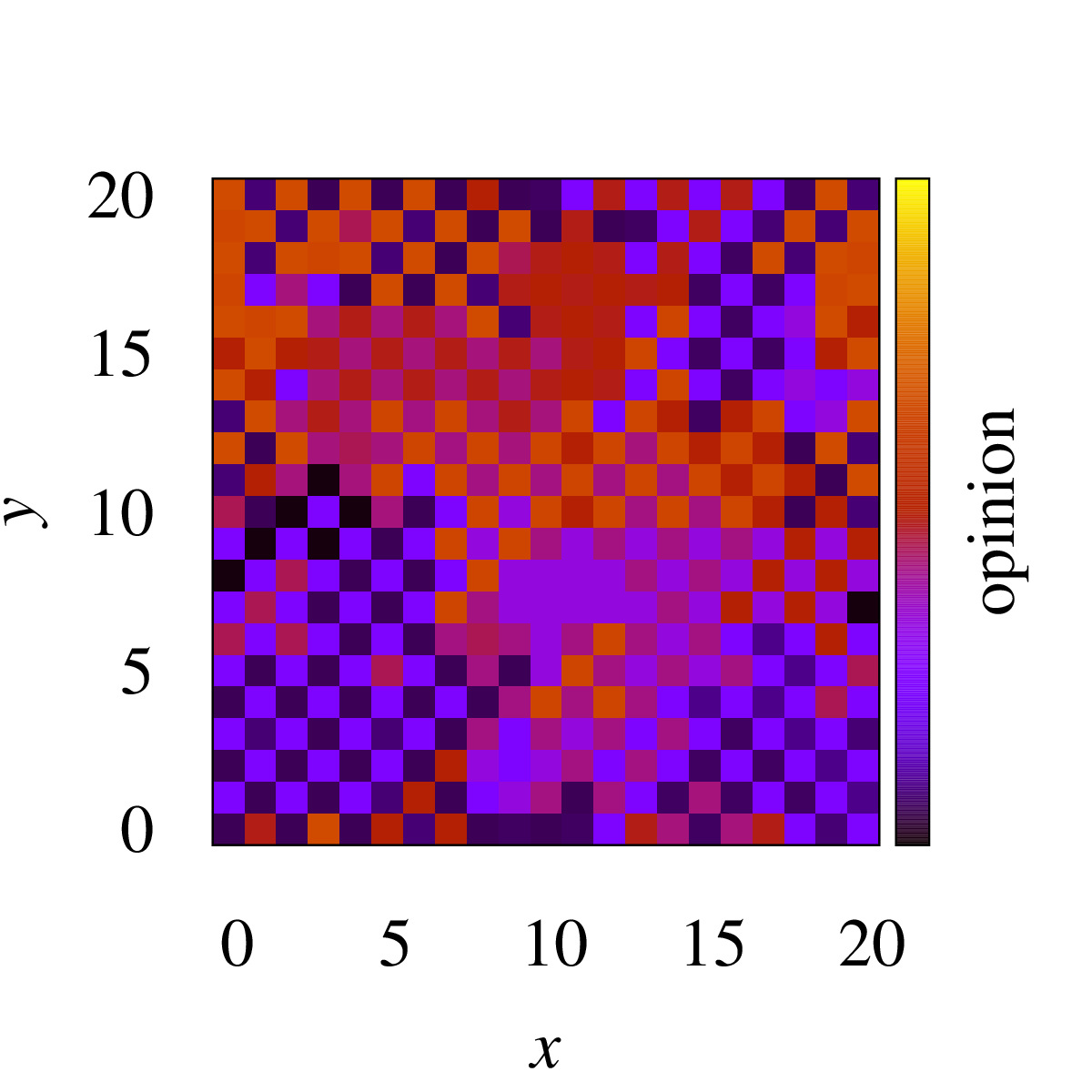}
\end{subfigure}
\hfill%% ---------------------------------------------------------------
\begin{subfigure}[t]{0.23\textwidth}
\caption{\label{subfig:V_even_t=0100}}
\includegraphics[trim={0mm 4mm 12mm 31mm},clip,width=.99\textwidth]{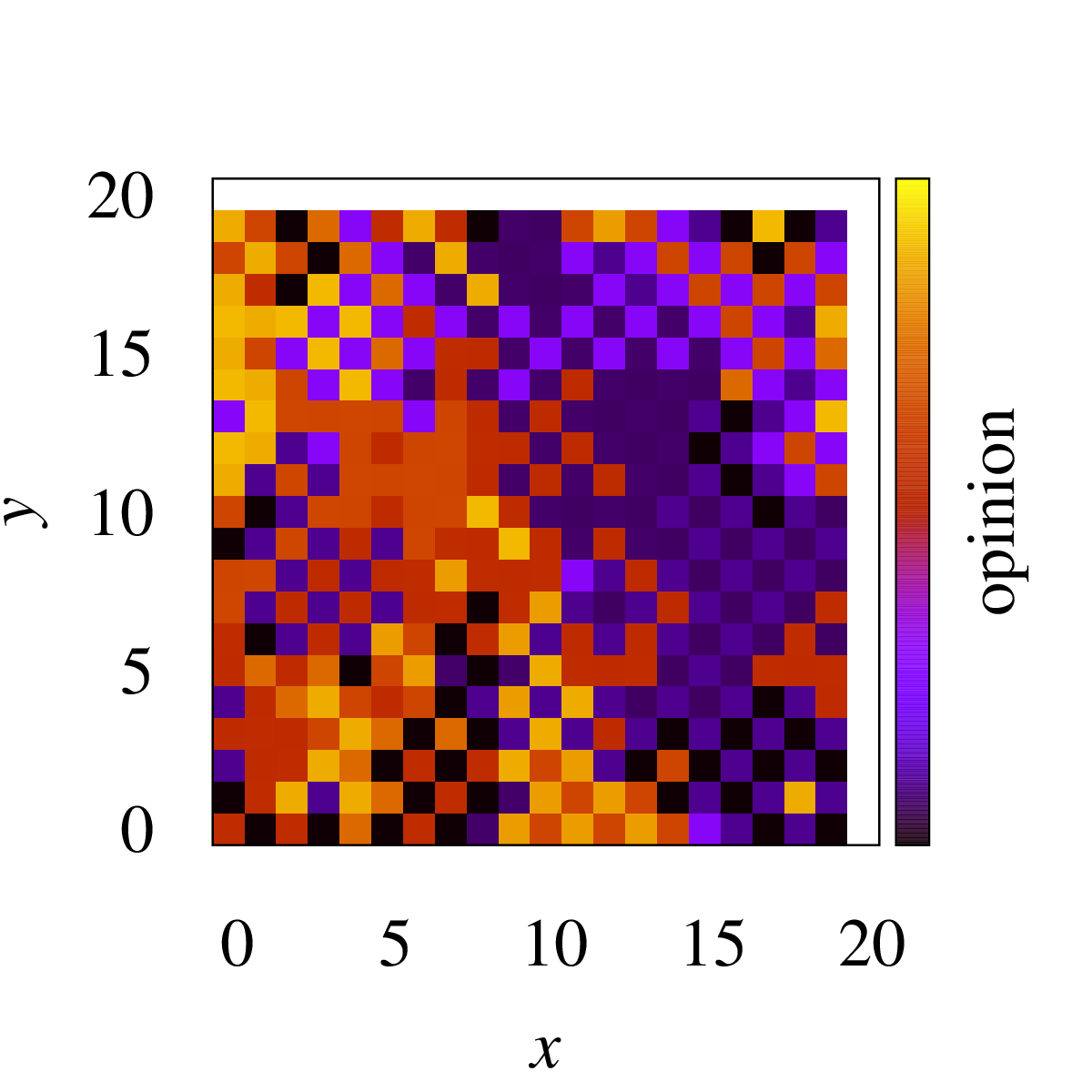}
\end{subfigure}
\hfill%% ---------------------------------------------------------------
\begin{subfigure}[t]{0.23\textwidth}
\caption{\label{subfig:V_seql_t=0100}}
\includegraphics[trim={0mm 4mm 12mm 31mm},clip,width=.99\textwidth]{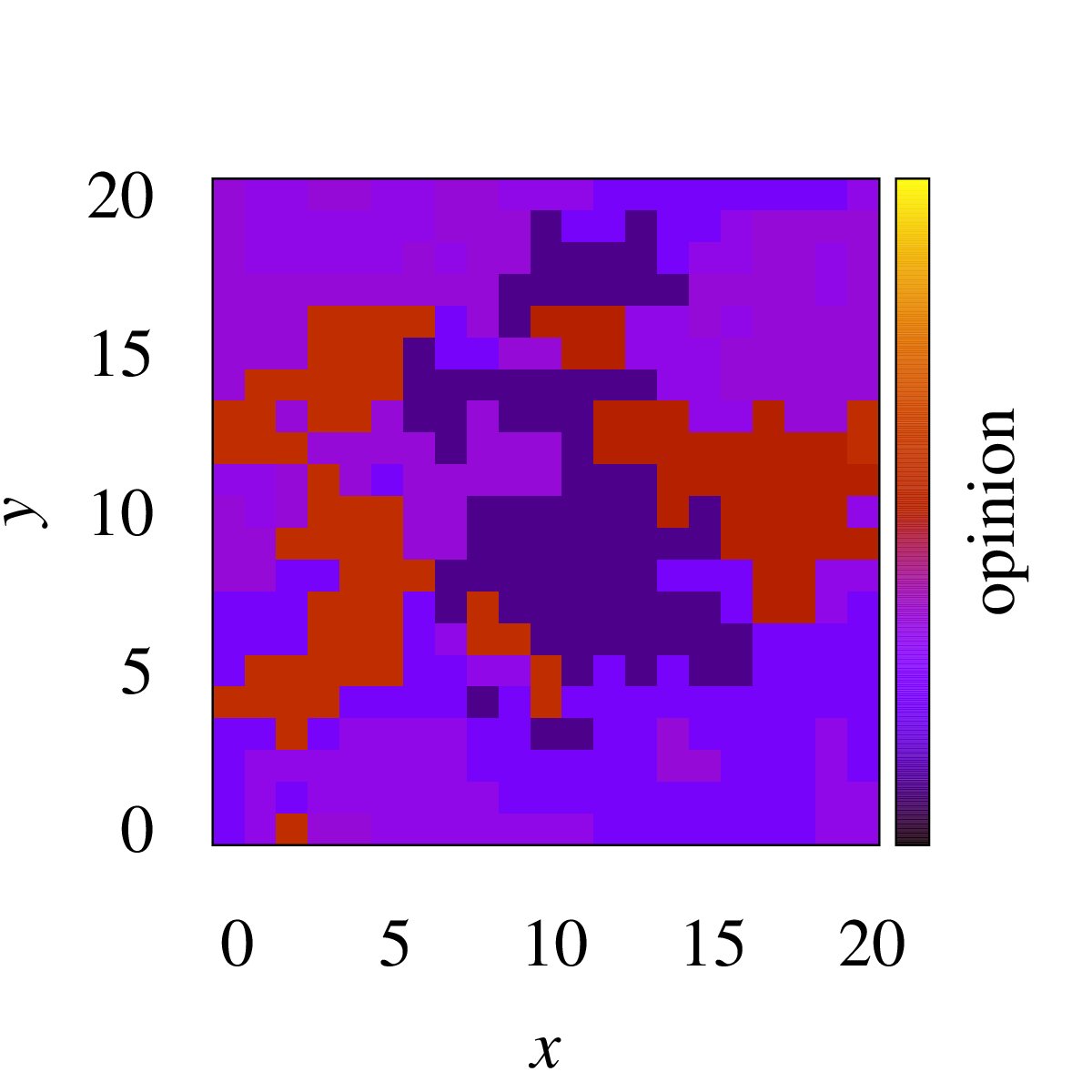}
\end{subfigure}
\hfill%% ---------------------------------------------------------------
\begin{subfigure}[t]{0.23\textwidth}
\caption{\label{subfig:V_rand_t=0100}}
\includegraphics[trim={0mm 4mm 12mm 31mm},clip,width=.99\textwidth]{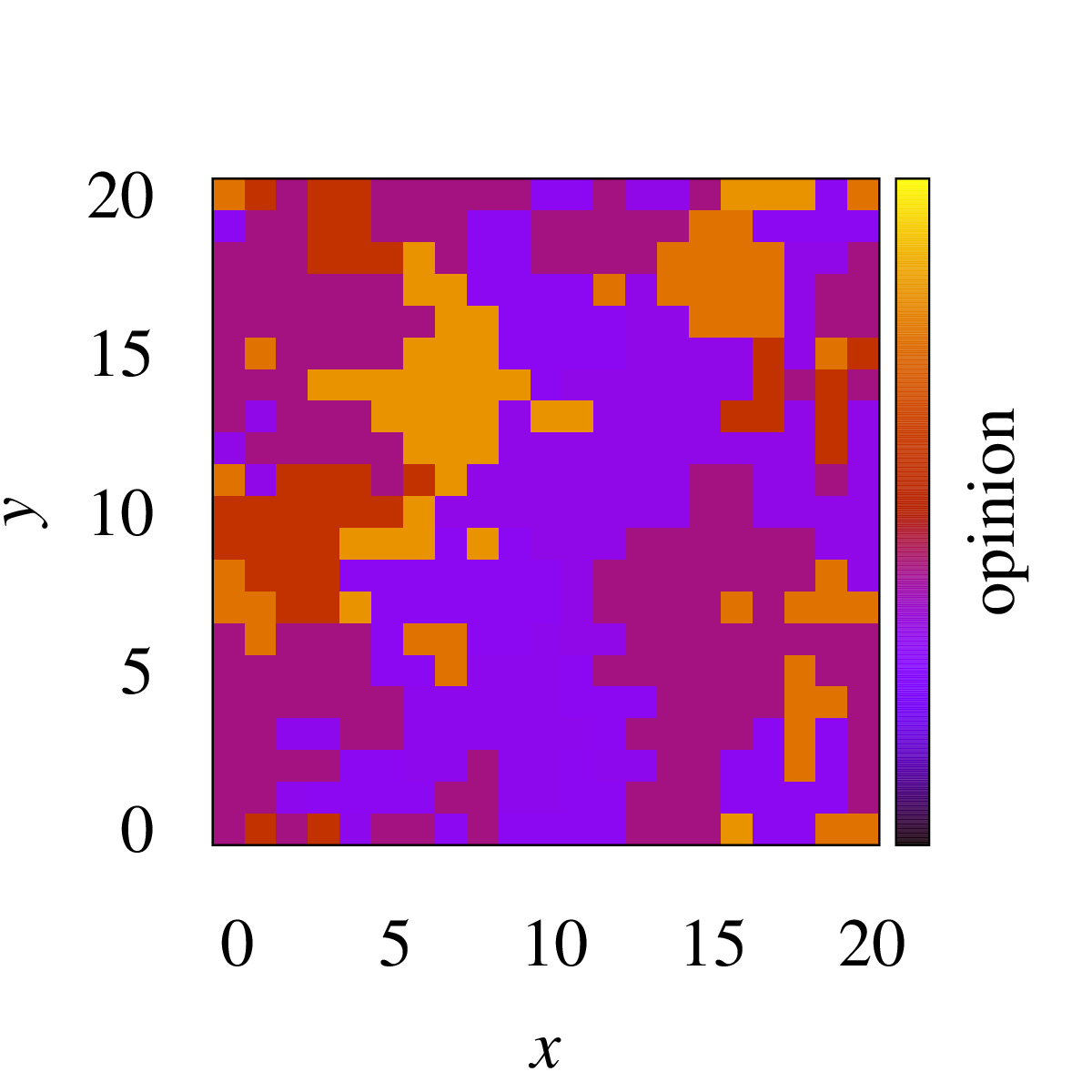}
\end{subfigure}
%% ---------------------------------------------------------------
%% ---------------------------------------------------------------
\begin{subfigure}[t]{0.23\textwidth}
\caption{\label{subfig:V_sync_t=1000}$t=10^3$}
\includegraphics[trim={0mm 4mm 12mm 31mm},clip,width=.99\textwidth]{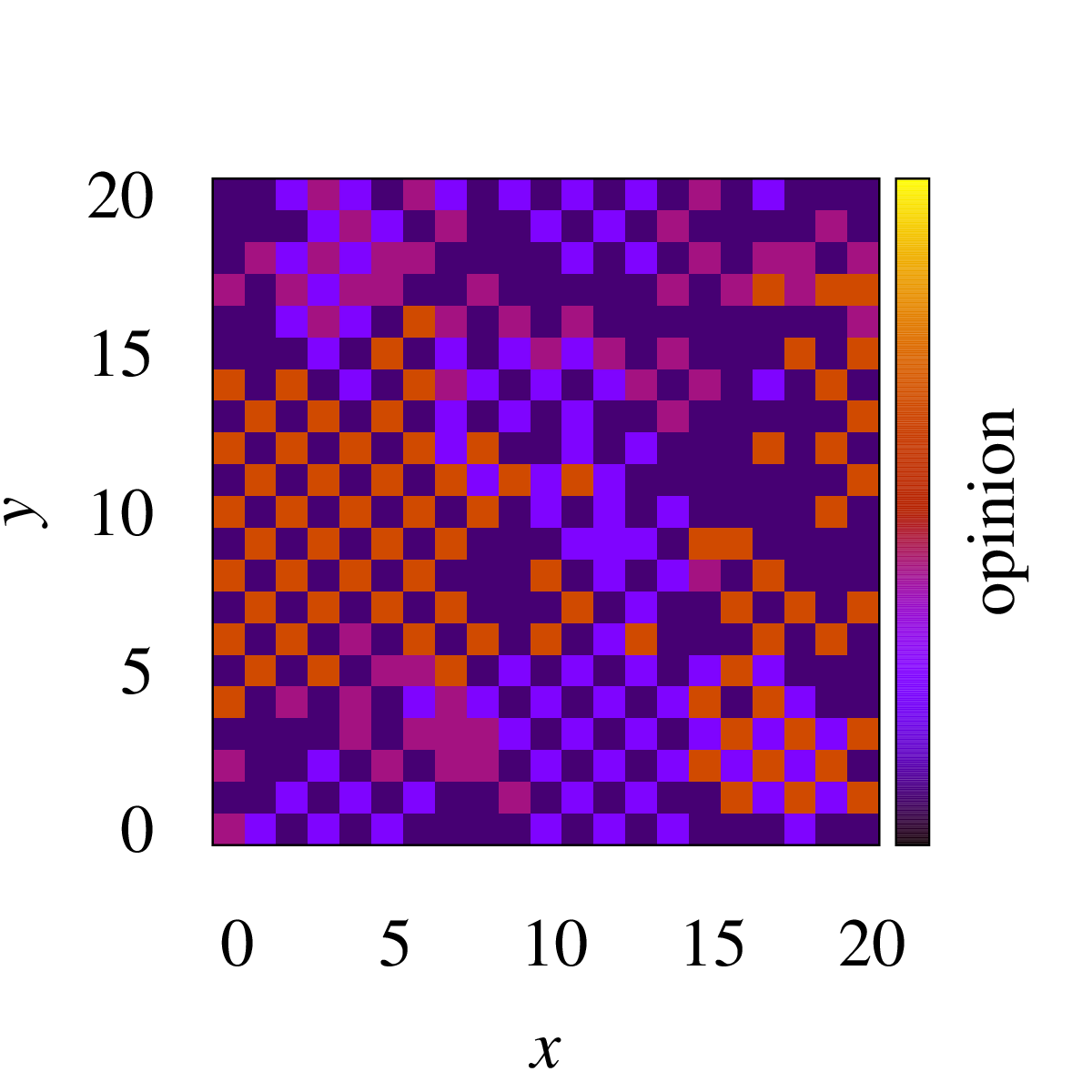}
\end{subfigure}
\hfill%% ---------------------------------------------------------------
\begin{subfigure}[t]{0.23\textwidth}
\caption{\label{subfig:V_even_t=1000}}
\includegraphics[trim={0mm 4mm 12mm 31mm},clip,width=.99\textwidth]{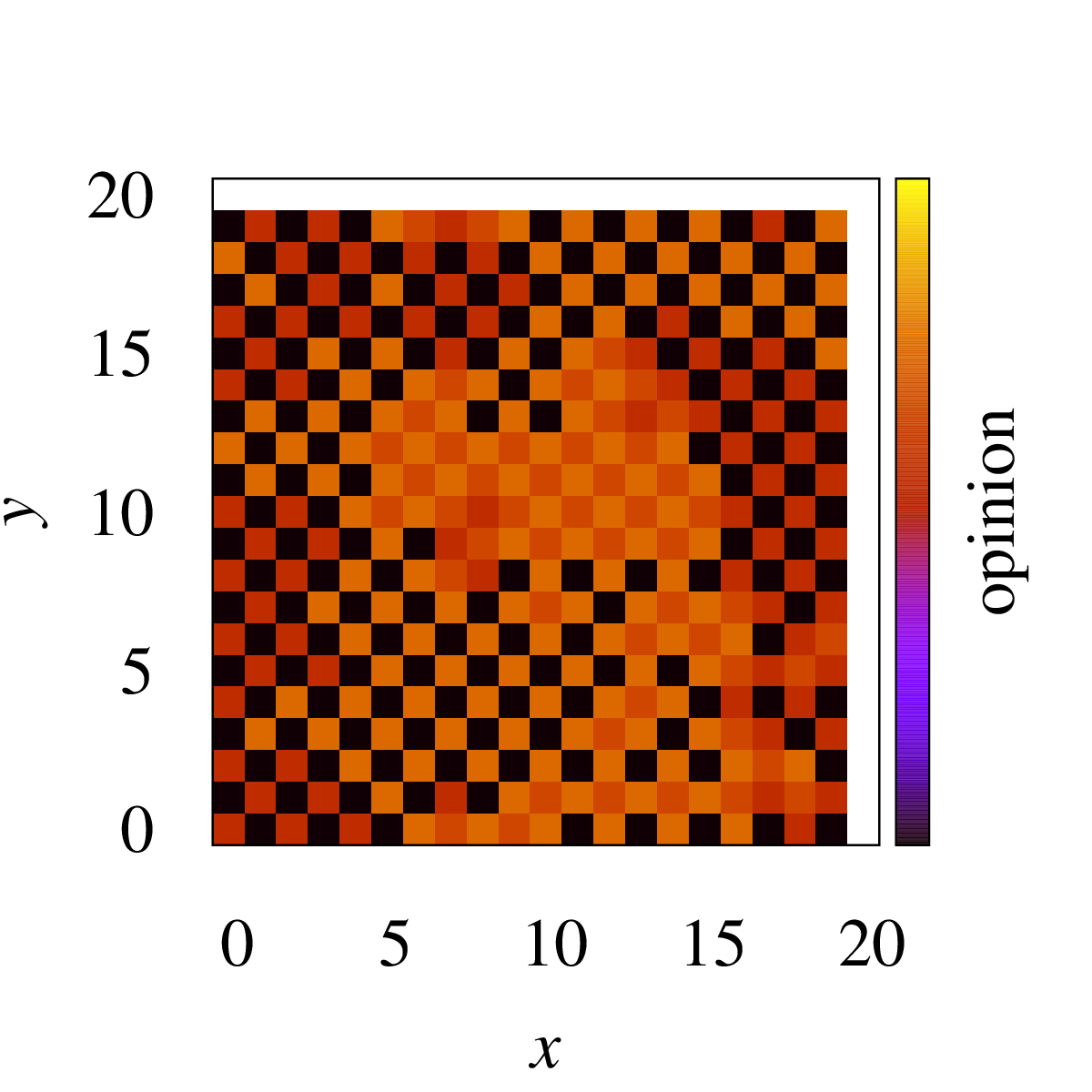}
\end{subfigure}
\hfill%% ---------------------------------------------------------------
\begin{subfigure}[t]{0.23\textwidth}
\caption{\label{subfig:V_seql_t=1000}}
\includegraphics[trim={0mm 4mm 12mm 31mm},clip,width=.99\textwidth]{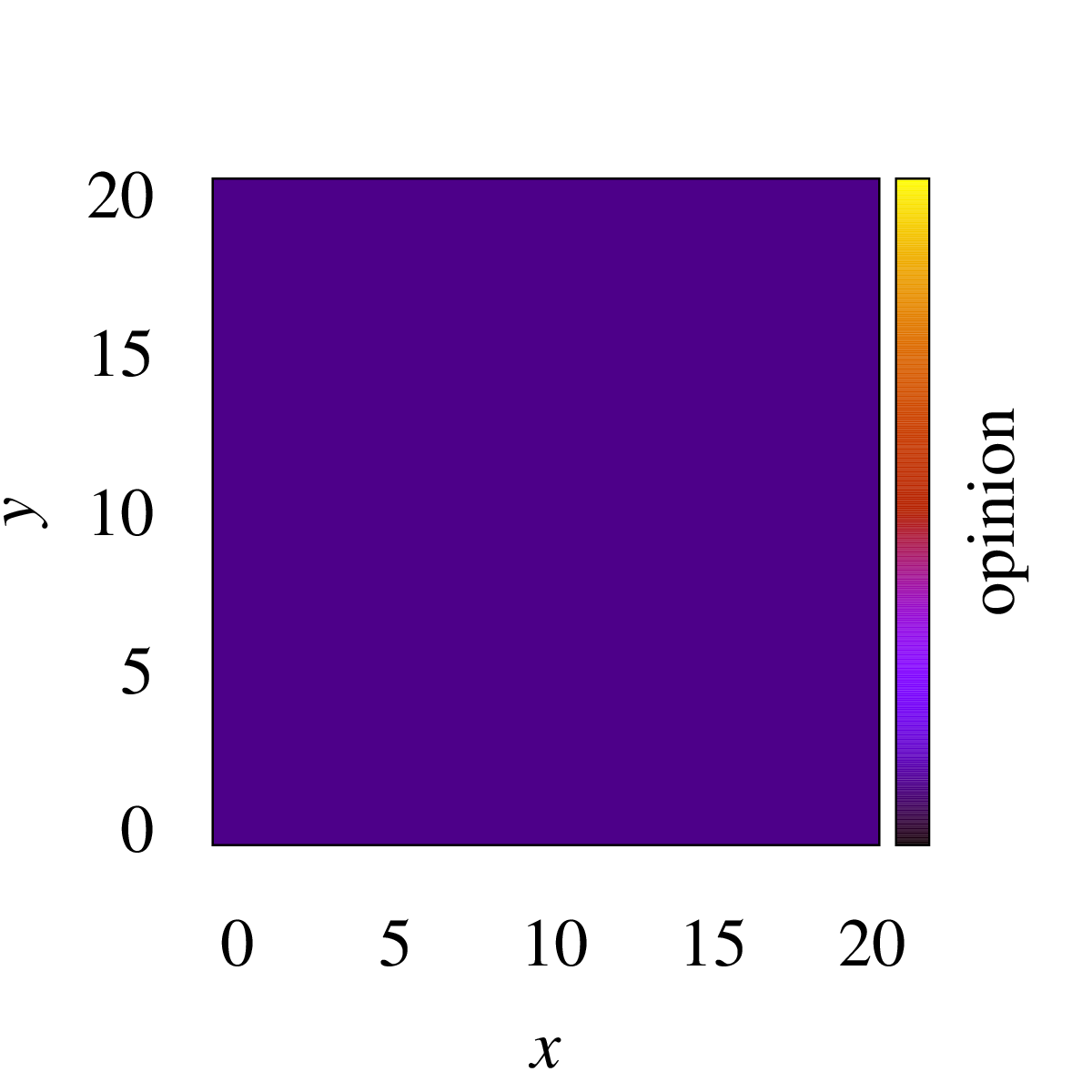}
\end{subfigure}
\hfill%% ---------------------------------------------------------------
\begin{subfigure}[t]{0.23\textwidth}
\caption{\label{subfig:V_rand_t=1000}}
\includegraphics[trim={0mm 4mm 12mm 31mm},clip,width=.99\textwidth]{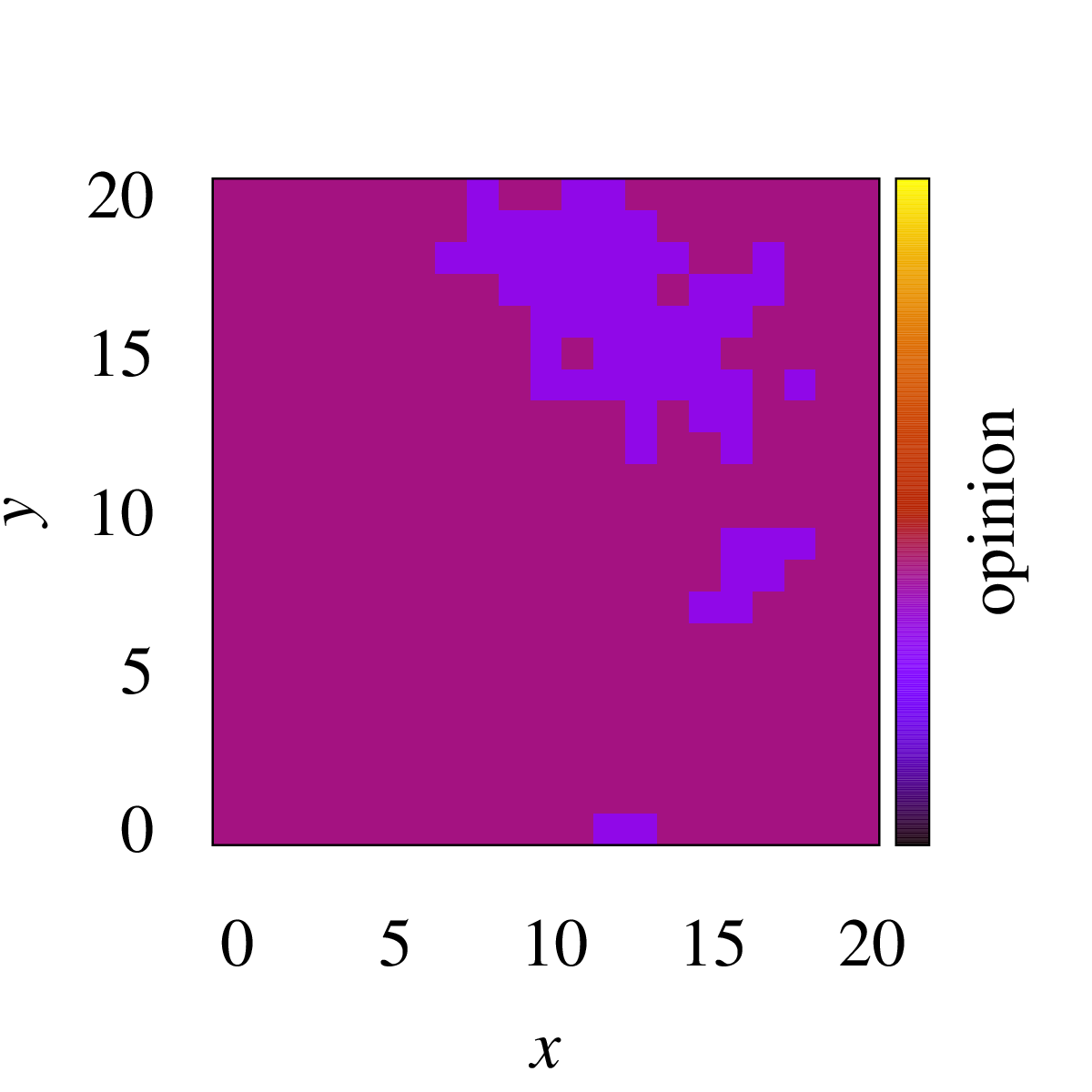}
\end{subfigure}
%% ---------------------------------------------------------------
%% ---------------------------------------------------------------
\begin{subfigure}[t]{0.23\textwidth}
\caption{\label{subfig:V_sync_t=end} $\tau=1495$}
\includegraphics[trim={0mm 4mm 12mm 31mm},clip,width=.99\textwidth]{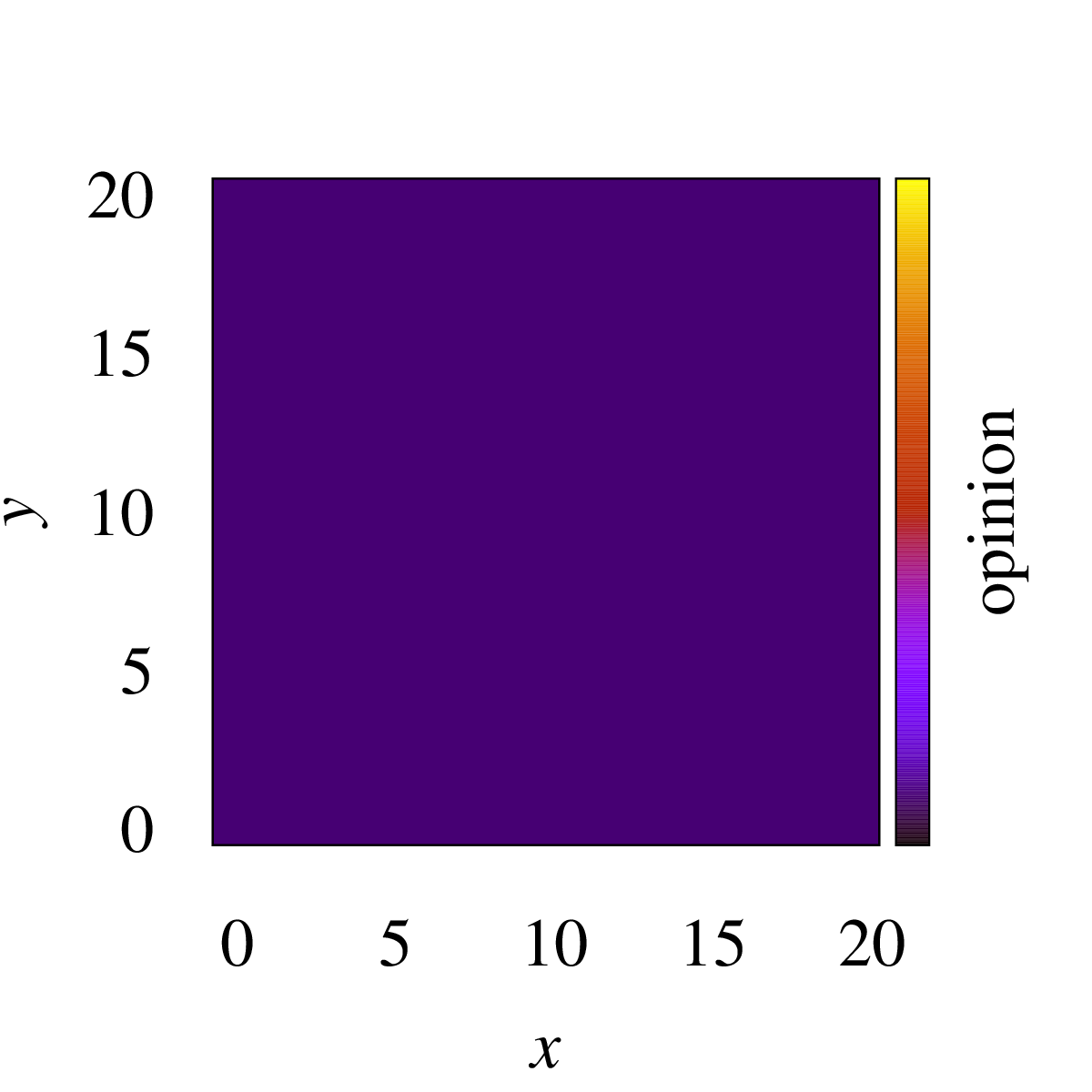}
\end{subfigure}
\hfill%% ---------------------------------------------------------------
\begin{subfigure}[t]{0.23\textwidth}
\caption{\label{subfig:V_even_t=end} $\tau=1322$}
\includegraphics[trim={0mm 4mm 12mm 31mm},clip,width=.99\textwidth]{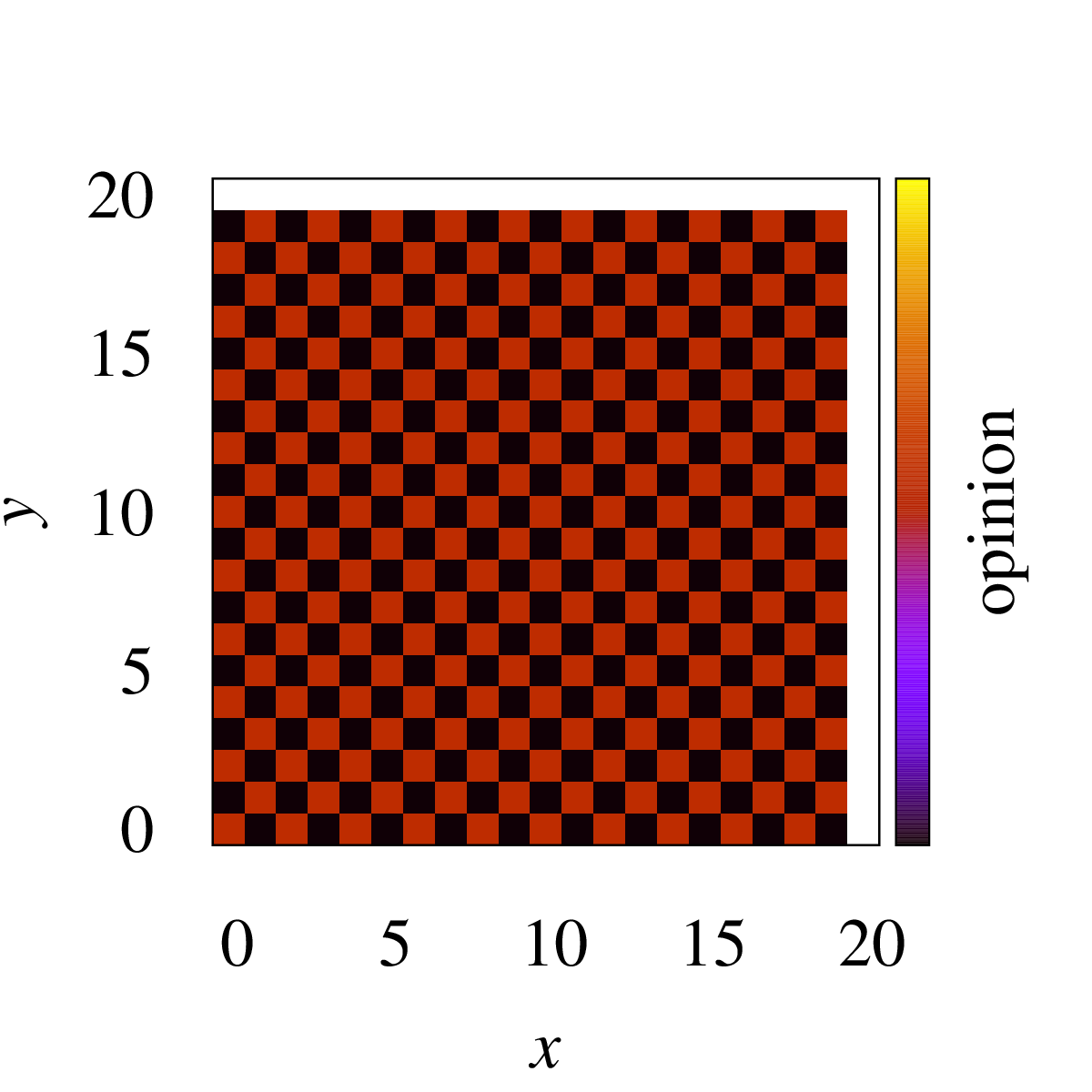}
\end{subfigure}
\hfill%% ---------------------------------------------------------------
\begin{subfigure}[t]{0.23\textwidth}
\caption{\label{subfig:V_seql_t=end}  $\tau=492$}
\includegraphics[trim={0mm 4mm 12mm 31mm},clip,width=.99\textwidth]{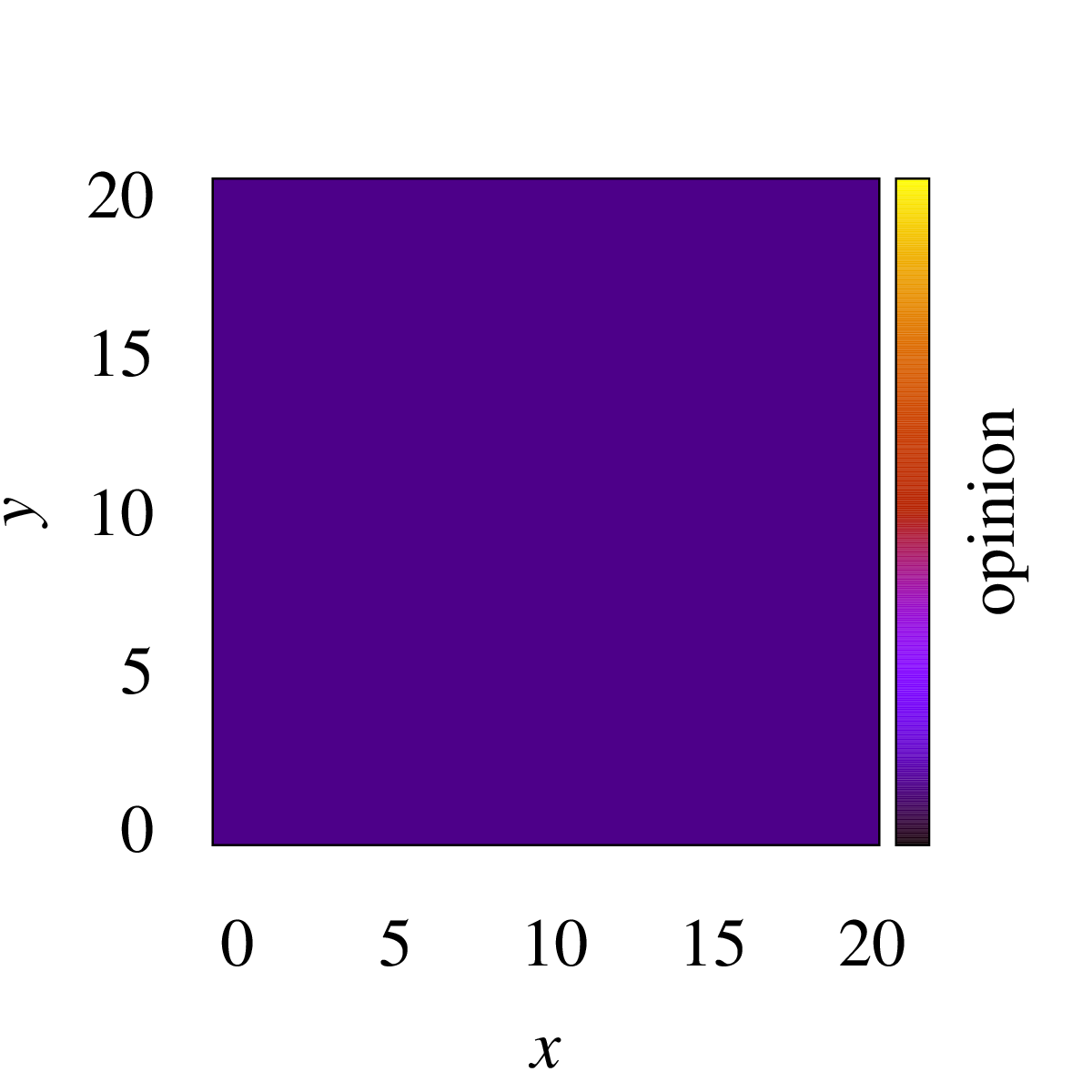}
\end{subfigure}
\hfill%% ---------------------------------------------------------------
\begin{subfigure}[t]{0.23\textwidth}
\caption{\label{subfig:V_rand_t=end} $\tau=1237$}
\includegraphics[trim={0mm 4mm 12mm 31mm},clip,width=.99\textwidth]{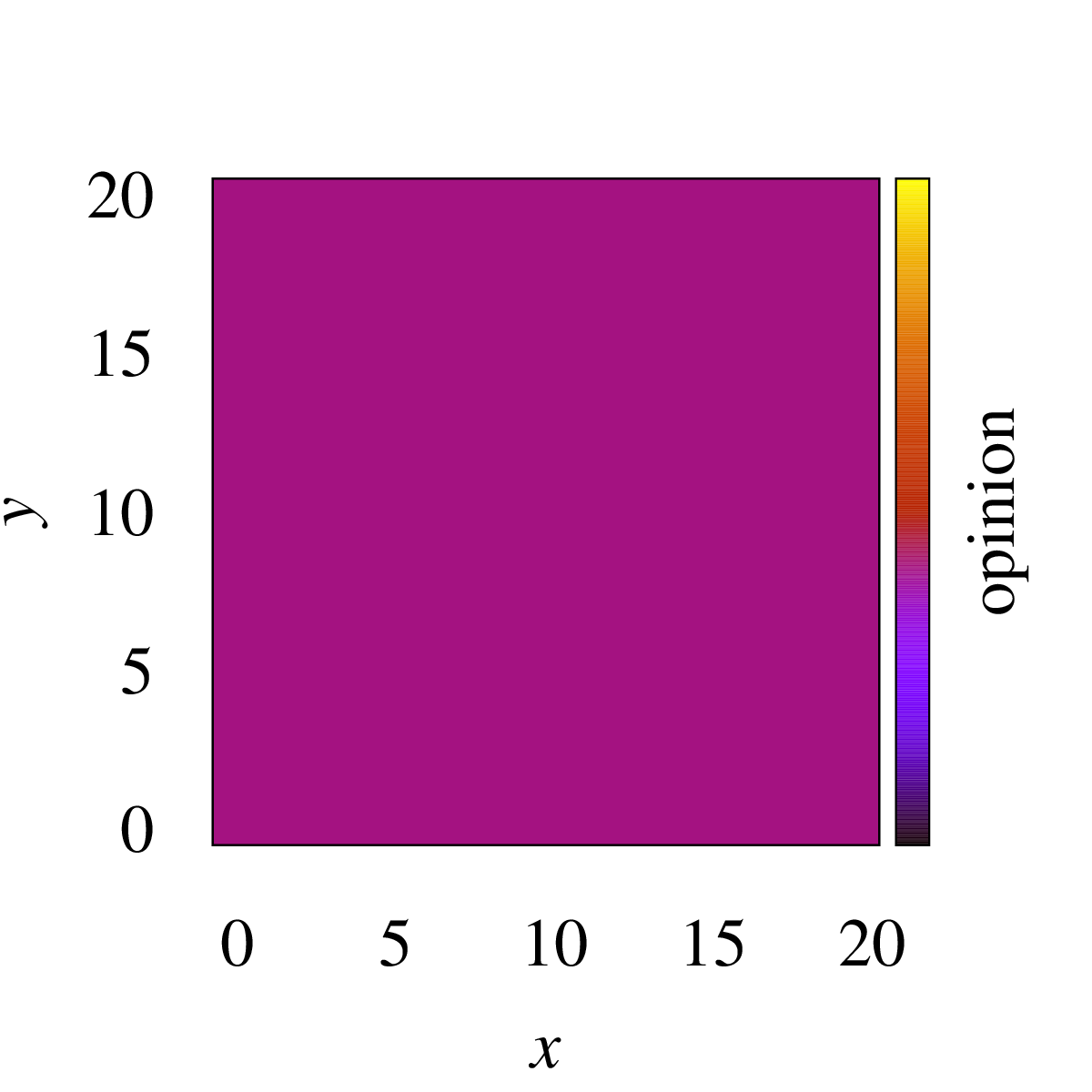}
\end{subfigure}
%% ---------------------------------------------------------------
\caption{\label{fig:V_snap}Snapshots for system evolution with voter model.
\subref{subfig:V_sync_t=0000}%%, \subref{subfig:V_even_t=0000}, \subref{subfig:V_seql_t=0000}, 
--\subref{subfig:V_rand_t=0000} $t=0$, 
\subref{subfig:V_sync_t=0010}%%, \subref{subfig:V_even_t=0010}, \subref{subfig:V_seql_t=0010}, 
--\subref{subfig:V_rand_t=0010} $t=10$, 
\subref{subfig:V_sync_t=0100}%%, \subref{subfig:V_even_t=0100}, \subref{subfig:V_seql_t=0100}, 
--\subref{subfig:V_rand_t=0100} $t=10^2$, 
\subref{subfig:V_sync_t=1000}%%, \subref{subfig:V_even_t=1000}, \subref{subfig:V_seql_t=1000}, 
--\subref{subfig:V_rand_t=1000} $t=10^3$, 
\subref{subfig:V_sync_t=end}%%,  \subref{subfig:V_even_t=end},  \subref{subfig:V_seql_t=end},  
--\subref{subfig:V_rand_t=end}  $t\to\infty$.
Various actors' opinion update schemes: \subref{subfig:V_sync_t=0000}, \subref{subfig:V_sync_t=0010}, \subref{subfig:V_sync_t=0100}, \subref{subfig:V_sync_t=1000}, \subref{subfig:V_sync_t=end} synchronous (odd $L=21$) 
\subref{subfig:V_even_t=0000}, \subref{subfig:V_even_t=0010}, \subref{subfig:V_even_t=0100}, \subref{subfig:V_even_t=1000}, \subref{subfig:V_even_t=end} synchronous (even $L=20$),
\subref{subfig:V_seql_t=0000}, \subref{subfig:V_seql_t=0010}, \subref{subfig:V_seql_t=0100}, \subref{subfig:V_seql_t=1000}, \subref{subfig:V_seql_t=end} asynchronous sequential, 
\subref{subfig:V_rand_t=0000}, \subref{subfig:V_rand_t=0010}, \subref{subfig:V_rand_t=0100}, \subref{subfig:V_rand_t=1000}, \subref{subfig:V_rand_t=end} asynchronous random}
%% ---------------------------------------------------------------
\end{figure*}

\begin{figure*}[htbp]
%% ===============================================================
\begin{subfigure}[t]{0.23\textwidth}
\caption{\label{subfig:SZ_neumann_t=0}von Neumann}
\includegraphics[trim={0mm 4mm 12mm 31mm},clip,width=.99\textwidth]{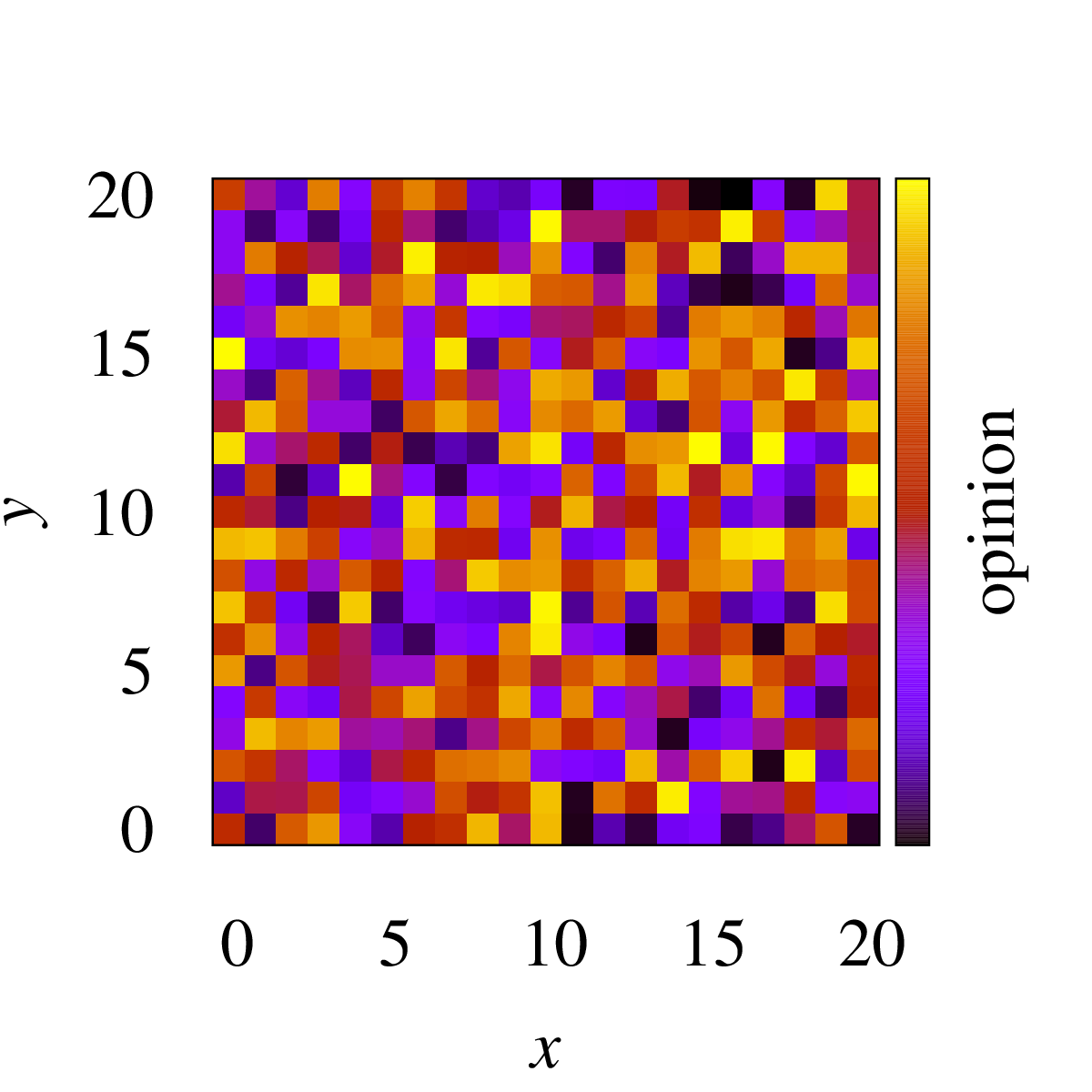}
\end{subfigure}
\hfill %% ---------------------------------------------------------------
\begin{subfigure}[t]{0.23\textwidth}
\caption{\label{subfig:SZ_moore_t=0}Moore}
\includegraphics[trim={0mm 4mm 12mm 31mm},clip,width=.99\textwidth]{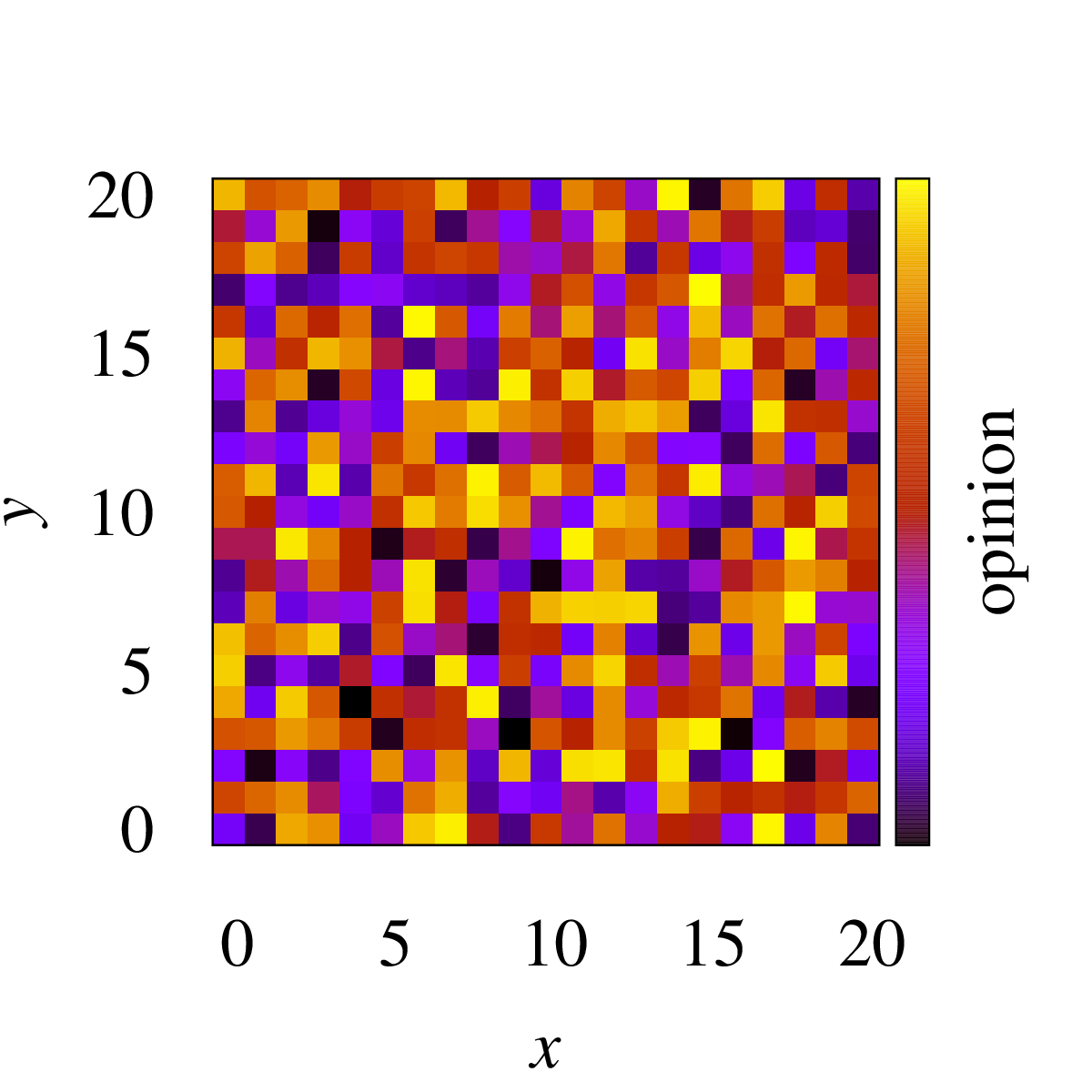}
\end{subfigure}
\hfill %% ---------------------------------------------------------------
\begin{subfigure}[t]{0.23\textwidth}
\caption{\label{subfig:SZ_triangular_t=0}triangular}
\includegraphics[trim={0mm 4mm 12mm 31mm},clip,width=.99\textwidth]{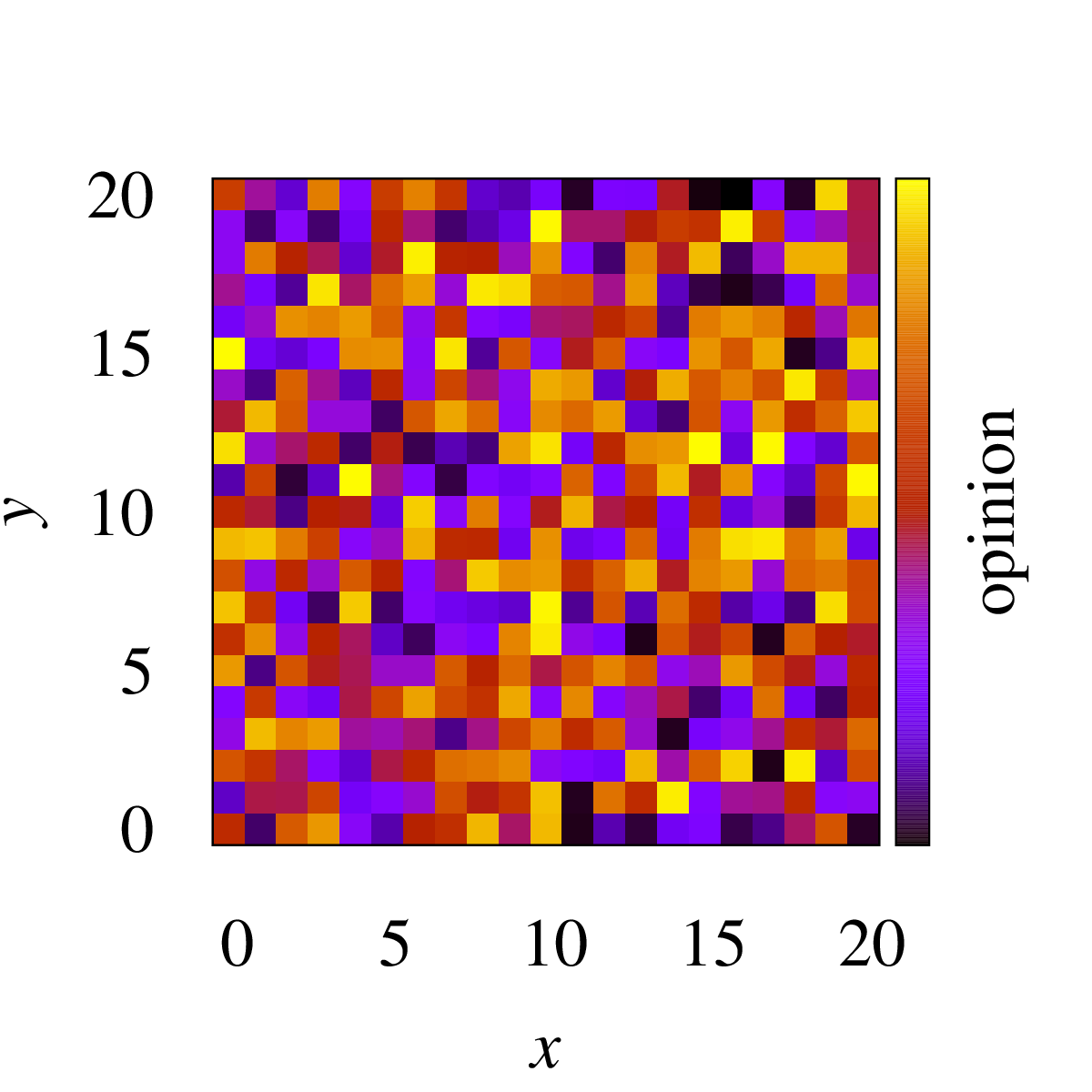}
\end{subfigure}
\hfill %% ---------------------------------------------------------------
\begin{subfigure}[t]{0.23\textwidth}
\caption{\label{subfig:SZ_honeycomb_t=0}honeycomb}
\includegraphics[trim={0mm 4mm 12mm 31mm},clip,width=.99\textwidth]{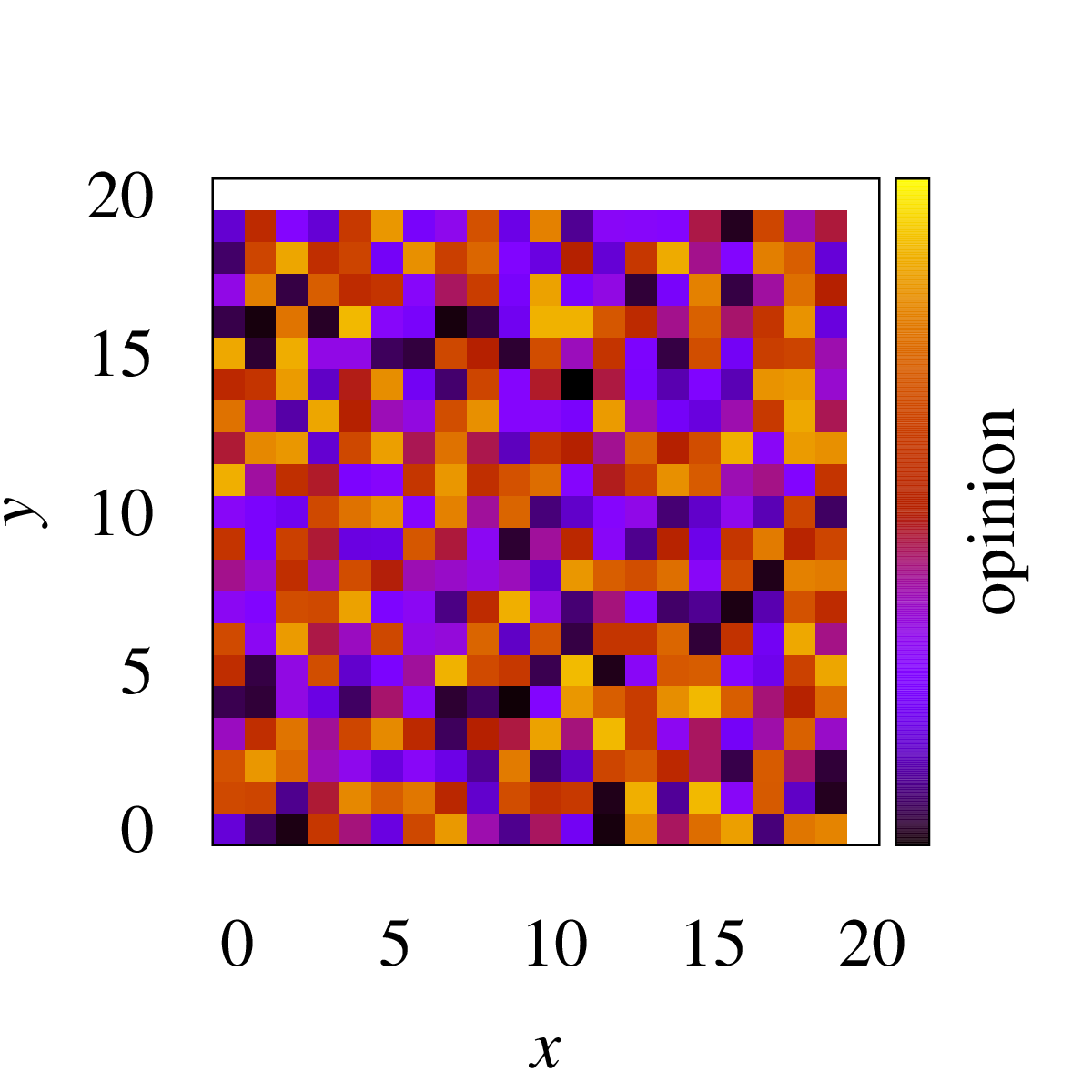}
\end{subfigure}
%% ===============================================================
\begin{subfigure}[t]{0.23\textwidth}
\caption{\label{subfig:SZ_neumann_t=100}$t=100$}
\includegraphics[trim={0mm 4mm 12mm 31mm},clip,width=.99\textwidth]{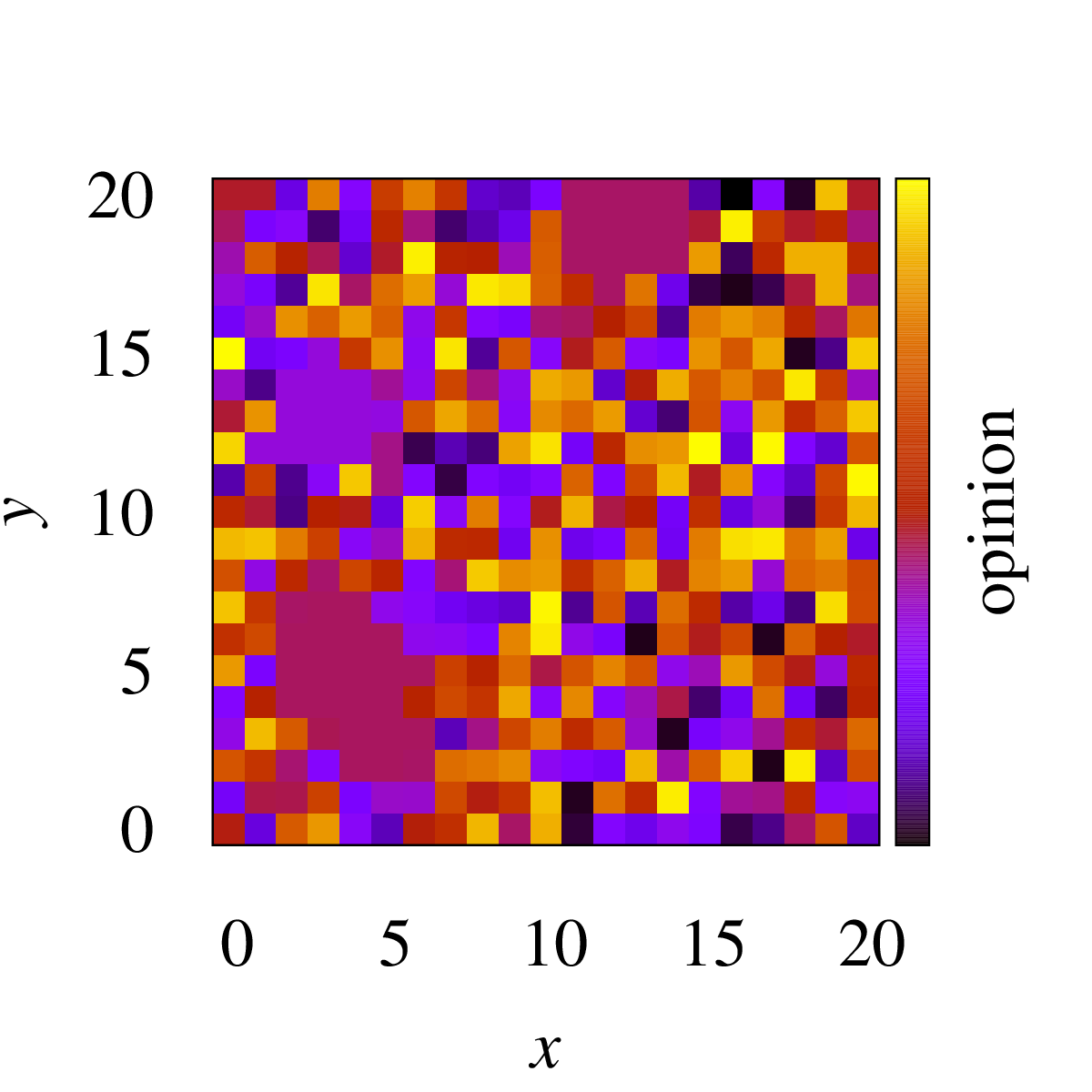}
\end{subfigure}
\hfill %% ---------------------------------------------------------------
\begin{subfigure}[t]{0.23\textwidth}
\caption{\label{subfig:SZ_moore_t=100}}
\includegraphics[trim={0mm 4mm 12mm 31mm},clip,width=.99\textwidth]{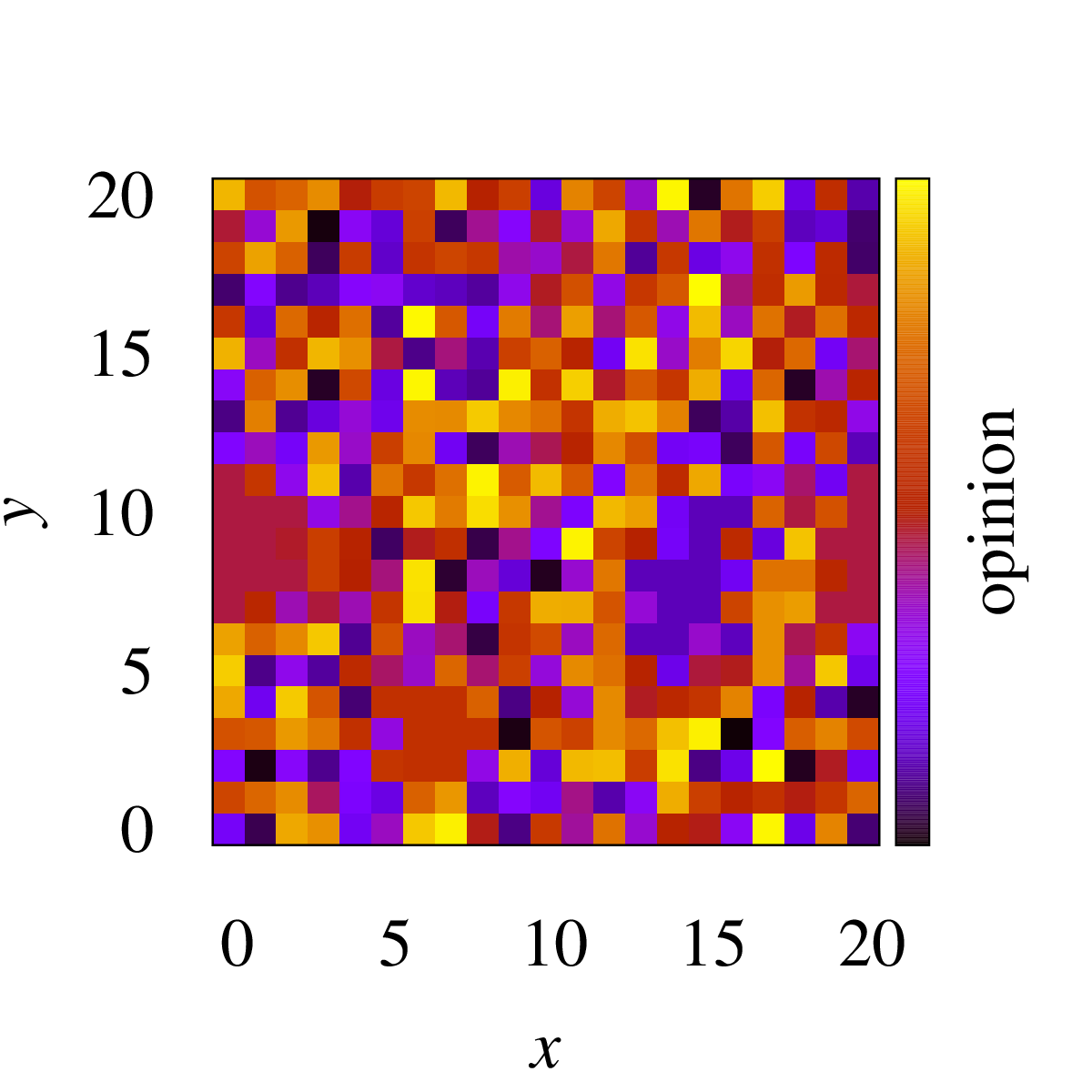}
\end{subfigure}
\hfill %% ---------------------------------------------------------------
\begin{subfigure}[t]{0.23\textwidth}
\caption{\label{subfig:SZ_triangular_t=100}}
\includegraphics[trim={0mm 4mm 12mm 31mm},clip,width=.99\textwidth]{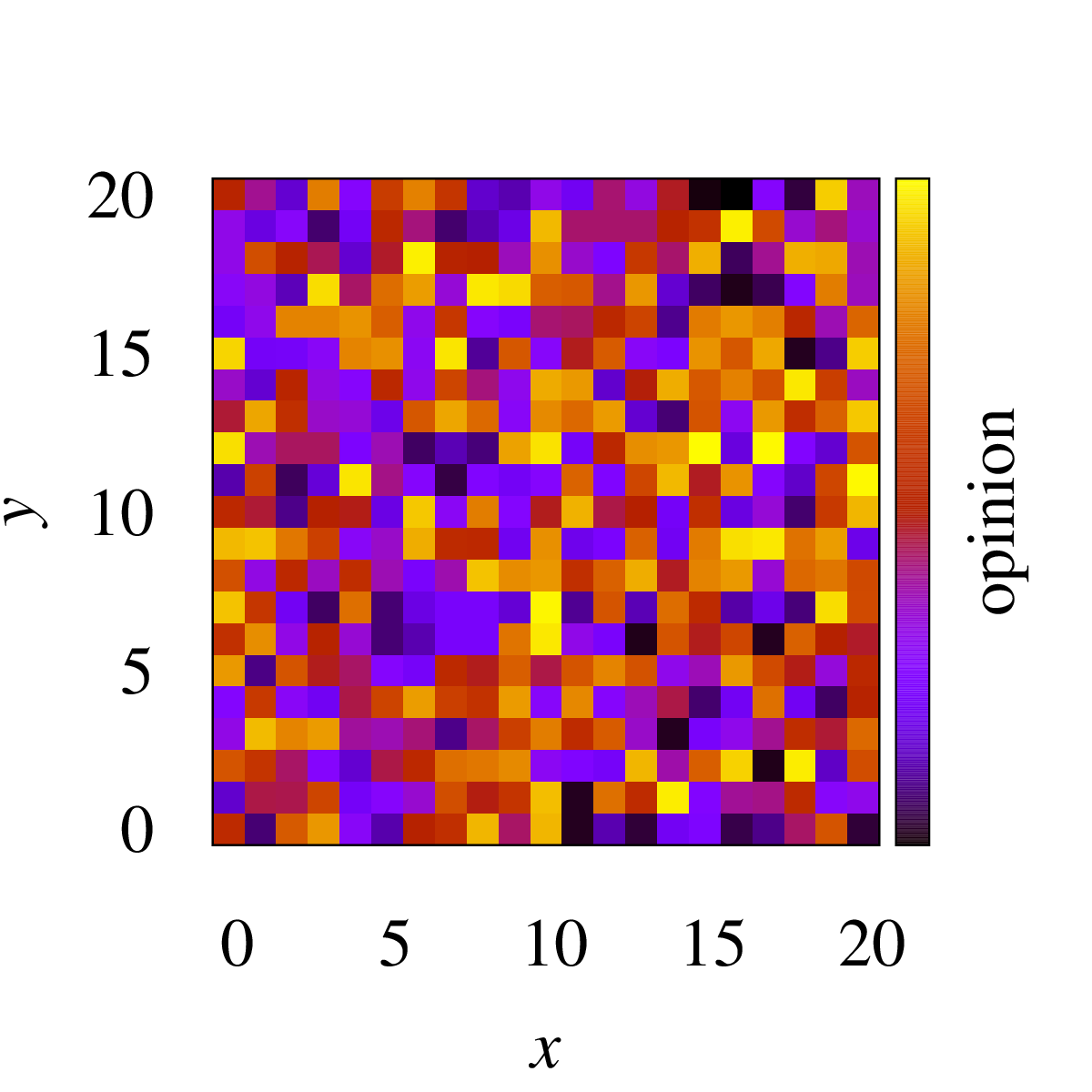}
\end{subfigure}
\hfill %% ---------------------------------------------------------------
\begin{subfigure}[t]{0.23\textwidth}
\caption{\label{subfig:SZ_honeycomb_t=100}}
\includegraphics[trim={0mm 4mm 12mm 31mm},clip,width=.99\textwidth]{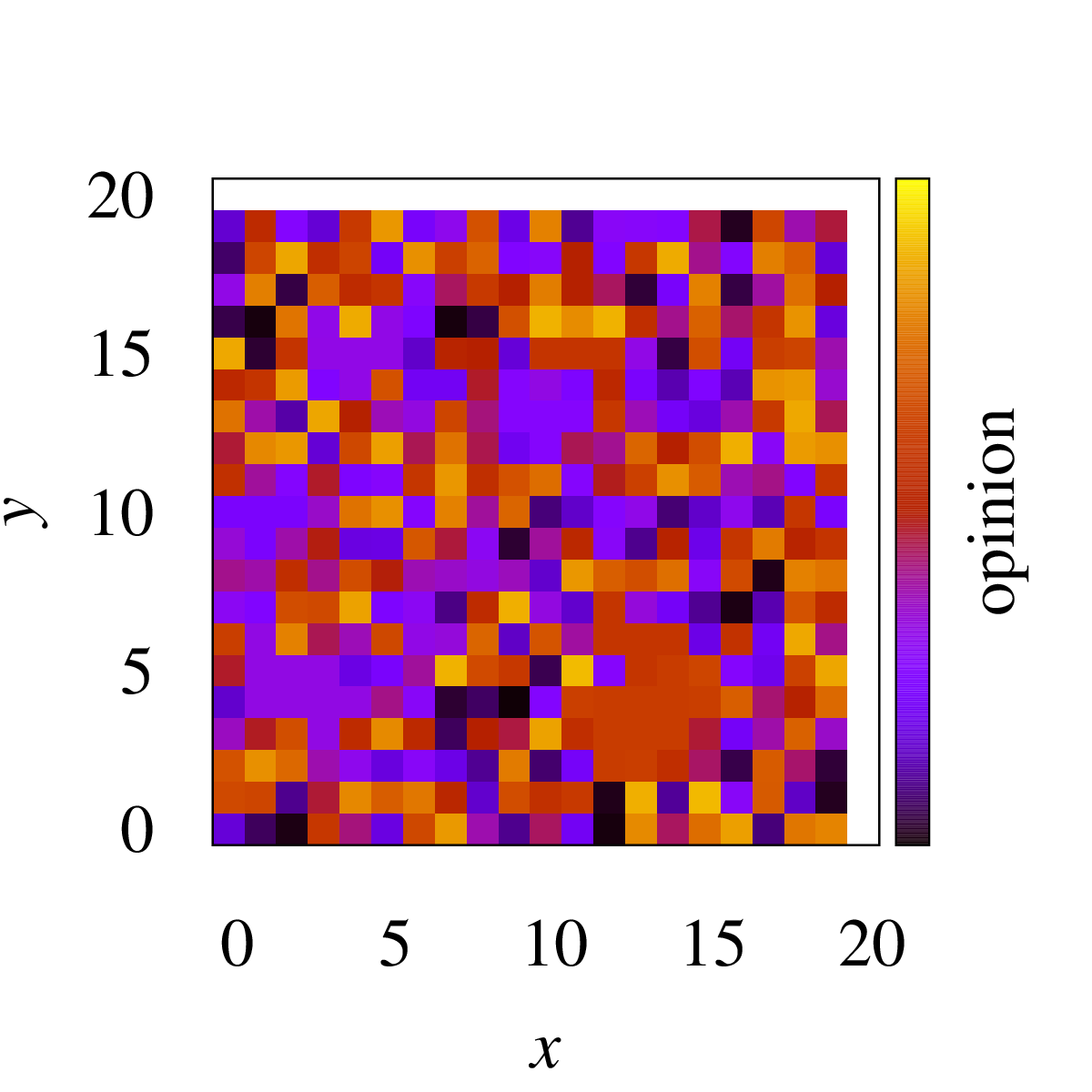}
\end{subfigure}
%% ===============================================================
\begin{subfigure}[t]{0.23\textwidth}
\caption{\label{subfig:SZ_neumann_t=200}$t=200$}
\includegraphics[trim={0mm 4mm 12mm 31mm},clip,width=.99\textwidth]{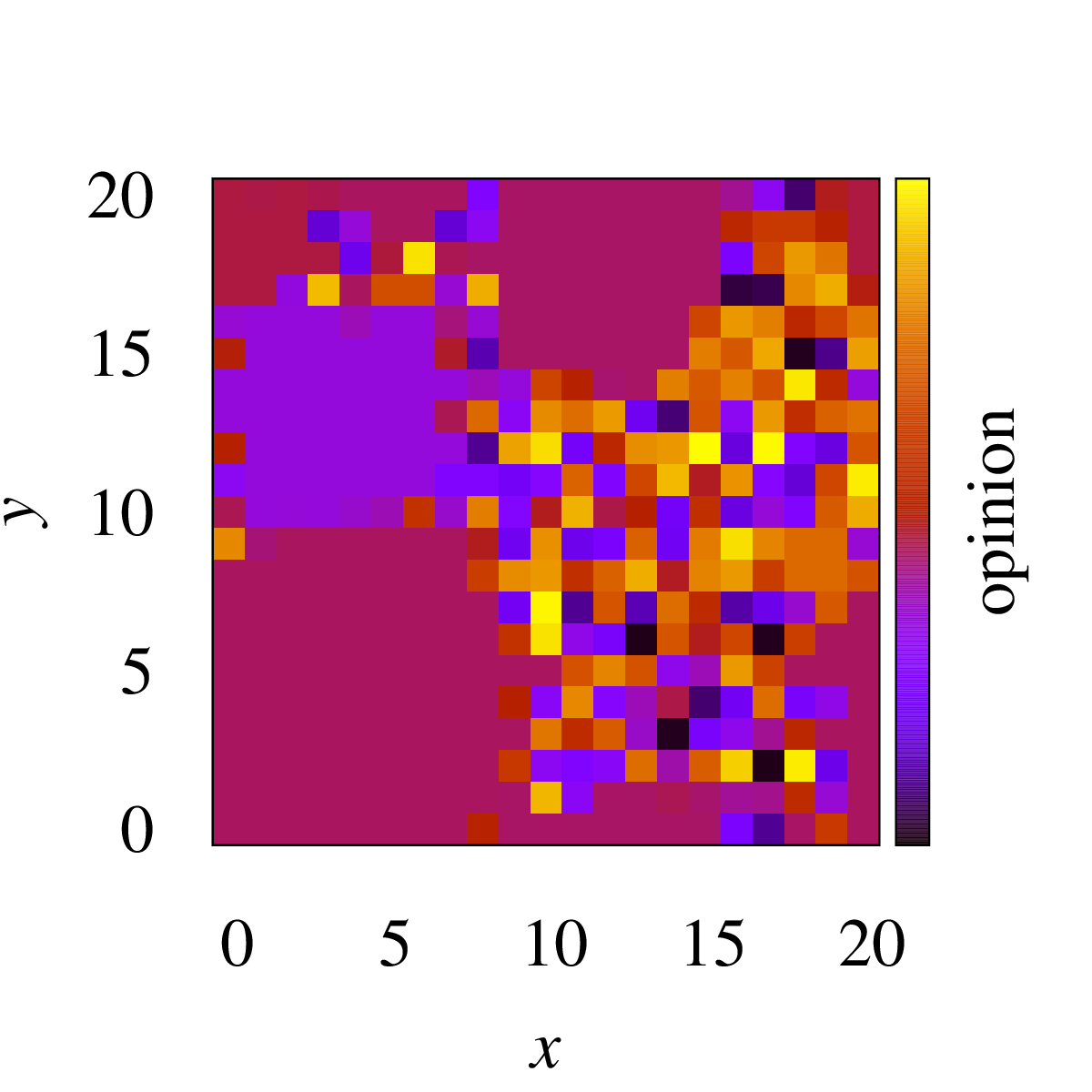}
\end{subfigure}
\hfill %% ---------------------------------------------------------------
\begin{subfigure}[t]{0.23\textwidth}
\caption{\label{subfig:SZ_moore_t=200}}
\includegraphics[trim={0mm 4mm 12mm 31mm},clip,width=.99\textwidth]{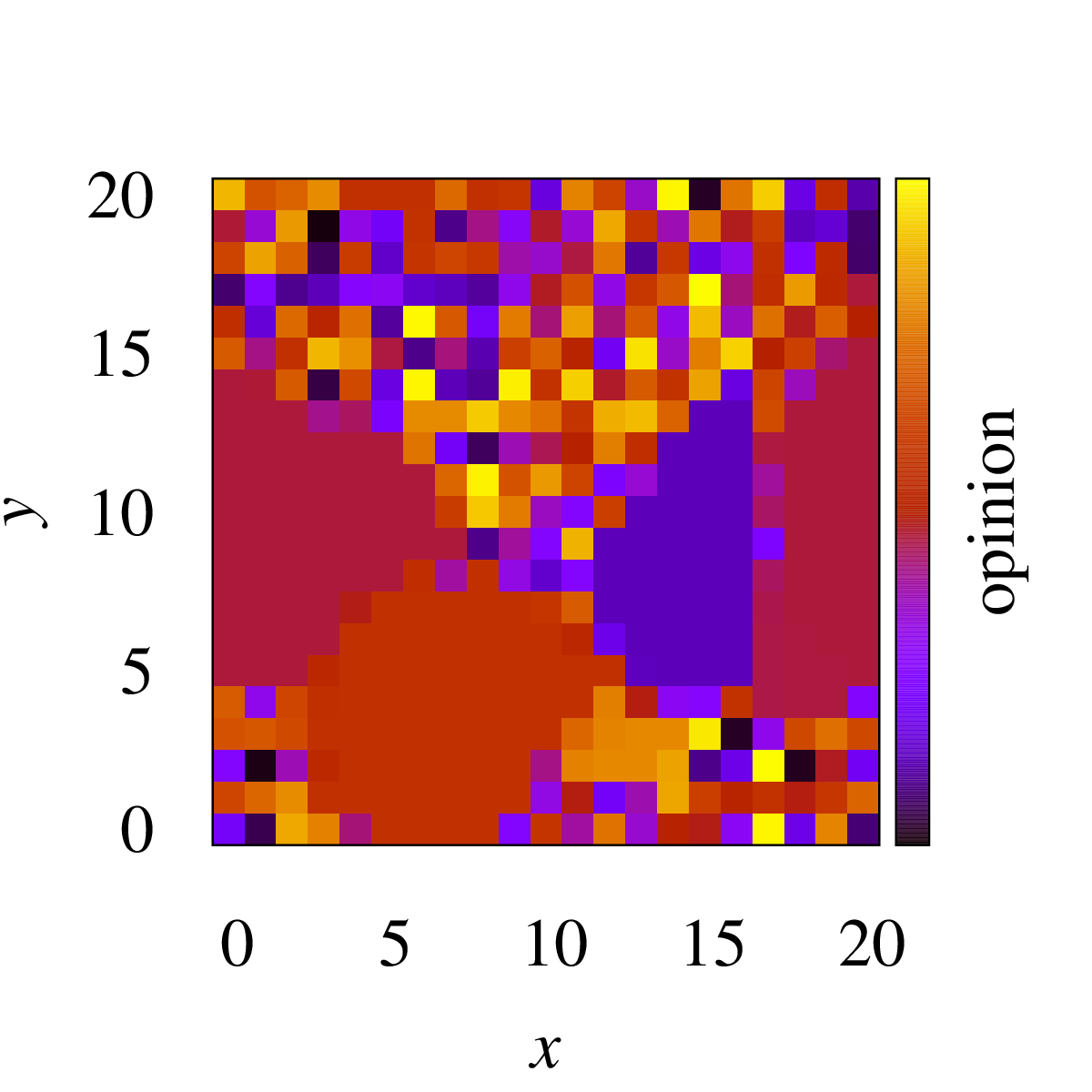}
\end{subfigure}
\hfill %% ---------------------------------------------------------------
\begin{subfigure}[t]{0.23\textwidth}
\caption{\label{subfig:SZ_triangular_t=200}}
\includegraphics[trim={0mm 4mm 12mm 31mm},clip,width=.99\textwidth]{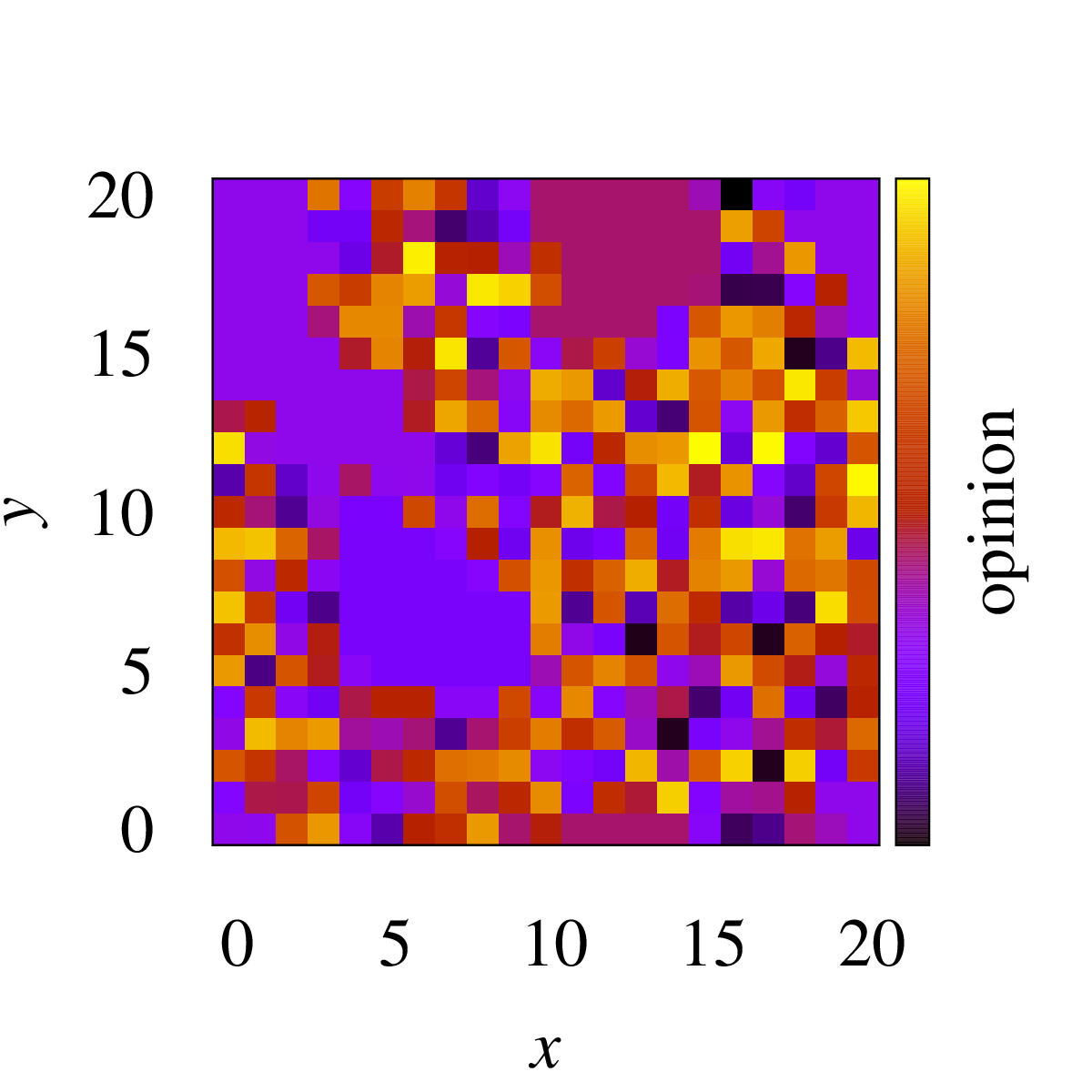}
\end{subfigure}
\hfill %% ---------------------------------------------------------------
\begin{subfigure}[t]{0.23\textwidth}
\caption{\label{subfig:SZ_honeycomb_t=200}}
\includegraphics[trim={0mm 4mm 12mm 31mm},clip,width=.99\textwidth]{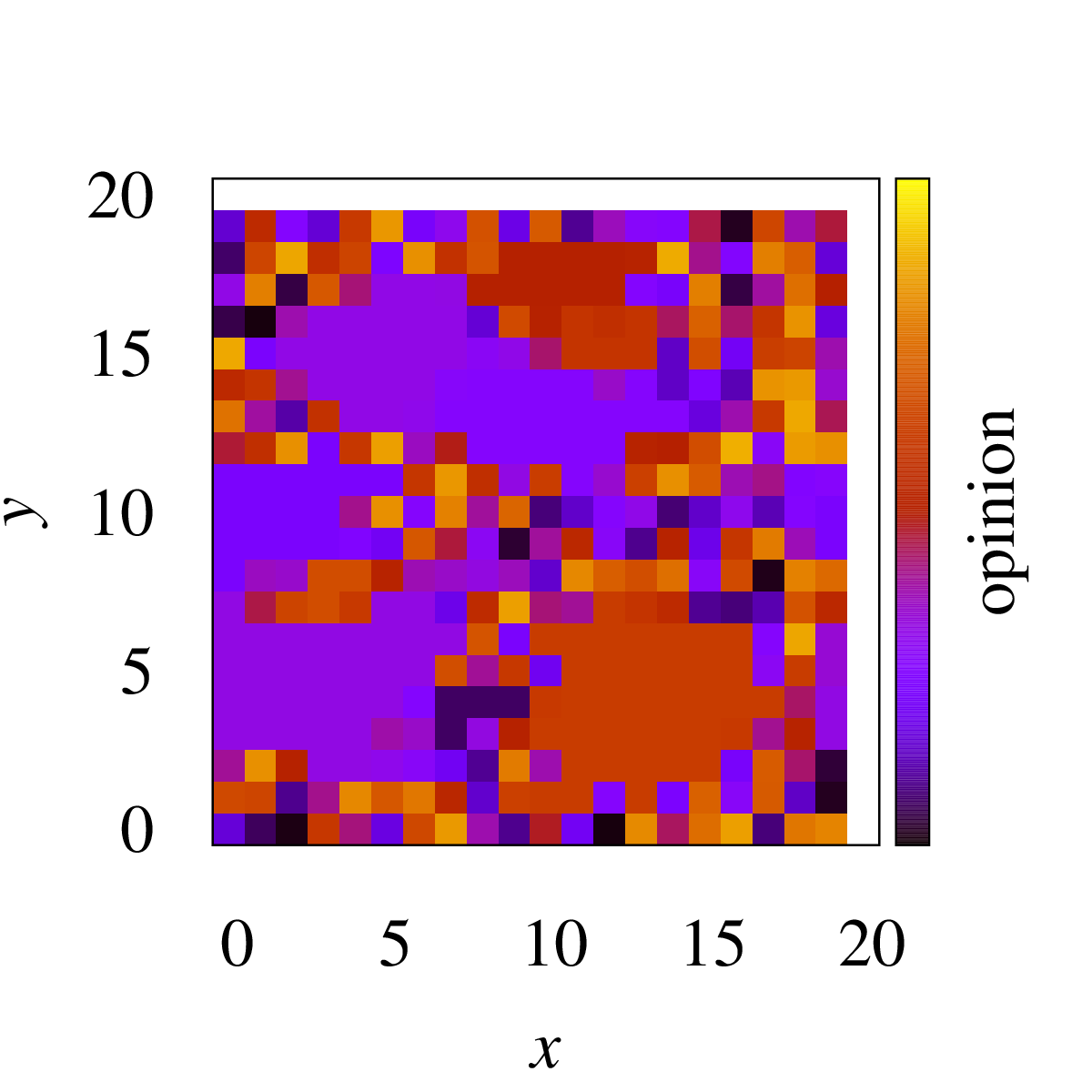}
\end{subfigure}
%% ===============================================================
\begin{subfigure}[t]{0.23\textwidth}
\caption{\label{subfig:SZ_neumann_t=400}$t=400$}
\includegraphics[trim={0mm 4mm 12mm 31mm},clip,width=.99\textwidth]{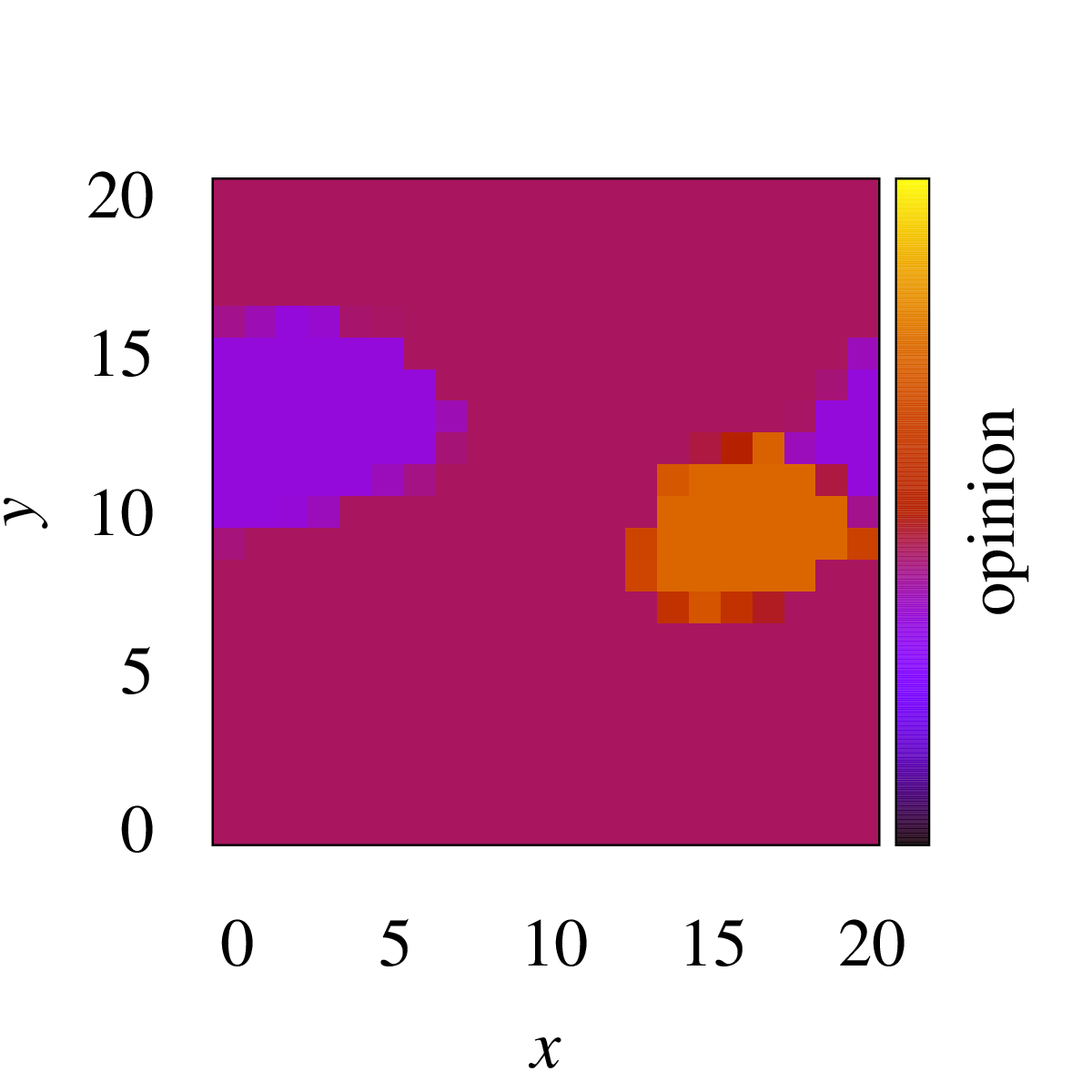}
\end{subfigure}
\hfill %% ---------------------------------------------------------------
\begin{subfigure}[t]{0.23\textwidth}
\caption{\label{subfig:SZ_moore_t=400}}
\includegraphics[trim={0mm 4mm 12mm 31mm},clip,width=.99\textwidth]{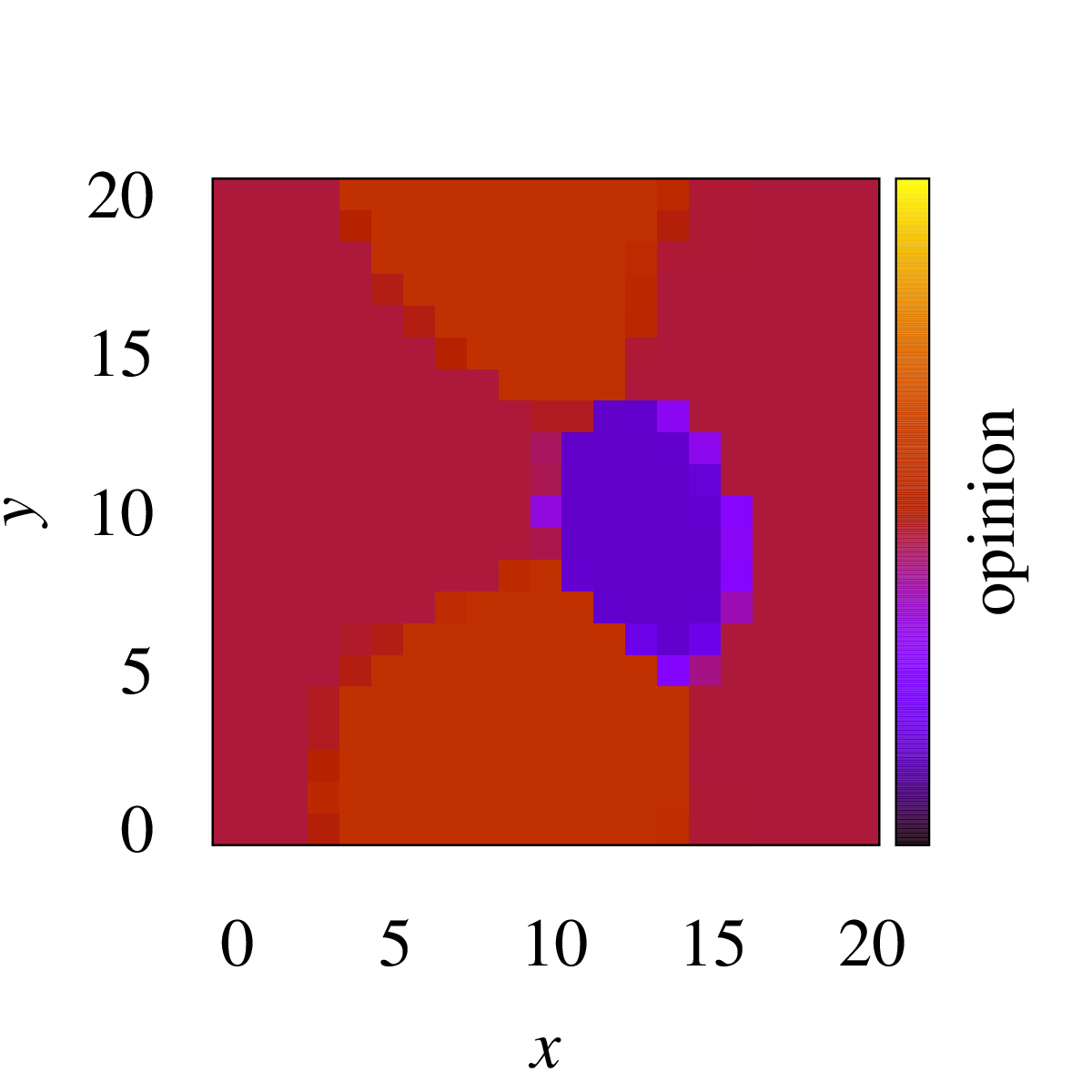}
\end{subfigure}
\hfill %% ---------------------------------------------------------------
\begin{subfigure}[t]{0.23\textwidth}
\caption{\label{subfig:SZ_triangular_t=400}}
\includegraphics[trim={0mm 4mm 12mm 31mm},clip,width=.99\textwidth]{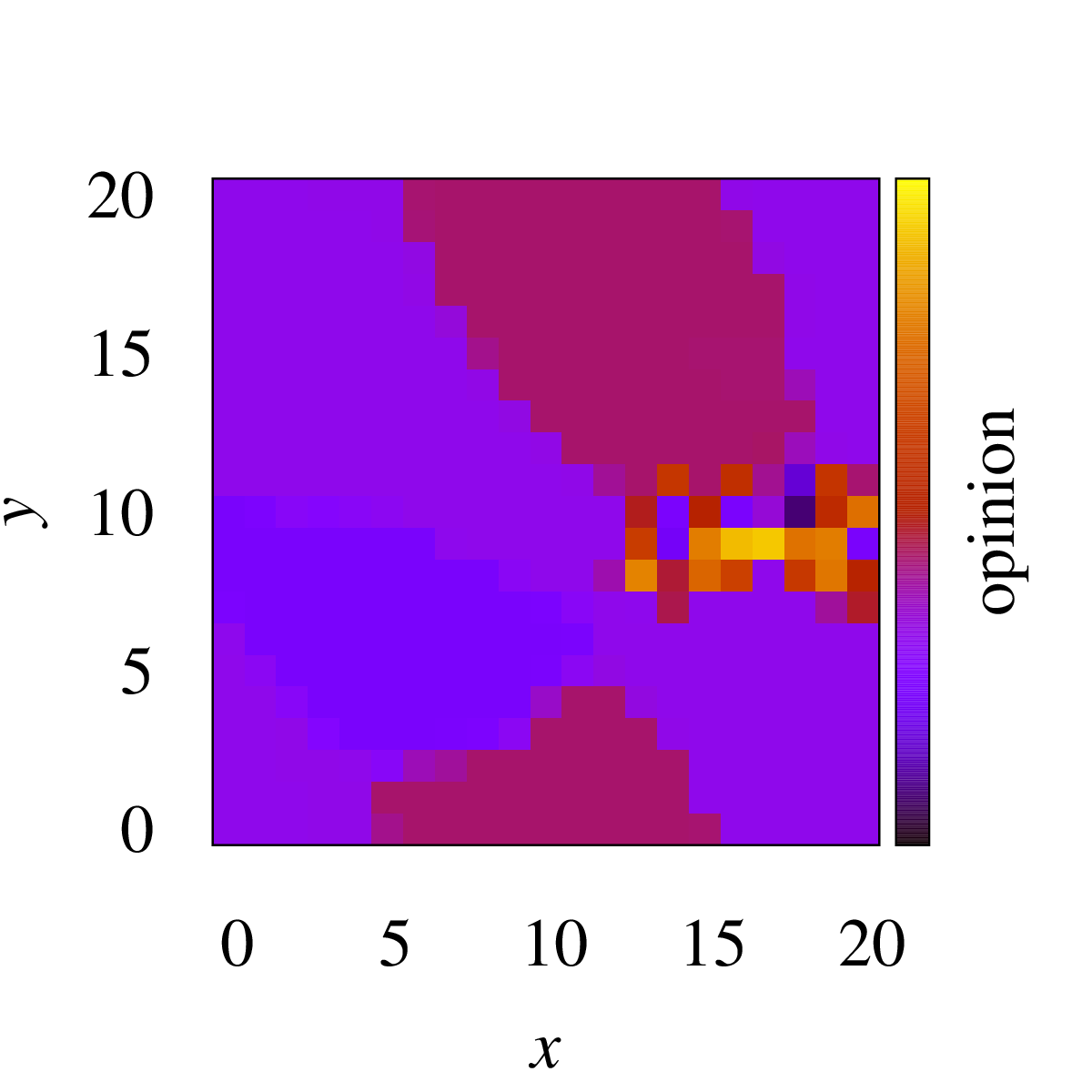}
\end{subfigure}
\hfill %% ---------------------------------------------------------------
\begin{subfigure}[t]{0.23\textwidth}
\caption{\label{subfig:SZ_honeycomb_t=400}}
\includegraphics[trim={0mm 4mm 12mm 31mm},clip,width=.99\textwidth]{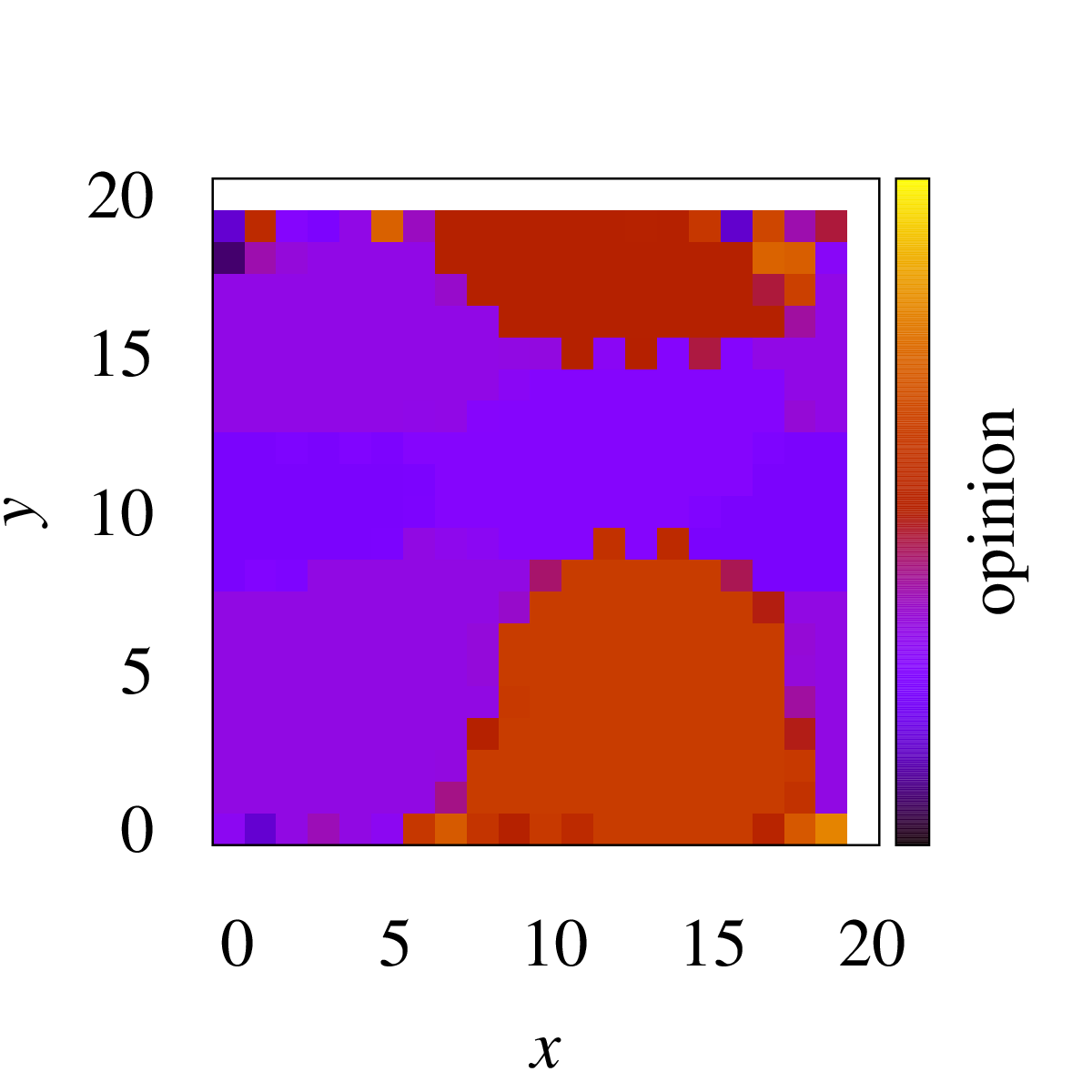}
\end{subfigure}
%% ===============================================================
\begin{subfigure}[t]{0.23\textwidth}
\caption{\label{subfig:SZ_neumann_t=end}$\tau=566$}
\includegraphics[trim={0mm 4mm 12mm 31mm},clip,width=.99\textwidth]{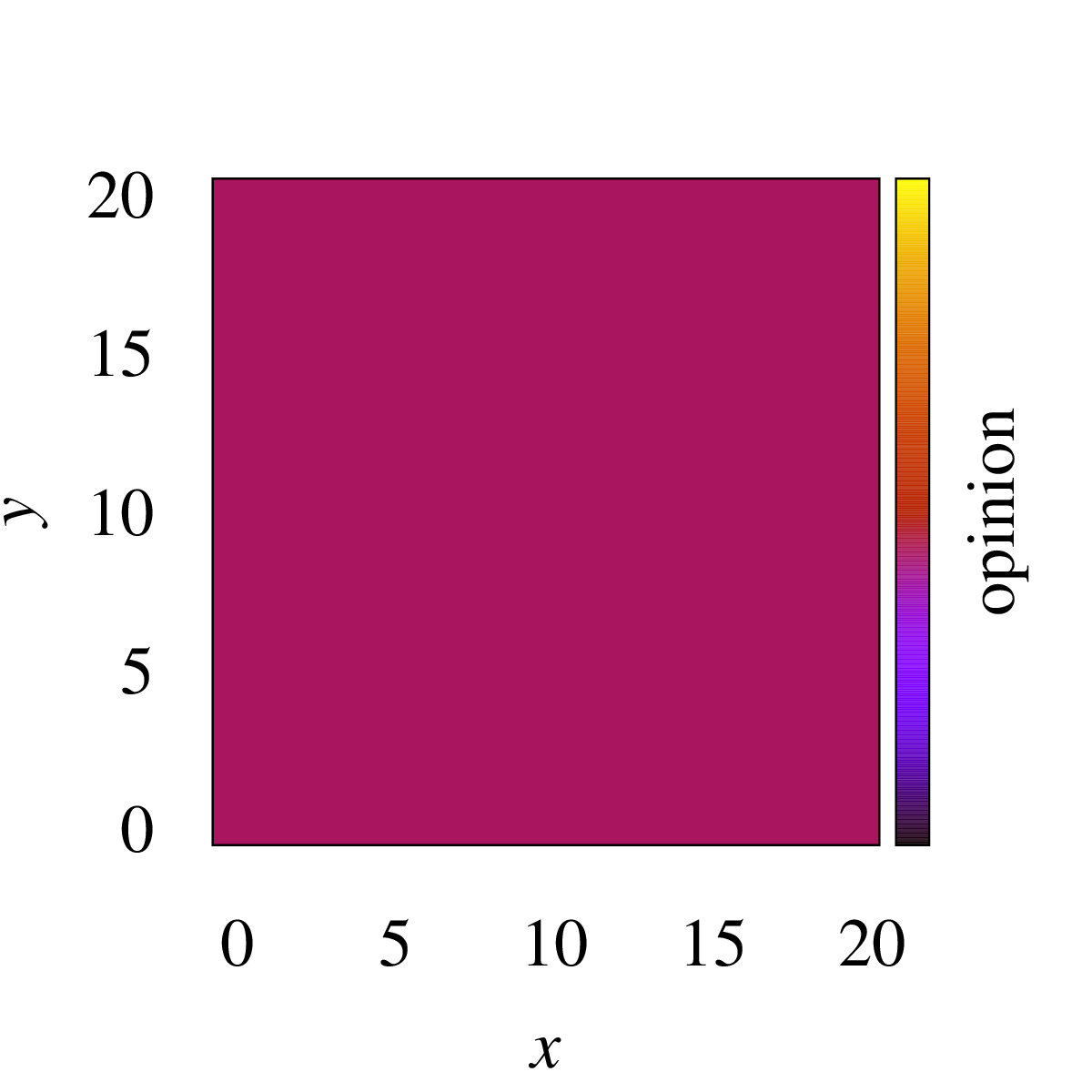}
\end{subfigure}
\hfill %% ---------------------------------------------------------------
\begin{subfigure}[t]{0.23\textwidth}
\caption{\label{subfig:SZ_moore_t=end}$\tau=657$}
\includegraphics[trim={0mm 4mm 12mm 31mm},clip,width=.99\textwidth]{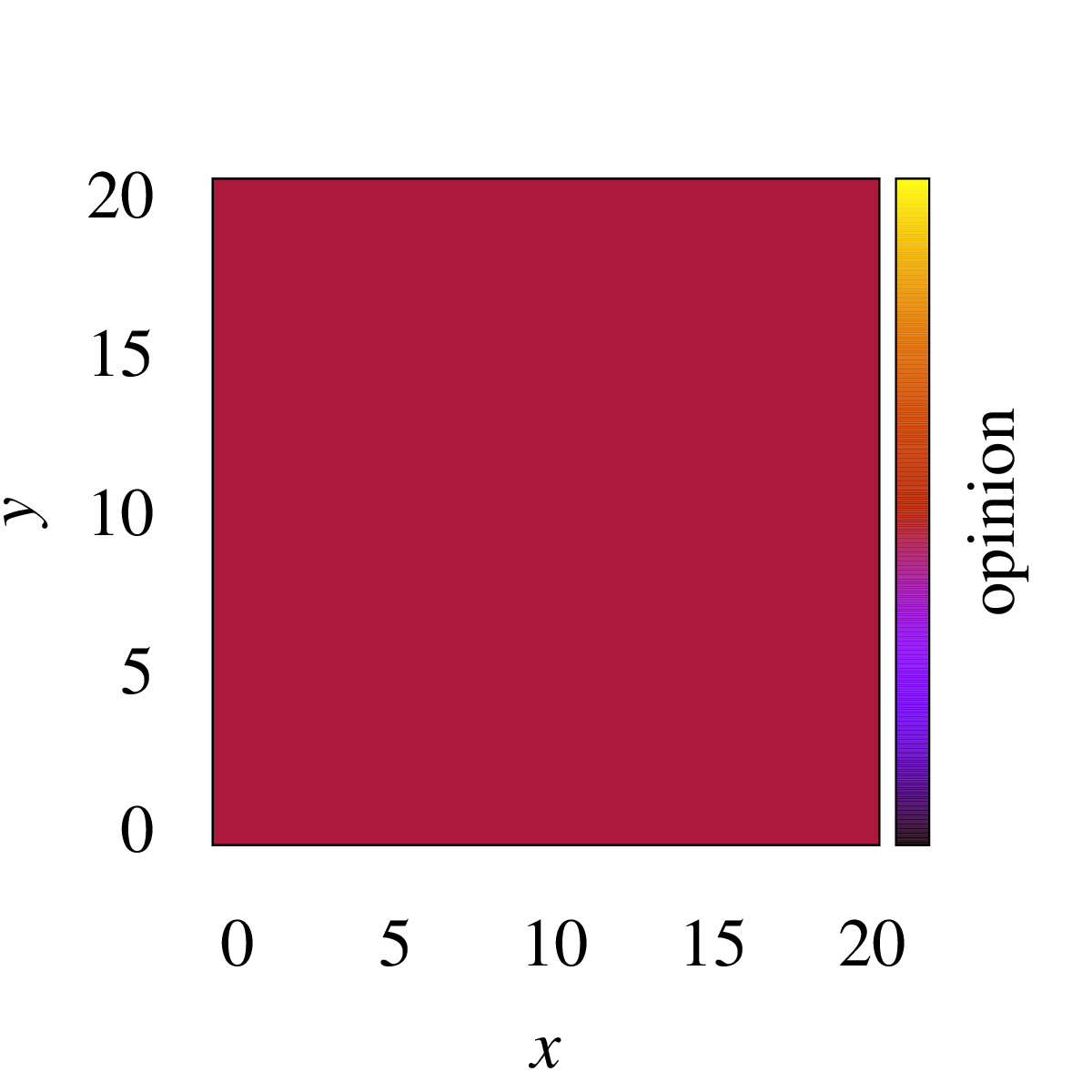}
\end{subfigure}
\hfill %% ---------------------------------------------------------------
\begin{subfigure}[t]{0.23\textwidth}
\caption{\label{subfig:SZ_triangular_t=end}$\tau=790$}
\includegraphics[trim={0mm 4mm 12mm 31mm},clip,width=.99\textwidth]{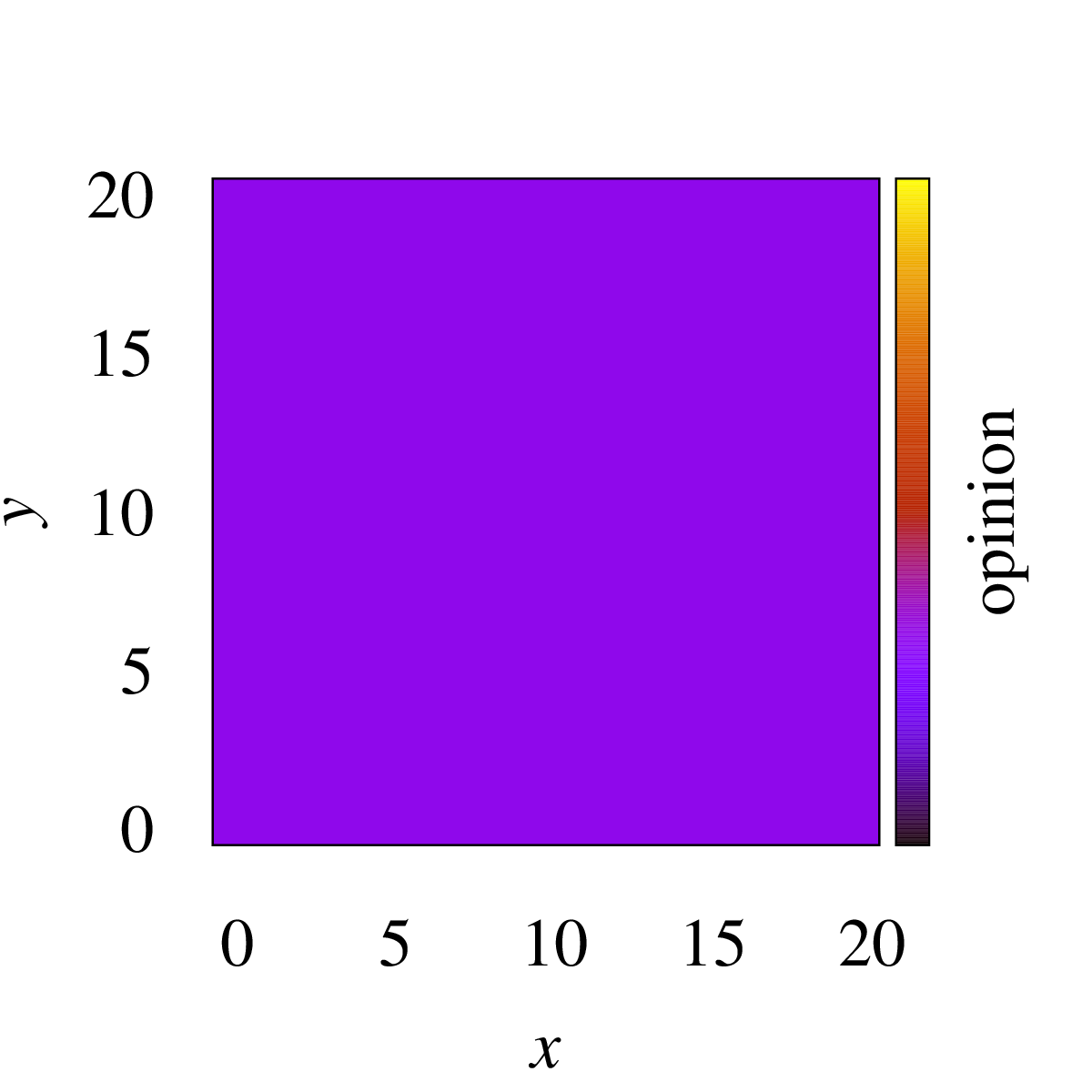}
\end{subfigure}
\hfill %% ---------------------------------------------------------------
\begin{subfigure}[t]{0.23\textwidth}
\caption{\label{subfig:SZ_honeycomb_t=end}$\tau=2160$}
\includegraphics[trim={0mm 4mm 12mm 31mm},clip,width=.99\textwidth]{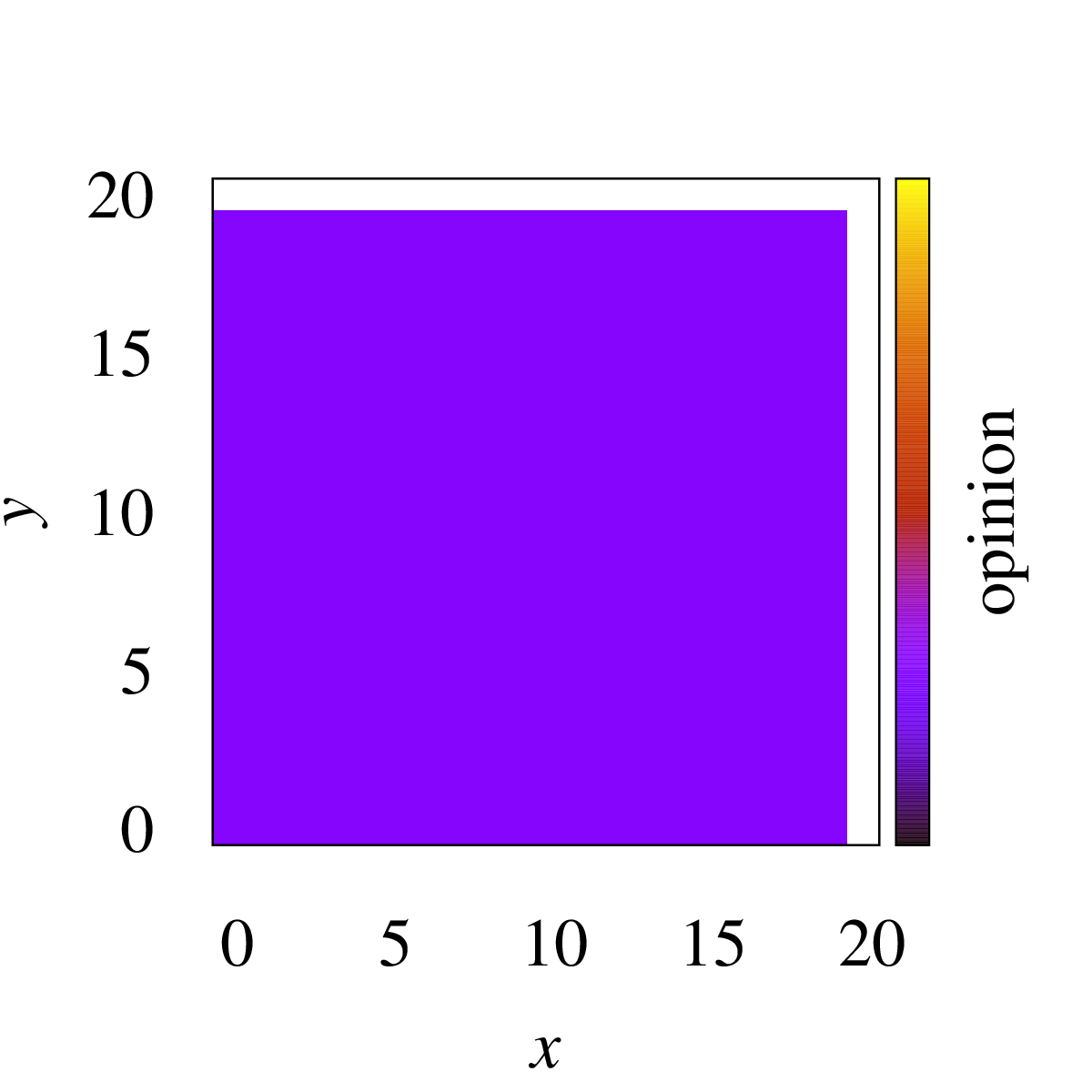}
\end{subfigure}
%% ===============================================================
\caption{\label{fig:S_snap}Snapshots for system evolution with Sznajd model. Asynchronous random update. $K=N/2$. 
\subref{subfig:SZ_neumann_t=0}--%%\subref{subfig:SZ_moore_t=0}, \subref{subfig:SZ_triangular_t=0}, 
\subref{subfig:SZ_honeycomb_t=0} $t=0$,
\subref{subfig:SZ_neumann_t=100}--%%, \subref{subfig:SZ_moore_t=100}, \subref{subfig:SZ_triangular_t=100}, 
\subref{subfig:SZ_honeycomb_t=100} $t=100$,
\subref{subfig:SZ_neumann_t=200}--%%, \subref{subfig:SZ_moore_t=200}, \subref{subfig:SZ_triangular_t=200}, 
\subref{subfig:SZ_honeycomb_t=200} $t=200$,
\subref{subfig:SZ_neumann_t=400}--%%, \subref{subfig:SZ_moore_t=400}, \subref{subfig:SZ_triangular_t=400}, 
\subref{subfig:SZ_honeycomb_t=400} $t=400$,
\subref{subfig:SZ_neumann_t=end}--%%, \subref{subfig:SZ_moore_t=end}, \subref{subfig:SZ_triangular_t=end}, 
\subref{subfig:SZ_honeycomb_t=end} $t\to\infty$.
Square lattice with \subref{subfig:SZ_neumann_t=0}, \subref{subfig:SZ_neumann_t=100}, \subref{subfig:SZ_neumann_t=200}, \subref{subfig:SZ_neumann_t=400}, \subref{subfig:SZ_neumann_t=end} the von Neumann neighborhood, 
square lattice with \subref{subfig:SZ_moore_t=0}, \subref{subfig:SZ_moore_t=100}, \subref{subfig:SZ_moore_t=200}, \subref{subfig:SZ_moore_t=400}, \subref{subfig:SZ_moore_t=end} Moore's neighborhood, 
\subref{subfig:SZ_triangular_t=0}, \subref{subfig:SZ_triangular_t=100}, \subref{subfig:SZ_triangular_t=200}, \subref{subfig:SZ_triangular_t=400}, \subref{subfig:SZ_triangular_t=end} triangular 
and \subref{subfig:SZ_honeycomb_t=0}, \subref{subfig:SZ_honeycomb_t=100}, \subref{subfig:SZ_honeycomb_t=200}, \subref{subfig:SZ_honeycomb_t=400}, \subref{subfig:SZ_honeycomb_t=end} honeycomb lattice.}
%% ===============================================================
\end{figure*}

\begin{figure*}[htbp]
%% ===============================================================
\begin{subfigure}[t]{0.23\textwidth}
\caption{\label{subfig:L_a30T00_t=0000}$\alpha=3$, $T=0$}
\includegraphics[trim={0mm 4mm 12mm 31mm},clip,width=.99\textwidth]{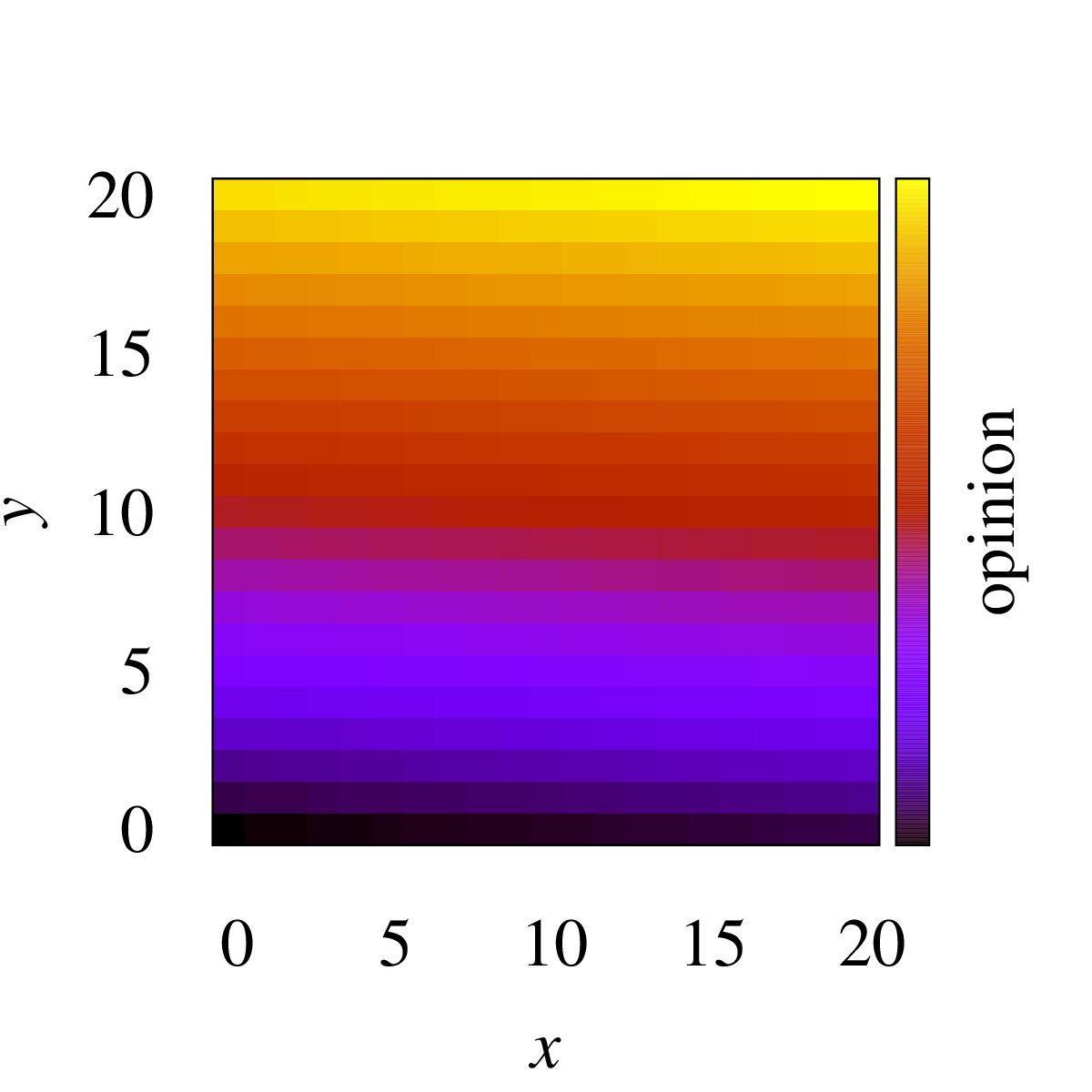}
\end{subfigure}
\hfill%% ---------------------------------------------------------------
\begin{subfigure}[t]{0.23\textwidth}
\caption{\label{subfig:L_a40T00_t=0000}$\alpha=4$, $T=0$}
\includegraphics[trim={0mm 4mm 12mm 31mm},clip,width=.99\textwidth]{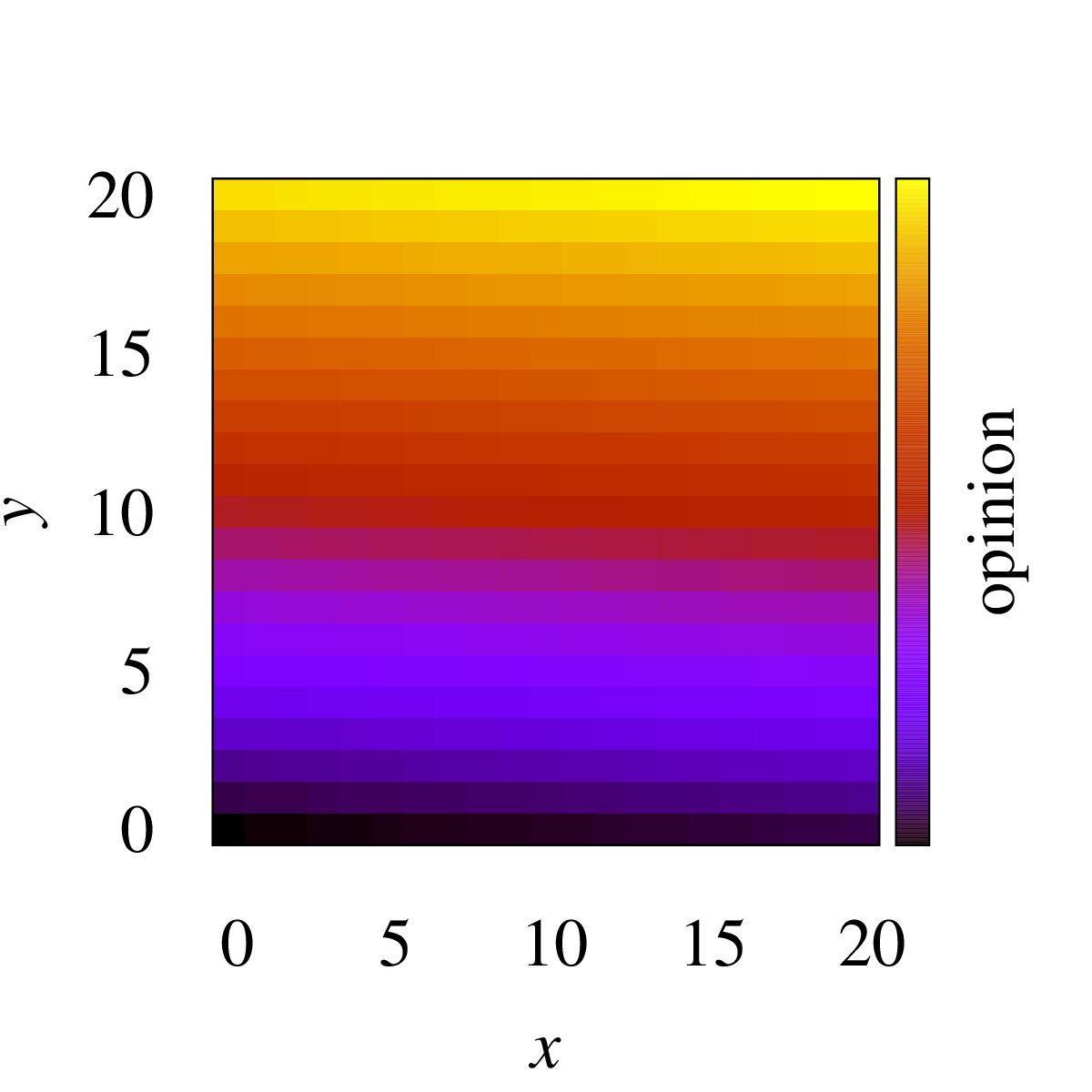}
\end{subfigure}
\hfill%% ---------------------------------------------------------------
\begin{subfigure}[t]{0.23\textwidth}
\caption{\label{subfig:L_a50T00_t=0000}$\alpha=5$, $T=0$}
\includegraphics[trim={0mm 4mm 12mm 31mm},clip,width=.99\textwidth]{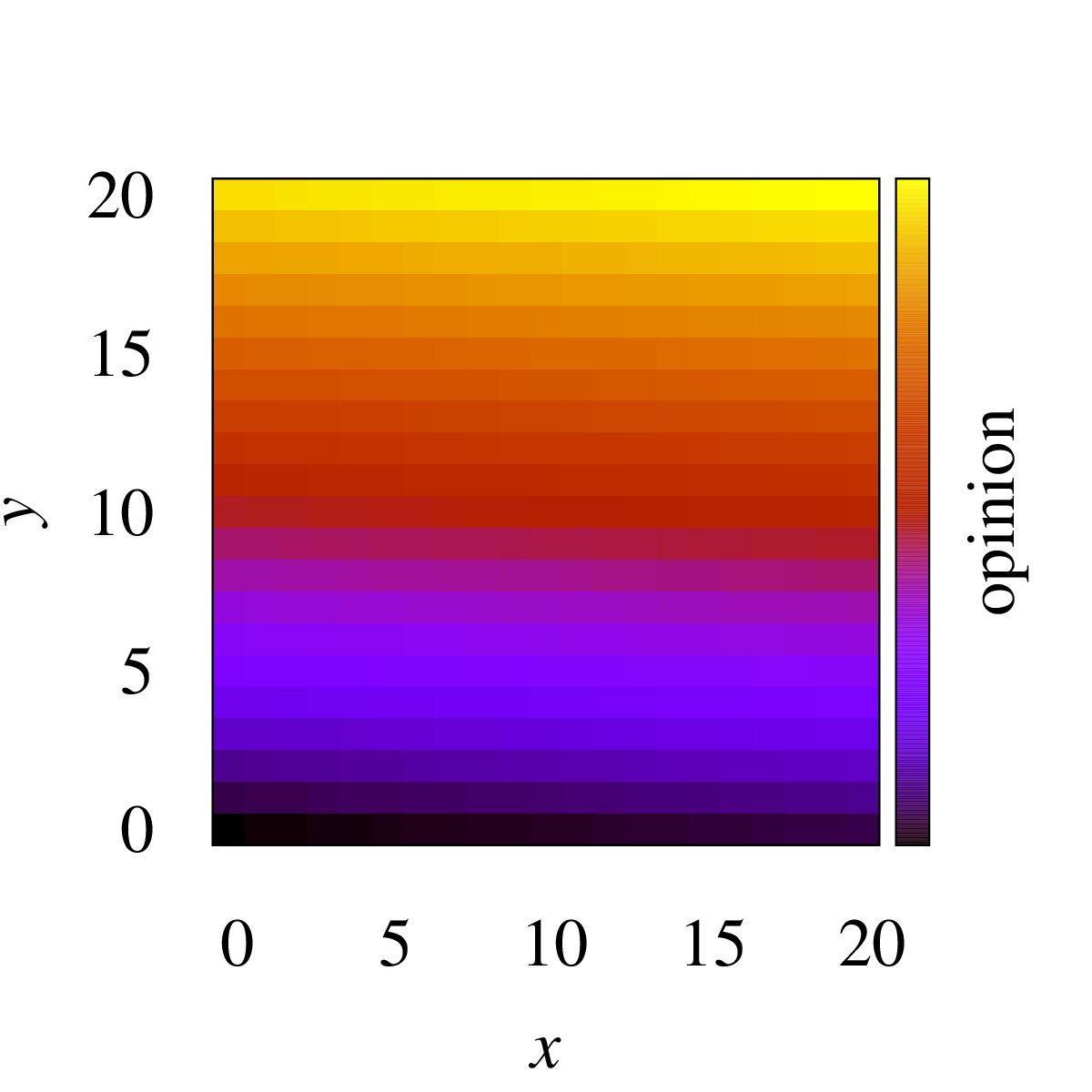}
\end{subfigure}
\hfill%% ---------------------------------------------------------------
\begin{subfigure}[t]{0.23\textwidth}
\caption{\label{subfig:L_a60T00_t=0000}$\alpha=6$, $T=0$}
\includegraphics[trim={0mm 4mm 12mm 31mm},clip,width=.99\textwidth]{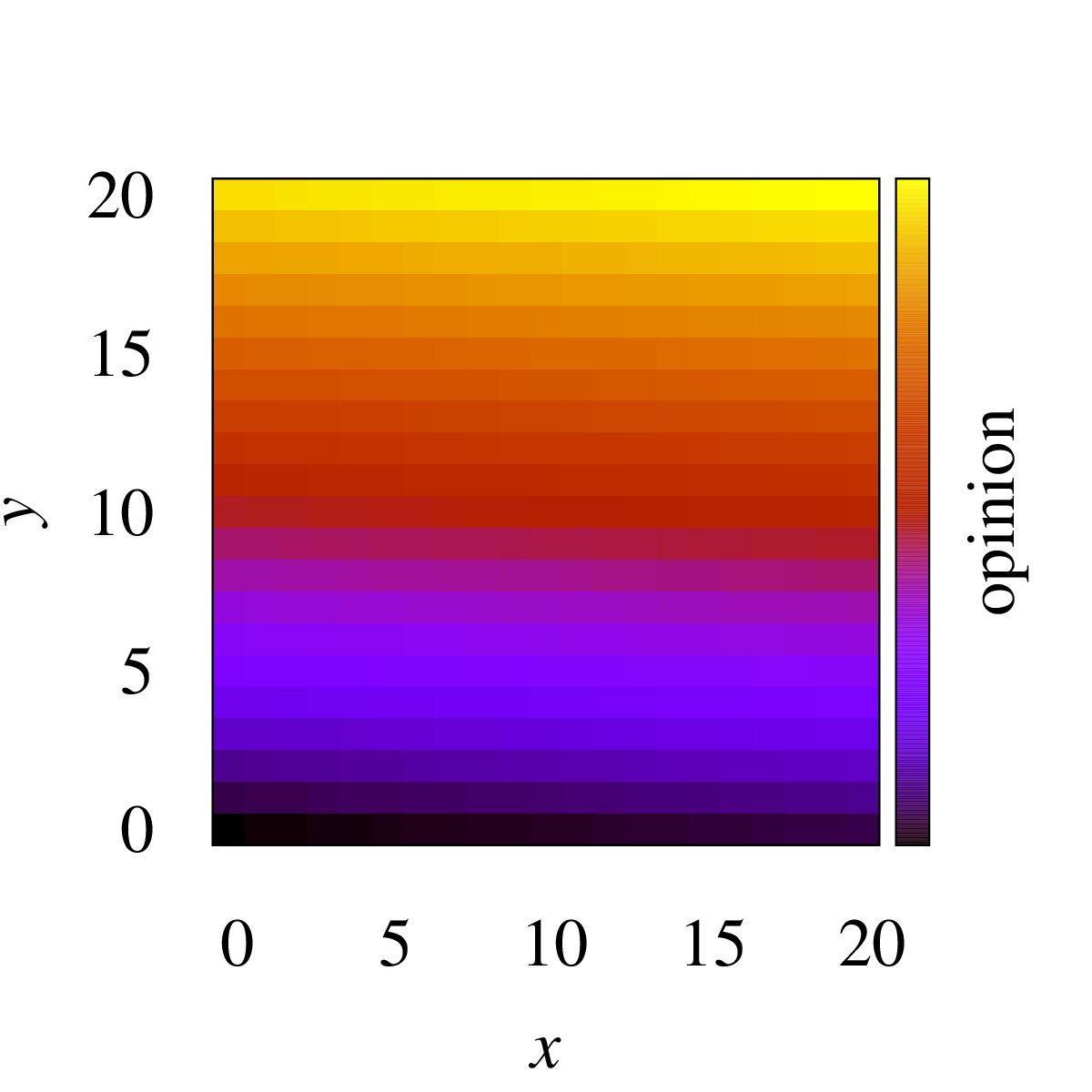}
\end{subfigure}
%% ---------------------------------------------------------------
%% ===============================================================
%% ---------------------------------------------------------------
\begin{subfigure}[t]{0.23\textwidth}
\caption{\label{subfig:L_a30T00_t=0010}$t=10$}
\includegraphics[trim={0mm 4mm 12mm 31mm},clip,width=.99\textwidth]{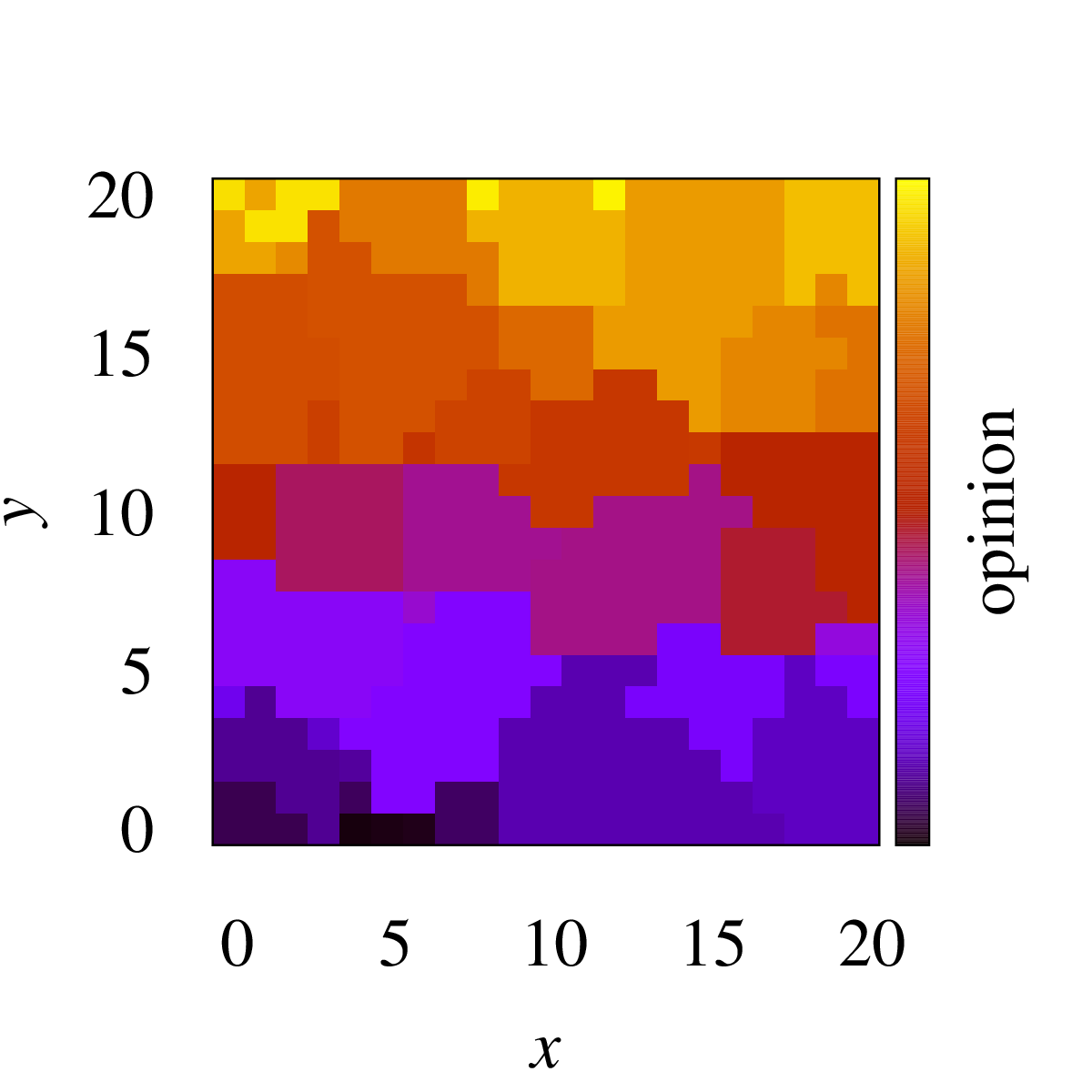}
\end{subfigure}
\hfill%% ---------------------------------------------------------------
\begin{subfigure}[t]{0.23\textwidth}
\caption{\label{subfig:L_a40T00_t=0010}}
\includegraphics[trim={0mm 4mm 12mm 31mm},clip,width=.99\textwidth]{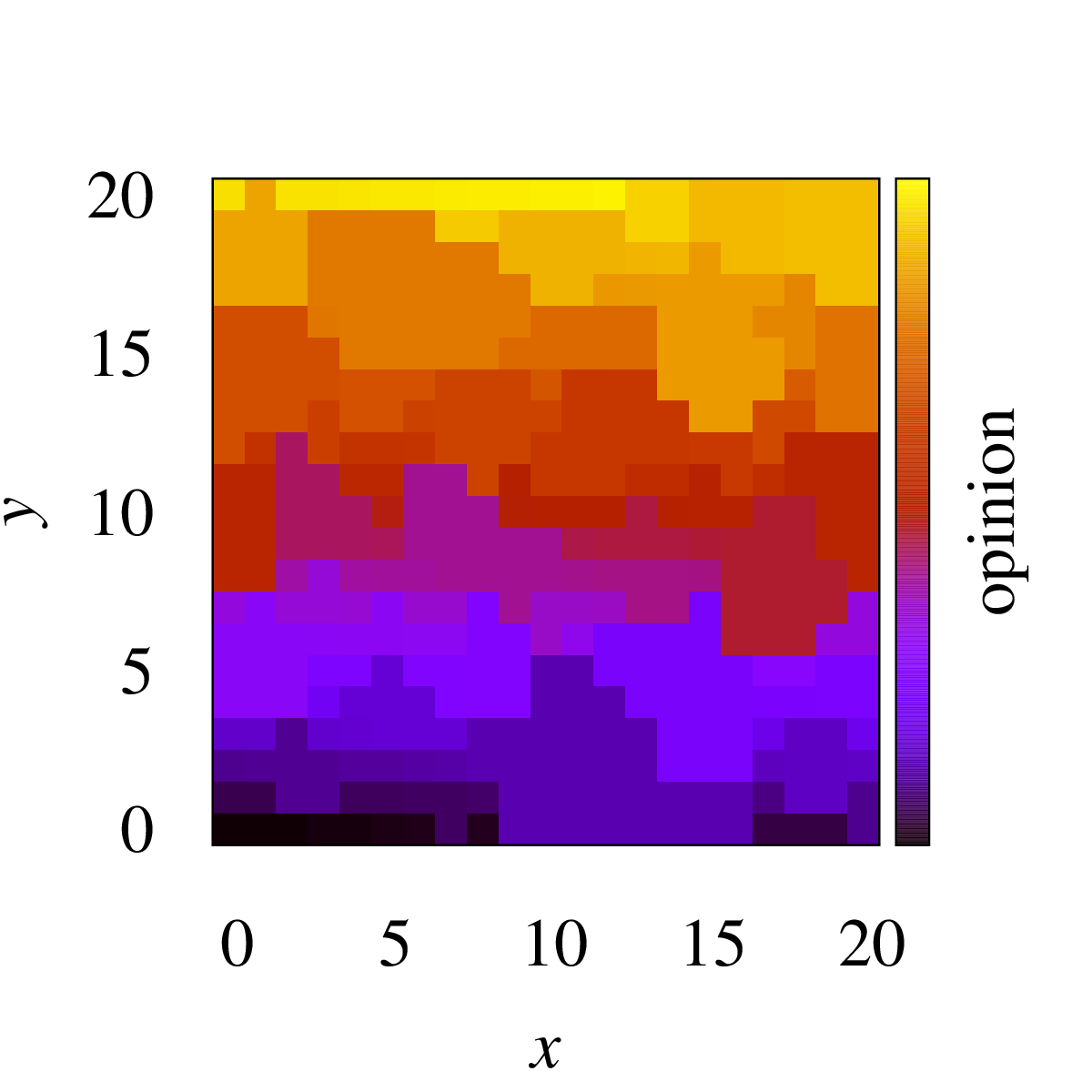}
\end{subfigure}
\hfill%% ---------------------------------------------------------------
\begin{subfigure}[t]{0.23\textwidth}
\caption{\label{subfig:L_a50T00_t=0010}}
\includegraphics[trim={0mm 4mm 12mm 31mm},clip,width=.99\textwidth]{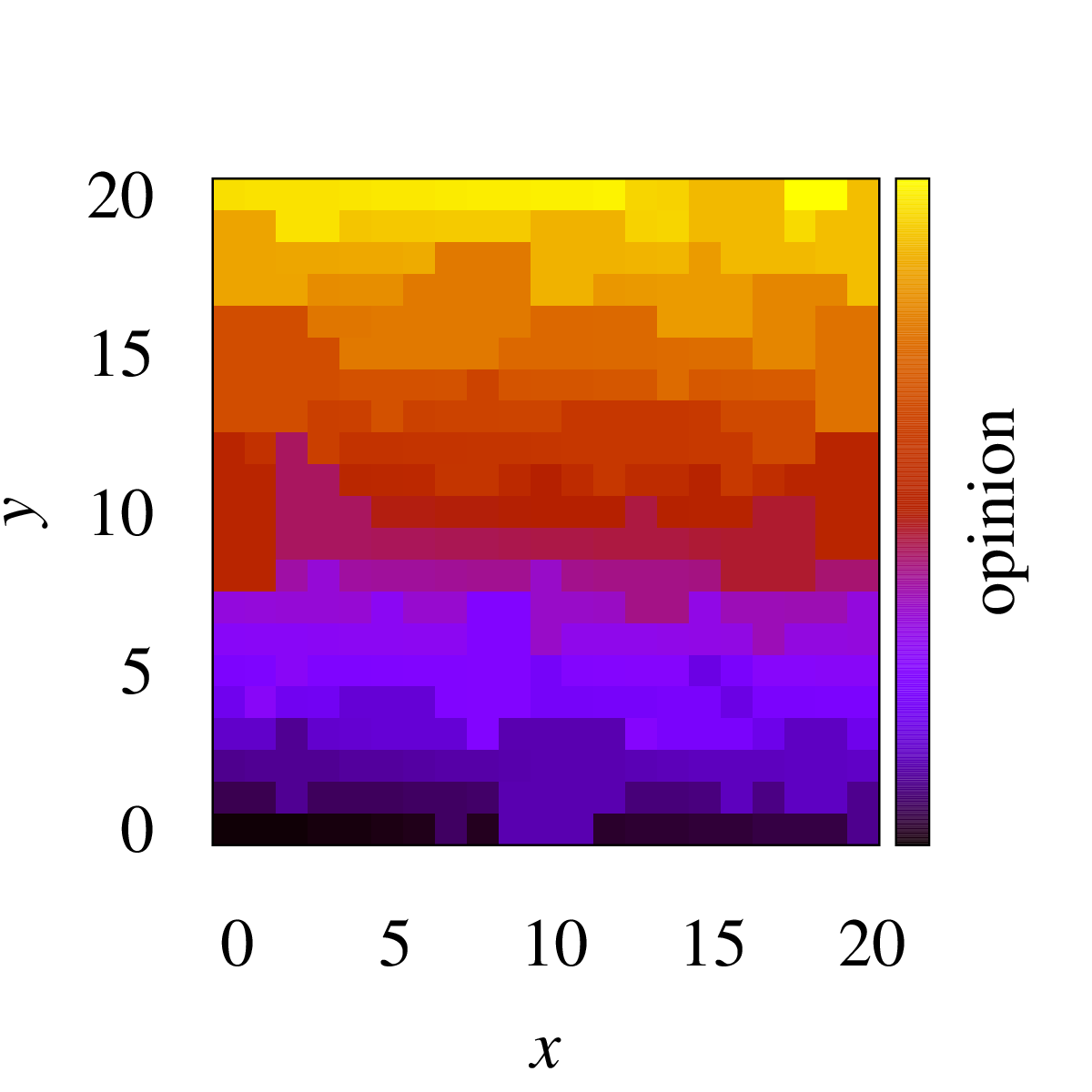}
\end{subfigure}
\hfill%% ---------------------------------------------------------------
\begin{subfigure}[t]{0.23\textwidth}
\caption{\label{subfig:L_a60T00_t=0010}}
\includegraphics[trim={0mm 4mm 12mm 31mm},clip,width=.99\textwidth]{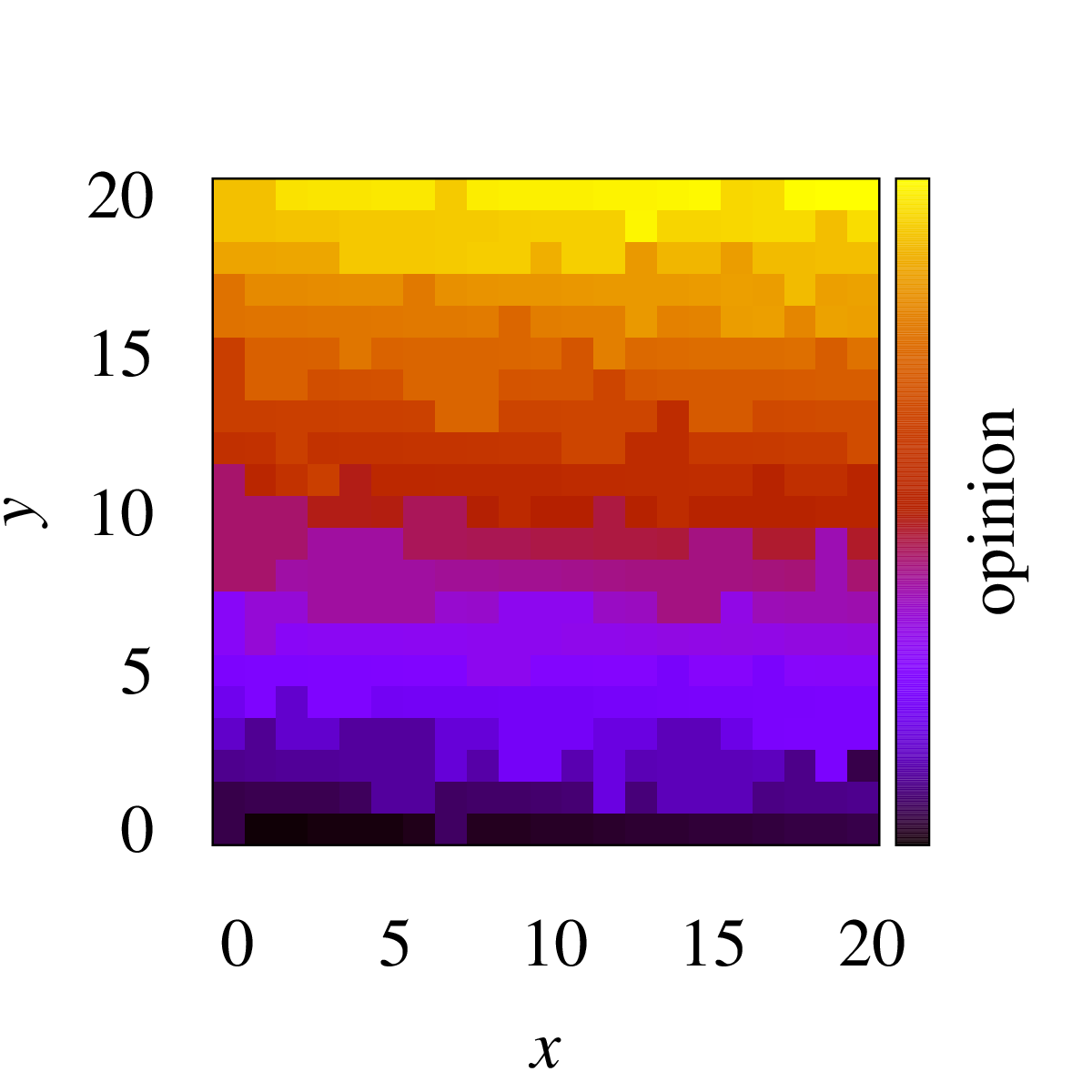}
\end{subfigure}
%% ---------------------------------------------------------------
%% ===============================================================
%% ---------------------------------------------------------------
\begin{subfigure}[t]{0.23\textwidth}
\caption{\label{subfig:L_a30T00_t=0100}$t=10^2$}
\includegraphics[trim={0mm 4mm 12mm 31mm},clip,width=.99\textwidth]{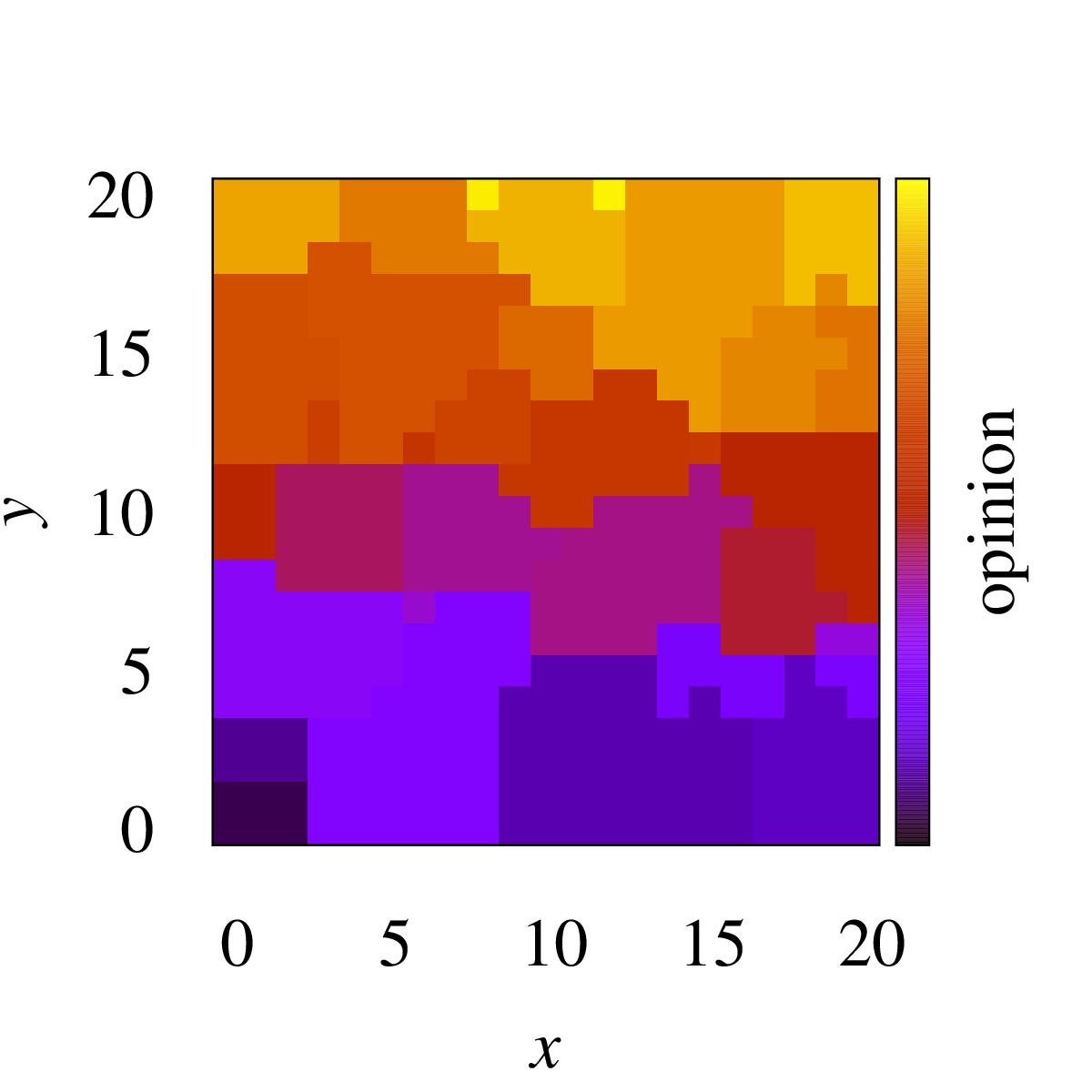}
\end{subfigure}
\hfill%% ---------------------------------------------------------------
\begin{subfigure}[t]{0.23\textwidth}
\caption{\label{subfig:L_a40T00_t=0100}}
\includegraphics[trim={0mm 4mm 12mm 31mm},clip,width=.99\textwidth]{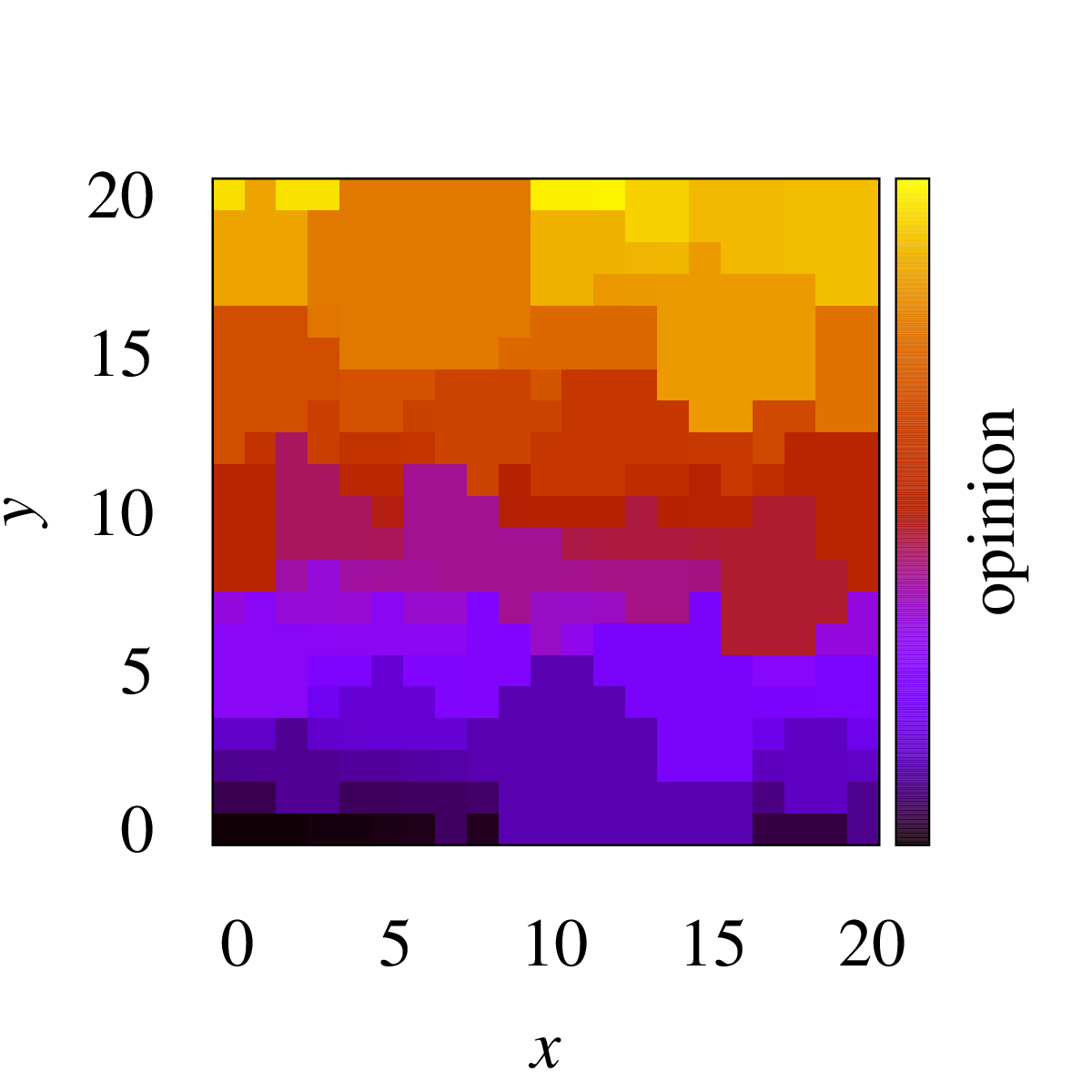}
\end{subfigure}
\hfill%% ---------------------------------------------------------------
\begin{subfigure}[t]{0.23\textwidth}
\caption{\label{subfig:L_a50T00_t=0100}}
\includegraphics[trim={0mm 4mm 12mm 31mm},clip,width=.99\textwidth]{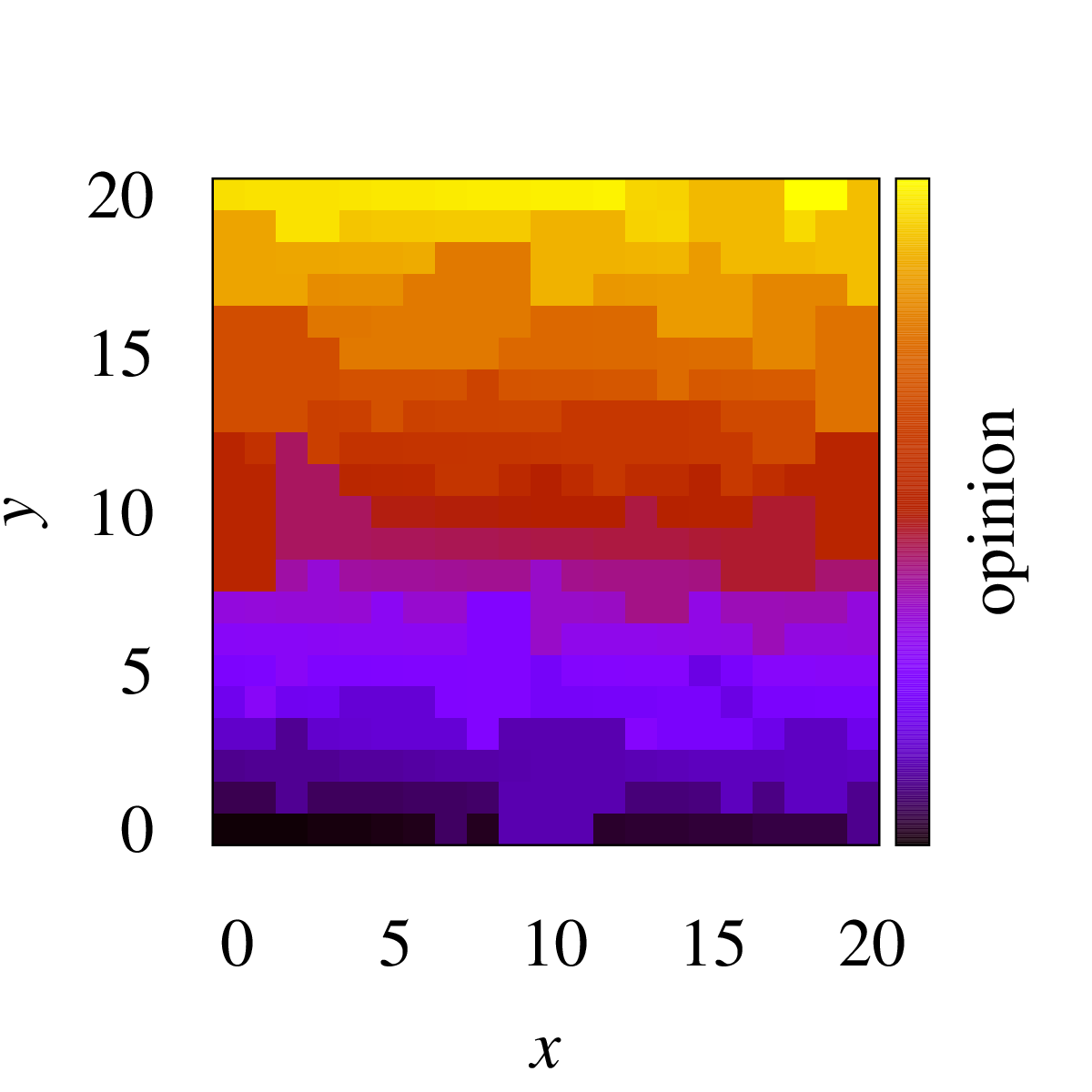}
\end{subfigure}
\hfill%% ---------------------------------------------------------------
\begin{subfigure}[t]{0.23\textwidth}
\caption{\label{subfig:L_a60T00_t=0100}}
\includegraphics[trim={0mm 4mm 12mm 31mm},clip,width=.99\textwidth]{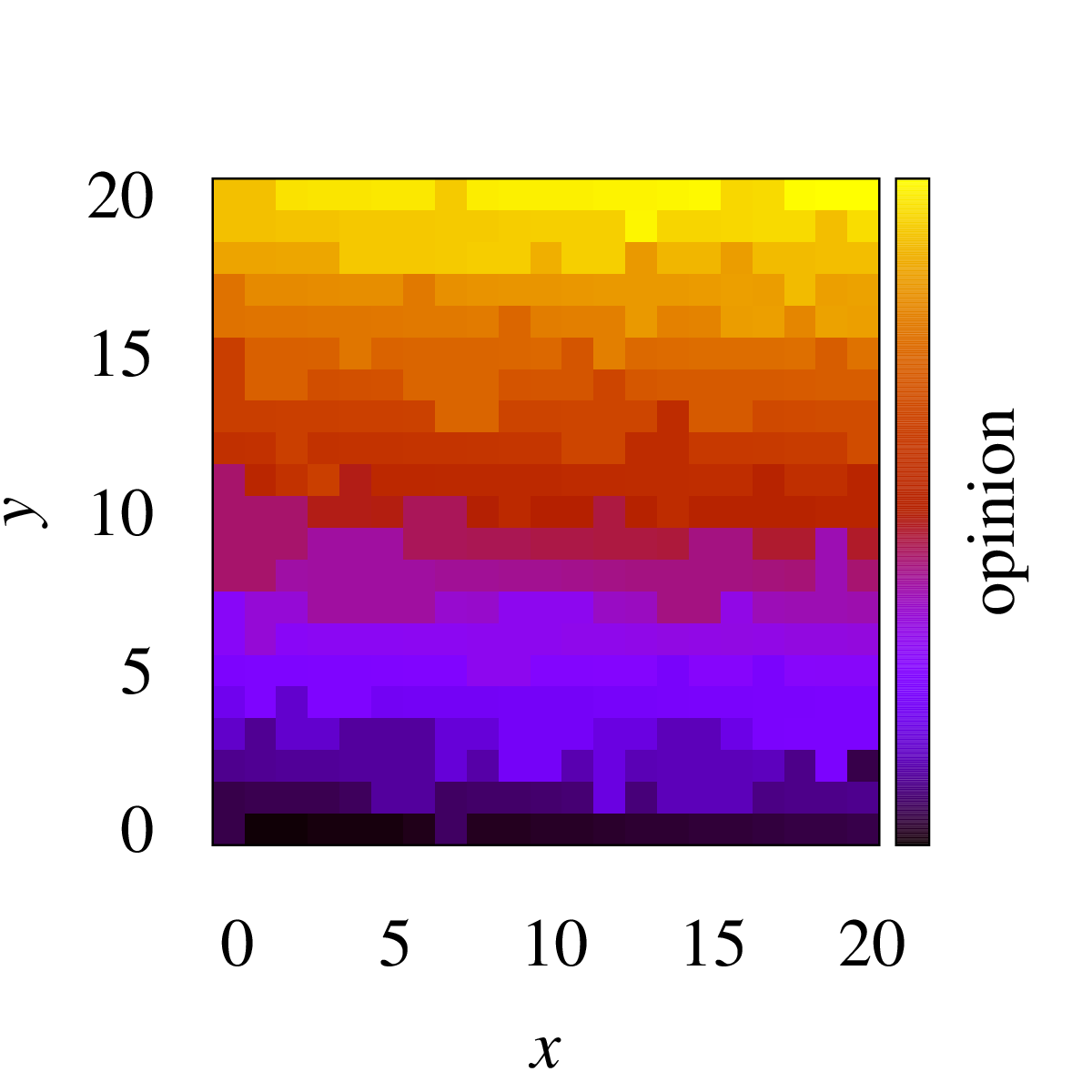}
\end{subfigure}
%% ---------------------------------------------------------------
%% ===============================================================
%% ---------------------------------------------------------------
\begin{subfigure}[t]{0.23\textwidth}
\caption{\label{subfig:L_a30T00_t=1000}$t=10^3$}
\includegraphics[trim={0mm 4mm 12mm 31mm},clip,width=.99\textwidth]{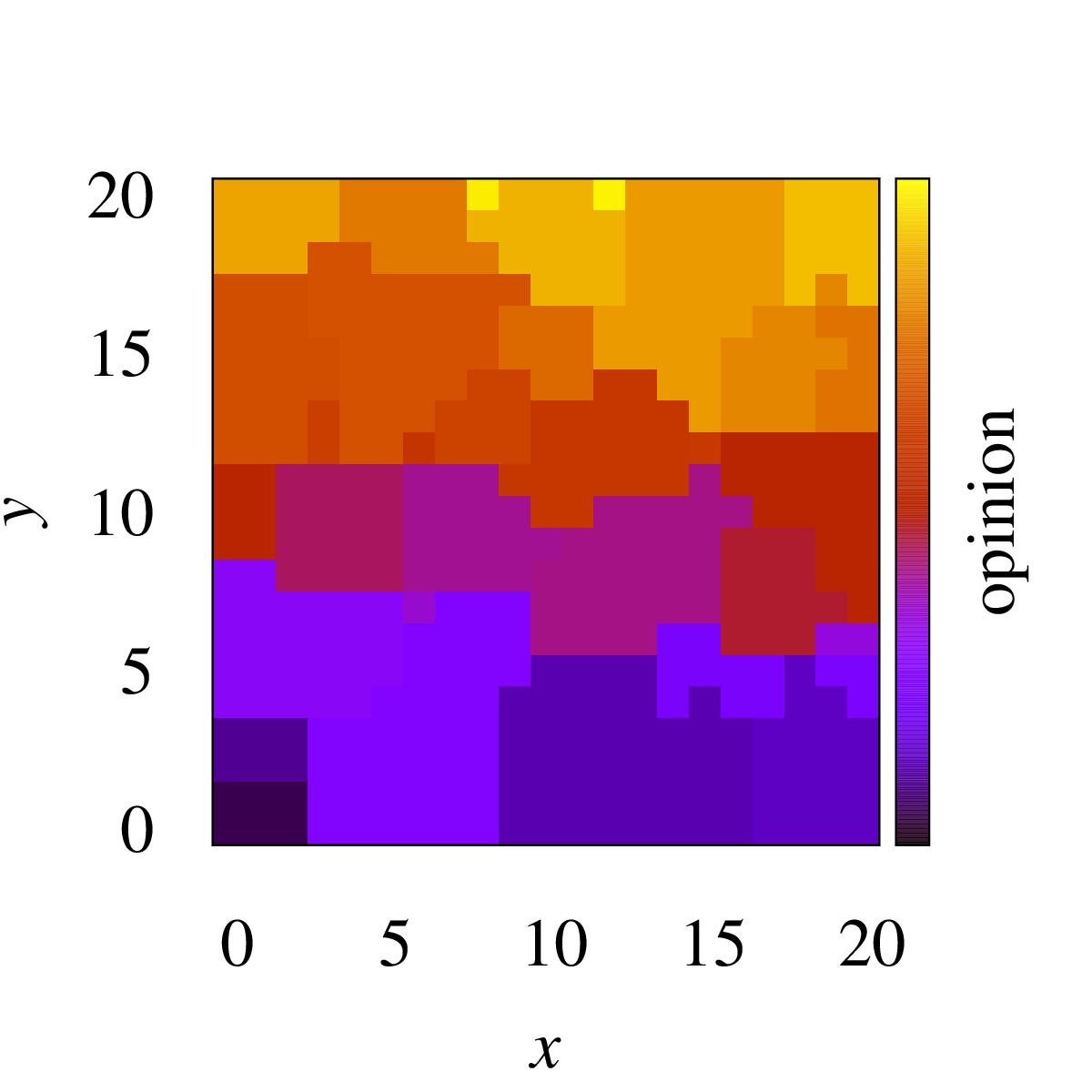}
\end{subfigure}
\hfill%% ---------------------------------------------------------------
\begin{subfigure}[t]{0.23\textwidth}
\caption{\label{subfig:L_a40T00_t=1000}}
\includegraphics[trim={0mm 4mm 12mm 31mm},clip,width=.99\textwidth]{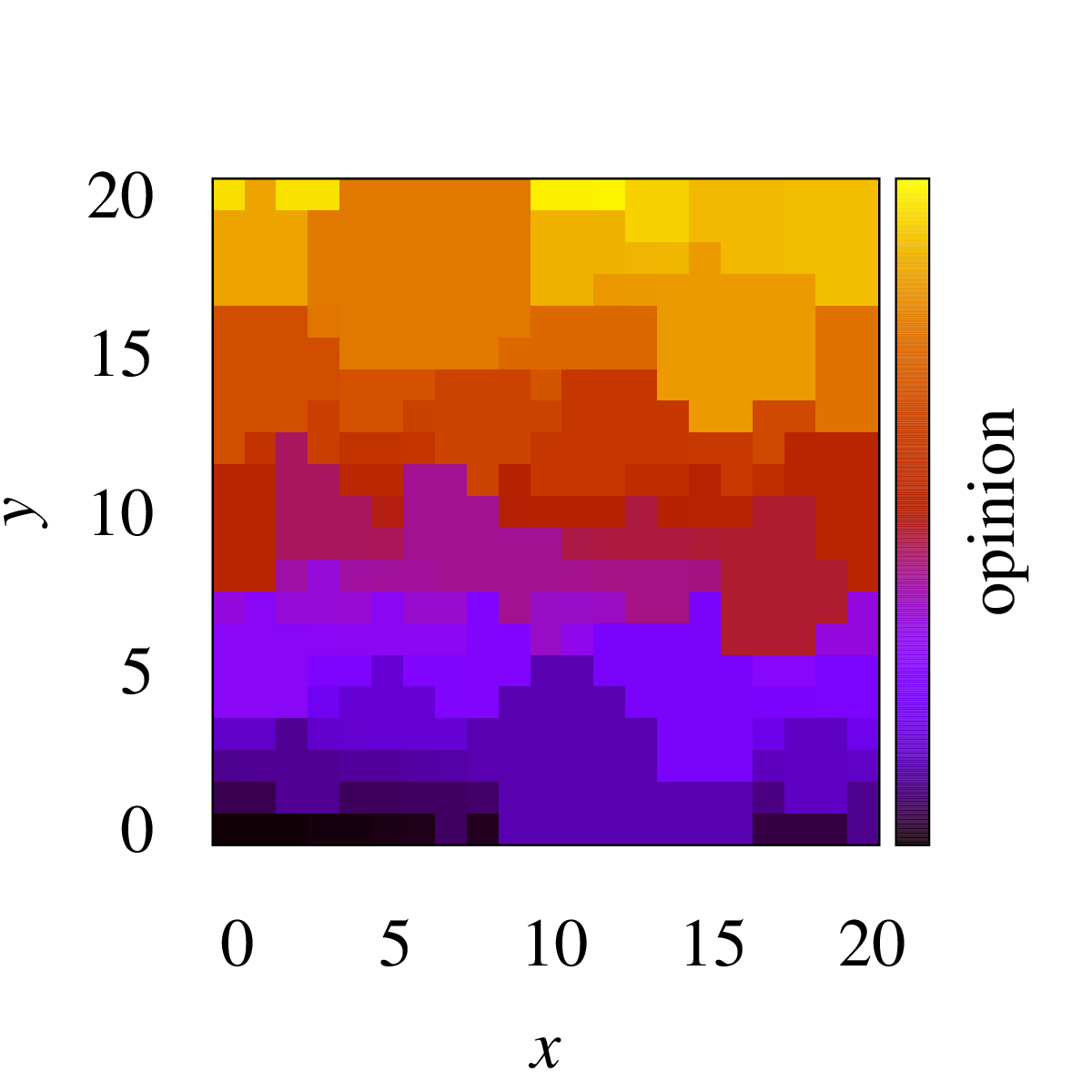}
\end{subfigure}
\hfill%% ---------------------------------------------------------------
\begin{subfigure}[t]{0.23\textwidth}
\caption{\label{subfig:L_a50T00_t=1000}}
\includegraphics[trim={0mm 4mm 12mm 31mm},clip,width=.99\textwidth]{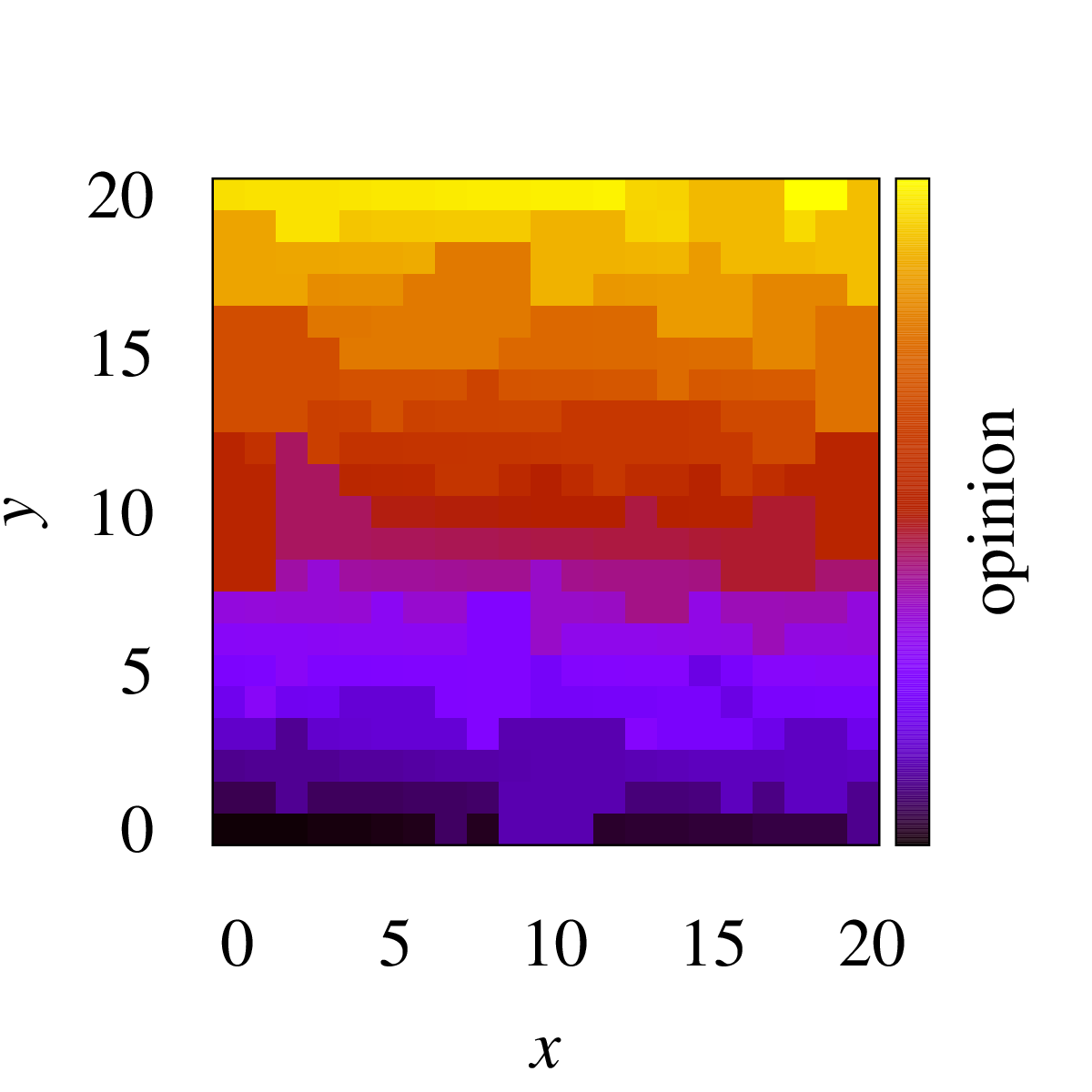}
\end{subfigure}
\hfill%% ---------------------------------------------------------------
\begin{subfigure}[t]{0.23\textwidth}
\caption{\label{subfig:L_a60T00_t=1000}}
\includegraphics[trim={0mm 4mm 12mm 31mm},clip,width=.99\textwidth]{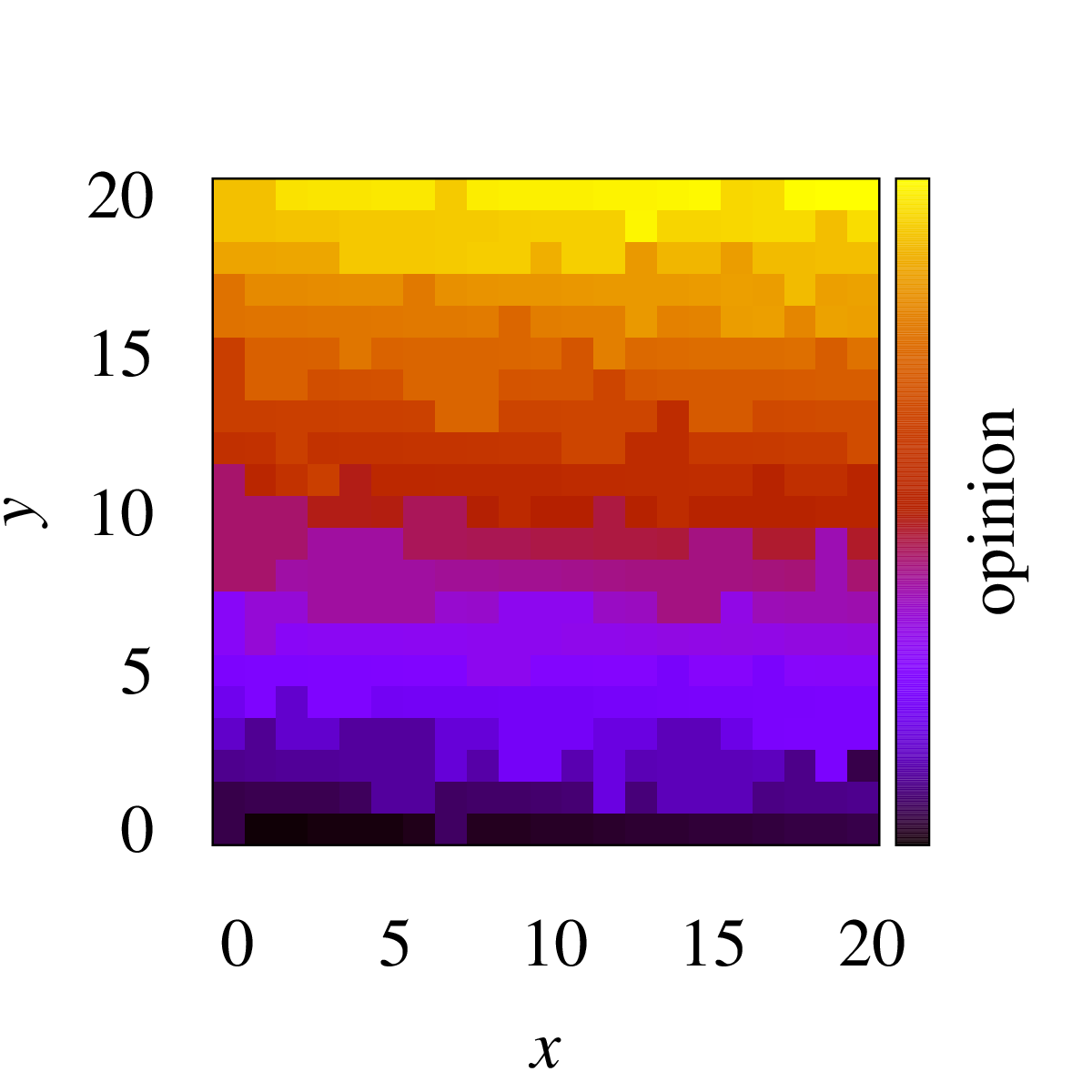}
\end{subfigure}
%% ---------------------------------------------------------------
%% ===============================================================
%% ---------------------------------------------------------------
\begin{subfigure}[t]{0.23\textwidth}
\caption{\label{subfig:L_a30T00_t=end} $t=10^5$}
\includegraphics[trim={0mm 4mm 12mm 31mm},clip,width=.99\textwidth]{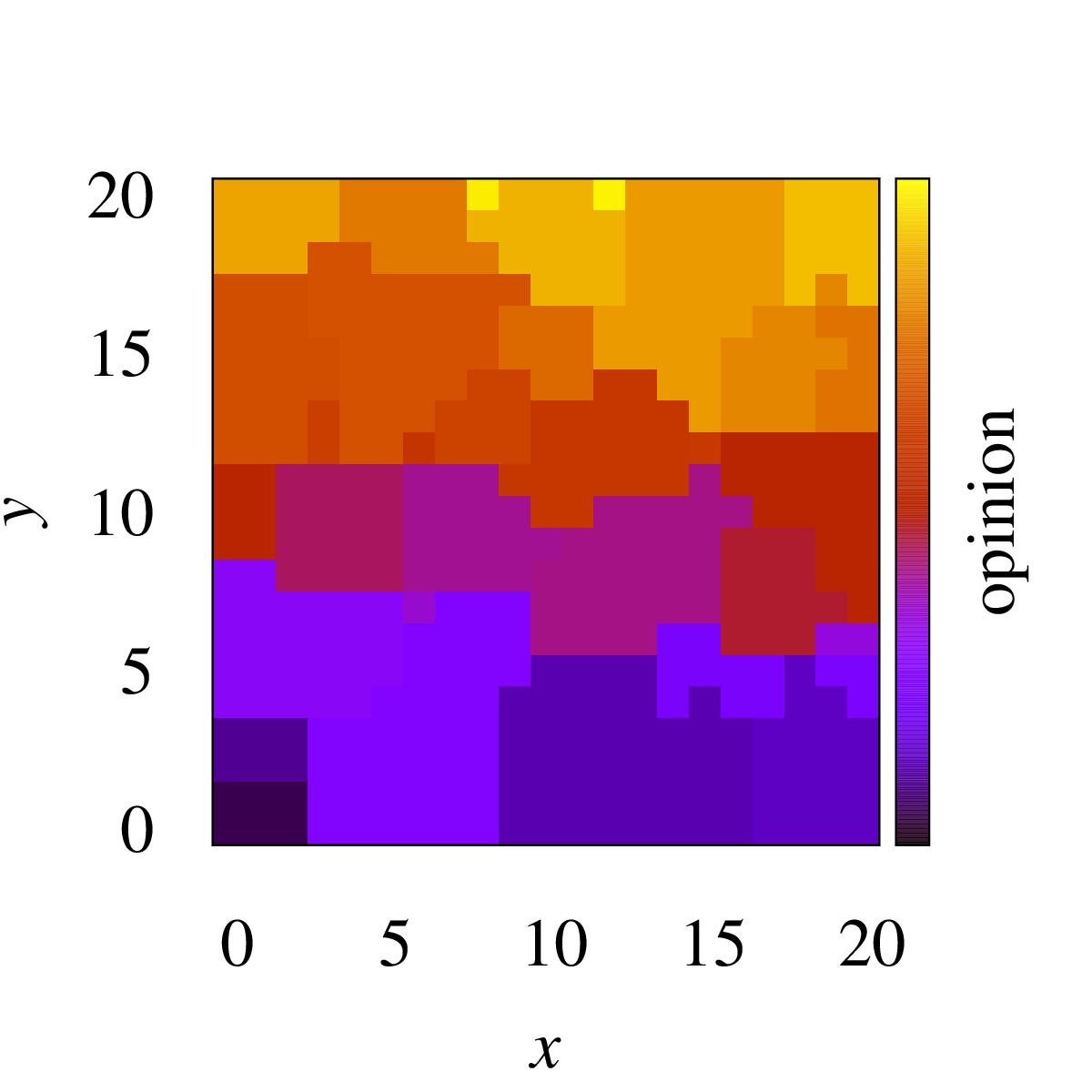}
\end{subfigure}
\hfill%% ---------------------------------------------------------------
\begin{subfigure}[t]{0.23\textwidth}
\caption{\label{subfig:L_a40T00_t=end}}
\includegraphics[trim={0mm 4mm 12mm 31mm},clip,width=.99\textwidth]{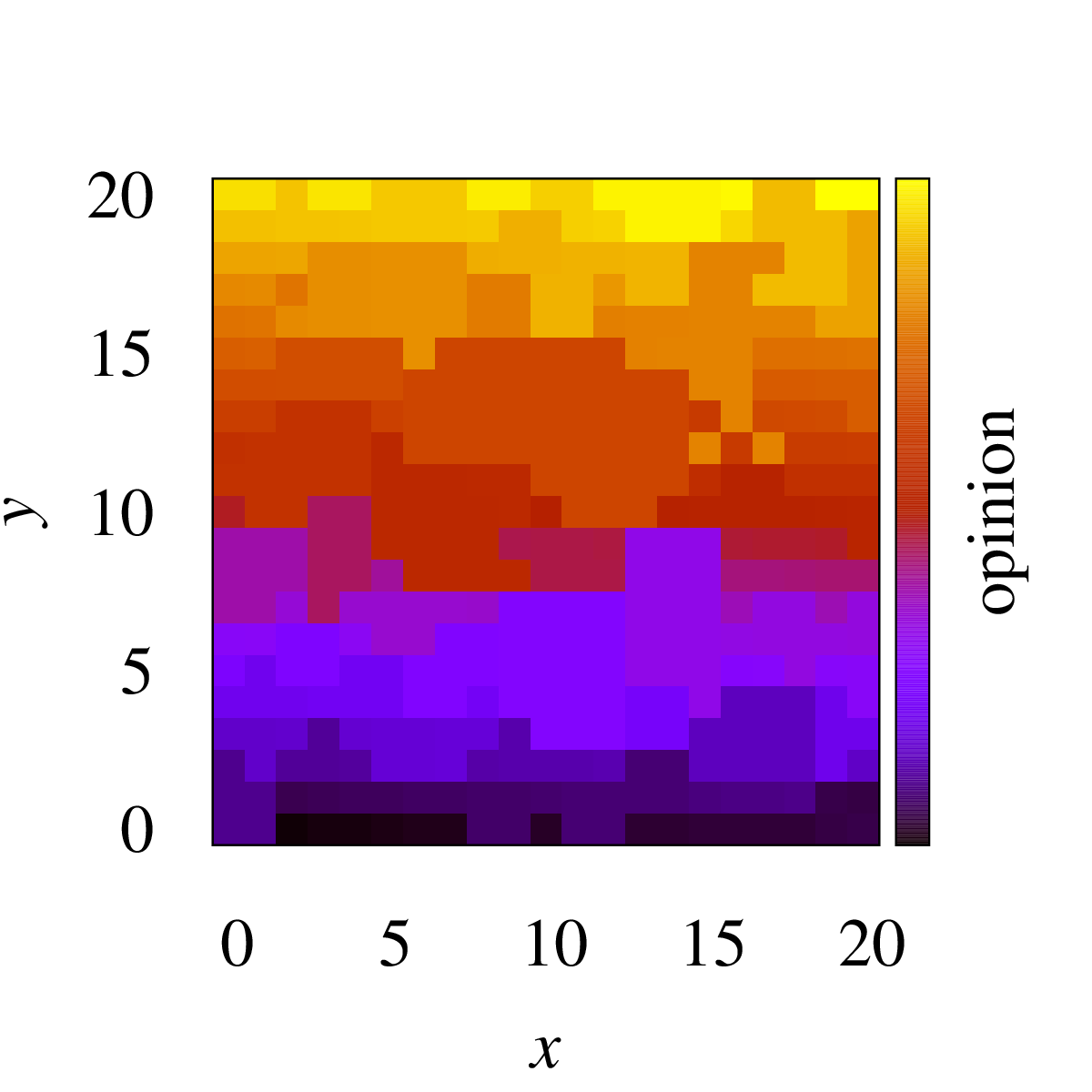}
\end{subfigure}
\hfill%% ---------------------------------------------------------------
\begin{subfigure}[t]{0.23\textwidth}
\caption{\label{subfig:L_a50T00_t=end}}
\includegraphics[trim={0mm 4mm 12mm 31mm},clip,width=.99\textwidth]{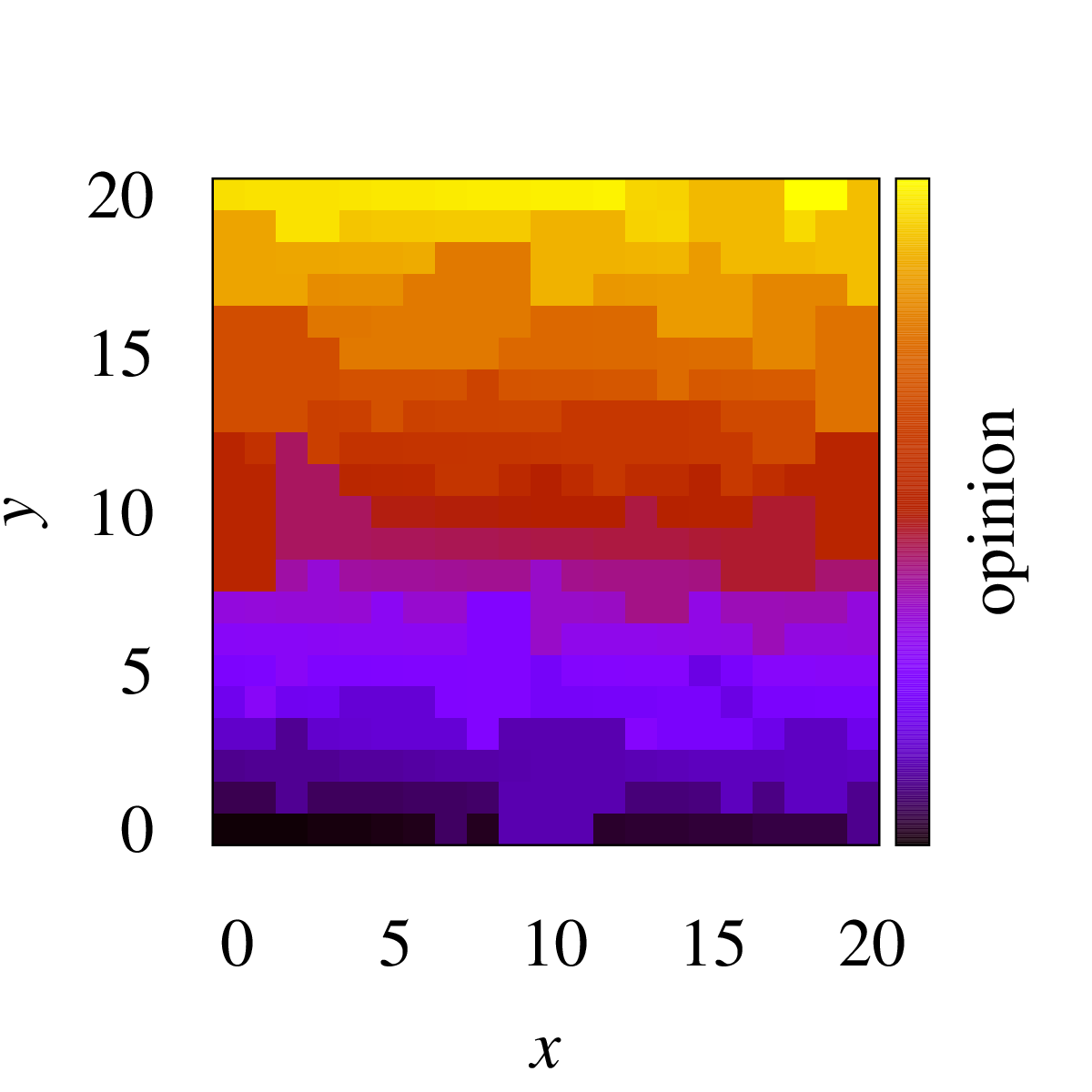}
\end{subfigure}
\hfill%% ---------------------------------------------------------------
\begin{subfigure}[t]{0.23\textwidth}
\caption{\label{subfig:L_a60T00_t=end}}
\includegraphics[trim={0mm 4mm 12mm 31mm},clip,width=.99\textwidth]{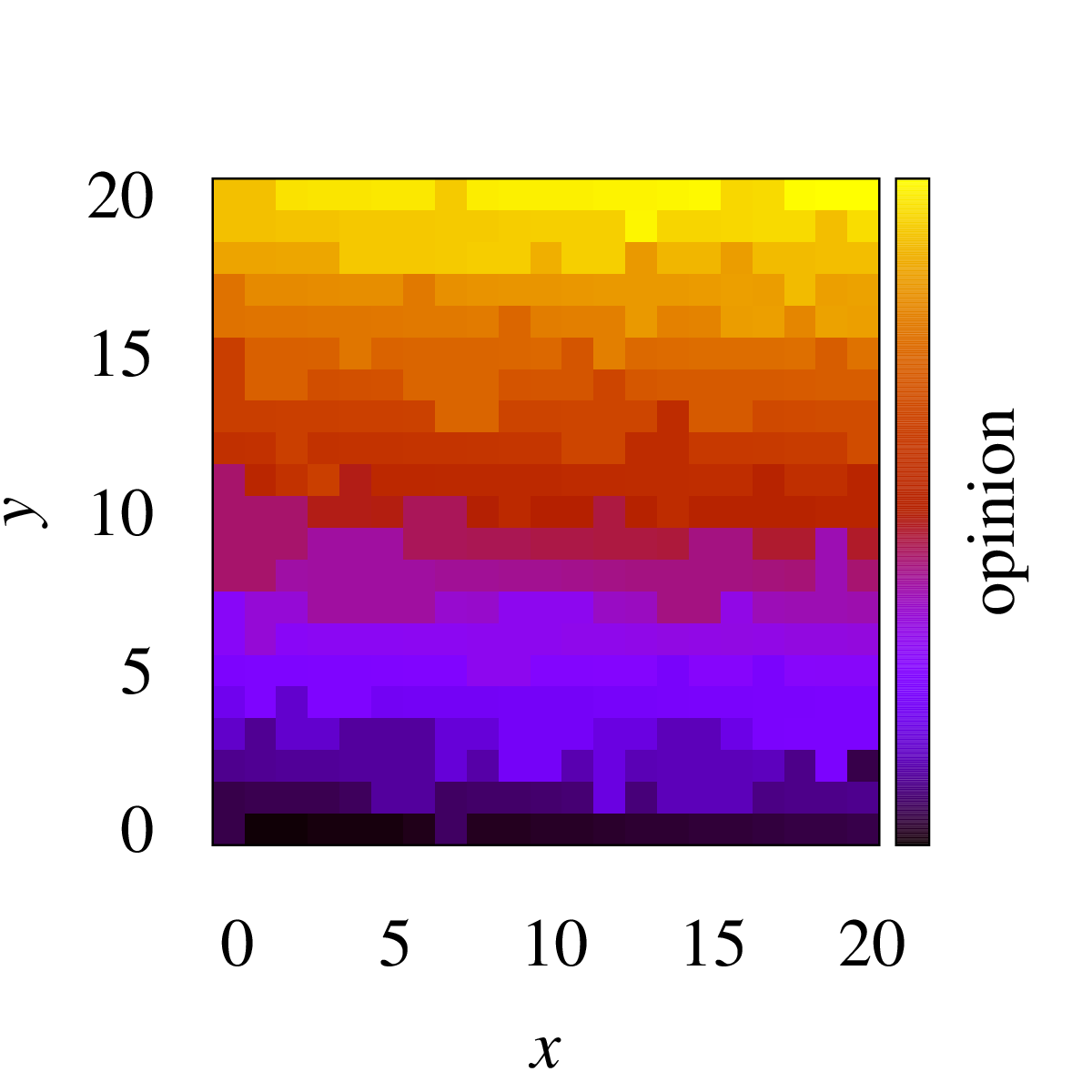}
\end{subfigure}
%% ---------------------------------------------------------------
\caption{\label{fig:L_snap_T0}Snapshots for system evolution with Latan\'e model. Synchronous update. $T=0$. $K=N$. 
\subref{subfig:L_a30T00_t=0000}%%, \subref{subfig:L_a40T00_t=0000}, \subref{subfig:L_a50T00_t=0000}, 
--\subref{subfig:L_a60T00_t=0000} $t=0$.
\subref{subfig:L_a30T00_t=0010}%%, \subref{subfig:L_a40T00_t=0010}, \subref{subfig:L_a50T00_t=0010}, 
--\subref{subfig:L_a60T00_t=0010} $t=10$,
\subref{subfig:L_a30T00_t=0100}%%, \subref{subfig:L_a40T00_t=0100}, \subref{subfig:L_a50T00_t=0100}, 
--\subref{subfig:L_a60T00_t=0100} $t=10^2$,
\subref{subfig:L_a30T00_t=1000}%%, \subref{subfig:L_a40T00_t=1000}, \subref{subfig:L_a50T00_t=1000}, 
--\subref{subfig:L_a60T00_t=1000} $t=10^3$,
\subref{subfig:L_a30T00_t=end}%%, \subref{subfig:L_a40T00_t=end}, \subref{subfig:L_a50T00_t=end}, 
--\subref{subfig:L_a60T00_t=end} $t\to\infty$.
Various effective ranges of interaction
\subref{subfig:L_a30T00_t=0000}, \subref{subfig:L_a30T00_t=0010}, \subref{subfig:L_a30T00_t=0100}, \subref{subfig:L_a30T00_t=1000}, \subref{subfig:L_a30T00_t=end} $\alpha=3$, 
\subref{subfig:L_a40T00_t=0000}, \subref{subfig:L_a40T00_t=0010}, \subref{subfig:L_a40T00_t=0100}, \subref{subfig:L_a40T00_t=1000}, \subref{subfig:L_a40T00_t=end} $\alpha=4$, 
\subref{subfig:L_a50T00_t=0000}, \subref{subfig:L_a50T00_t=0010}, \subref{subfig:L_a50T00_t=0100}, \subref{subfig:L_a50T00_t=1000}, \subref{subfig:L_a50T00_t=end} $\alpha=5$, 
\subref{subfig:L_a60T00_t=0000}, \subref{subfig:L_a60T00_t=0010}, \subref{subfig:L_a60T00_t=0100}, \subref{subfig:L_a60T00_t=1000}, \subref{subfig:L_a60T00_t=end} $\alpha=6$}
%% ---------------------------------------------------------------
\end{figure*}

\begin{figure*}[htbp]
%% ---------------------------------------------------------------
\begin{subfigure}[t]{0.23\textwidth}
\caption{\label{subfig:L_a20T10_t=0000}$\alpha=2$, $T=1$}
\includegraphics[trim={0mm 4mm 12mm 31mm},clip,width=.99\textwidth]{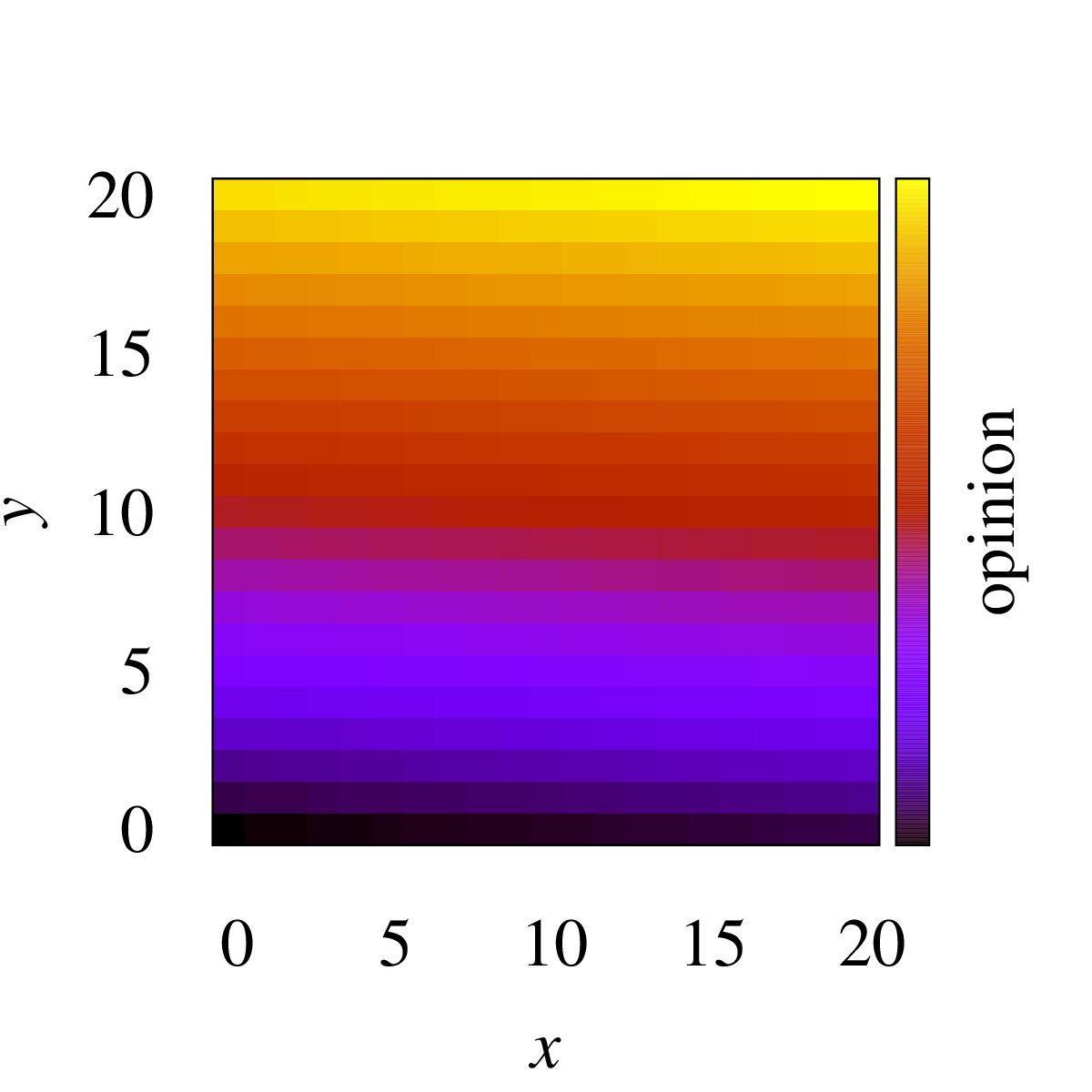}
\end{subfigure}
\hfill%% ---------------------------------------------------------------
\begin{subfigure}[t]{0.23\textwidth}
\caption{\label{subfig:L_a30T10_t=0000}$\alpha=3$, $T=1$}
\includegraphics[trim={0mm 4mm 12mm 31mm},clip,width=.99\textwidth]{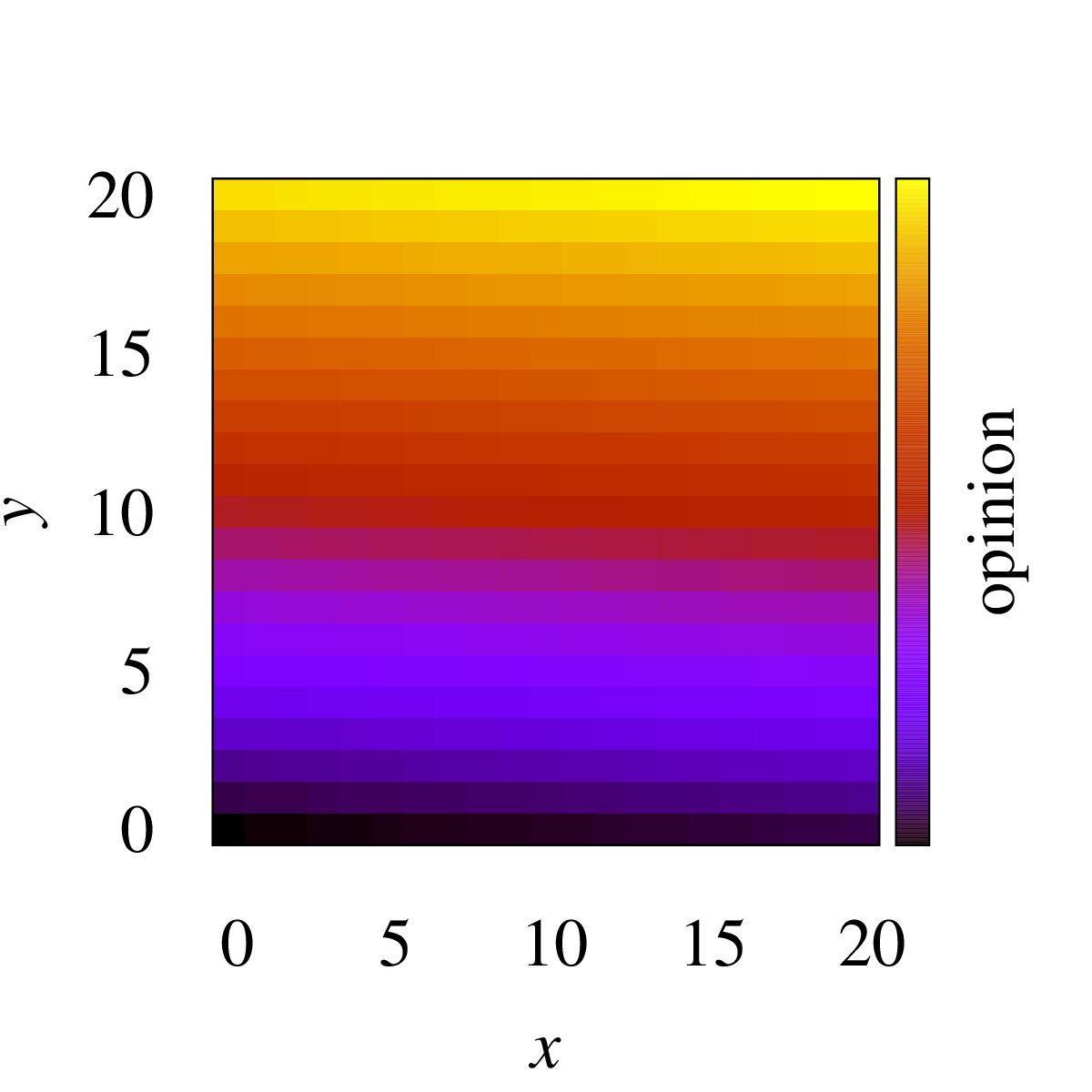}
\end{subfigure}
\hfill%% ---------------------------------------------------------------
\begin{subfigure}[t]{0.23\textwidth}
\caption{\label{subfig:L_a50T10_t=0000}$\alpha=5$, $T=1$}
\includegraphics[trim={0mm 4mm 12mm 31mm},clip,width=.99\textwidth]{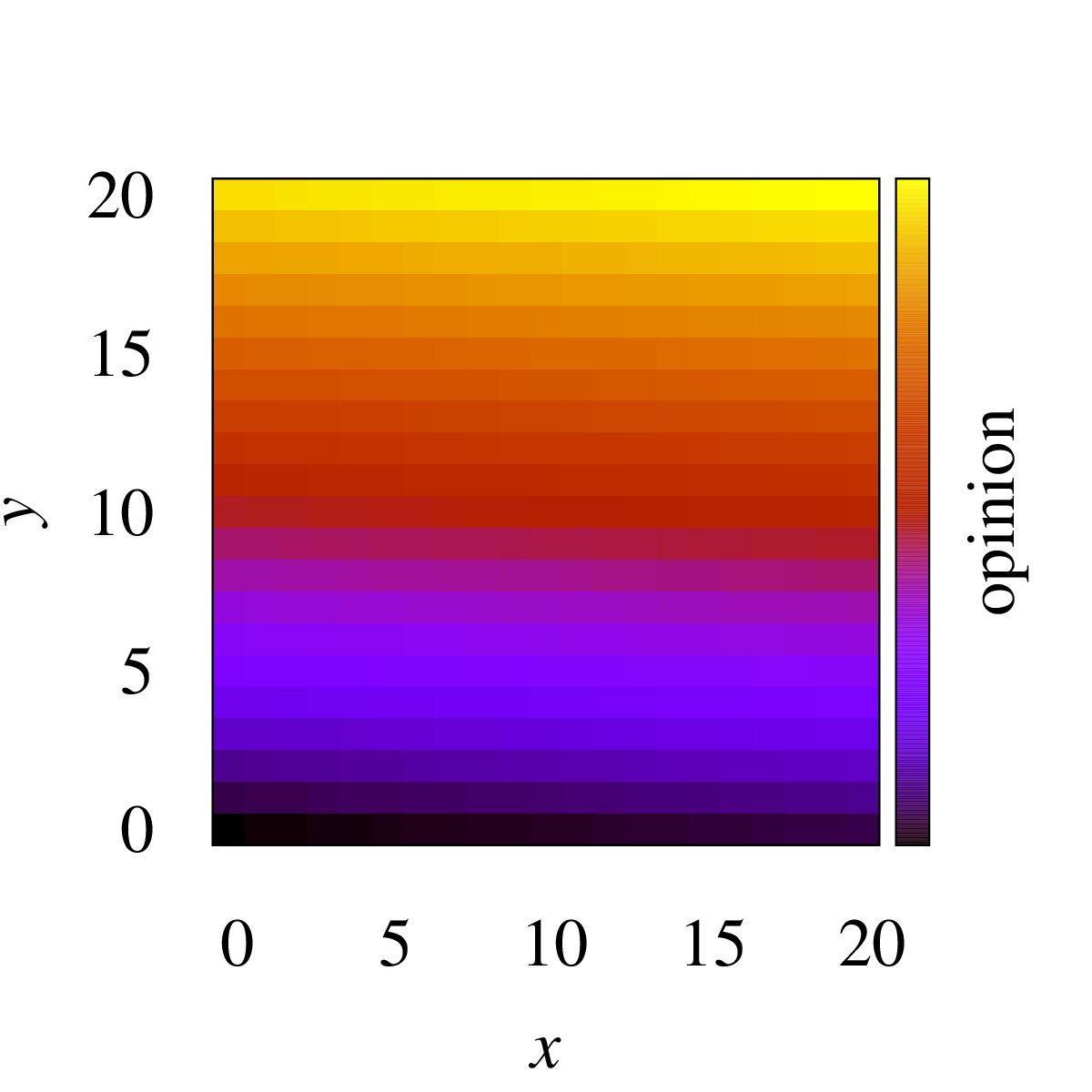}
\end{subfigure}
\hfill%% ---------------------------------------------------------------
\begin{subfigure}[t]{0.23\textwidth}
\caption{\label{subfig:L_a60T10_t=0000}$\alpha=6$, $T=1$}
\includegraphics[trim={0mm 4mm 12mm 31mm},clip,width=.99\textwidth]{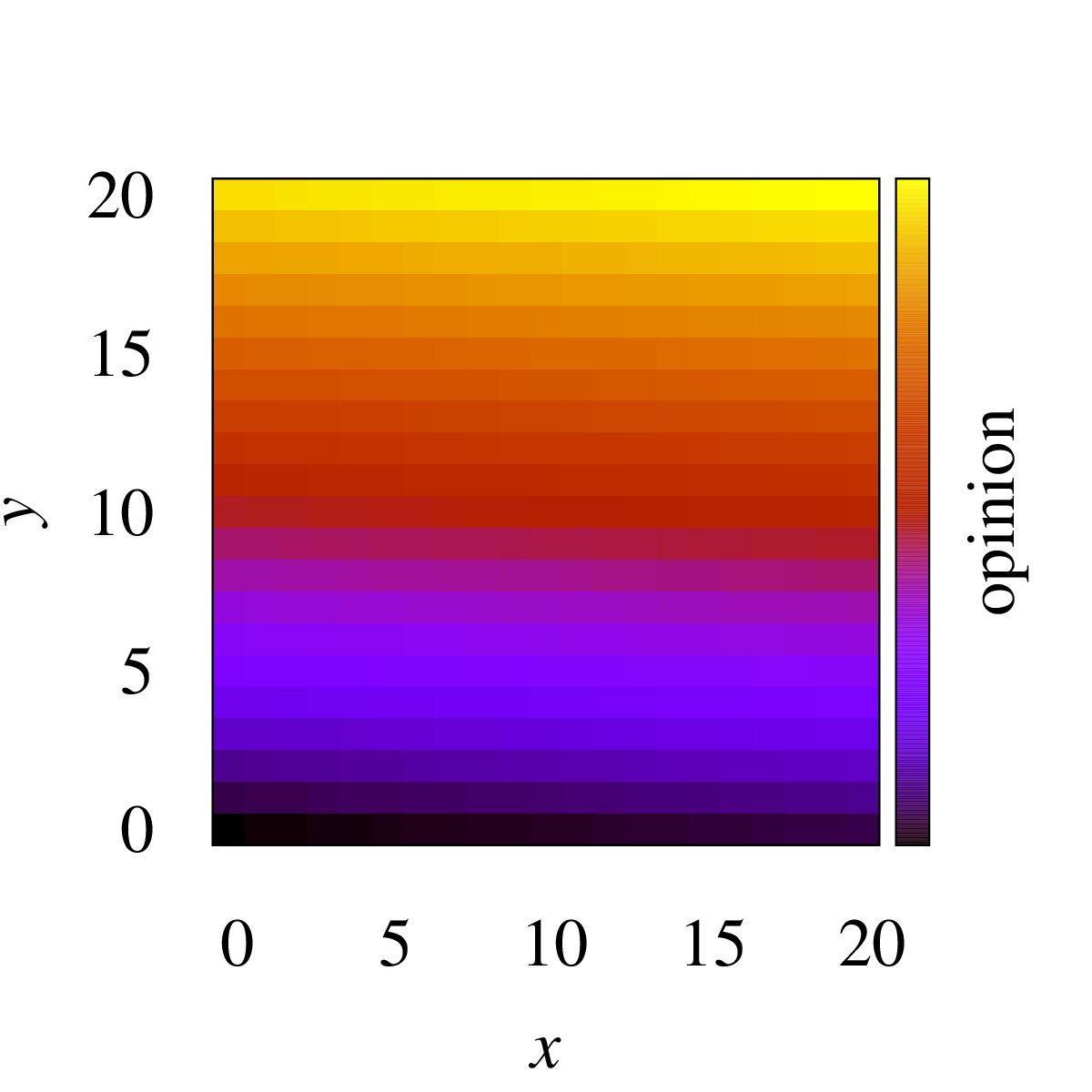}
\end{subfigure}
%% ---------------------------------------------------------------
%% ===============================================================
%% ---------------------------------------------------------------
\begin{subfigure}[t]{0.23\textwidth}
\caption{\label{subfig:L_a20T10_t=0005}$t=5$}
\includegraphics[trim={0mm 4mm 12mm 31mm},clip,width=.99\textwidth]{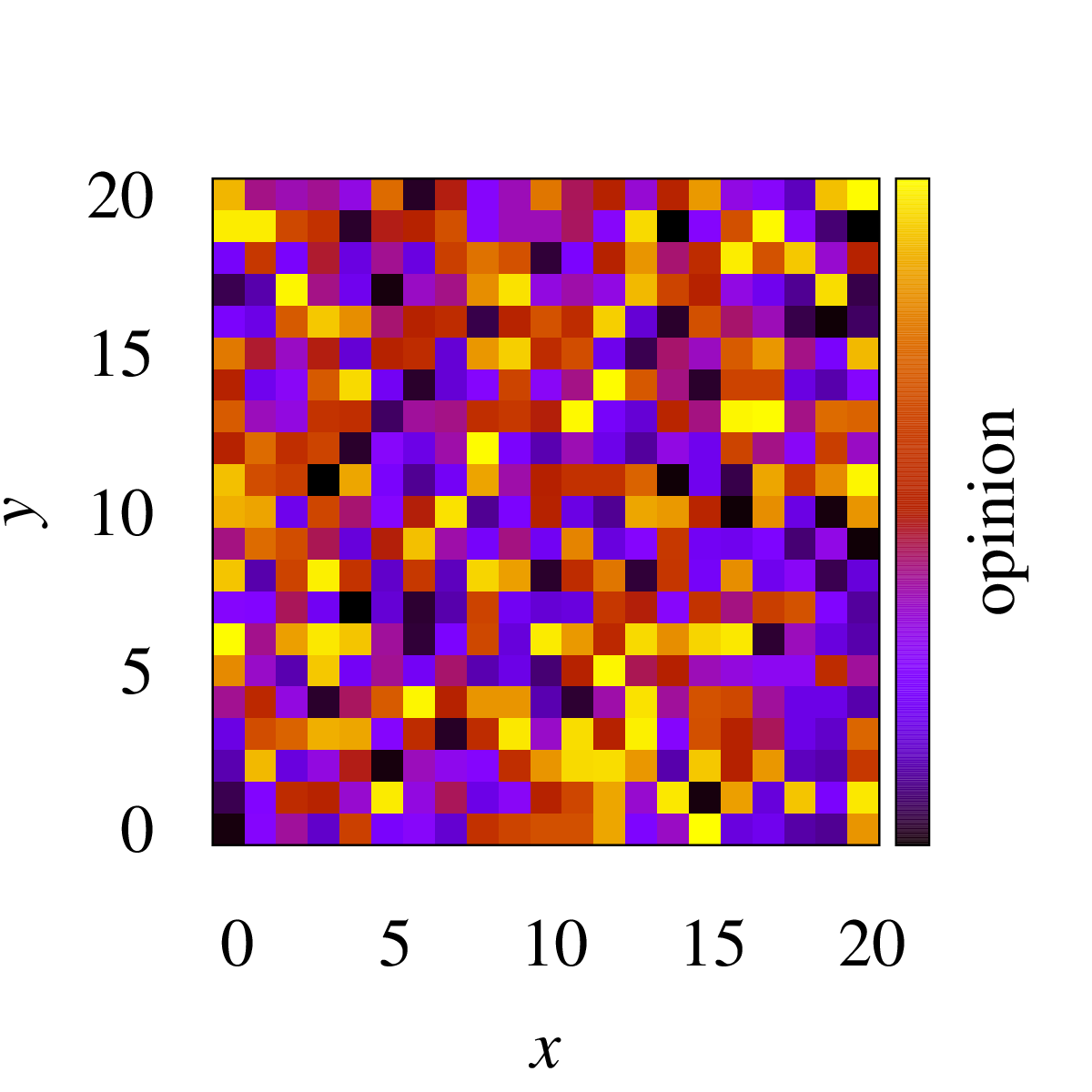}
\end{subfigure}
\hfill%% ---------------------------------------------------------------
\begin{subfigure}[t]{0.23\textwidth}
\caption{\label{subfig:L_a30T10_t=0005}$t=5$}
\includegraphics[trim={0mm 4mm 12mm 31mm},clip,width=.99\textwidth]{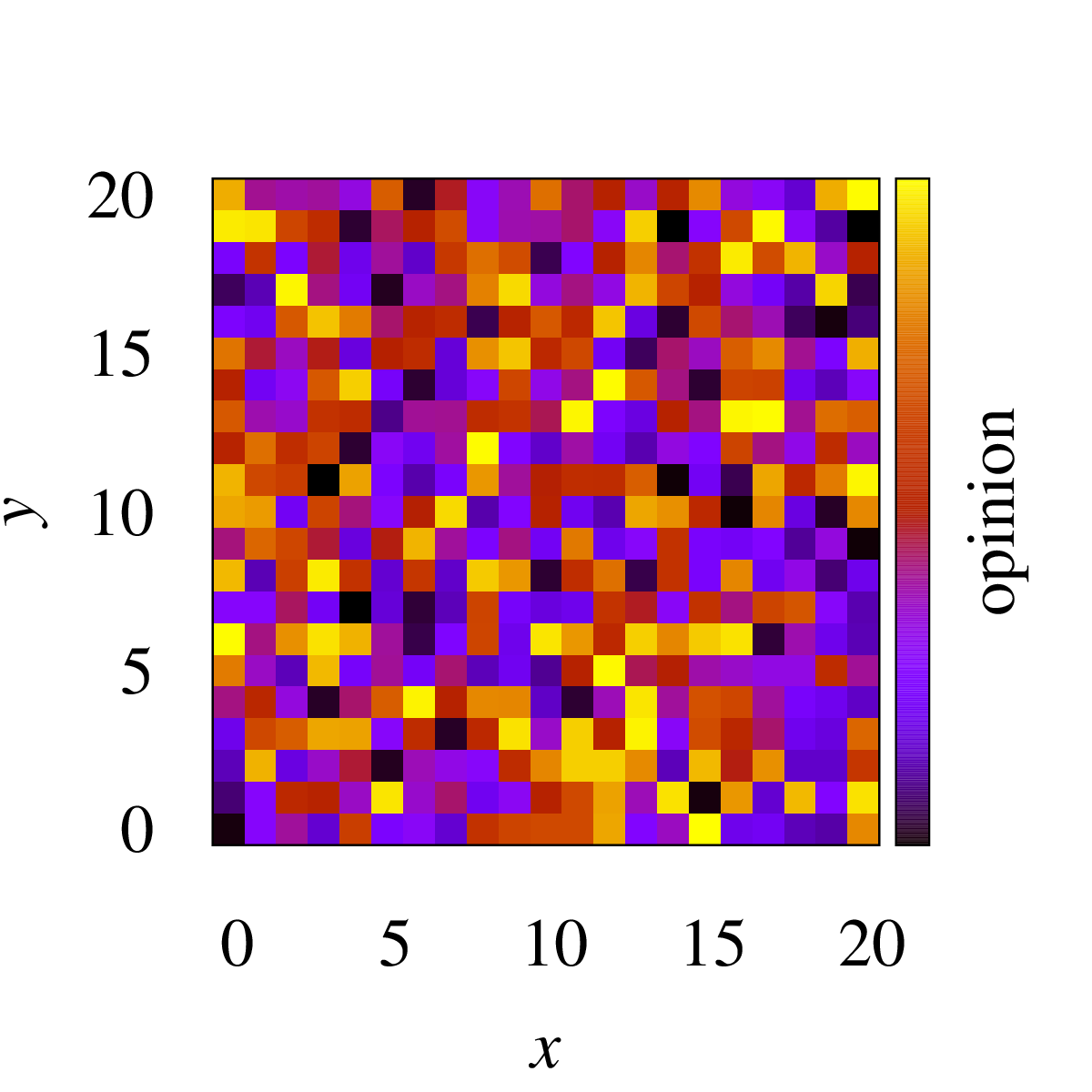}
\end{subfigure}
\hfill%% ---------------------------------------------------------------
\begin{subfigure}[t]{0.23\textwidth}
\caption{\label{subfig:L_a50T10_t=0100}$t=10^2$}
\includegraphics[trim={0mm 4mm 12mm 31mm},clip,width=.99\textwidth]{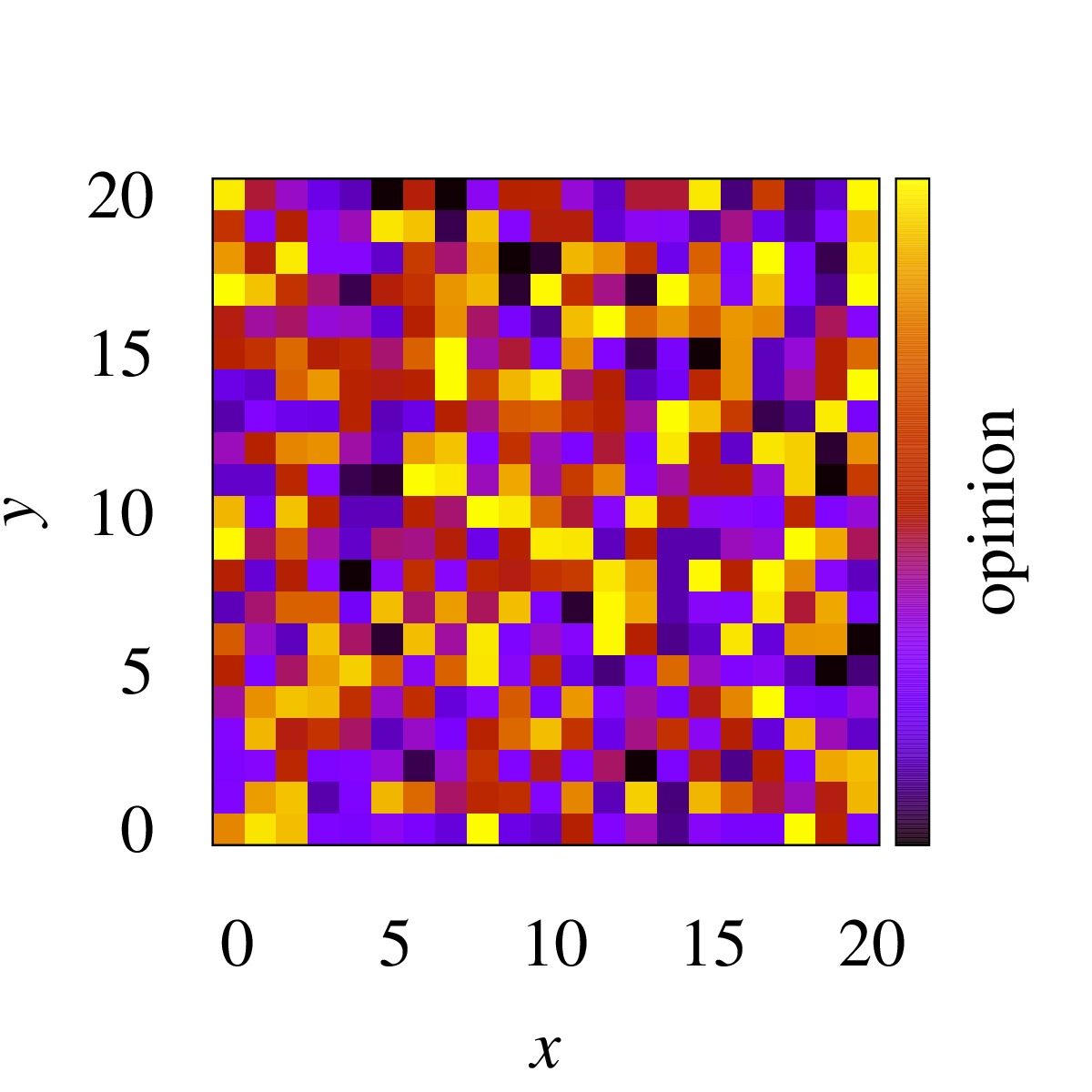}
\end{subfigure}
\hfill%% ---------------------------------------------------------------
\begin{subfigure}[t]{0.23\textwidth}
\caption{\label{subfig:L_a60T10_t=5e3}$t=5\times 10^3$}
\includegraphics[trim={0mm 4mm 12mm 31mm},clip,width=.99\textwidth]{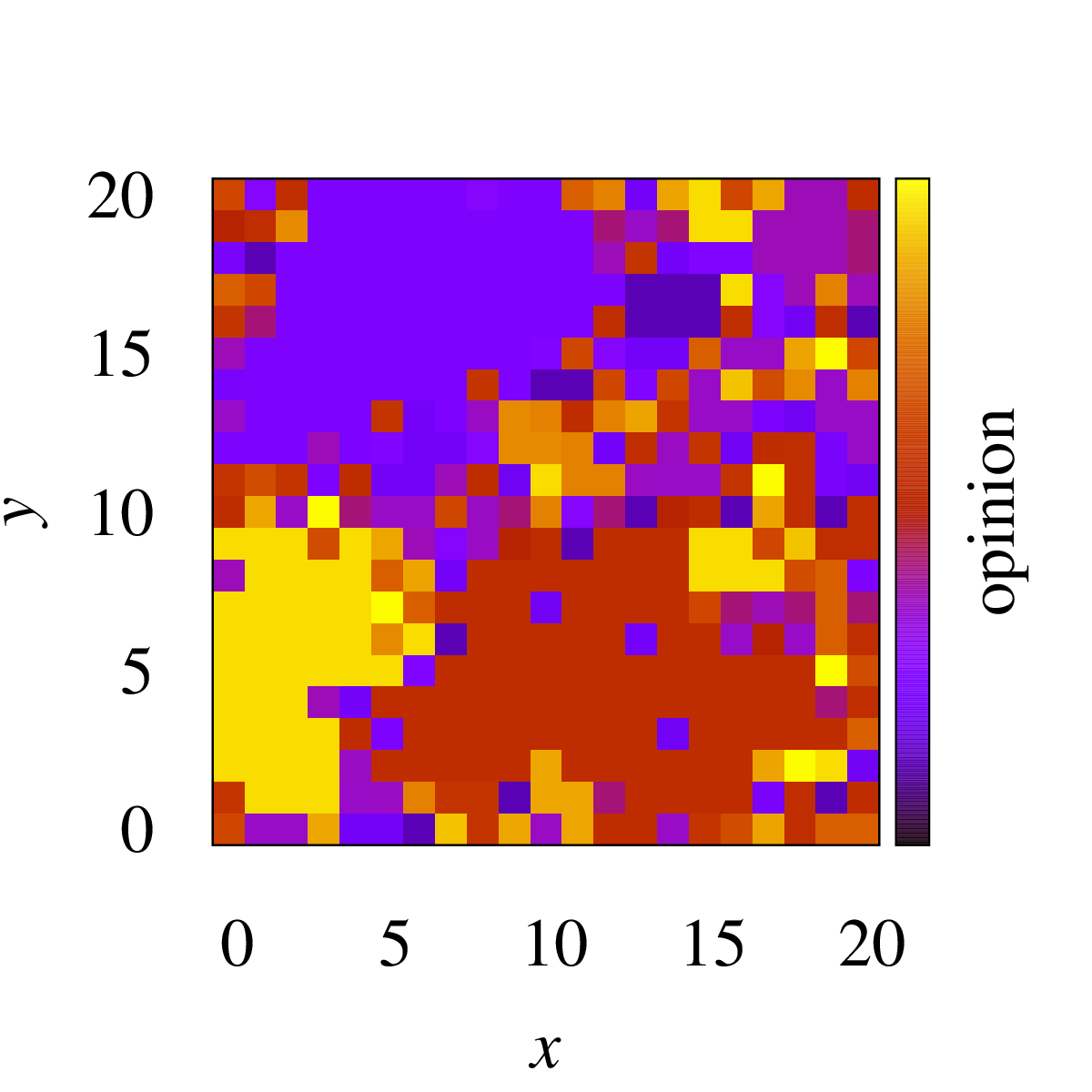}
\end{subfigure}
%% ---------------------------------------------------------------
%% ===============================================================
%% ---------------------------------------------------------------
\begin{subfigure}[t]{0.23\textwidth}
\caption{\label{subfig:L_a20T10_t=0010}$t=10$}
\includegraphics[trim={0mm 4mm 12mm 31mm},clip,width=.99\textwidth]{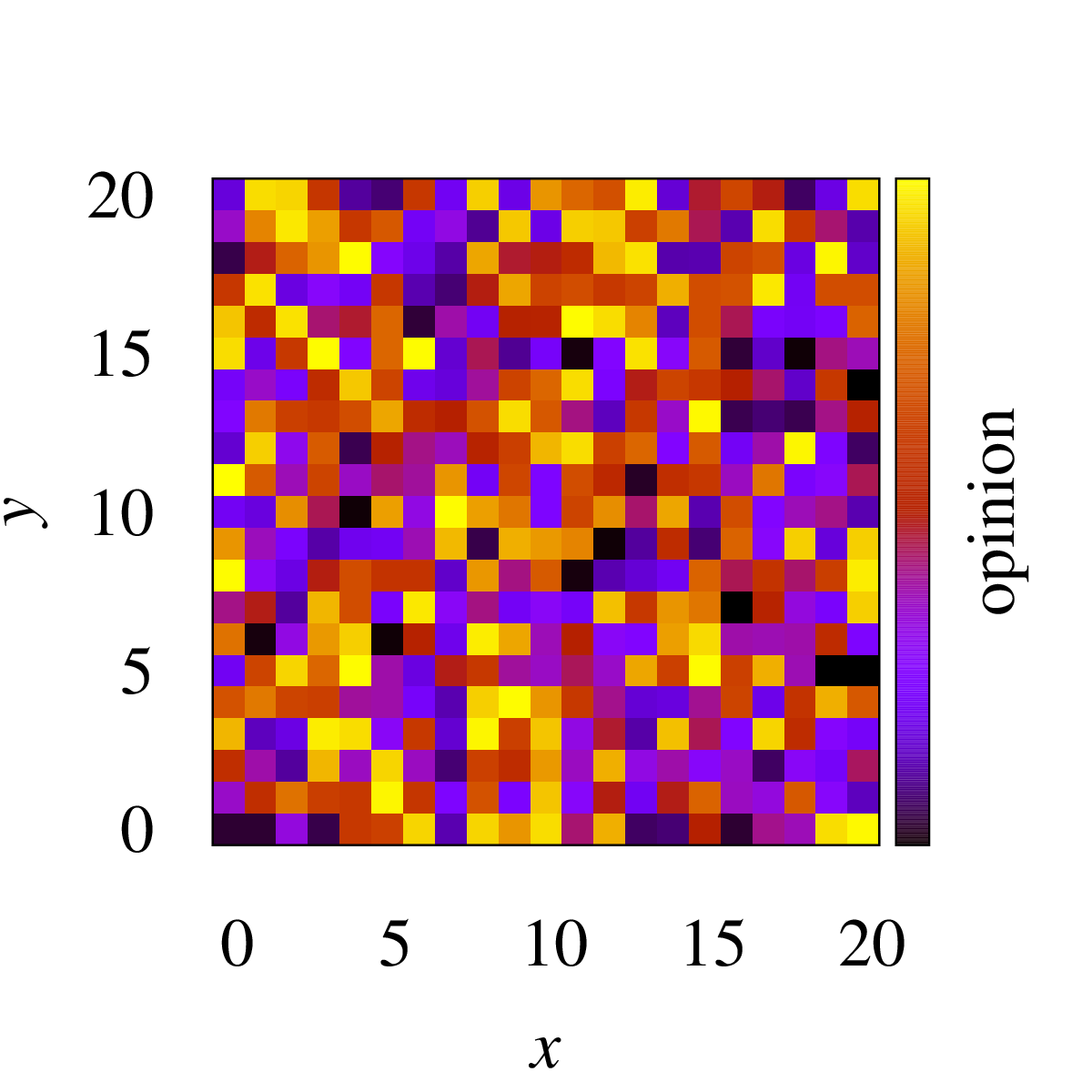}
\end{subfigure}
\hfill%% ---------------------------------------------------------------
\begin{subfigure}[t]{0.23\textwidth}
\caption{\label{subfig:L_a30T10_t=0010}$t=10$}
\includegraphics[trim={0mm 4mm 12mm 31mm},clip,width=.99\textwidth]{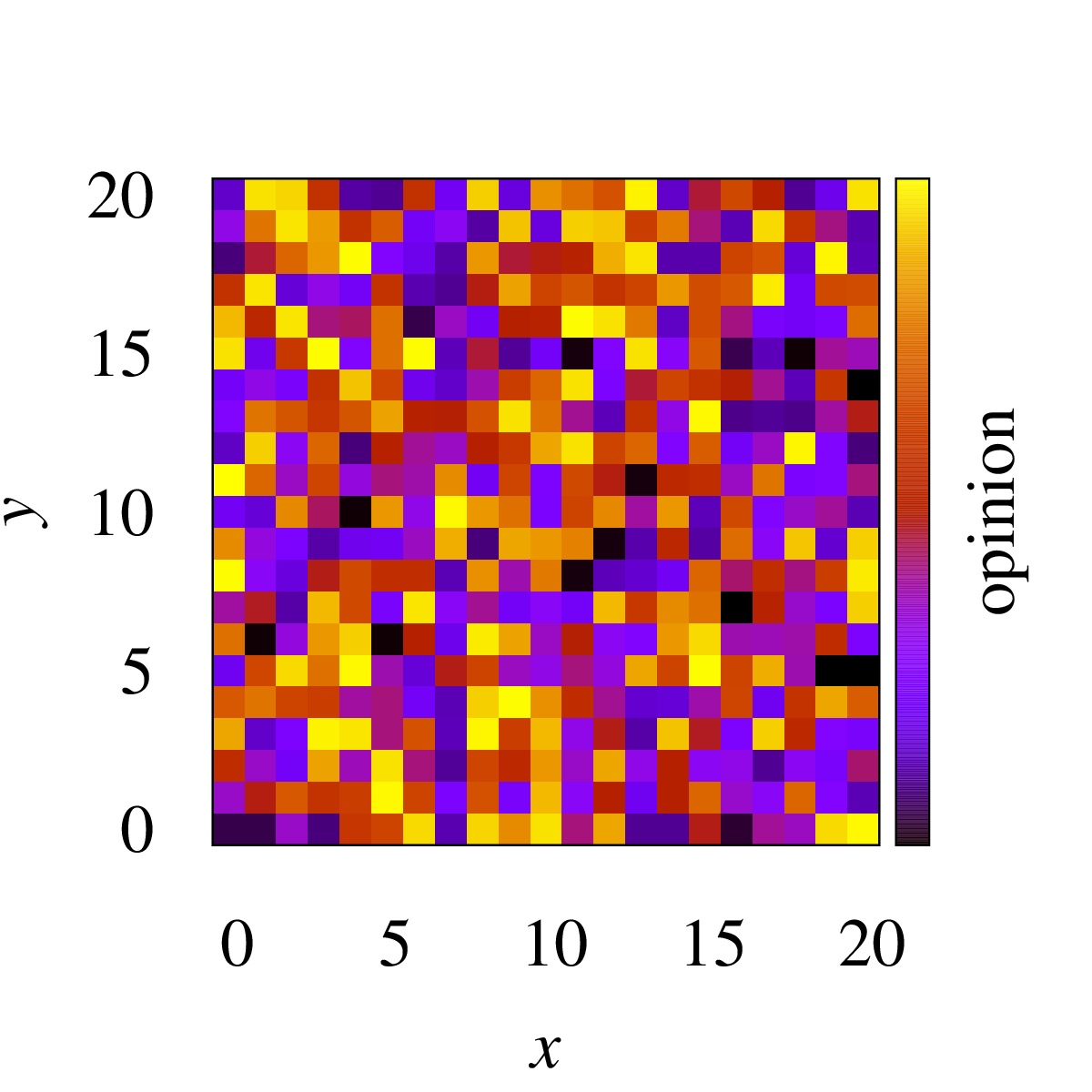}
\end{subfigure}
\hfill%% ---------------------------------------------------------------
\begin{subfigure}[t]{0.23\textwidth}
\caption{\label{subfig:L_a50T10_t=1000}$t=10^3$}
\includegraphics[trim={0mm 4mm 12mm 31mm},clip,width=.99\textwidth]{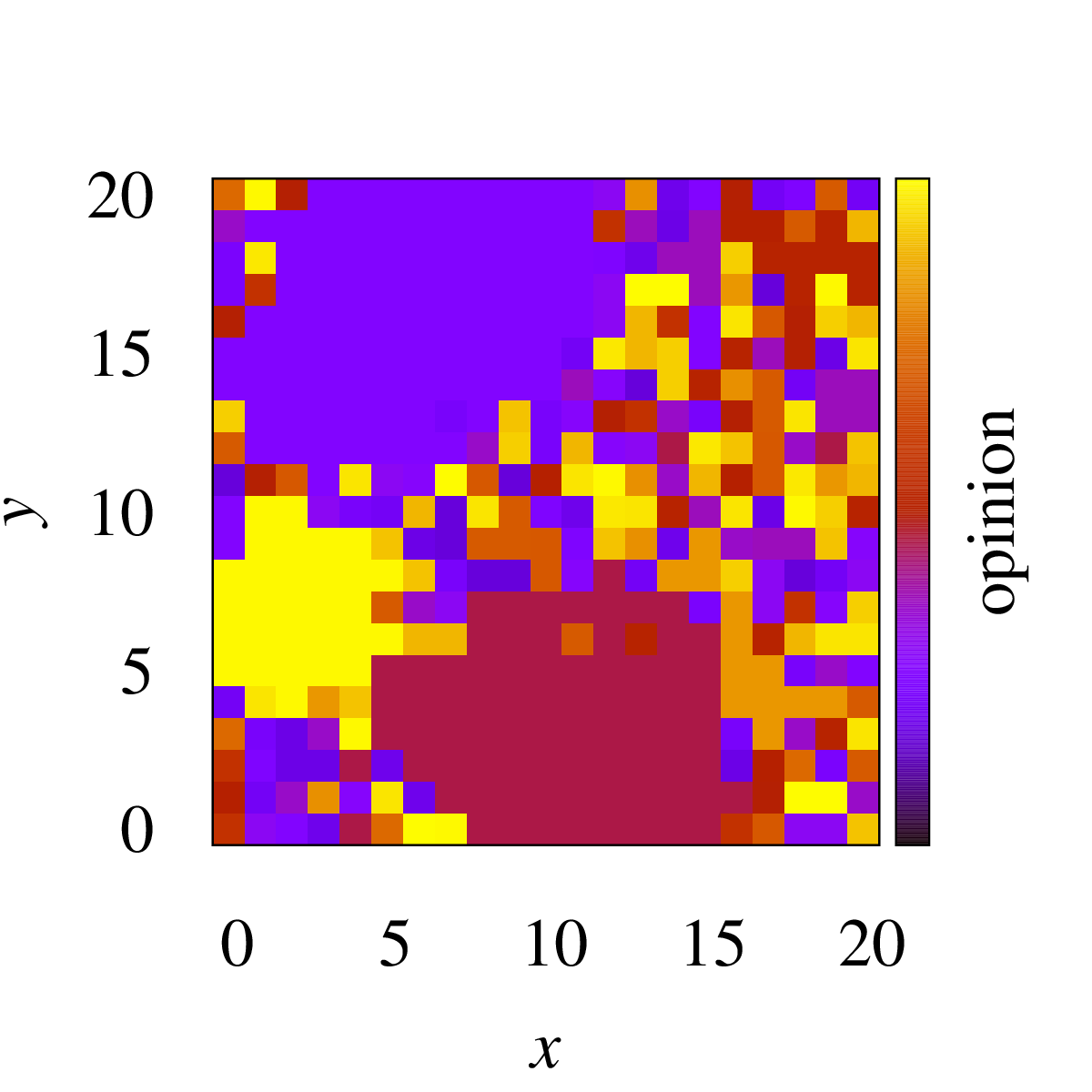}
\end{subfigure}
\hfill%% ---------------------------------------------------------------
\begin{subfigure}[t]{0.23\textwidth}
\caption{\label{subfig:L_a60T10_t=1e4}$t=10^4$}
\includegraphics[trim={0mm 4mm 12mm 31mm},clip,width=.99\textwidth]{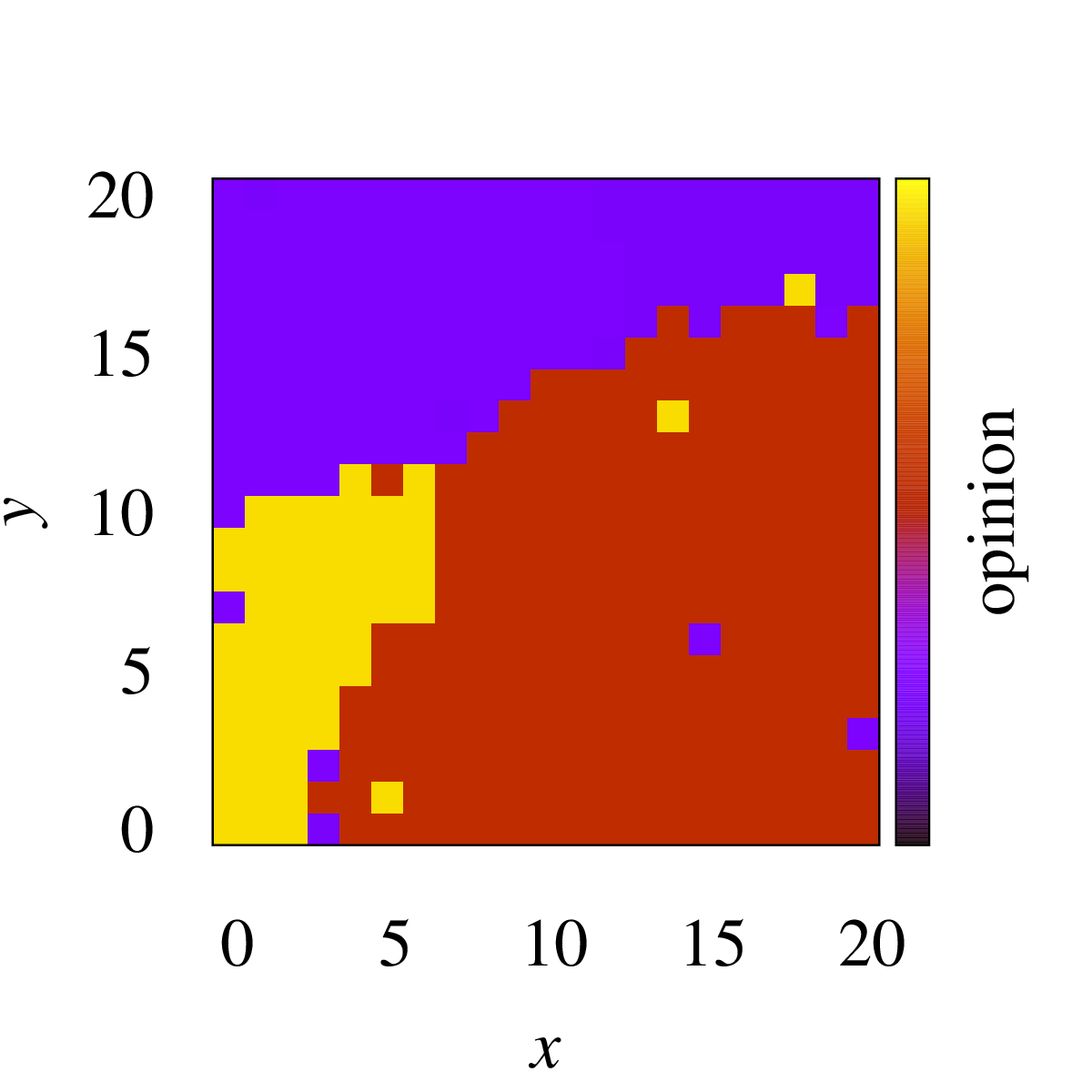}
\end{subfigure}
%% ---------------------------------------------------------------
%% ===============================================================
%% ---------------------------------------------------------------
\begin{subfigure}[t]{0.23\textwidth}
\caption{\label{subfig:L_a20T10_t=0050}$t=50$}
\includegraphics[trim={0mm 4mm 12mm 31mm},clip,width=.99\textwidth]{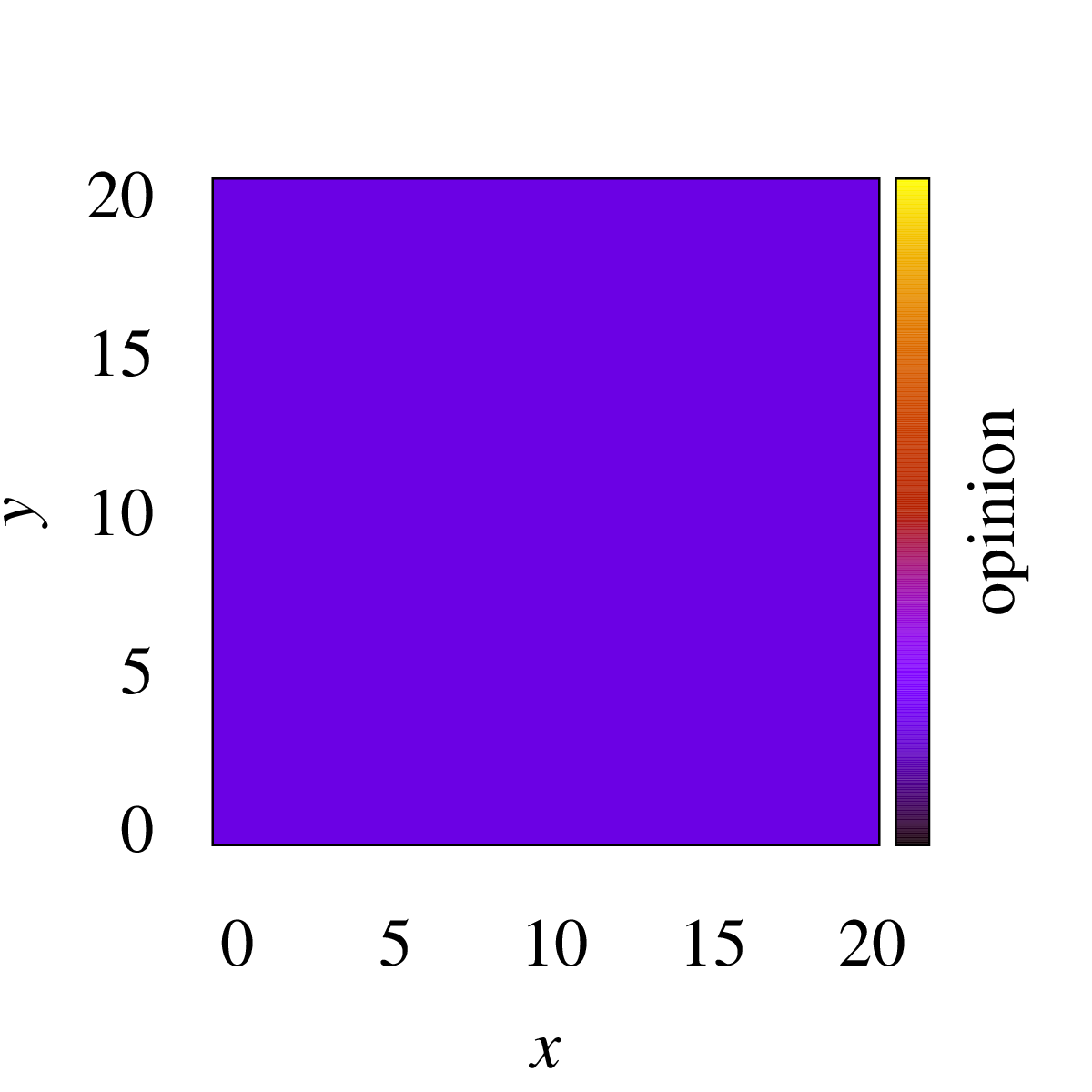}
\end{subfigure}
\hfill%% ---------------------------------------------------------------
\begin{subfigure}[t]{0.23\textwidth}
\caption{\label{subfig:L_a30T10_t=0050}$t=50$}
\includegraphics[trim={0mm 4mm 12mm 31mm},clip,width=.99\textwidth]{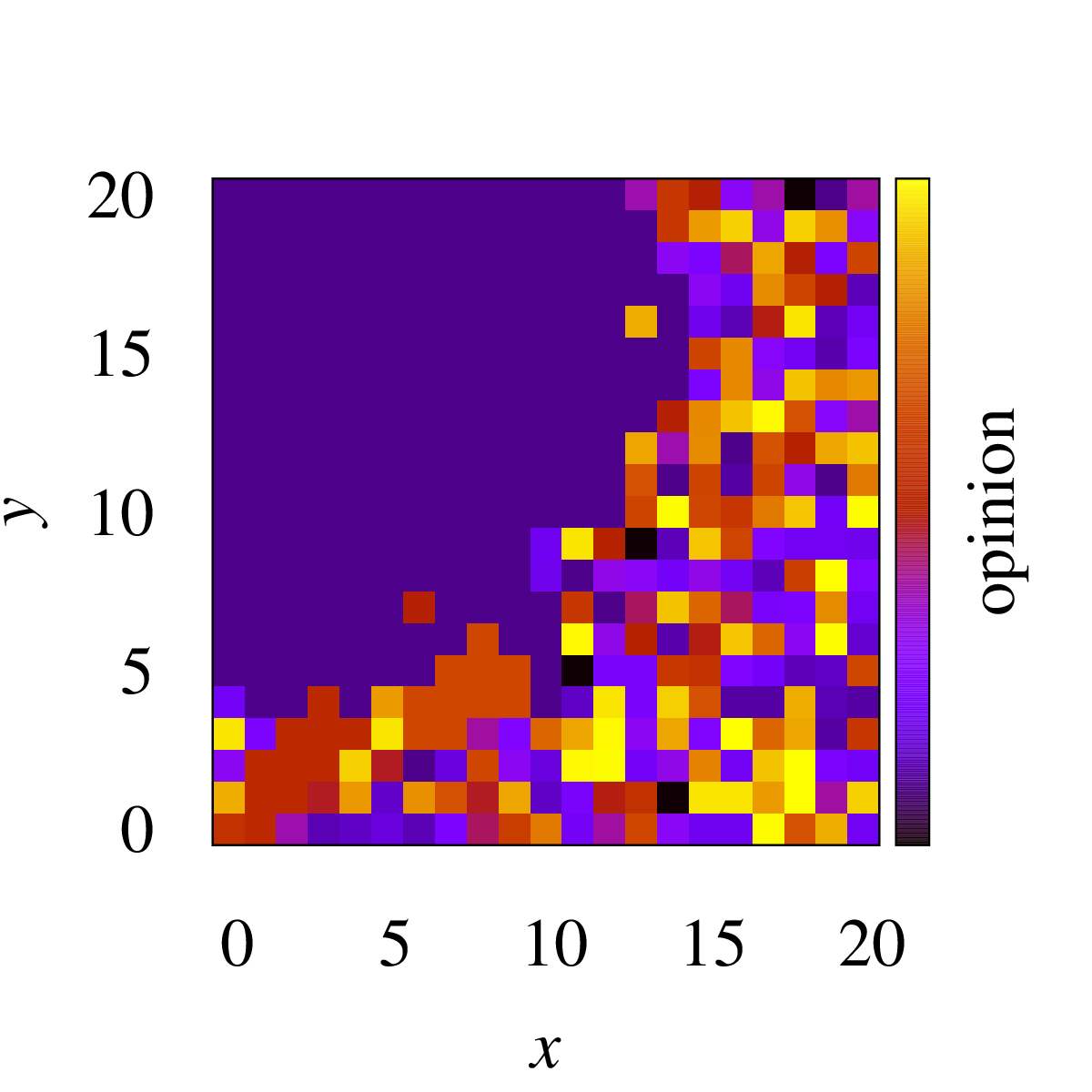}
\end{subfigure}
\hfill%% ---------------------------------------------------------------
\begin{subfigure}[t]{0.23\textwidth}
\caption{\label{subfig:L_a50T10_t=5000}$t=5\times 10^3$}
\includegraphics[trim={0mm 4mm 12mm 31mm},clip,width=.99\textwidth]{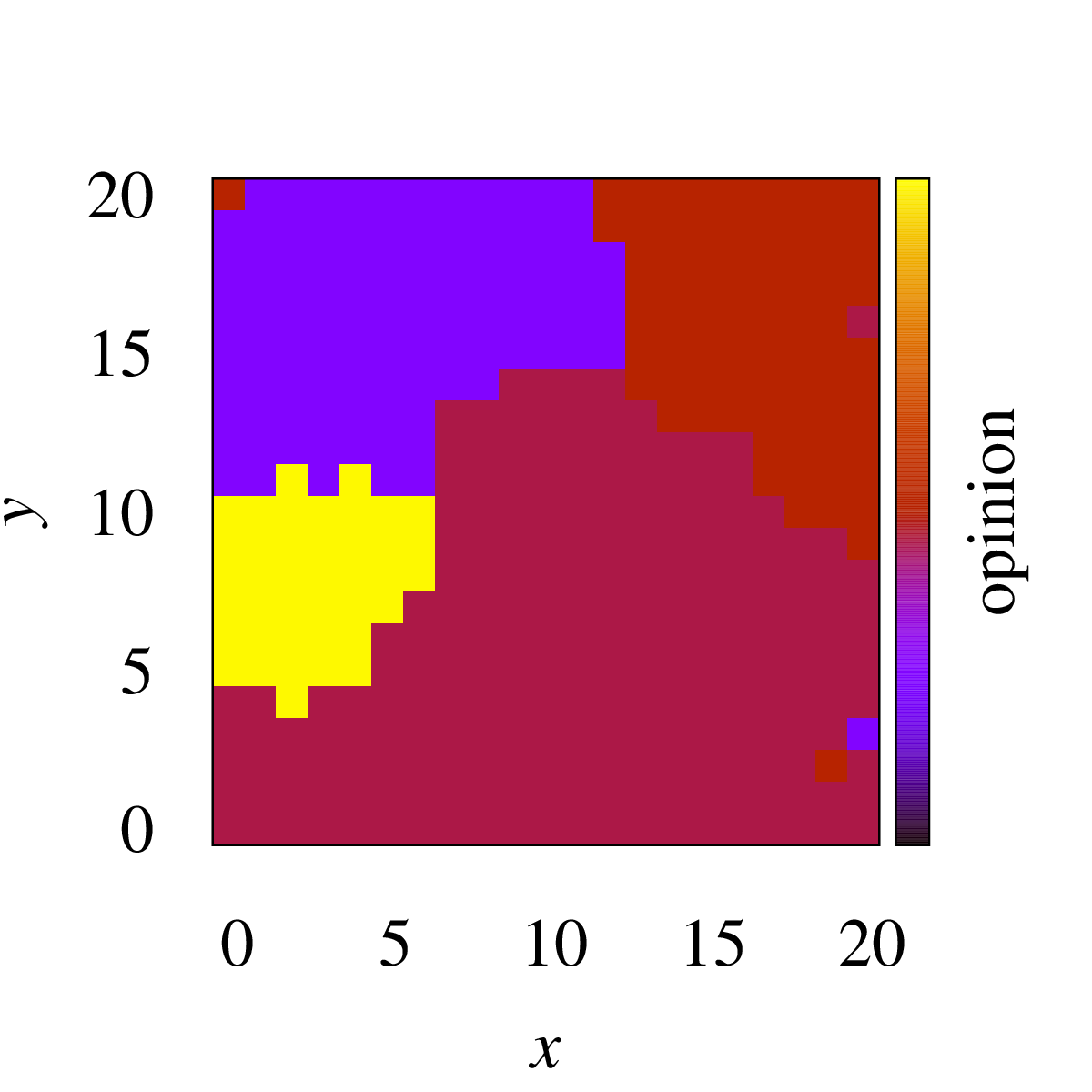}
\end{subfigure}
\hfill%% ---------------------------------------------------------------
\begin{subfigure}[t]{0.23\textwidth}
\caption{\label{subfig:L_a60T10_t=5e4}$t=5\times 10^4$}
\includegraphics[trim={0mm 4mm 12mm 31mm},clip,width=.99\textwidth]{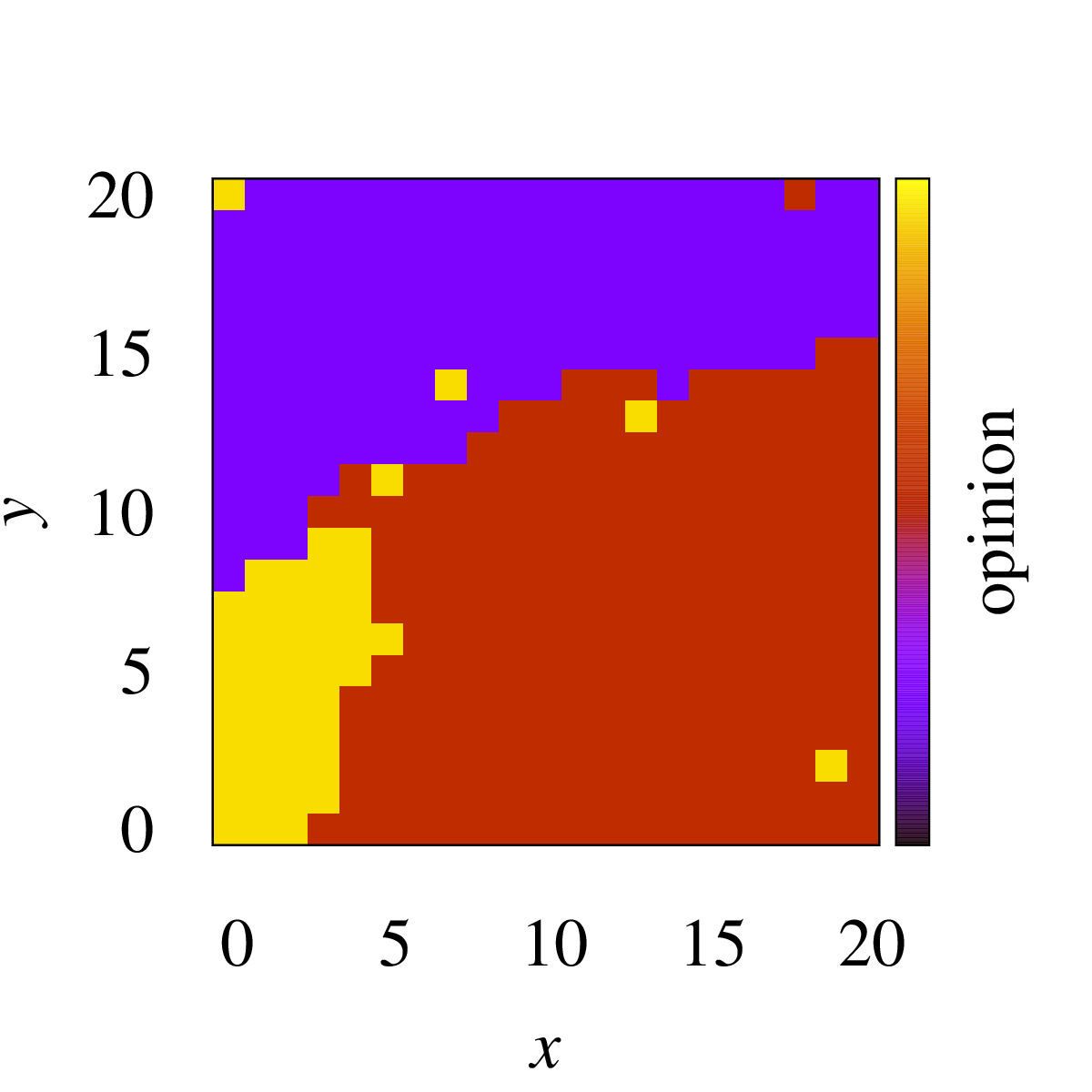}
\end{subfigure}
%% ---------------------------------------------------------------
%% ===============================================================
%% ---------------------------------------------------------------
\begin{subfigure}[t]{0.23\textwidth}
\caption{\label{subfig:L_a20T10_t=end}$\tau=25$}
\includegraphics[trim={0mm 4mm 12mm 31mm},clip,width=.99\textwidth]{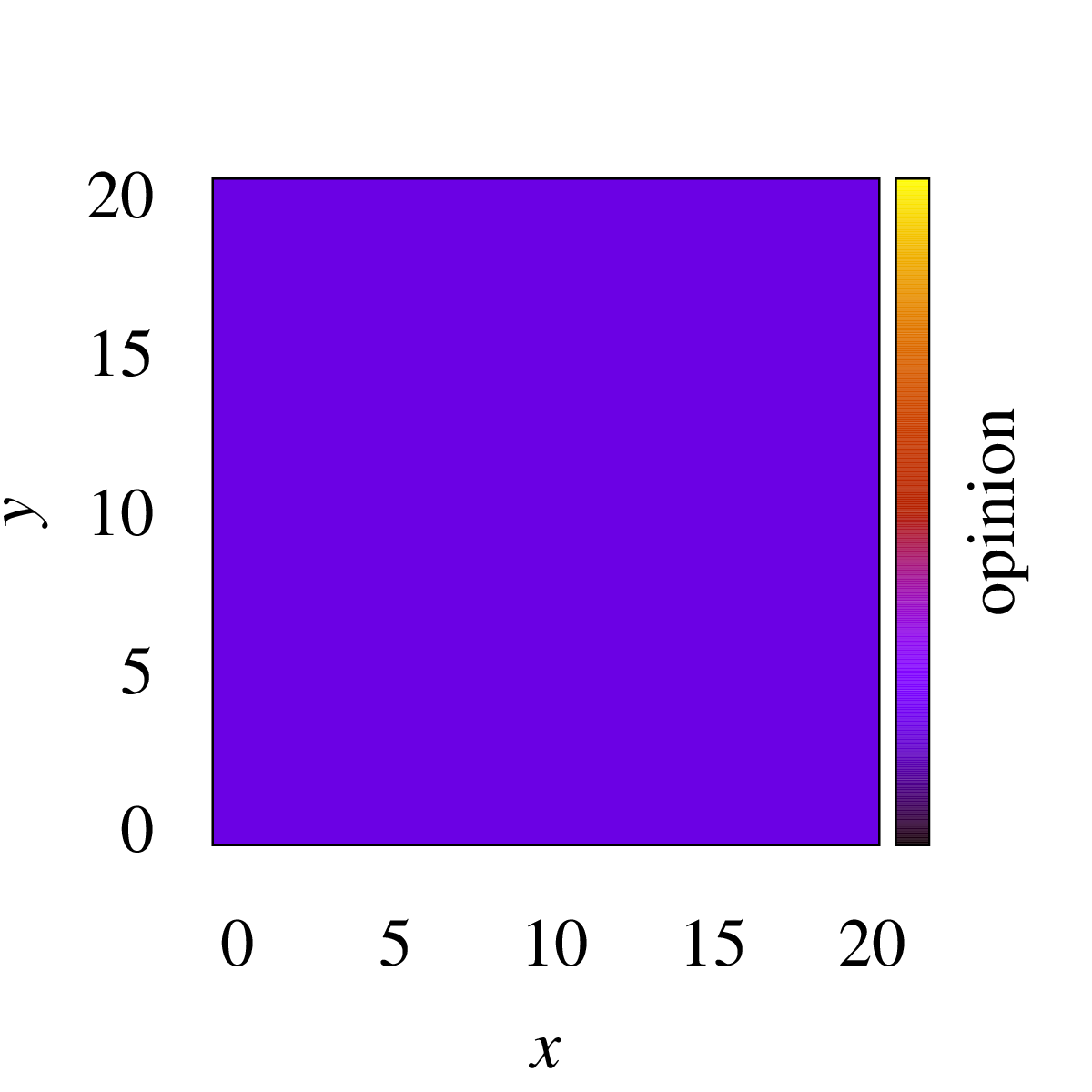}
\end{subfigure}
\hfill%% ---------------------------------------------------------------
\begin{subfigure}[t]{0.23\textwidth}
\caption{\label{subfig:L_a30T10_t=end}$\tau=103$}
\includegraphics[trim={0mm 4mm 12mm 31mm},clip,width=.99\textwidth]{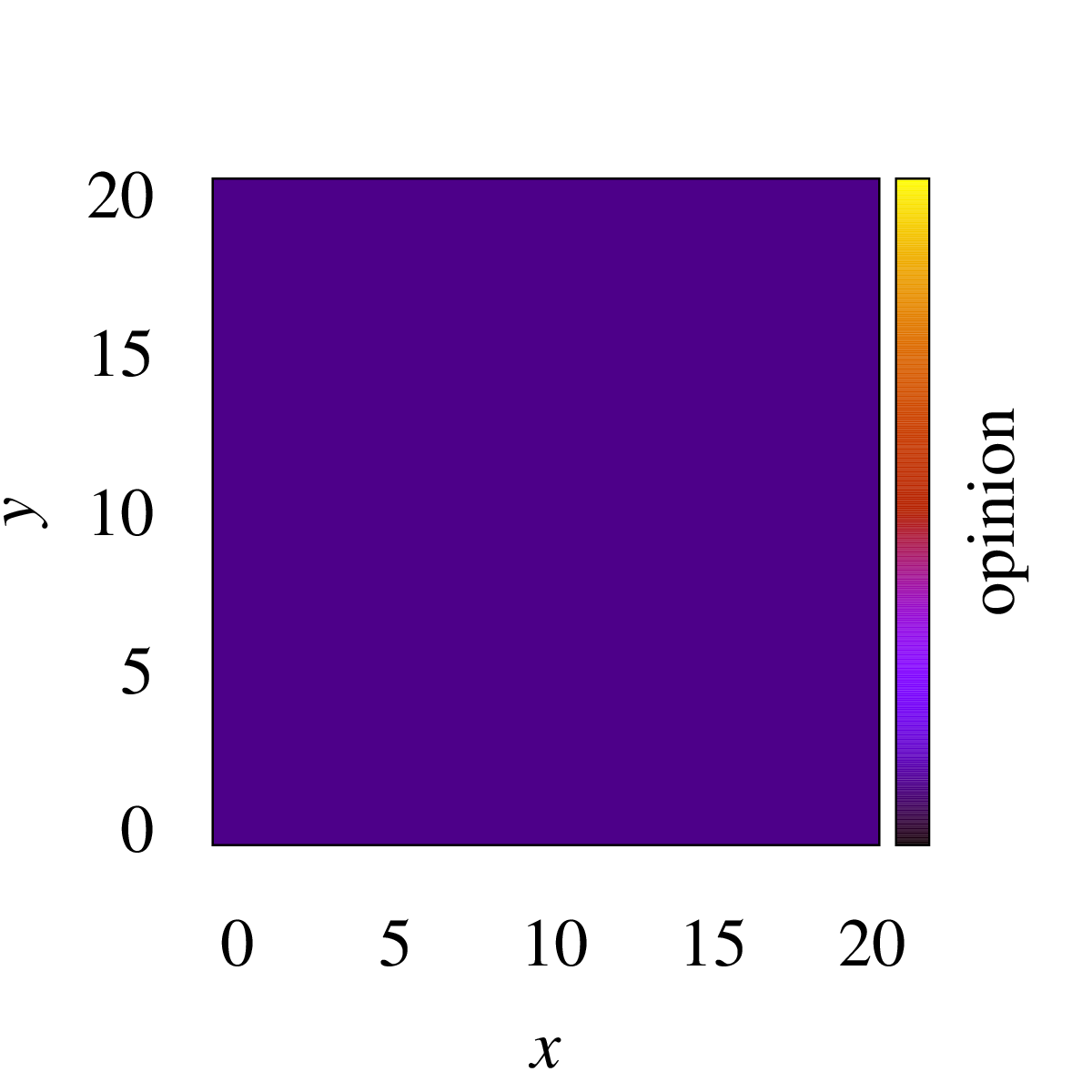}
\end{subfigure}
\hfill%% ---------------------------------------------------------------
\begin{subfigure}[t]{0.23\textwidth}
\caption{\label{subfig:L_a50T10_t=end}$t=10^5$}
\includegraphics[trim={0mm 4mm 12mm 31mm},clip,width=.99\textwidth]{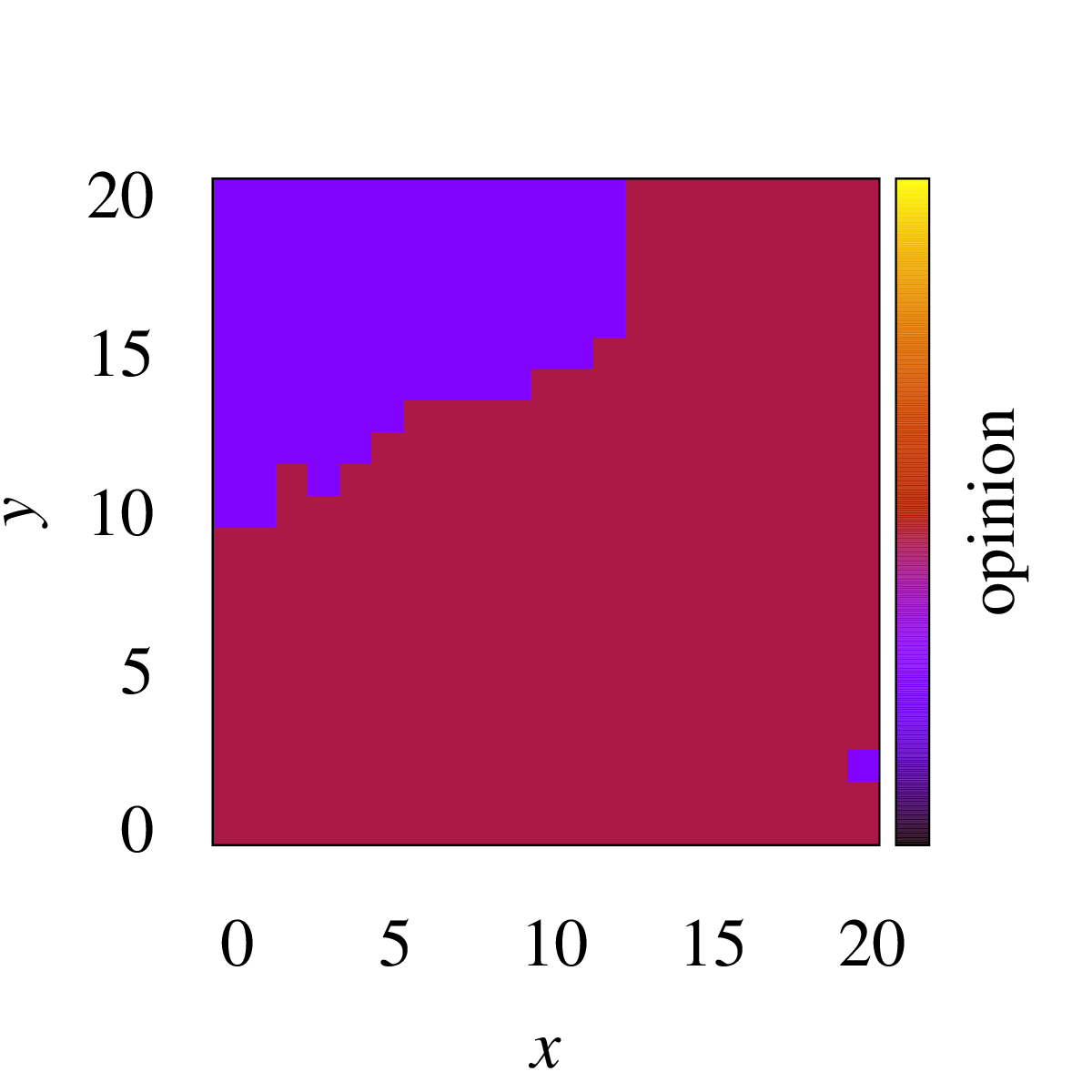}
\end{subfigure}
\hfill%% ---------------------------------------------------------------
\begin{subfigure}[t]{0.23\textwidth}
\caption{\label{subfig:L_a60T10_t=end}$t=10^5$}
\includegraphics[trim={0mm 4mm 12mm 31mm},clip,width=.99\textwidth]{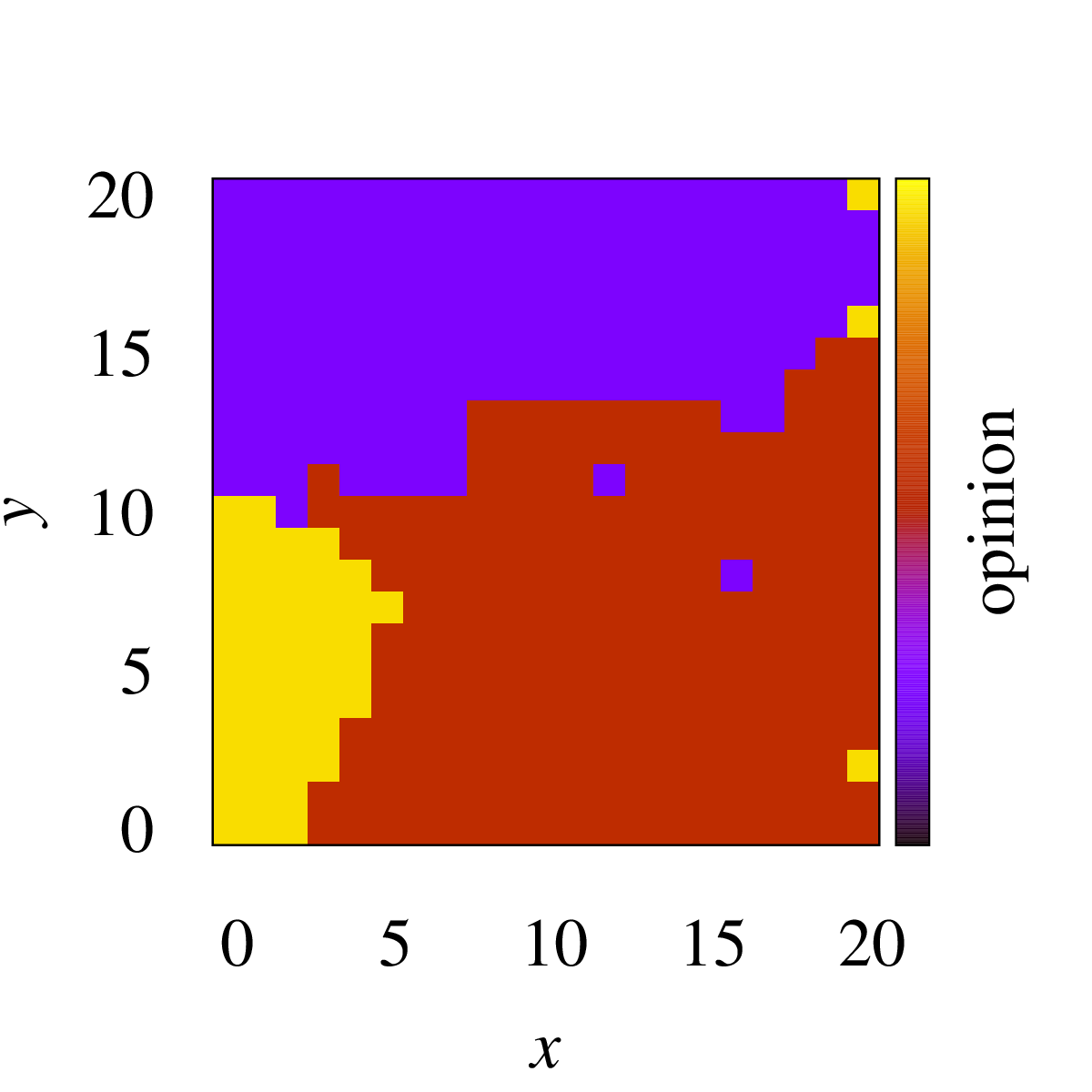}
\end{subfigure}
%% ---------------------------------------------------------------
\caption{\label{fig:L_snap_T1}Snapshots for system evolution with Latan\'e model. Synchronous update. $T=1$. $K=N$. 
\subref{subfig:L_a20T10_t=0000}%%, \subref{subfig:L_a30T00_t=0000}, \subref{subfig:L_a50T00_t=0000}, 
--\subref{subfig:L_a60T10_t=0000} $t=0$.
%\subref{subfig:L_a20T10_t=0005}%%, \subref{subfig:L_a30T00_t=0005}, \subref{subfig:L_a50T00_t=0005}, 
%--\subref{subfig:L_a60T10_t=0005} $t=5$,
%\subref{subfig:L_a20T10_t=0010}%%, \subref{subfig:L_a30T00_t=0010}, \subref{subfig:L_a50T00_t=0010}, 
%--\subref{subfig:L_a60T10_t=0010} $t=10$,
%\subref{subfig:L_a20T10_t=0050}%%, \subref{subfig:L_a30T00_t=0050}, \subref{subfig:L_a50T00_t=0050}, 
%--\subref{subfig:L_a60T10_t=0050} $t=50$,
\subref{subfig:L_a20T10_t=end}%%, \subref{subfig:L_a30T00_t=end}, \subref{subfig:L_a50T00_t=end}, 
--\subref{subfig:L_a60T10_t=end} $t\to\infty$.
Various effective ranges of interaction
\subref{subfig:L_a20T10_t=0000}, \subref{subfig:L_a20T10_t=0005}, \subref{subfig:L_a20T10_t=0010}, \subref{subfig:L_a20T10_t=0050}, \subref{subfig:L_a20T10_t=end} $\alpha=2$, 
\subref{subfig:L_a30T10_t=0000}, \subref{subfig:L_a30T10_t=0005}, \subref{subfig:L_a30T10_t=0010}, \subref{subfig:L_a30T10_t=0050}, \subref{subfig:L_a30T10_t=end} $\alpha=3$, 
\subref{subfig:L_a50T10_t=0000}, \subref{subfig:L_a50T10_t=0100}, \subref{subfig:L_a50T10_t=1000}, \subref{subfig:L_a50T10_t=5000}, \subref{subfig:L_a50T10_t=end} $\alpha=5$, 
\subref{subfig:L_a60T10_t=0000}, \subref{subfig:L_a60T10_t=5e3}, \subref{subfig:L_a60T10_t=1e4}, \subref{subfig:L_a60T10_t=5e4}, \subref{subfig:L_a60T10_t=end} $\alpha=6$}
%% ---------------------------------------------------------------
\end{figure*}

In \Cref{fig:V_snap,fig:S_snap,fig:L_snap_T0,fig:L_snap_T1} examples of snapshots from the evolution of the system for the voter, Sznajd, Latan\'e model (for $T=0$ and $T=1$) are presented, respectively.

%% ################################################################
%% \bibliography{Bib/km,Bib/opiniondynamics}

\begin{thebibliography}{60}%
\makeatletter
\providecommand \@ifxundefined [1]{%
 \@ifx{#1\undefined}
}%
\providecommand \@ifnum [1]{%
 \ifnum #1\expandafter \@firstoftwo
 \else \expandafter \@secondoftwo
 \fi
}%
\providecommand \@ifx [1]{%
 \ifx #1\expandafter \@firstoftwo
 \else \expandafter \@secondoftwo
 \fi
}%
\providecommand \natexlab [1]{#1}%
\providecommand \enquote  [1]{``#1''}%
\providecommand \bibnamefont  [1]{#1}%
\providecommand \bibfnamefont [1]{#1}%
\providecommand \citenamefont [1]{#1}%
\providecommand \href@noop [0]{\@secondoftwo}%
\providecommand \href [0]{\begingroup \@sanitize@url \@href}%
\providecommand \@href[1]{\@@startlink{#1}\@@href}%
\providecommand \@@href[1]{\endgroup#1\@@endlink}%
\providecommand \@sanitize@url [0]{\catcode `\\12\catcode `\$12\catcode
  `\&12\catcode `\#12\catcode `\^12\catcode `\_12\catcode `\%12\relax}%
\providecommand \@@startlink[1]{}%
\providecommand \@@endlink[0]{}%
\providecommand \url  [0]{\begingroup\@sanitize@url \@url }%
\providecommand \@url [1]{\endgroup\@href {#1}{\urlprefix }}%
\providecommand \urlprefix  [0]{URL }%
\providecommand \Eprint [0]{\href }%
\providecommand \doibase [0]{https://doi.org/}%
\providecommand \selectlanguage [0]{\@gobble}%
\providecommand \bibinfo  [0]{\@secondoftwo}%
\providecommand \bibfield  [0]{\@secondoftwo}%
\providecommand \translation [1]{[#1]}%
\providecommand \BibitemOpen [0]{}%
\providecommand \bibitemStop [0]{}%
\providecommand \bibitemNoStop [0]{.\EOS\space}%
\providecommand \EOS [0]{\spacefactor3000\relax}%
\providecommand \BibitemShut  [1]{\csname bibitem#1\endcsname}%
\let\auto@bib@innerbib\@empty
%</preamble>
\bibitem [{\citenamefont {Galam}(2012)}]{Galam_2012}%
  \BibitemOpen
  \bibfield  {author} {\bibinfo {author} {\bibfnamefont {S.}~\bibnamefont
  {Galam}},\ }\href {https://doi.org/10.1007/978-1-4614-2032-3} {\emph
  {\bibinfo {title} {Sociophysics: A Physicist's Modeling of Psycho-political
  Phenomena}}}\ (\bibinfo  {publisher} {Springer},\ \bibinfo {address} {New
  York, NY},\ \bibinfo {year} {2012})\BibitemShut {NoStop}%
\bibitem [{\citenamefont {Stauffer}(2013)}]{Stauffer_2013}%
  \BibitemOpen
  \bibfield  {author} {\bibinfo {author} {\bibfnamefont {D.}~\bibnamefont
  {Stauffer}},\ }\bibfield  {title} {\bibinfo {title} {A biased review of
  sociophysics},\ }\href {https://doi.org/10.1007/s10955-012-0604-9} {\bibfield
   {journal} {\bibinfo  {journal} {Journal of Statistical Physics}\ }\textbf
  {\bibinfo {volume} {151}},\ \bibinfo {pages} {9} (\bibinfo {year}
  {2013})}\BibitemShut {NoStop}%
\bibitem [{\citenamefont {Sen}\ and\ \citenamefont
  {Chakrabarti}(2014)}]{Sen_2014}%
  \BibitemOpen
  \bibfield  {author} {\bibinfo {author} {\bibfnamefont {P.}~\bibnamefont
  {Sen}}\ and\ \bibinfo {author} {\bibfnamefont {B.~K.}\ \bibnamefont
  {Chakrabarti}},\ }\href@noop {} {\emph {\bibinfo {title} {Sociophysics: {A}n
  Introduction}}}\ (\bibinfo  {publisher} {Oxford Univeristy Press},\ \bibinfo
  {address} {Oxford},\ \bibinfo {year} {2014})\BibitemShut {NoStop}%
\bibitem [{\citenamefont {Schweitzer}(2018)}]{Schweitzer_Sociophysics}%
  \BibitemOpen
  \bibfield  {author} {\bibinfo {author} {\bibfnamefont {F.}~\bibnamefont
  {Schweitzer}},\ }\bibfield  {title} {\bibinfo {title} {Sociophysics},\ }\href
  {https://doi.org/10.1063/PT.3.3845} {\bibfield  {journal} {\bibinfo
  {journal} {Physics Today}\ }\textbf {\bibinfo {volume} {71}},\ \bibinfo
  {pages} {40} (\bibinfo {year} {2018})}\BibitemShut {NoStop}%
\bibitem [{\citenamefont {Perc}(2019)}]{Matjaz_2019}%
  \BibitemOpen
  \bibfield  {author} {\bibinfo {author} {\bibfnamefont {M.}~\bibnamefont
  {Perc}},\ }\bibfield  {title} {\bibinfo {title} {The social physics
  collective},\ }\href@noop {} {\bibfield  {journal} {\bibinfo  {journal}
  {Scientific Reports}\ }\textbf {\bibinfo {volume} {9}},\ \bibinfo {pages}
  {16549} (\bibinfo {year} {2019})}\BibitemShut {NoStop}%
\bibitem [{\citenamefont {Jusup}\ \emph {et~al.}(2022)\citenamefont {Jusup},
  \citenamefont {Holme}, \citenamefont {Kanazawa}, \citenamefont {Takayasu},
  \citenamefont {Romić}, \citenamefont {Wang}, \citenamefont {Geček},
  \citenamefont {Lipić}, \citenamefont {Podobnik}, \citenamefont {Wang},
  \citenamefont {Luo}, \citenamefont {Klanjšček}, \citenamefont {Fan},
  \citenamefont {Boccaletti},\ and\ \citenamefont {Perc}}]{Jusup_2022}%
  \BibitemOpen
  \bibfield  {author} {\bibinfo {author} {\bibfnamefont {M.}~\bibnamefont
  {Jusup}}, \bibinfo {author} {\bibfnamefont {P.}~\bibnamefont {Holme}},
  \bibinfo {author} {\bibfnamefont {K.}~\bibnamefont {Kanazawa}}, \bibinfo
  {author} {\bibfnamefont {M.}~\bibnamefont {Takayasu}}, \bibinfo {author}
  {\bibfnamefont {I.}~\bibnamefont {Romić}}, \bibinfo {author} {\bibfnamefont
  {Z.}~\bibnamefont {Wang}}, \bibinfo {author} {\bibfnamefont {S.}~\bibnamefont
  {Geček}}, \bibinfo {author} {\bibfnamefont {T.}~\bibnamefont {Lipić}},
  \bibinfo {author} {\bibfnamefont {B.}~\bibnamefont {Podobnik}}, \bibinfo
  {author} {\bibfnamefont {L.}~\bibnamefont {Wang}}, \bibinfo {author}
  {\bibfnamefont {W.}~\bibnamefont {Luo}}, \bibinfo {author} {\bibfnamefont
  {T.}~\bibnamefont {Klanjšček}}, \bibinfo {author} {\bibfnamefont
  {J.}~\bibnamefont {Fan}}, \bibinfo {author} {\bibfnamefont {S.}~\bibnamefont
  {Boccaletti}},\ and\ \bibinfo {author} {\bibfnamefont {M.}~\bibnamefont
  {Perc}},\ }\bibfield  {title} {\bibinfo {title} {Social physics},\ }\href
  {https://doi.org/10.1016/j.physrep.2021.10.005} {\bibfield  {journal}
  {\bibinfo  {journal} {Physics Reports}\ }\textbf {\bibinfo {volume} {948}},\
  \bibinfo {pages} {1} (\bibinfo {year} {2022})}\BibitemShut {NoStop}%
\bibitem [{SI:(2023{\natexlab{a}})}]{SI:Trends_Sociophysics}%
  \BibitemOpen
  \href
  {https://www.mdpi.com/journal/entropy/special_issues/Trends_Sociophysics}
  {\bibinfo {title} {Modern trends in sociophysics}} (\bibinfo {year}
  {accessed: Dec. 23, 2023}{\natexlab{a}})\BibitemShut {NoStop}%
\bibitem [{\citenamefont {{da Luz}}\ \emph {et~al.}(2023)\citenamefont {{da
  Luz}}, \citenamefont {Anteneodo}, \citenamefont {Crokidakis},\ and\
  \citenamefont {Perc}}]{da_Luz_2023}%
  \BibitemOpen
  \bibfield  {author} {\bibinfo {author} {\bibfnamefont {M.~G.}\ \bibnamefont
  {{da Luz}}}, \bibinfo {author} {\bibfnamefont {C.}~\bibnamefont {Anteneodo}},
  \bibinfo {author} {\bibfnamefont {N.}~\bibnamefont {Crokidakis}},\ and\
  \bibinfo {author} {\bibfnamefont {M.}~\bibnamefont {Perc}},\ }\bibfield
  {title} {\bibinfo {title} {Sociophysics: {S}ocial collective behavior from
  the physics point of view},\ }\href
  {https://doi.org/10.1016/j.chaos.2023.113379} {\bibfield  {journal} {\bibinfo
   {journal} {Chaos, Solitons \& Fractals}\ }\textbf {\bibinfo {volume}
  {170}},\ \bibinfo {pages} {113379} (\bibinfo {year} {2023})}\BibitemShut
  {NoStop}%
\bibitem [{SI:(2023{\natexlab{b}})}]{SI:SergeGalam70}%
  \BibitemOpen
  \href {https://www.mdpi.com/journal/physics/special_issues/SergeGalam70}
  {\bibinfo {title} {In honor of {P}rofessor {S}erge {G}alam for his 70th
  birthday and forty years of sociophysics}} (\bibinfo {year} {accessed: Dec.
  23, 2023}{\natexlab{b}})\BibitemShut {NoStop}%
\bibitem [{\citenamefont {Castellano}\ \emph {et~al.}(2009)\citenamefont
  {Castellano}, \citenamefont {Fortunato},\ and\ \citenamefont
  {Loreto}}]{Castellano-2009}%
  \BibitemOpen
  \bibfield  {author} {\bibinfo {author} {\bibfnamefont {C.}~\bibnamefont
  {Castellano}}, \bibinfo {author} {\bibfnamefont {S.}~\bibnamefont
  {Fortunato}},\ and\ \bibinfo {author} {\bibfnamefont {V.}~\bibnamefont
  {Loreto}},\ }\bibfield  {title} {\bibinfo {title} {Statistical physics of
  social dynamics},\ }\href {https://doi.org/10.1103/RevModPhys.81.591}
  {\bibfield  {journal} {\bibinfo  {journal} {Reviews of Modern Physics}\
  }\textbf {\bibinfo {volume} {81}},\ \bibinfo {pages} {591} (\bibinfo {year}
  {2009})}\BibitemShut {NoStop}%
\bibitem [{\citenamefont {Galam}(2006)}]{Galam_2006}%
  \BibitemOpen
  \bibfield  {author} {\bibinfo {author} {\bibfnamefont {S.}~\bibnamefont
  {Galam}},\ }\bibinfo {title} {Opinion dynamics, minority spreading and
  heterogeneous beliefs},\ in\ \href
  {https://doi.org/10.1002/9783527610006.ch13} {\emph {\bibinfo {booktitle}
  {Econophysics and Sociophysics}}}\ (\bibinfo  {publisher} {John Wiley \&
  Sons, Ltd},\ \bibinfo {year} {2006})\ Chap.~\bibinfo {chapter} {13}, pp.\
  \bibinfo {pages} {367--391}\BibitemShut {NoStop}%
\bibitem [{\citenamefont {Weisbuch}(2006)}]{Weisbuch_2006}%
  \BibitemOpen
  \bibfield  {author} {\bibinfo {author} {\bibfnamefont {G.}~\bibnamefont
  {Weisbuch}},\ }\bibinfo {title} {Social opinion dynamics},\ in\ \href
  {https://doi.org/10.1002/9783527610006.ch12} {\emph {\bibinfo {booktitle}
  {Econophysics and Sociophysics}}}\ (\bibinfo  {publisher} {John Wiley \&
  Sons, Ltd},\ \bibinfo {year} {2006})\ Chap.~\bibinfo {chapter} {12}, pp.\
  \bibinfo {pages} {339--366}\BibitemShut {NoStop}%
\bibitem [{\citenamefont {Sobkowicz}(2019)}]{Sobkowicz_2019}%
  \BibitemOpen
  \bibfield  {author} {\bibinfo {author} {\bibfnamefont {P.}~\bibnamefont
  {Sobkowicz}},\ }\bibfield  {title} {\bibinfo {title} {Social simulation
  models at the ethical crossroads},\ }\href
  {https://doi.org/10.1007/s11948-017-9993-0} {\bibfield  {journal} {\bibinfo
  {journal} {Science and Engineering Ethics}\ }\textbf {\bibinfo {volume}
  {25}},\ \bibinfo {pages} {143} (\bibinfo {year} {2019})}\BibitemShut
  {NoStop}%
\bibitem [{\citenamefont {Grabisch}\ and\ \citenamefont
  {Rusinowska}(2020)}]{Grabisch_2020}%
  \BibitemOpen
  \bibfield  {author} {\bibinfo {author} {\bibfnamefont {M.}~\bibnamefont
  {Grabisch}}\ and\ \bibinfo {author} {\bibfnamefont {A.}~\bibnamefont
  {Rusinowska}},\ }\bibfield  {title} {\bibinfo {title} {A survey on
  nonstrategic models of opinion dynamics},\ }\href
  {https://doi.org/10.3390/g11040065} {\bibfield  {journal} {\bibinfo
  {journal} {Games}\ }\textbf {\bibinfo {volume} {11}},\ \bibinfo {pages} {65}
  (\bibinfo {year} {2020})}\BibitemShut {NoStop}%
\bibitem [{\citenamefont {Ellero}\ \emph {et~al.}(2023)\citenamefont {Ellero},
  \citenamefont {Fasano},\ and\ \citenamefont {Favaretto}}]{Ellero_2023}%
  \BibitemOpen
  \bibfield  {author} {\bibinfo {author} {\bibfnamefont {A.}~\bibnamefont
  {Ellero}}, \bibinfo {author} {\bibfnamefont {G.}~\bibnamefont {Fasano}},\
  and\ \bibinfo {author} {\bibfnamefont {D.}~\bibnamefont {Favaretto}},\
  }\bibfield  {title} {\bibinfo {title} {Mathematical programming for the
  dynamics of opinion diffusion},\ }\href
  {https://doi.org/10.3390/physics5030061} {\bibfield  {journal} {\bibinfo
  {journal} {Physics}\ }\textbf {\bibinfo {volume} {5}},\ \bibinfo {pages}
  {936} (\bibinfo {year} {2023})}\BibitemShut {NoStop}%
\bibitem [{\citenamefont {Sousa}(2005)}]{ISI:000226629700050}%
  \BibitemOpen
  \bibfield  {author} {\bibinfo {author} {\bibfnamefont {A.}~\bibnamefont
  {Sousa}},\ }\bibfield  {title} {\bibinfo {title} {Consensus formation on a
  triad scale-free network},\ }\href
  {https://doi.org/10.1016/j.physa.2004.09.027} {\bibfield  {journal} {\bibinfo
   {journal} {Physica A}\ }\textbf {\bibinfo {volume} {348}},\ \bibinfo {pages}
  {701} (\bibinfo {year} {2005})}\BibitemShut {NoStop}%
\bibitem [{\citenamefont {Rodrigues}\ and\ \citenamefont
  {Da~F.~Costa}(2005)}]{Rodrigues_2005}%
  \BibitemOpen
  \bibfield  {author} {\bibinfo {author} {\bibfnamefont {F.~A.}\ \bibnamefont
  {Rodrigues}}\ and\ \bibinfo {author} {\bibfnamefont {L.}~\bibnamefont
  {Da~F.~Costa}},\ }\bibfield  {title} {\bibinfo {title} {Surviving opinions in
  {S}znajd models on complex networks},\ }\href
  {https://doi.org/10.1142/S0129183105008278} {\bibfield  {journal} {\bibinfo
  {journal} {International Journal of Modern Physics C}\ }\textbf {\bibinfo
  {volume} {16}},\ \bibinfo {pages} {1785} (\bibinfo {year}
  {2005})}\BibitemShut {NoStop}%
\bibitem [{\citenamefont {Malarz}\ and\ \citenamefont
  {Ku{\l}akowski}(2010)}]{Kulakowski2010}%
  \BibitemOpen
  \bibfield  {author} {\bibinfo {author} {\bibfnamefont {K.}~\bibnamefont
  {Malarz}}\ and\ \bibinfo {author} {\bibfnamefont {K.}~\bibnamefont
  {Ku{\l}akowski}},\ }\bibfield  {title} {\bibinfo {title} {Indifferents as an
  interface between contra and pro},\ }\href
  {https://doi.org/10.12693/APhysPolA.117.695} {\bibfield  {journal} {\bibinfo
  {journal} {Acta Physica Polonica A}\ }\textbf {\bibinfo {volume} {117}},\
  \bibinfo {pages} {695} (\bibinfo {year} {2010})}\BibitemShut {NoStop}%
\bibitem [{\citenamefont {\"{O}zt\"{u}rk}(2013)}]{Ozturk_2013}%
  \BibitemOpen
  \bibfield  {author} {\bibinfo {author} {\bibfnamefont {M.~K.}\ \bibnamefont
  {\"{O}zt\"{u}rk}},\ }\bibfield  {title} {\bibinfo {title} {Dynamics of
  discrete opinions without compromise},\ }\href
  {https://doi.org/10.1142/S0219525913500100} {\bibfield  {journal} {\bibinfo
  {journal} {Advances in Complex Systems}\ }\textbf {\bibinfo {volume} {16}},\
  \bibinfo {pages} {1350010} (\bibinfo {year} {2013})}\BibitemShut {NoStop}%
\bibitem [{\citenamefont {Ba\'ncerowski}\ and\ \citenamefont
  {Malarz}(2019)}]{1902.03454}%
  \BibitemOpen
  \bibfield  {author} {\bibinfo {author} {\bibfnamefont {P.}~\bibnamefont
  {Ba\'ncerowski}}\ and\ \bibinfo {author} {\bibfnamefont {K.}~\bibnamefont
  {Malarz}},\ }\bibfield  {title} {\bibinfo {title} {Multi-choice opinion
  dynamics model based on {L}atan\'e theory},\ }\href
  {https://doi.org/10.1140/epjb/e2019-90533-0} {\bibfield  {journal} {\bibinfo
  {journal} {The European Physical Journal B}\ }\textbf {\bibinfo {volume}
  {92}},\ \bibinfo {pages} {219} (\bibinfo {year} {2019})}\BibitemShut
  {NoStop}%
\bibitem [{\citenamefont {Martins}(2020)}]{Martins_2020}%
  \BibitemOpen
  \bibfield  {author} {\bibinfo {author} {\bibfnamefont {A.~C.~R.}\
  \bibnamefont {Martins}},\ }\bibfield  {title} {\bibinfo {title} {Discrete
  opinion dynamics with {$M$} choices},\ }\href
  {https://doi.org/10.1140/epjb/e2019-100298-3} {\bibfield  {journal} {\bibinfo
   {journal} {The European Physical Journal B}\ }\textbf {\bibinfo {volume}
  {93}},\ \bibinfo {pages} {1} (\bibinfo {year} {2020})}\BibitemShut {NoStop}%
\bibitem [{\citenamefont {Zubillaga}\ \emph {et~al.}(2022)\citenamefont
  {Zubillaga}, \citenamefont {Vilela}, \citenamefont {Wang}, \citenamefont
  {Du}, \citenamefont {Dong},\ and\ \citenamefont {Stanley}}]{Zubillaga_2022}%
  \BibitemOpen
  \bibfield  {author} {\bibinfo {author} {\bibfnamefont {B.}~\bibnamefont
  {Zubillaga}}, \bibinfo {author} {\bibfnamefont {A.}~\bibnamefont {Vilela}},
  \bibinfo {author} {\bibfnamefont {M.}~\bibnamefont {Wang}}, \bibinfo {author}
  {\bibfnamefont {R.}~\bibnamefont {Du}}, \bibinfo {author} {\bibfnamefont
  {G.}~\bibnamefont {Dong}},\ and\ \bibinfo {author} {\bibfnamefont
  {H.}~\bibnamefont {Stanley}},\ }\bibfield  {title} {\bibinfo {title}
  {Three-state majority-vote model on small-world networks},\ }\href
  {https://doi.org/10.1038/s41598-021-03467-6} {\bibfield  {journal} {\bibinfo
  {journal} {Scientific Reports}\ }\textbf {\bibinfo {volume} {12}},\ \bibinfo
  {pages} {282} (\bibinfo {year} {2022})}\BibitemShut {NoStop}%
\bibitem [{\citenamefont {Li}\ \emph {et~al.}(2022)\citenamefont {Li},
  \citenamefont {Zeng}, \citenamefont {Fan},\ and\ \citenamefont
  {Di}}]{Li_2022}%
  \BibitemOpen
  \bibfield  {author} {\bibinfo {author} {\bibfnamefont {L.}~\bibnamefont
  {Li}}, \bibinfo {author} {\bibfnamefont {A.}~\bibnamefont {Zeng}}, \bibinfo
  {author} {\bibfnamefont {Y.}~\bibnamefont {Fan}},\ and\ \bibinfo {author}
  {\bibfnamefont {Z.}~\bibnamefont {Di}},\ }\bibfield  {title} {\bibinfo
  {title} {Modeling multi-opinion propagation in complex systems with
  heterogeneous relationships via {P}otts model on signed networks},\ }\href
  {https://doi.org/10.1063/5.0084525} {\bibfield  {journal} {\bibinfo
  {journal} {Chaos}\ }\textbf {\bibinfo {volume} {32}},\ \bibinfo {pages}
  {083101} (\bibinfo {year} {2022})}\BibitemShut {NoStop}%
\bibitem [{\citenamefont {Doniec}\ \emph {et~al.}(2022)\citenamefont {Doniec},
  \citenamefont {Lipiecki},\ and\ \citenamefont {Sznajd-Weron}}]{Doniec_2022}%
  \BibitemOpen
  \bibfield  {author} {\bibinfo {author} {\bibfnamefont {M.}~\bibnamefont
  {Doniec}}, \bibinfo {author} {\bibfnamefont {A.}~\bibnamefont {Lipiecki}},\
  and\ \bibinfo {author} {\bibfnamefont {K.}~\bibnamefont {Sznajd-Weron}},\
  }\bibfield  {title} {\bibinfo {title} {Consensus, polarization and hysteresis
  in the three-state noisy $q$-voter model with bounded confidence},\ }\href
  {https://doi.org/10.3390/e24070983} {\bibfield  {journal} {\bibinfo
  {journal} {Entropy}\ }\textbf {\bibinfo {volume} {24}},\ \bibinfo {pages}
  {983} (\bibinfo {year} {2022})}\BibitemShut {NoStop}%
\bibitem [{\citenamefont {Xiong}\ \emph {et~al.}(2017)\citenamefont {Xiong},
  \citenamefont {Liu}, \citenamefont {Wang},\ and\ \citenamefont
  {Wang}}]{000409112600017}%
  \BibitemOpen
  \bibfield  {author} {\bibinfo {author} {\bibfnamefont {F.}~\bibnamefont
  {Xiong}}, \bibinfo {author} {\bibfnamefont {Y.}~\bibnamefont {Liu}}, \bibinfo
  {author} {\bibfnamefont {L.}~\bibnamefont {Wang}},\ and\ \bibinfo {author}
  {\bibfnamefont {X.}~\bibnamefont {Wang}},\ }\bibfield  {title} {\bibinfo
  {title} {Analysis and application of opinion model with multiple topic
  interactions},\ }\href {https://doi.org/10.1063/1.4998736} {\bibfield
  {journal} {\bibinfo  {journal} {Chaos}\ }\textbf {\bibinfo {volume} {27}},\
  \bibinfo {pages} {083113} (\bibinfo {year} {2017})}\BibitemShut {NoStop}%
\bibitem [{\citenamefont {Galam}(2013)}]{000316891200004}%
  \BibitemOpen
  \bibfield  {author} {\bibinfo {author} {\bibfnamefont {S.}~\bibnamefont
  {Galam}},\ }\bibfield  {title} {\bibinfo {title} {The drastic outcomes from
  voting alliances in three-party democratic voting (1990--2013)},\ }\href
  {https://doi.org/10.1007/s10955-012-0641-4} {\bibfield  {journal} {\bibinfo
  {journal} {Journal of Statistical Physics}\ }\textbf {\bibinfo {volume}
  {151}},\ \bibinfo {pages} {46} (\bibinfo {year} {2013})}\BibitemShut
  {NoStop}%
\bibitem [{\citenamefont {Wu}\ and\ \citenamefont
  {Szeto}(2018)}]{000432967700004}%
  \BibitemOpen
  \bibfield  {author} {\bibinfo {author} {\bibfnamefont {D.}~\bibnamefont
  {Wu}}\ and\ \bibinfo {author} {\bibfnamefont {K.~Y.}\ \bibnamefont {Szeto}},\
  }\bibfield  {title} {\bibinfo {title} {Analysis of timescale to consensus in
  voting dynamics with more than two options},\ }\href
  {https://doi.org/10.1103/PhysRevE.97.042320} {\bibfield  {journal} {\bibinfo
  {journal} {Physical Review E}\ }\textbf {\bibinfo {volume} {{97}}},\ \bibinfo
  {pages} {042320} (\bibinfo {year} {{2018}})}\BibitemShut {NoStop}%
\bibitem [{\citenamefont {Mobilia}(2023)}]{Mobilia_2023}%
  \BibitemOpen
  \bibfield  {author} {\bibinfo {author} {\bibfnamefont {M.}~\bibnamefont
  {Mobilia}},\ }\bibfield  {title} {\bibinfo {title} {Polarization and
  consensus in a voter model under time-fluctuating influences},\ }\href
  {https://doi.org/10.3390/physics5020037} {\bibfield  {journal} {\bibinfo
  {journal} {Physics}\ }\textbf {\bibinfo {volume} {5}},\ \bibinfo {pages}
  {517} (\bibinfo {year} {2023})}\BibitemShut {NoStop}%
\bibitem [{\citenamefont {Malarz}\ and\ \citenamefont
  {Mas{\l}yk}(2023)}]{Maslyk_2023}%
  \BibitemOpen
  \bibfield  {author} {\bibinfo {author} {\bibfnamefont {K.}~\bibnamefont
  {Malarz}}\ and\ \bibinfo {author} {\bibfnamefont {T.}~\bibnamefont
  {Mas{\l}yk}},\ }\bibfield  {title} {\bibinfo {title} {Phase diagram for
  social impact theory in initially fully differentiated society},\ }\href
  {https://doi.org/10.3390/physics5040067} {\bibfield  {journal} {\bibinfo
  {journal} {Physics}\ }\textbf {\bibinfo {volume} {5}},\ \bibinfo {pages}
  {1031} (\bibinfo {year} {2023})}\BibitemShut {NoStop}%
\bibitem [{\citenamefont {Dworak}\ and\ \citenamefont
  {Malarz}(2023)}]{2211.04183}%
  \BibitemOpen
  \bibfield  {author} {\bibinfo {author} {\bibfnamefont {M.}~\bibnamefont
  {Dworak}}\ and\ \bibinfo {author} {\bibfnamefont {K.}~\bibnamefont
  {Malarz}},\ }\bibfield  {title} {\bibinfo {title} {Vanishing opinions in
  {L}atan\'e model of opinion formation},\ }\href
  {https://doi.org/10.3390/e25010058} {\bibfield  {journal} {\bibinfo
  {journal} {Entropy}\ }\textbf {\bibinfo {volume} {25}},\ \bibinfo {pages}
  {58} (\bibinfo {year} {2023})}\BibitemShut {NoStop}%
\bibitem [{\citenamefont {Kowalska-Stycze\'n}\ and\ \citenamefont
  {Malarz}(2020)}]{2002.05451}%
  \BibitemOpen
  \bibfield  {author} {\bibinfo {author} {\bibfnamefont {A.}~\bibnamefont
  {Kowalska-Stycze\'n}}\ and\ \bibinfo {author} {\bibfnamefont
  {K.}~\bibnamefont {Malarz}},\ }\bibfield  {title} {\bibinfo {title} {Noise
  induced unanimity and disorder in opinion formation},\ }\href
  {https://doi.org/10.1371/journal.pone.0235313} {\bibfield  {journal}
  {\bibinfo  {journal} {Plos One}\ }\textbf {\bibinfo {volume} {15}},\ \bibinfo
  {pages} {e0235313} (\bibinfo {year} {2020})}\BibitemShut {NoStop}%
\bibitem [{\citenamefont {Holley}\ and\ \citenamefont
  {Liggett}(1975)}]{Holley_1975}%
  \BibitemOpen
  \bibfield  {author} {\bibinfo {author} {\bibfnamefont {R.~A.}\ \bibnamefont
  {Holley}}\ and\ \bibinfo {author} {\bibfnamefont {T.~M.}\ \bibnamefont
  {Liggett}},\ }\bibfield  {title} {\bibinfo {title} {Ergodic theorems for
  weakly interacting infinite systems and voter model},\ }\href
  {https://doi.org/10.1214/aop/1176996306} {\bibfield  {journal} {\bibinfo
  {journal} {Annals of Probability}\ }\textbf {\bibinfo {volume} {3}},\
  \bibinfo {pages} {643} (\bibinfo {year} {1975})}\BibitemShut {NoStop}%
\bibitem [{\citenamefont {Liggett}(1999)}]{Liggett_1999}%
  \BibitemOpen
  \bibfield  {author} {\bibinfo {author} {\bibfnamefont {T.~M.}\ \bibnamefont
  {Liggett}},\ }\href {https://doi.org/10.1007/978-3-662-03990-8} {\emph
  {\bibinfo {title} {Stochastic Interacting Systems: Contact, Voter and
  Exclusion Processes}}}\ (\bibinfo  {publisher} {Springer},\ \bibinfo
  {address} {Berlin, Heidelberg},\ \bibinfo {year} {1999})\BibitemShut
  {NoStop}%
\bibitem [{\citenamefont {Sznajd-Weron}\ and\ \citenamefont
  {Sznajd}(2000)}]{Sznajd-2000}%
  \BibitemOpen
  \bibfield  {author} {\bibinfo {author} {\bibfnamefont {K.}~\bibnamefont
  {Sznajd-Weron}}\ and\ \bibinfo {author} {\bibfnamefont {J.}~\bibnamefont
  {Sznajd}},\ }\bibfield  {title} {\bibinfo {title} {Opinion evolution in
  closed community},\ }\href {https://doi.org/10.1142/S0129183100000936}
  {\bibfield  {journal} {\bibinfo  {journal} {International Journal of Modern
  Physics C}\ }\textbf {\bibinfo {volume} {11}},\ \bibinfo {pages} {1157}
  (\bibinfo {year} {2000})}\BibitemShut {NoStop}%
\bibitem [{\citenamefont {Latan\'e}(1981)}]{Latane-1981}%
  \BibitemOpen
  \bibfield  {author} {\bibinfo {author} {\bibfnamefont {B.}~\bibnamefont
  {Latan\'e}},\ }\bibfield  {title} {\bibinfo {title} {The psychology of social
  impact},\ }\href {https://doi.org/10.1037/0003-066X.36.4.343} {\bibfield
  {journal} {\bibinfo  {journal} {American Psychologist}\ }\textbf {\bibinfo
  {volume} {36}},\ \bibinfo {pages} {343} (\bibinfo {year} {1981})}\BibitemShut
  {NoStop}%
\bibitem [{\citenamefont {Herrer\'{\i}as-Azcu\'e}\ and\ \citenamefont
  {Galla}(2019)}]{PhysRevE.100.022304}%
  \BibitemOpen
  \bibfield  {author} {\bibinfo {author} {\bibfnamefont {F.}~\bibnamefont
  {Herrer\'{\i}as-Azcu\'e}}\ and\ \bibinfo {author} {\bibfnamefont
  {T.}~\bibnamefont {Galla}},\ }\bibfield  {title} {\bibinfo {title} {Consensus
  and diversity in multistate noisy voter models},\ }\href
  {https://doi.org/10.1103/PhysRevE.100.022304} {\bibfield  {journal} {\bibinfo
   {journal} {Physical Review E}\ }\textbf {\bibinfo {volume} {100}},\ \bibinfo
  {pages} {022304} (\bibinfo {year} {2019})}\BibitemShut {NoStop}%
\bibitem [{\citenamefont {Vendeville}\ \emph {et~al.}(2024)\citenamefont
  {Vendeville}, \citenamefont {Zhou},\ and\ \citenamefont
  {Guedj}}]{PhysRevE.109.024312}%
  \BibitemOpen
  \bibfield  {author} {\bibinfo {author} {\bibfnamefont {A.}~\bibnamefont
  {Vendeville}}, \bibinfo {author} {\bibfnamefont {S.}~\bibnamefont {Zhou}},\
  and\ \bibinfo {author} {\bibfnamefont {B.}~\bibnamefont {Guedj}},\ }\bibfield
   {title} {\bibinfo {title} {Discord in the voter model for complex
  networks},\ }\href {https://doi.org/10.1103/PhysRevE.109.024312} {\bibfield
  {journal} {\bibinfo  {journal} {Physical Review E}\ }\textbf {\bibinfo
  {volume} {109}},\ \bibinfo {pages} {024312} (\bibinfo {year}
  {2024})}\BibitemShut {NoStop}%
\bibitem [{\citenamefont {Cialdini}(1984)}]{Cialdini_1984}%
  \BibitemOpen
  \bibfield  {author} {\bibinfo {author} {\bibfnamefont {R.~B.}\ \bibnamefont
  {Cialdini}},\ }\href@noop {} {\emph {\bibinfo {title} {Influence: The
  Psychology of Persuasion}}}\ (\bibinfo  {publisher} {William Morrow and
  Company},\ \bibinfo {address} {New York},\ \bibinfo {year}
  {1984})\BibitemShut {NoStop}%
\bibitem [{\citenamefont {Morgan}\ and\ \citenamefont
  {Laland}(2012)}]{Morgan_2012}%
  \BibitemOpen
  \bibfield  {author} {\bibinfo {author} {\bibfnamefont {T.}~\bibnamefont
  {Morgan}}\ and\ \bibinfo {author} {\bibfnamefont {K.}~\bibnamefont
  {Laland}},\ }\bibfield  {title} {\bibinfo {title} {The biological bases of
  conformity},\ }\href {https://doi.org/10.3389/fnins.2012.00087} {\bibfield
  {journal} {\bibinfo  {journal} {Frontiers in Neuroscience}\ }\textbf
  {\bibinfo {volume} {6}},\ \bibinfo {pages} {87} (\bibinfo {year}
  {2012})}\BibitemShut {NoStop}%
\bibitem [{\citenamefont {Asch}(1956)}]{Asch_1956}%
  \BibitemOpen
  \bibfield  {author} {\bibinfo {author} {\bibfnamefont {S.~E.}\ \bibnamefont
  {Asch}},\ }\bibfield  {title} {\bibinfo {title} {Studies of independence and
  conformity: {A} minority of one against a unanimous majority},\ }\href
  {https://doi.org/10.1037/h0093718} {\bibfield  {journal} {\bibinfo  {journal}
  {Psychological Monographs: General and Applied}\ }\textbf {\bibinfo {volume}
  {70}},\ \bibinfo {pages} {1} (\bibinfo {year} {1956})}\BibitemShut {NoStop}%
\bibitem [{\citenamefont {Deutsch}\ and\ \citenamefont
  {Gerard}(1955)}]{Deutsch_1955}%
  \BibitemOpen
  \bibfield  {author} {\bibinfo {author} {\bibfnamefont {M.}~\bibnamefont
  {Deutsch}}\ and\ \bibinfo {author} {\bibfnamefont {H.~B.}\ \bibnamefont
  {Gerard}},\ }\bibfield  {title} {\bibinfo {title} {A study of normative and
  informational social influences upon individual judgment},\ }\href
  {https://doi.org/10.1037/h0046408} {\bibfield  {journal} {\bibinfo  {journal}
  {Journal of Abnormal and Social Psychology}\ }\textbf {\bibinfo {volume}
  {51}},\ \bibinfo {pages} {629} (\bibinfo {year} {1955})}\BibitemShut
  {NoStop}%
\bibitem [{\citenamefont {Milgram}(1963)}]{Milgram_1963}%
  \BibitemOpen
  \bibfield  {author} {\bibinfo {author} {\bibfnamefont {S.}~\bibnamefont
  {Milgram}},\ }\bibfield  {title} {\bibinfo {title} {Behavioral study of
  obedience},\ }\href {https://doi.org/10.1037/h0040525} {\bibfield  {journal}
  {\bibinfo  {journal} {Journal of Abnormal and Social Psychology}\ }\textbf
  {\bibinfo {volume} {67}},\ \bibinfo {pages} {371} (\bibinfo {year}
  {1963})}\BibitemShut {NoStop}%
\bibitem [{\citenamefont {Blass}(1999)}]{Blass_1999}%
  \BibitemOpen
  \bibfield  {author} {\bibinfo {author} {\bibfnamefont {T.}~\bibnamefont
  {Blass}},\ }\bibfield  {title} {\bibinfo {title} {The {M}ilgram paradigm
  after 35 years: {S}ome things we now know about obedience to authority},\
  }\href {https://doi.org/10.1111/j.1559-1816.1999.tb00134.x} {\bibfield
  {journal} {\bibinfo  {journal} {Journal of Applied Social Psychology}\
  }\textbf {\bibinfo {volume} {29}},\ \bibinfo {pages} {955} (\bibinfo {year}
  {1999})}\BibitemShut {NoStop}%
\bibitem [{\citenamefont {Cialdini}\ and\ \citenamefont
  {Goldstein}(2004)}]{Cialdini_2004}%
  \BibitemOpen
  \bibfield  {author} {\bibinfo {author} {\bibfnamefont {R.~B.}\ \bibnamefont
  {Cialdini}}\ and\ \bibinfo {author} {\bibfnamefont {N.~J.}\ \bibnamefont
  {Goldstein}},\ }\bibfield  {title} {\bibinfo {title} {Social influence:
  Compliance and conformity},\ }\href
  {https://doi.org/10.1146/annurev.psych.55.090902.142015} {\bibfield
  {journal} {\bibinfo  {journal} {Annual Review of Psychology}\ }\textbf
  {\bibinfo {volume} {55}},\ \bibinfo {pages} {591} (\bibinfo {year}
  {2004})}\BibitemShut {NoStop}%
\bibitem [{\citenamefont {Moscovici}\ and\ \citenamefont
  {Nemeth}(1974)}]{Moscovici-Nemeth_1974}%
  \BibitemOpen
  \bibfield  {author} {\bibinfo {author} {\bibfnamefont {S.}~\bibnamefont
  {Moscovici}}\ and\ \bibinfo {author} {\bibfnamefont {C.}~\bibnamefont
  {Nemeth}},\ }\bibinfo {title} {Social influence: {II}. {M}inority
  influence.},\ in\ \href@noop {} {\emph {\bibinfo {booktitle} {Social
  psychology: {C}lassic and contemporary integrations}}},\ \bibinfo {editor}
  {edited by\ \bibinfo {editor} {\bibfnamefont {C.}~\bibnamefont {Nemeth}}}\
  (\bibinfo  {publisher} {Rand Mcnally},\ \bibinfo {year} {1974})\BibitemShut
  {NoStop}%
\bibitem [{\citenamefont {Stauffer}\ \emph {et~al.}(2000)\citenamefont
  {Stauffer}, \citenamefont {Sousa},\ and\ \citenamefont
  {De~Oliveira}}]{ISI:000166141900014}%
  \BibitemOpen
  \bibfield  {author} {\bibinfo {author} {\bibfnamefont {D.}~\bibnamefont
  {Stauffer}}, \bibinfo {author} {\bibfnamefont {A.}~\bibnamefont {Sousa}},\
  and\ \bibinfo {author} {\bibfnamefont {S.}~\bibnamefont {De~Oliveira}},\
  }\bibfield  {title} {\bibinfo {title} {Generalization to square lattice of
  {S}znajd sociophysics model},\ }\href
  {https://doi.org/10.1142/S012918310000105X} {\bibfield  {journal} {\bibinfo
  {journal} {International Journal of Modern Physics C}\ }\textbf {\bibinfo
  {volume} {11}},\ \bibinfo {pages} {1239} (\bibinfo {year}
  {2000})}\BibitemShut {NoStop}%
\bibitem [{\citenamefont {Sznajd-Weron}\ and\ \citenamefont
  {Sznajd}(2005)}]{Sznajd-2005a}%
  \BibitemOpen
  \bibfield  {author} {\bibinfo {author} {\bibfnamefont {K.}~\bibnamefont
  {Sznajd-Weron}}\ and\ \bibinfo {author} {\bibfnamefont {J.}~\bibnamefont
  {Sznajd}},\ }\bibfield  {title} {\bibinfo {title} {Who is left, who is
  right?},\ }\href {https://doi.org/10.1016/j.physa.2004.12.038} {\bibfield
  {journal} {\bibinfo  {journal} {Physica A}\ }\textbf {\bibinfo {volume}
  {351}},\ \bibinfo {pages} {593} (\bibinfo {year} {2005})}\BibitemShut
  {NoStop}%
\bibitem [{\citenamefont {Sznajd-Weron}(2005)}]{Sznajd-2005}%
  \BibitemOpen
  \bibfield  {author} {\bibinfo {author} {\bibfnamefont {K.}~\bibnamefont
  {Sznajd-Weron}},\ }\bibfield  {title} {\bibinfo {title} {Sznajd model and its
  applications},\ }\href
  {http://www.actaphys.uj.edu.pl/fulltext?series=Reg&vol=36&page=2537}
  {\bibfield  {journal} {\bibinfo  {journal} {Acta Physica Polonica B}\
  }\textbf {\bibinfo {volume} {36}},\ \bibinfo {pages} {2537} (\bibinfo {year}
  {2005})}\BibitemShut {NoStop}%
\bibitem [{\citenamefont {Sznajd-Weron}\ \emph {et~al.}(2021)\citenamefont
  {Sznajd-Weron}, \citenamefont {Sznajd},\ and\ \citenamefont
  {Weron}}]{Sznajd-Sznajd-Sznajd}%
  \BibitemOpen
  \bibfield  {author} {\bibinfo {author} {\bibfnamefont {K.}~\bibnamefont
  {Sznajd-Weron}}, \bibinfo {author} {\bibfnamefont {J.}~\bibnamefont
  {Sznajd}},\ and\ \bibinfo {author} {\bibfnamefont {T.}~\bibnamefont
  {Weron}},\ }\bibfield  {title} {\bibinfo {title} {A review on the {S}znajd
  model---20 years after},\ }\href
  {https://doi.org/10.1016/j.physa.2020.125537} {\bibfield  {journal} {\bibinfo
   {journal} {Physica A}\ }\textbf {\bibinfo {volume} {565}},\ \bibinfo {pages}
  {125537} (\bibinfo {year} {2021})}\BibitemShut {NoStop}%
\bibitem [{\citenamefont {Asch}(1955)}]{Asch1955}%
  \BibitemOpen
  \bibfield  {author} {\bibinfo {author} {\bibfnamefont {S.~E.}\ \bibnamefont
  {Asch}},\ }\bibfield  {title} {\bibinfo {title} {Opinions and social
  pressure},\ }\href {https://doi.org/10.1038/scientificamerican1155-31}
  {\bibfield  {journal} {\bibinfo  {journal} {Scientific American}\ }\textbf
  {\bibinfo {volume} {193}},\ \bibinfo {pages} {31} (\bibinfo {year}
  {1955})}\BibitemShut {NoStop}%
\bibitem [{\citenamefont {Milgram}\ \emph {et~al.}(1969)\citenamefont
  {Milgram}, \citenamefont {Bickman},\ and\ \citenamefont
  {Berkowitz}}]{Milgram1969}%
  \BibitemOpen
  \bibfield  {author} {\bibinfo {author} {\bibfnamefont {S.}~\bibnamefont
  {Milgram}}, \bibinfo {author} {\bibfnamefont {L.}~\bibnamefont {Bickman}},\
  and\ \bibinfo {author} {\bibfnamefont {L.}~\bibnamefont {Berkowitz}},\
  }\bibfield  {title} {\bibinfo {title} {Note on drawing power of crowds of
  different size},\ }\href {https://doi.org/10.1037/h0028070} {\bibfield
  {journal} {\bibinfo  {journal} {Journal of Personality and Social
  Psychology}\ }\textbf {\bibinfo {volume} {13}},\ \bibinfo {pages} {79}
  (\bibinfo {year} {1969})}\BibitemShut {NoStop}%
\bibitem [{\citenamefont {Darley}\ and\ \citenamefont
  {Latan\'e}(1968)}]{Darley1968}%
  \BibitemOpen
  \bibfield  {author} {\bibinfo {author} {\bibfnamefont {J.~M.}\ \bibnamefont
  {Darley}}\ and\ \bibinfo {author} {\bibfnamefont {B.}~\bibnamefont
  {Latan\'e}},\ }\bibfield  {title} {\bibinfo {title} {Bystander intervention
  in emergencies---{D}iffusion of responsibility},\ }\href
  {https://doi.org/10.1037/h0025589} {\bibfield  {journal} {\bibinfo  {journal}
  {Journal of Personality and Social Psychology}\ }\textbf {\bibinfo {volume}
  {8}},\ \bibinfo {pages} {377} (\bibinfo {year} {1968})}\BibitemShut {NoStop}%
\bibitem [{\citenamefont {Latan{\'e}}\ and\ \citenamefont
  {Harkins}(1976)}]{Latane-1976}%
  \BibitemOpen
  \bibfield  {author} {\bibinfo {author} {\bibfnamefont {B.}~\bibnamefont
  {Latan{\'e}}}\ and\ \bibinfo {author} {\bibfnamefont {S.}~\bibnamefont
  {Harkins}},\ }\bibfield  {title} {\bibinfo {title} {Cross-modality matches
  suggest anticipated stage fright a multiplicative power function of audience
  size and status},\ }\href {https://doi.org/10.3758/BF03208286} {\bibfield
  {journal} {\bibinfo  {journal} {Perception {\&} Psychophysics}\ }\textbf
  {\bibinfo {volume} {20}},\ \bibinfo {pages} {482} (\bibinfo {year}
  {1976})}\BibitemShut {NoStop}%
\bibitem [{\citenamefont {Latan\'e}\ and\ \citenamefont
  {Nida}(1981)}]{Latane-1981a}%
  \BibitemOpen
  \bibfield  {author} {\bibinfo {author} {\bibfnamefont {B.}~\bibnamefont
  {Latan\'e}}\ and\ \bibinfo {author} {\bibfnamefont {S.}~\bibnamefont
  {Nida}},\ }\bibfield  {title} {\bibinfo {title} {Ten years of research on
  group size and helping},\ }\href {https://doi.org/10.1037/0033-2909.89.2.308}
  {\bibfield  {journal} {\bibinfo  {journal} {Psychological Bulletin}\ }\textbf
  {\bibinfo {volume} {89}},\ \bibinfo {pages} {308} (\bibinfo {year}
  {1981})}\BibitemShut {NoStop}%
\bibitem [{\citenamefont {Nowak}\ \emph {et~al.}(1990)\citenamefont {Nowak},
  \citenamefont {Szamrej},\ and\ \citenamefont {Latan\'e}}]{Nowak-1990}%
  \BibitemOpen
  \bibfield  {author} {\bibinfo {author} {\bibfnamefont {A.}~\bibnamefont
  {Nowak}}, \bibinfo {author} {\bibfnamefont {J.}~\bibnamefont {Szamrej}},\
  and\ \bibinfo {author} {\bibfnamefont {B.}~\bibnamefont {Latan\'e}},\
  }\bibfield  {title} {\bibinfo {title} {From private attitude to public
  opinion: {A} dynamic theory of social impact},\ }\href
  {https://doi.org/10.1037/0033-295X.97.3.362} {\bibfield  {journal} {\bibinfo
  {journal} {Psychological Review}\ }\textbf {\bibinfo {volume} {97}},\
  \bibinfo {pages} {362} (\bibinfo {year} {1990})}\BibitemShut {NoStop}%
\bibitem [{\citenamefont {Ho{\l}yst}\ \emph {et~al.}(2011)\citenamefont
  {Ho{\l}yst}, \citenamefont {Kacperski},\ and\ \citenamefont
  {Schweitzer}}]{ARCPIX253}%
  \BibitemOpen
  \bibfield  {author} {\bibinfo {author} {\bibfnamefont {J.~A.}\ \bibnamefont
  {Ho{\l}yst}}, \bibinfo {author} {\bibfnamefont {K.}~\bibnamefont
  {Kacperski}},\ and\ \bibinfo {author} {\bibfnamefont {F.}~\bibnamefont
  {Schweitzer}},\ }\bibfield  {title} {\bibinfo {title} {Social impact models
  of opinion dynamics},\ }in\ \href
  {https://doi.org/10.1142/9789812811578_0005} {\emph {\bibinfo {booktitle}
  {Annual Reviews of Computational Physics IX}}},\ \bibinfo {editor} {edited
  by\ \bibinfo {editor} {\bibfnamefont {D.}~\bibnamefont {Stauffer}}}\
  (\bibinfo  {publisher} {World Scientific},\ \bibinfo {address} {Singapore},\
  \bibinfo {year} {2011})\ pp.\ \bibinfo {pages} {253--273}\BibitemShut
  {NoStop}%
\bibitem [{\citenamefont {Noelle-Neumann}(1984)}]{Noelle-Neumann_1984}%
  \BibitemOpen
  \bibfield  {author} {\bibinfo {author} {\bibfnamefont {E.}~\bibnamefont
  {Noelle-Neumann}},\ }\href@noop {} {\emph {\bibinfo {title} {The Spiral of
  Silence. Public Opinion---Our Social Skin}}}\ (\bibinfo  {publisher} {Chicago
  University Press},\ \bibinfo {address} {Chicago},\ \bibinfo {year}
  {1984})\BibitemShut {NoStop}%
\bibitem [{\citenamefont {Eady}\ \emph {et~al.}(2019)\citenamefont {Eady},
  \citenamefont {Nagler}, \citenamefont {Guess}, \citenamefont {Zilinsky},\
  and\ \citenamefont {Tucker}}]{Eady_2019}%
  \BibitemOpen
  \bibfield  {author} {\bibinfo {author} {\bibfnamefont {G.}~\bibnamefont
  {Eady}}, \bibinfo {author} {\bibfnamefont {J.}~\bibnamefont {Nagler}},
  \bibinfo {author} {\bibfnamefont {A.}~\bibnamefont {Guess}}, \bibinfo
  {author} {\bibfnamefont {J.}~\bibnamefont {Zilinsky}},\ and\ \bibinfo
  {author} {\bibfnamefont {J.~A.}\ \bibnamefont {Tucker}},\ }\bibfield  {title}
  {\bibinfo {title} {How many people live in political bubbles on social media?
  {E}vidence from linked survey and twitter data},\ }\href
  {https://doi.org/10.1177/2158244019832705} {\bibfield  {journal} {\bibinfo
  {journal} {Sage Open}\ }\textbf {\bibinfo {volume} {9}},\ \bibinfo {pages}
  {2158244019832705} (\bibinfo {year} {2019})}\BibitemShut {NoStop}%
\bibitem [{\citenamefont {Lin}(2008)}]{Lin_2008}%
  \BibitemOpen
  \bibfield  {author} {\bibinfo {author} {\bibfnamefont {N.}~\bibnamefont
  {Lin}},\ }\bibinfo {title} {A network theory of social capital},\ in\
  \href@noop {} {\emph {\bibinfo {booktitle} {The Handbook of Social
  Capital}}},\ \bibinfo {editor} {edited by\ \bibinfo {editor} {\bibfnamefont
  {J.~W.}\ \bibnamefont {van Deth}}\ and\ \bibinfo {editor} {\bibfnamefont
  {G.}~\bibnamefont {Wolleb}}}\ (\bibinfo  {publisher} {Oxford University Press
  Inc.},\ \bibinfo {address} {New York},\ \bibinfo {year} {2008})\
  Chap.~\bibinfo {chapter} {2}, pp.\ \bibinfo {pages} {50--69}\BibitemShut
  {NoStop}%
\bibitem [{\citenamefont {Goldman}\ and\ \citenamefont
  {Mutz}(2011)}]{Goldman_2011}%
  \BibitemOpen
  \bibfield  {author} {\bibinfo {author} {\bibfnamefont {S.~K.}\ \bibnamefont
  {Goldman}}\ and\ \bibinfo {author} {\bibfnamefont {D.~C.}\ \bibnamefont
  {Mutz}},\ }\bibfield  {title} {\bibinfo {title} {The friendly media
  phenomenon: {A} cross-national analysis of cross-cutting exposure},\ }\href
  {https://doi.org/10.1080/10584609.2010.544280} {\bibfield  {journal}
  {\bibinfo  {journal} {Political Communication}\ }\textbf {\bibinfo {volume}
  {28}},\ \bibinfo {pages} {42} (\bibinfo {year} {2011})}\BibitemShut {NoStop}%
\end{thebibliography}

%% ===============================================================
%% ===============================================================

%% ===============================================================
%% ===============================================================

%% ===============================================================
%% ===============================================================

%% ===============================================================
%% ===============================================================

\end{document}